%% file: MechAWN.tex
\RequirePackage{fix-cm}
\documentclass[smallcondensed,envcountsect,envcountsame]{svjour3b} 
\pdfoutput=1
\smartqed  
\usepackage{stfloats}
\fnbelowfloat
\usepackage[T1]{fontenc}
\usepackage[utf8]{inputenc}
\usepackage{textcomp}
\usepackage{cite}
\usepackage{xcolor}

\usepackage{tikz}
\usetikzlibrary{positioning,chains,matrix,shapes.arrows,scopes,
                shapes.misc,arrows,automata,calc,decorations.markings}

\usepackage{graphicx}
\graphicspath{{figures/}}
\usepackage{hyperref}
\hypersetup{%
    final=true,
    colorlinks=true,
    urlcolor=black,
    linkcolor=black,
    citecolor=black,
    pdfborder={0 0 0},
    pdfauthor={Timothy Bourke \& Robert J.\ van Glabbeek \& Peter H\"ofner},
    pdftitle={Mechanizing a Process Algebra for Network Protocols},
    pdfsubject={Mechanized compositional invariant proofs of process 
    algebra models},
    pdfkeywords={Interactive Theorem Proving, Isabelle/HOL,
                 Process Algebra, Compositional Invariant Proofs,
                 Wireless Mesh Networks, Mobile Ad hoc Networks},
}
\usepackage{amssymb}
\usepackage{latexsym}
\usepackage{stmaryrd}

\usepackage{mathpartir}
\mprset{sep=1.10em}  

\usepackage{url}
\usepackage{extarrows}
\usepackage{relsize}

\usepackage[printonlyused]{acronym}
\newcommand{\acworkaround}[1]{{\acsfont {#1}}}

\usepackage{isabelle, isabellesym}
\isabellestyle{tim}
\renewcommand{\isasymAnd}{\isamath{{\mathsmaller{\bigwedge}}}}

\definecolor{gray}{rgb}{0.4,0.4,0.4}
\newcommand{\gray}[1]{\textcolor{gray}{#1}}

\newcommand{\snip}[4]
  {\expandafter\newcommand\csname snippet--#1\endcsname{#4}}
\newcommand{\snippet}[1]{\csname snippet--#1\endcsname}
\newcommand{\msnippet}[1]{\mbox{\csname snippet--#1\endcsname}}

\input{snippets}

\usepackage[inline]{enumitem}
\newlist{inparaenum}{enumerate*}{1}
\setlist[inparaenum,1]{label=(\arabic*),ref=\arabic*}

\newcommand{\refsec}[1]{\hyperref[sec:#1]{Section~\ref*{sec:#1}}}
\newcommand{\refsecs}[2]{\hyperref[sec:#1]{Sections~\ref*{sec:#1}}
                         \hyperref[sec:#2]{and~\ref*{sec:#2}}}
\newcommand{\reffig}[1]{\hyperref[fig:#1]{Figure~\ref*{fig:#1}}}
\newcommand{\reffigs}[2]{\hyperref[fig:#1]{Figures~\ref*{fig:#1}}
                         \hyperref[fig:#2]{and~\ref*{fig:#2}}}
\newcommand{\reffigss}[3]{\hyperref[fig:#1]{Figures~\ref*{fig:#1}},
                          \hyperref[fig:#2]{\ref*{fig:#2}},
                          \hyperref[fig:#3]{and~\ref*{fig:#3}}}
\newcommand{\reffigsss}[4]{\hyperref[fig:#1]{Figures~\ref*{fig:#1}},
                           \hyperref[fig:#2]{\ref*{fig:#2}},
                           \hyperref[fig:#3]{\ref*{fig:#3}},
                           \hyperref[fig:#4]{and~\ref*{fig:#4}}}
\newcommand{\reffigr}[2]{\hyperref[fig:#1]{Figures~\ref*{fig:#1}}%
                         \hyperref[fig:#1]{--\ref*{fig:#2}}}
\newcommand{\refthm}[1]{\hyperref[thm:#1]{Theorem~\ref*{thm:#1}}}
\newcommand{\refthms}[2]{\hyperref[thm:#1]{Theorems~\ref*{thm:#1}}
                         \hyperref[thm:#2]{and~\ref*{thm:#2}}}
\newcommand{\reflem}[1]{\hyperref[thm:#1]{Lemma~\ref*{thm:#1}}}
\newcommand{\reflems}[2]{\hyperref[thm:#1]{Lemmas~\ref*{thm:#1}}
                         \hyperref[thm:#2]{and~\ref*{thm:#2}}}
\newcommand{\refcoro}[1]{\hyperref[thm:#1]{Corollary~\ref*{thm:#1}}}
\newcommand{\refcoros}[2]{\hyperref[thm:#1]{Corollaries~\ref*{thm:#1}}
                         \hyperref[thm:#2]{and~\ref*{thm:#2}}}
\newcommand{\refcoross}[3]{\hyperref[thm:#1]{Corollaries~\ref*{thm:#1}},
                         \hyperref[thm:#2]{\ref*{thm:#2}},
                         \hyperref[thm:#3]{and~\ref*{thm:#3}}}
\newcommand{\refdef}[1]{\hyperref[def:#1]{Definition~\ref*{def:#1}}}
\newcommand{\refdefs}[2]{\hyperref[def:#1]{Definitions~\ref*{def:#1}}
                         \hyperref[def:#1]{and~\ref*{def:#2}}}
\newcommand{\refdefr}[2]{\hyperref[def:#1]{Definitions~\ref*{def:#1}}%
                         \hyperref[def:#1]{--\ref*{def:#2}}}
\newcommand{\refeq}[1]{\hyperref[eq:#1]{(\ref*{eq:#1})}}
\newcommand{\refeqs}[2]{\hyperref[eq:#1]{(\ref*{eq:#1}}\hyperref[eq:#1]{, 
\ref*{eq:#2})}}
\newcommand{\refeqsss}[2]{\hyperref[eq:#1]{(\ref*{eq:#1})} 
and~\hyperref[eq:#1]{(\ref*{eq:#2})}}
\newcommand{\refeqss}[2]{\hyperref[eq:#1]{(\ref*{eq:#1}}--\hyperref[eq:#1]{\ref*{eq:#2})}}
\newcommand{\refstep}[1]{\hyperref[step:#1]{Step~\ref*{step:#1}}}

\newcommand{\refitem}[1]{\hyperref[item:#1]{Item~\ref*{item:#1}}}
\newcommand{\refitemr}[2]{\hyperref[item:#1]{Items~\ref*{item:#1}}
                          \hyperref[item:#2]{to~\ref*{item:#2}}}
\newcommand{\refitems}[2]{\hyperref[item:#1]{Items~\ref*{item:#1}}
                          \hyperref[item:#2]{and~\ref*{item:#2}}}

\newcommand{\puncgap}{\hskip.1em}

\newcommand{\seqpsos}{\isa{seqp{\isacharunderscore}sos}}
\newcommand{\parpsos}{\isa{parp{\isacharunderscore}sos}}
\newcommand{\nodesos}{\isa{node{\isacharunderscore}sos}}
\newcommand{\pnetsos}{\isa{pnet{\isacharunderscore}sos}}
\newcommand{\cnetsos}{\isa{cnet{\isacharunderscore}sos}}
\newcommand{\oseqpsos}{\isa{oseqp{\isacharunderscore}sos}}
\newcommand{\oparpsos}{\isa{oparp{\isacharunderscore}sos}}
\newcommand{\onodesos}{\isa{onode{\isacharunderscore}sos}}
\newcommand{\opnetsos}{\isa{opnet{\isacharunderscore}sos}}
\newcommand{\ocnetsos}{\isa{ocnet{\isacharunderscore}sos}}
\newcommand{\Ri}{R\isactrlsub 0}

\def\isacharasterisk{\ensuremath{\ast}}

\newcommand{\nodes}[3]{%
    $\mbox{#2}^{\mbox{\,\isastylescript{#1}}}_{\mbox{\isastylescript{#3}}}$}
\newcommand{\subnetsless}[2]{{#1}{\isamath{\,\shortparallel\hspace{.3pt}}}{#2}}
\newcommand{\subnets}[2]{{#1}{\isamath{\,\shortparallel\,}}{#2}}
\newcommand{\subnetsadj}[2]{{#1}{\isamath{\hspace{.35pt}\shortparallel\,}}{#2}}

\def\NodeS\ #1\ #2\ #3{\nodes{#1}{#2}{#3}}
\renewcommand{\isasymin}{\isamath{\,\in\,}}

\newcommand{\listconcat}{{\scriptsize\ensuremath{+\!+}}}
\newcommand{\parallelcomp}{{\scriptsize\raisebox{.19ex}{\ensuremath{\,\parallel\,}}}}
\newcommand{\gammatoy}{{{\isasymGamma}\isactrlsub {\mbox{\sf\scriptsize Toy}}}}
\newcommand{\gammax}{{{\isasymGamma}\isactrlsub {\mbox{\sf\scriptsize x}}}}
\newcommand{\gammaqmsg}{{{\isasymGamma}\isactrlsub {\textsc{qmsg}}}}
\newcommand{\sigmatoy}{{{\isasymsigma}\isactrlsub {\textsc{toy}}}}

\newcommand{\selmsg}{{s}\isactrlsub {\mathit{msg}}}
\newcommand{\updmsg}{{u}\isactrlsub {\mathit{msg}}}
\newcommand{\selips}{{s}\isactrlsub {\mathit{ids}}}
\newcommand{\selip}{{s}\isactrlsub {\mathit{id}}}
\newcommand{\seldata}{{s}\isactrlsub {\mathit{data}}}

\newcommand{\Pkt}[2]{\isa{Pkt {#1} {#2}}}

\newcommand{\Newpkt}[2]{\isa{Newpkt {#1} {#2}}}

\newcommand{\microstep}[1]{\isa{{\isasymleadsto}\isactrlbsub{#1}\isactrlesub}}

\newcommand{\microsteprtcl}[1]{\isa{{\isasymleadsto}\isactrlbsub{#1}\isactrlesub}\isactrlisup{\isacharasterisk}}

\newcommand{\stepinv}{\isa{\isamath{\mid\!\mid\hspace{-2pt}\equiv}}}
\newcommand{\ostepinv}{\isa{\isamath{\mid\hspace{-2pt}\equiv}}}

\acrodef{AODV}{Ad hoc On-demand Distance Vector}
\acrodef{AFP}{Archive of Formal Proofs}
\acrodef{AWN}{Algebra for Wireless Networks}
\acrodef{MANET}{Mobile Ad hoc Network}
\acrodef{ITP}{Interactive Theorem Prover}
\acrodef{HOL}{Higher Order Logic}
\acrodef{WMN}{Wireless Mesh Network}
\acrodef{SOS}{Structural Operational Semantics}

\begin{document}

\title{Mechanizing a Process Algebra for Network Protocols%
\thanks{NICTA is funded by the Australian Government through the Department of 
Communications and the Australian Research Council through the ICT Centre of 
Excellence Program.}
}

\author{
	Timothy Bourke
\and
        Robert~J.~van~Glabbeek
\and 
	Peter H\"ofner
}

\authorrunning{Bourke, van Glabbeek, and H\"ofner} 
\institute{T.\ Bourke \at
              Inria Paris and \'Ecole normale supérieure, Paris, France\\
              \email{Timothy.Bourke@inria.fr}           
           \and
              R.\ J.\ van Glabbeek \at
              NICTA and UNSW, Sydney, Australia\\
              \email{rvg@cs.stanford.edu}
           \and
              P.\ H\"ofner \at
              NICTA and UNSW, Sydney, Australia\\
              \email{Peter.Hoefner@nicta.com.au}
}

\date{~}

\maketitle

\begin{abstract}
This paper presents the mechanization of a process algebra for Mobile Ad hoc 
Networks and Wireless Mesh Networks, and the development of a compositional 
framework for proving invariant properties.
Mechanizing the core process algebra in Isabelle/HOL is relatively standard, 
but its layered structure necessitates special treatment.
The control states of reactive processes, such as nodes in a network, are 
modelled by terms of the process algebra.
We propose a technique based on these terms to streamline proofs of 
inductive invariance.
This is not sufficient, however, to state and prove invariants that relate 
states across multiple processes (entire networks).
To this end, we propose a novel compositional technique for lifting global 
invariants stated at the level of individual nodes to networks of nodes.
\end{abstract}

\section{Introduction and related work}\label{sec:intro}

The \ac{AWN} is a process algebra developed in particular for modelling and 
analysing protocols for \acp{MANET} and 
\acp{WMN}~\cite{ESOP12,FehnkerEtAl:AWN:2013}, but that can be used for 
reasoning about routing and communication protocols in general.
This paper reports on both its mechanization in 
Isabelle/HOL~\cite{NipkowPauWen:IsabelleTut:2002} and the development of a 
compositional framework for showing invariant properties of models.%
\footnote{The Isabelle/HOL source files can be found in the \ac{AFP}~\cite{Bourke14}.}
The techniques we describe are a response to problems encountered during the 
mechanization of a model and proof of a crucial correctness property for the 
\ac{AODV} routing protocol, a widely used protocol, standardized by the 
IETF~\cite{RFC3561}.
The \ac{AODV} case study is described in detail 
elsewhere~\cite{ATVA14} and we only refer to it briefly in this paper.
The property we study is \emph{loop freedom}, meaning that no data packet is 
sent in cycles forever. Such a property can only be expressed by relating 
states of different (neighbouring) network nodes.
Encoding such inter-node properties in an \ac{ITP} proved quite 
challenging, since the proof is performed inductively for an arbitrary 
number of nodes and the base case is a single node whose neighbours do not 
yet exist.
We develop a novel compositional technique to address this challenge.

Despite extensive research on related 
problems~\cite{deRoeverEtAl:ConcVer:2001} and several mechanized frameworks 
for reactive systems~\cite{HeydCre:Unity:1996, ChaudhuriEtAl:TLA+:2010, 
Muller:PhD:1998}, we are not aware of other solutions that allow the 
compositional statement and proof of properties relating the states of 
different nodes in a message-passing model---at least not within the 
strictures imposed by an \ac{ITP}.

\paragraph{Related work.} 
\ac{AWN} is a process algebra, but for the purposes of proving properties we 
treat it essentially as a structured programming language and employ a
technique originally proposed by Floyd~\cite{Floyd:MeaningsToPrograms:1967} 
and later developed by Manna and Pnueli~\cite{MannaPnu:Safety}, whereby
a set of semantic rules is defined to link the syntax of a program to an 
induced transition system. Safety properties are then shown to hold for
all reachable states by induction from a set of initial states over
the set of transitions. Rather than define the induced transition system
in terms of labels and (virtual) program counters~\cite[Chapter 1]{MannaPnu:Safety},
we use term derivatives and \ac{SOS} rules \cite{Pl04}.

This separation between language and model differs from the approach taken 
in formalisms like UNITY~\cite{ChandyMis:Unity:1988} and I/O 
Automata~\cite{LynchTut:IOAutoIntro:1989}, where initial states and sets of 
transitions are specified directly, and also from that of 
TLA\textsuperscript{+}~\cite{Lamport:TLA+:2002}, where the initial states 
and transition relation are written as a formula of first-order logic.
The advantage of the language-plus-semantics approach is that sequencing and 
branching in models is expressed by syntactic operators with the implied 
changes in the underlying control state being managed by the semantic rules.
Arguably, this permits models that are easier to understand by experts in 
the system being modelled.
The disadvantage is some extra complexity and layers of definitions.
We find, however, that these details are well managed by \acp{ITP} 
and---once defined---intrude little on the verification task.

\ac{AWN} provides a unique mix of communication primitives and a treatment 
of data structures that are essential for studying \ac{MANET} and \ac{WMN} 
protocols with dynamic topologies and sophisticated routing 
logic~\cite[\textsection 
1]{FehnkerEtAl:AWN:2013}. 
It supports communication primitives for one-to-one (\isa{unicast}), 
one-to-many (\isa{groupcast}), and one-to-all (\isa{broadcast}) message 
passing.
\ac{AWN} comprises distinct layers for expressing the structure of nodes and 
networks.
We exploit this structure critically in our proofs, and we
expect the techniques proposed in \refsecs{proof-base}{proof-comp} to
also apply to similar layered modelling 
languages~\cite{NH06,CWS,CMAN,CMN,SRS10,RBPT}.

Besides this, our work differs from other mechanizations for verifying 
reactive systems, like UNITY~\cite{HeydCre:Unity:1996}, 
TLA\textsuperscript{+}~\cite{ChaudhuriEtAl:TLA+:2010}, or I/O 
Automata~\cite{Muller:PhD:1998} (from which we drew the most inspiration), 
in its explicit treatment of control states, in the form of process algebra 
terms, as distinct from data states.
In this respect, our approach is close to that of 
Isabelle/Circus~\cite{FeliachiGauWol:Circus:2012}, but it differs in
\begin{inparaenum}
\item
the treatment of operators for composing nodes, which we model directly as 
functions on automata,
\item
the treatment of recursive invocations, which we do not permit, and
\item
our inclusion of a framework for compositional proofs.
\end{inparaenum}

Within the process algebraic tradition,
other work in \acp{ITP} focuses on showing properties of process algebras, 
such as the treatment of binders~\cite{BengtsonParrow09}, that bisimulation 
equivalence is a congruence~\cite{GothelGle:TimedCSP:2010, 
Hirschkoff:picalc:1997}, or properties of fix-point 
induction~\cite{TejWolff97}, while we focus on what has been termed `proof 
methodology'~\cite{FokkinkGroRen:ProcAlgProof:2004},
and develop a compositional method for showing correctness properties of 
protocols specified in a process algebra.

As an alternative to the frameworks cited above, and the work we present,
Paulson's inductive approach~\cite{Paulson:Inductive:1998} can be applied to 
show properties of protocols specified with less generic infrastructure.
In fact, it has also been applied to model the \ac{AODV} 
protocol~\cite{ZYZW09}; a detailed comparison is given 
elsewhere~\cite[\textsection 9]{ATVA14}.
But we think this approach to be better suited to systems 
specified in a `declarative' style as opposed to the strongly operational 
models we consider.
The question of style has practical implications.
It determines the `distance' between the original specification and the 
formal model---perhaps surprisingly protocol descriptions are often quite 
operational (this is the case for \ac{AODV}~\cite{RFC3561}).
It also likely influences proofs of refinement between abstract and 
implementation models.

\paragraph{Structure and contributions.} 
\refsec{awn} describes the mechanization of \ac{AWN}.
The basic definitions are routine but the layered structure of the language 
and the treatment of operators on networks as functions on automata are 
relatively novel and essential to understanding later sections.
\refsec{proof-base} describes our mechanization of the theory of inductive 
invariants, closely following~\cite{MannaPnu:Safety}.
We exploit the structure of \ac{AWN} to generate verification conditions 
corresponding to those of pen-and-paper proofs~\cite[\textsection 
7]{FehnkerEtAl:AWN:2013}.
\refsec{proof-comp} presents a compositional technique for stating and 
proving invariants that relate states across multiple nodes.
Basically, we substitute `open' \ac{SOS} rules over the global state for the 
standard rules over local states (\refsec{omodel}), show the property over a 
single sequential process (\refsec{oinv}), `lift' it successively over 
layers that model message queueing and network communication
(\refsec{lift}), and, ultimately, `transfer' it to the original model 
(\refsec{transfer}).

\paragraph{Note.} This paper is an extended version of \cite{ITP14}. It presents 
all details with regards to the mechanization---many of which were skipped 
in \cite{ITP14} due to lack of space.
We also present more details about the novel compositional technique for 
lifting global invariants, including motivation and examples.
As a case study, the framework we present in this paper was successfully 
applied in the mechanization of a proof of \ac{AODV}'s loop freedom, 
the details of which are available in the \ac{AFP}~\cite{BourkeHof14} and 
presented elsewhere~\cite{ATVA14}.

\section{The process algebra \acs{AWN}} \label{sec:awn} 

The \acf{AWN} comprises five layers~\cite[\textsection 
4]{FehnkerEtAl:AWN:2013}:
\begin{inparaenum}
\item
\emph{sequential processes} for encoding the 
protocol logic as a recursive specification;
\item
\emph{parallel composition} of sequential processes for running multiple 
processes simultaneously on a single node;
\item
\emph{node expressions} for encapsulating processes running on a node and 
tracking a node's address and neighbours (other nodes within 
transmission range);
\item
\emph{partial network expressions} for describing networks 
as parallel compositions of nodes and
\item
\emph{complete network expressions} for closing partial networks to further interactions with 
an environment.
\end{inparaenum}
We treat each layer as an automaton with states of a specific form and a 
given set of transition rules.
We describe the layers from the bottom up over the following sections.

\subsection{Sequential processes}\label{sec:awn:seq} 

Sequential processes are used to encode protocol logic.
Each is modelled by a \emph{(recursive) specification} \isa{\isasymGamma} of 
type \isa{'p {\isasymRightarrow} ('k, 'p, 'l) seqp}, which maps 
process names of type \isa{'p} to terms of type \isa{('k, 'p, 'l) 
seqp}, also
parameterized by \isa{'k}, data states, and~\isa{'l}, labels.
States of sequential processes have the form \isa{(\isasymxi, p)} where 
\isa{\isasymxi} is a data state of type \isa{'k} and \isa{p} is a control 
term of type \isa{('k, 'p, 'l) seqp}.\footnote{In fact, control terms are 
also parameterized by the type of messages, which are specific to a given 
protocol, but we prefer to omit this detail from the presentation given 
here.}

\begin{figure}[t]
\tiny\centering 
  \begin{tabular}{@{}l@{\ }l@{}}
    \snippet{lseqp_assign} &
        \isa{'l {\isasymRightarrow}
             ('k {\isasymRightarrow} 'k) {\isasymRightarrow}
             ('k, 'p, 'l) seqp {\isasymRightarrow}
             ('k, 'p, 'l) seqp}
    \\[.5em]
    \snippet{lseqp_guard} &
        \isa{'l {\isasymRightarrow}
             ('k {\isasymRightarrow} 'k set) {\isasymRightarrow}
             ('k, 'p, 'l) seqp {\isasymRightarrow}
             ('k, 'p, 'l) seqp}
    \\[.5em]
    \snippet{lseqp_ucast}\hspace{1em} &
        \isa{'l {\isasymRightarrow}
             ('k {\isasymRightarrow} ip) {\isasymRightarrow}
             ('k {\isasymRightarrow} msg) {\isasymRightarrow}
             ('k, 'p, 'l) seqp {\isasymRightarrow}
           }\\[.2em] &\isa{
             ('k, 'p, 'l) seqp {\isasymRightarrow}
             ('k, 'p, 'l) seqp}
    \\[.5em]
    \snippet{lseqp_bcast} &
        \isa{'l {\isasymRightarrow}
             ('k {\isasymRightarrow} msg) {\isasymRightarrow}
             ('k, 'p, 'l) seqp {\isasymRightarrow}
             ('k, 'p, 'l) seqp}
    \\[.5em]
    \snippet{lseqp_gcast} &
        \isa{'l {\isasymRightarrow}
             ('k {\isasymRightarrow} ip set) {\isasymRightarrow}
             ('k {\isasymRightarrow} msg) {\isasymRightarrow}
             ('k, 'p, 'l) seqp {\isasymRightarrow}
          }\\[.2em] &\isa{
             ('k, 'p, 'l) seqp}
    \\[.5em]

    \snippet{lseqp_send} &
        \isa{'l {\isasymRightarrow}
             ('k {\isasymRightarrow} msg) {\isasymRightarrow}
             ('k, 'p, 'l) seqp {\isasymRightarrow}
             ('k, 'p, 'l) seqp}
    \\[.5em]
    \snippet{lseqp_receive} &
        \isa{'l {\isasymRightarrow}
             (msg {\isasymRightarrow} \!'k {\isasymRightarrow} \!'k) 
             {\isasymRightarrow}
             ('k, 'p, 'l) seqp {\isasymRightarrow}
             ('k, 'p, 'l) seqp}
    \\[.5em]
    \snippet{lseqp_deliver} &
        \isa{'l {\isasymRightarrow}
             ('k {\isasymRightarrow} data) {\isasymRightarrow}
             ('k, 'p, 'l) seqp {\isasymRightarrow}
             ('k, 'p, 'l) seqp}
    \\[.5em]

    \snippet{seqp_choice} &
        \isa{('k, 'p, 'l) seqp {\isasymRightarrow}
             ('k, 'p, 'l) seqp {\isasymRightarrow}
             ('k, 'p, 'l) seqp}
    \\[.5em]
    \snippet{seqp_call} &
        \isa{'p {\isasymRightarrow}
             ('k, 'p, 'l) seqp}
  \end{tabular}
\caption{Term constructors for sequential processes: {\sf ('k, 'p, 'l) 
seqp}.\newline%
\null\hspace{3.5em}(Leading \isa{\isasymlambda}-abstractions are omitted, for example,
 \isa{\isasymlambda l\ u\ p.\ }\snippet{lseqp_assign} is written 
 \snippet{lseqp_assign}.)}\label{fig:seqp:terms}
\end{figure}

\begin{figure}[t]
\tiny\begin{mathpar} 
        \msnippet{assignT'}\and
        \msnippet{guardT} \and
        \msnippet{unicastT}\and
        \msnippet{notunicastT}\and
        \msnippet{broadcastT}\and
        \msnippet{groupcastT}\and
        \msnippet{sendT}\and
        \msnippet{receiveT}\and
        \msnippet{deliverT}\and
        \msnippet{choiceT1}\and
        \msnippet{choiceT2}\and
        \msnippet{callT}
    \end{mathpar} 
\caption{\ac{SOS} rules for sequential processes: 
\seqpsos.\label{fig:seqp:sos}}
\end{figure}

Process terms are built from the constructors that are shown with their 
types in \reffig{seqp:terms}.
Here we make use of types \isa{data}, \isa{msg}, and \isa{ip} of 
\emph{application layer data}, \emph{messages} and \emph{IP addresses} (or 
any other node identifiers).
These are to be defined separately for any application of \ac{AWN}.
Furthermore, for any type \isa{'t}, the type of sets of objects of type 
\isa{'t} is denoted \isa{'t set}.
The inductive set \seqpsos, shown in \reffig{seqp:sos}, contains 
\ac{SOS} rules for each constructor.
It is parameterized by a specification \isa{\isasymGamma} and relates 
triples of source states, actions, and destination states.

The `prefix' constructors are each labelled with an 
\isa{\gray{{\isacharbraceleft}l{\isacharbraceright}}}.
Labels are used to strengthen invariants when a property is only true 
in or between certain states; they have no influence on control flow (unlike 
in~\cite{MannaPnu:Safety}).
The prefix constructors are {assignment}, {guard/bind}, {network 
synchronizations} 
\isa{unicast}/\isa{broadcast}/\isa{groupcast}/\isa{receive}, and
{internal communications} \isa{send}/\isa{receive}/\isa{deliver}.

The \emph{assignment} \snippet{lseqp_assign} transforms the data state 
\isa{\isasymxi} deterministically into the data state 
\isa{\isasymxi\isacharprime}, according to the function \isa{u}, and then 
acts as \isa{p}. `During' the update a \isa{\isasymtau}-action is performed.
In the original \ac{AWN} \cite{ESOP12,FehnkerEtAl:AWN:2013},
the data state \isa{\isasymxi} was defined as a partial function from
data variables to values of the appropriate type, and the assignment \isa{u} 
modified or extended this
partial function by (re)mapping a specific variable to a new value, which 
could depend on the current
data state.  In our mechanization the type of data states is given as an 
abstract parameter of the language that
is not yet instantiated in any particular way. Consequently, 
\isa{u} is taken to be any
function of type \isa{'k {\isasymRightarrow} 'k}, modifying the data state.
In comparison with \cite{ESOP12,FehnkerEtAl:AWN:2013}, our 
current treatment is less syntactic and more general.

The \emph{guard/bind} statement \snippet{lseqp_guard} encodes both guards 
and variable bindings. Here \isa{g} is of type \isa{'k {\isasymRightarrow} 
'k set}, a
  function from data states to sets of data states.
  Executing a guard amounts to making a nondeterministic choice of one of 
  the
  data states obtainable from the current state \isa{\isasymxi} by applying
  \isa{g}; in case \isa{g(\isasymxi)} is empty no transition is possible.
 For a valuation function \isa{h} of type \isa{'k\ {\isasymRightarrow}\ 
 bool} the guard statement is implemented as   
 \isa{\gray{{\isacharbraceleft}l{\isacharbraceright}}\snippet{stmt_guard}},
 which has no outgoing transition if \isa{h} evaluates to \isa{false}.
Variable binding like \mbox{\snippet{stmt_bind}} returns
all possible states that satisfy the binding constraint.
In the original \ac{AWN} \cite{ESOP12,FehnkerEtAl:AWN:2013},
where the data state \isa{\isasymxi} was a partial function from
data variables to values, the execution of a
guard/bind construct could only extend the domain of~\isa{\isasymxi},
thereby assigning values to previously unbound variables.
In our more abstract approach to data states, we must allow
any manipulation of the (as of yet unspecified) data state.
As this includes changing values of already bound variables, the guard/bind 
construct strictly subsumes assignment.
Since this `misuse' of a guard as assignment is not allowed in the original 
semantics of \ac{AWN}~\cite{ESOP12,FehnkerEtAl:AWN:2013},
we prefer to keep both.

The sequential process \snippet{lseqp_ucast}
tries to unicast the message \isa{\selmsg} to the destination
\isa{\selip}; 
if successful it continues to act as \isa{p} and otherwise
as \isa{q}. In other words, \isa{unicast{\isacharparenleft}\selip,\ 
\selmsg{\isacharparenright}\ {\isachardot}\ p} is
prioritized over \isa{q}, which is only considered when the unicast action 
is not possible (\isa{{\isasymnot}unicast\ {\isacharparenleft}\selip\ 
{\isasymxi}{\isacharparenright}}).
Which of the actions \isa{unicast} or \isa{{\isasymnot}unicast} will occur 
depends on whether the
destination \isa{\selip} is in transmission range of the current node;
this is implemented by the first two rules of Figure~\ref{fig:node} 
(described later).
In \cite{ESOP12,FehnkerEtAl:AWN:2013} the message \isa{\selmsg} is an 
expression with variables that evaluates to a message depending
on the current values of those variables.
Here, more abstractly, it can be any function of type \isa{'k 
{\isasymRightarrow} msg} that
constructs a message from the current data state.
The sequential process \mbox{\snippet{lseqp_bcast}} broadcasts \isa{\selmsg} to the other network nodes within transmission 
range.\footnote{Whether a node is within transmission range or not is 
determined later on.}
The process \mbox{\snippet{lseqp_gcast}} tries
to transmit \isa{\selmsg} to all destinations \isa{\selips}, and proceeds
as \isa{p} regardless of whether any of the transmissions is successful.

The sequential process \snippet{lseqp_send} synchronously transmits a message to another
process running on the same network node; this action can occur only
when the other sequential process is able to receive the message.
The sequential process \snippet{lseqp_receive} receives any message 
\isa{\updmsg}
either from another node, from another
sequential process running on the same node, or from the client\footnote{The
application layer that initiates packet sending and awaits receipt of
a packet.} connected to the local node.  It then proceeds as 
\isa{p}, but 
with an updated data state (the state change is triggered by the message).
In the original syntax and semantics of \ac{AWN},~\isa{\updmsg} was a data 
variable of type \isa{msg};
here it is an abstract function of type \isa{msg {\isasymRightarrow} 'k 
{\isasymRightarrow} 'k}, which changes the data state.
The submission of data from a client
is modelled by the receipt of a special message (\snippet{newpkt}),
where the function \isa{Newpkt} generates a message containing the
data \isa{d} and the intended destination \isa{dst}. Data is delivered to 
the client by \snippet{lseqp_deliver}.

The other constructors are unlabelled and serve to `glue' processes 
together: The \emph{choice} construct \snippet{seqp_choice} takes the union 
of two transition sets and hence may act either as \isa{p} or as
\isa{q}.
The procedure call \snippet{seqp_call} affixes 
a term from the specification~(\isa{{\isasymGamma}\ pn}).
The behaviour of \snippet{seqp_call} is exactly the same as that of the 
sequential process that \isa{\isasymGamma} associates to the process name 
\isa{pn}.
In \cite{ESOP12,FehnkerEtAl:AWN:2013}, on the other hand, process 
names \isa{pn} are explicitly parameterized with a list of data 
variables which can be defined by arbitrary data expressions at 
the call site.
The semantics of the process call involves running the 
process \isa{{\isasymGamma}\ pn} on an \emph{updated} data state, obtained 
by evaluating the data expressions in the current state and assigning the 
resulting values to the corresponding variables, while clearing the values 
of all variables that do not occur as parameters of \isa{pn}, effectively 
making them undefined.
In the current treatment, this behaviour is recovered by preceding a 
\snippet{seqp_call} by an explicit assignment statement.
As variables cannot be made undefined, they are cleared by setting them to 
arbitrary values.
This change is the biggest departure from the original definition of 
\ac{AWN}; it simplifies the treatment of \isa{call}, as we show in 
\refsec{cterms}, and facilitates working with automata where variable 
locality makes little sense.
The drawback is that the atomic `assign and jump' semantics is lost, 
which is sometimes inconvenient (an example is given later 
in~\refsec{awn:par}).

\paragraph{An example sequential process.}

\newcommand{\myhyperlink}[1]{%
 \hyperlink{#1}{\ptoy{#1}}%
  }

\newcommand{\myhypertargetcol}[1]{%
 \hypertarget{#1}{& \ptoy{#1}}%
  }

\newcommand{\ptoy}[1]{\raisebox{.1ex}{%
    \isa{\scriptsize\gray{{\isacharbraceleft}PToy-:{#1}{\isacharbraceright}}}}}%

We give the specification of a simple `toy' protocol as a running 
example. The formal \ac{AWN} specification is presented in \reffig{toy}.
Nodes following the protocol broadcast messages 
containing an integer \isa{no}. Each remembers the largest integer it has 
received and drops messages containing smaller or equal values.

The protocol is defined by a process named \isa{PToy} that maintains three 
variables: the integer \isa{no};
an identifier \isa{id}---also an integer, which uniquely identifies a node 
(for example, the node's IP address);
and an identifier \isa{nhid} that stores a node address
(either that of the node itself, or the address of another node that
supplied the largest number in the last comparison it made).\footnote{The
protocol behaviour regarding \isa{nhid} is rather arbitrary; it only
serves to illustrate some forthcoming concepts.}
The initial values of \isa{nhid} and \isa{no} are \isa{id} and \isa{0}, 
respectively.
\begin{figure}[t]
\centering
\tiny

\mbox{\isa{\gammatoy\ PToy\ \isacharequal\ labelled\ PToy\  \hspace{-0.3pt}
{\isacharparenleft}}}\hspace{-6pt}
\mbox{%
\renewcommand{\arraystretch}{1.8}%
\begin{tabular}[t]{p{6.0cm}@{\hspace{.5cm}}p{1.5cm}}
\isa{
receive{\isacharparenleft}{\isasymlambda}msg{\isacharprime}\ 
{\isasymxi}{\isachardot}\ {\isasymxi}\ {\isasymlparr}\ msg\ 
{\isacharcolon}{\isacharequal}\ msg{\isacharprime}\ 
{\isasymrparr}{\isacharparenright}{\isachardot}} \myhypertargetcol{0} \\
\isa{
{\isasymlbrakk}{\isasymlambda}{\isasymxi}{\isachardot}\ {\isasymxi}\ 
{\isasymlparr}nhid\ {\isacharcolon}{\isacharequal}\ id\ 
{\isasymxi}{\isasymrparr}{\isasymrbrakk}} \myhypertargetcol{1} \\
\isa{
{\isacharparenleft}\ \ \ 
{\isasymlangle}is{\isacharunderscore}newpkt{\isasymrangle}} \myhypertargetcol{2} \\
\isa{
\ \ \ \ \ {\isasymlbrakk}{\isasymlambda}{\isasymxi}{\isachardot}\ 
{\isasymxi}\ {\isasymlparr}no\ {\isacharcolon}{\isacharequal}\ max\ 
{\isacharparenleft}no\ {\isasymxi}{\isacharparenright}\ 
{\isacharparenleft}num\ 
{\isasymxi}{\isacharparenright}{\isasymrparr}{\isasymrbrakk}} \myhypertargetcol{3} \\
\isa{
\ \ \ \ \ 
broadcast{\isacharparenleft}{\isasymlambda}{\isasymxi}{\isachardot}\ 
Pkt (no\ {\isasymxi})\ (id\ 
{\isasymxi}){\isacharparenright}{\isachardot}
} \myhypertargetcol{4} \\
\isa{
\ \ \ \ \ 
\isa{{\isasymlbrakk}clear{\isacharunderscore}locals{\isasymrbrakk}\ 
call(PToy)}
} \myhypertargetcol{5} \\
\isa{
\ {\isasymoplus}\ {\isasymlangle}is{\isacharunderscore}pkt{\isasymrangle}} & 
\ptoy{2} \\
\isa{
\ \ \ \ \ {\isacharparenleft}
\ {\isasymlangle}%
{\isasymlambda}{\isasymxi}{\isachardot}\ %
if\ %
num\ {\isasymxi}\ \isachargreater\ no\ {\isasymxi}%
\ then\ \isacharbraceleft\isasymxi\isacharbraceright\ else\ \isasymemptyset%
{\isasymrangle}} \myhypertargetcol{6} \\
\isa{
\ \ \ \ \ \ \ \ \ {\isasymlbrakk}{\isasymlambda}{\isasymxi}{\isachardot}\ 
{\isasymxi}\ {\isasymlparr}no\ {\isacharcolon}{\isacharequal}\ num\ 
{\isasymxi}{\isasymrparr}{\isasymrbrakk}} \myhypertargetcol{7} \\
\isa{
\ \ \ \ \ \ \ \ \ {\isasymlbrakk}{\isasymlambda}{\isasymxi}{\isachardot}\ 
{\isasymxi}\ {\isasymlparr}nhid\ {\isacharcolon}{\isacharequal}\ sid\ 
{\isasymxi}{\isasymrparr}{\isasymrbrakk}} \myhypertargetcol{8} \\
\isa{
\ \ \ \ \ \ \ \ \ 
broadcast{\isacharparenleft}{\isasymlambda}{\isasymxi}{\isachardot}\ 
Pkt\ (no\ {\isasymxi})\ (id\ 
{\isasymxi}){\isacharparenright}{\isachardot}} \myhypertargetcol{9} \\
\isa{
\ \ \ \ \ \ \ \ \ 
{\isasymlbrakk}clear{\isacharunderscore}locals{\isasymrbrakk}\ call(PToy)} 
\myhypertargetcol{10} \\
\isa{
\ \ \ \ \ {\isasymoplus}\ {\isasymlangle}%
{\isasymlambda}{\isasymxi}{\isachardot}\ %
if\ %
num\ {\isasymxi}\ \isasymle\ no\ {\isasymxi}%
\ then\ \isacharbraceleft\isasymxi\isacharbraceright\ else\ \isasymemptyset%
{\isasymrangle}} &\ptoy{6} \\
\isa{
\ \ \ \ \ \ \ \ \ 
{\isasymlbrakk}clear{\isacharunderscore}locals{\isasymrbrakk}\ call(PToy)%
{\isacharparenright}%
{\isacharparenright}{\isacharparenright}} \myhypertargetcol{11}
\end{tabular}} 
\caption{\ac{AWN}-specification of a toy protocol.\label{fig:toy}}
\end{figure}

The behaviour of a single node in our toy protocol is given by
  the recursive specification \isa{\gammatoy} and an initial state
  \isa{(\isasymxi, p)} consisting of a data state
  \isa{\isasymxi}---defined above---and a control term \isa{p}---here
  the process \isa{\gammatoy\ PToy}. The specification \isa{\gammatoy}, given in \reffig{toy},
  assigns a process term to each process name---here only to the name 
  \isa{PToy}.
  The process term \isa{\gammatoy\ PToy}
  is defined as the result of applying a function \isa{labelled} to two arguments:
an identifier and the actual process without labels.
The labels are supplied by the function \isa{labelled}:
  it associates its first argument paired with a number as a
  label to every prefix construct occurring as a subterm. We show these 
  labels on
  the right-hand side of \reffig{toy}. Note that the choice construct 
  \isa{\isasymoplus} and the
  subterms \isa{call(PToy)} do not receive a label. Moreover, the
  function \isa{labelled} is defined in such a way that both arguments
  of the \isa{\isasymoplus} receive the same label; this way labels
  correspond exactly to states that can be reached during the execution
  of the process.

A node \isa{id} running the protocol \isa{PToy} will wait until it receives 
a message \isa{msg\isacharprime}
(line~\myhyperlink{0}). The protocol then updates the local data
  state \isa{\isasymxi} by assigning the message \isa{msg{\isacharprime}} to the
variable \isa{msg} (\isa{{{\isasymlambda}\isasymxi}{\isachardot}\ {\isasymxi}\ {\isasymlparr}\ msg\ 
{\isacharcolon}{\isacharequal}\ msg{\isacharprime}\ {\isasymrparr}}). In our 
scenario, there are two message constructors \Pkt{d}{src} and 
\Newpkt{d}{dst};
both carry an identifier (\isa{src} and \isa{dst}) and an integer-payload \isa{d}.
Here, \isa{src} is, by design, the sender of the message.
We require that all messages from the client of a node must have the
form~\mbox{\snippet{newpkt}}. All messages sent by a node have the
  form \Pkt{d}{src}. The message type thus uniquely determines whether
  the message originated from the application layer or from another node.

\begin{figure}[t]
\centering
\newcommand{\assign}{\isa{{\isasymlbrakk}$\cdots${\isasymrbrakk}}}
\newcommand{\guard}{\isa{{\isasymlangle}$\cdots${\isasymrangle}}}
\begin{tikzpicture}[
    level distance=.9cm,
    sibling distance=.5cm,
    every node/.style={inner sep=2pt},
]
  \tikzstyle{ext}=[draw,-latex,thick,solid]
  \tikzstyle{int}=[draw,-latex,thin,dashed]
  \tikzstyle{extt}=[edge from parent/.style={ext}]
  \tikzstyle{intt}=[edge from parent/.style={int}]
  \tikzstyle{label}=[font=\tiny,right,inner sep=5pt,yshift=2]
  \node {\isa{PToy}} child[dotted,->] {
    node (n00) {\ptoy{0}} child[extt] {
    node (n01) {\ptoy{1}} { child[intt] {
    node (n02) {\ptoy{2}} { [sibling distance=3.0cm]
        child { node (n03) {\ptoy{3}} child {
                node (n04) {\ptoy{4}} { child[extt] {
                node (n05) {\ptoy{5}} edge from parent node[label] {\isa{broadcast}} }
                edge from parent node[label] {\assign} }
        } edge from parent node[label,left] {\guard} }
        child { node (n06) {\ptoy{6}} [sibling distance=1.5cm]
            child { node (n07) {\ptoy{7}} { child {
                    node (n08) {\ptoy{8}} { child {
                    node (n09) {\ptoy{9}} { child[extt] {
                    node (n10) (n10) {\ptoy{10}
            } edge from parent node[label] {\isa{broadcast}} }
            } edge from parent node[label] {\assign} }
            } edge from parent node[label] {\assign} }
            } edge from parent node[label,left] {\guard} }
            child { node (n11) {\ptoy{11}}
            edge from parent node[label] {\guard}
        } edge from parent node[label] {\guard} }
    } edge from parent node[label] {\assign} }
    } edge from parent node[label] {\isa{receive}}
    }};
  \path (n05.west)  edge[int,bend left=60] node[label,left,near start] {\assign} (n00.west)
        (n11.east) edge[int,bend right=40] node[label,near start] {\assign} ([yshift=-1]n00.east)
        (n10.east)  edge[int,bend right=90,looseness=1.8] node[label,near start] {\assign} ([yshift=1]n00.east)
  ; 
\end{tikzpicture}
\caption[]{Control state structure of \isa{\gammatoy}.\label{fig:auto}}
\end{figure}

The choice (\myhyperlink{2}) makes a case distinction based on
whether the message received is a new packet or a `standard' one. In the 
former case, the guard/bind statement \isa{is{\isacharunderscore}newpkt}
`evaluates to true'\footnote{By this we mean that when it is
  applied to the current data state it returns a non-empty set of
  updated data states.}
and copies the message content \isa{d} to the variable \isa{num}.
Formally, \isa{is{\isacharunderscore}newpkt} is defined as
\begin{center}
\begin{tabular}{lcl}
\isa{is{\isacharunderscore}newpkt\ {\isasymxi}}
    & \isa{\isacharequal}
    & \isa{\textsf{case}\ msg\ {\isasymxi}\ \textsf{of}}\\
& & \isa{\ \ \ Pkt\ d\ src\ {\isasymRightarrow} \ {\isasymemptyset}}\\
& & \isa{\ {\isacharbar}\ Newpkt\ d\ dst\ {\isasymRightarrow}\ 
{\isacharbraceleft}{\isasymxi}{\isasymlparr}num\ 
{\isacharcolon}{\isacharequal}\ 
d{\isasymrparr}{\isacharbraceright}}\puncgap.
\end{tabular}
\end{center}
Afterwards, the process
proceeds to execute the lines 
labelled
\myhyperlink{3}, \myhyperlink{4}, and \myhyperlink{5}.
In the case of a `standard' message, the statement 
\isa{is{\isacharunderscore}pkt}
evaluates to true, the local state is updated 
by copying  the message contents \isa{d} into \isa{num} and \isa{src} into 
\isa{sid},
and the protocol proceeds with lines \myhyperlink{6}--\myhyperlink{11}.

In line \myhyperlink{3} the protocol compares the stored integer \isa{no} 
with the integer \isa{num} that came from the incoming message,
determines and stores the larger one into the variable \isa{no}, and 
broadcasts this value to all its neighbours with itself listed as sender 
(line~\myhyperlink{4}). After that, in line \myhyperlink{5}, the process 
calls itself recursively,
after resetting the local variables \isa{msg}, \isa{num}, 
and \isa{sid} to arbitrary values.

Depending on the contents of the `standard' message, the protocol 
performs two different sequences of actions.
(1) If the integer taken from the message and stored in variable
\isa{num} is larger than the stored \isa{no} (line \myhyperlink{6}), then it 
is stored in variable \isa{no} (line \myhyperlink{7}) and the sender of the 
message is stored in \isa{nhid} (line~\myhyperlink{8}).
Before resetting the local variables and returning to the 
start of the protocol by a recursive call (line \myhyperlink{10}), the node 
sends out the just updated number \isa{no}, again identifying itself as 
sender (line \myhyperlink{9}).
(2) If the integer from the message is smaller than or equal to \isa{no} 
(line \myhyperlink{6}), the node considers the message content outdated, 
drops the message, and calls itself recursively. 

As mentioned before, every sequential process is modelled by an 
automa\-ton---a record\footnote{The generic record has type 
\snippet{automaton_type}, where the type \isa{'s} is the domain of states, 
here pairs of data records and control terms, and \isa{'a} is the domain of 
actions.
} of two fields: a set of initial states 
and a set of transitions---parameterized by an address \isa{i}:
\begin{center}
\snippet{ptoy_lhs} \isa{\isacharequal} \snippet{ptoy_rhs}\puncgap,
\end{center}
where \isa{toy-init\ i} yields the initial data state \snippet{toy_init}.
The last three variables are initialized to arbitrary values, as they are 
considered local.
A representation of the automaton \isa{toy-init\ i} that abstracts from the 
data state is depicted in \reffig{auto}.

\subsection{Local parallel composition} \label{sec:awn:par} 

\newcommand{\qname}{Qmsg}
\newcommand{\myhyperlinkq}[1]{\hyperlink{qmsg-#1}{\qmsg{#1}}}
\newcommand{\myhypertargetcolq}[1]{\hypertarget{qmsg-#1}{& \qmsg{#1}}}
\newcommand{\qmsg}[1]{\raisebox{.1ex}{%
    \isa{\scriptsize\gray{{\isacharbraceleft}\qname-:{#1}{\isacharbraceright}}}}}%

Message sending protocols must nearly always be input enabled, that is, 
nodes should always be in a state where they can receive 
messages.\footnote{The semantics of \ac{AWN} ensures that any message 
transmitted by a node \emph{will} be received by all intended destinations 
that are within transmission range---the reasons for this design decision 
are given in \cite{ESOP12,FehnkerEtAl:AWN:2013}. In this setting, the 
absence of input enabledness would give rise to the unrealistic phenomenon 
of \emph{blocking}, the situation where one node is unable to transmit a 
message simply because another one is not ready to receive it.}
To achieve this, and to model asynchronous message transmission, the 
protocol process is combined with a queue model.
A queue can be expressed in \ac{AWN} as the specification \gammaqmsg{} with 
a single process \isa{Qmsg} shown in \reffig{qmsg}.
\begin{figure}[t]
\tiny
\!\mbox{\isa{\gammaqmsg\ \qname\ {\isacharequal}\ labelled\ \qname \  \hspace{-0.3pt}
{\isacharparenleft}}}\\
\null\hfil\hfil\hfil\hfil\hfil\mbox{%
\renewcommand{\arraystretch}{1.8}%
\begin{tabular}[t]{p{8.0cm}@{\hspace{.3cm}}p{1.2cm}@{}}
\isa{\ \ \ \,receive{\isacharparenleft}{\isasymlambda}msg\ msgs{\isachardot}\ msgs\ {\isacharat}\ {\isacharbrackleft}msg{\isacharbrackright}{\isacharparenright}\ {\isachardot}}\ %
\isa{call{\isacharparenleft}\qname{\isacharparenright}} \myhypertargetcolq{0} \\
\isa{{\isasymoplus}}\ %
\isa{{\isasymlangle}{\isasymlambda}msgs{\isachardot}\ \textsf{if}\ msgs\ {\isasymnoteq}\ {\isacharbrackleft}\,{\isacharbrackright}\ \textsf{then}\ {\isacharbraceleft}msgs{\isacharbraceright}\ \textsf{else}\ {\isasymemptyset}{\isasymrangle}} & \qmsg{0} \\
\isa{\ \ \ {\isacharparenleft}\ \ \,\ send{\isacharparenleft}{\isasymlambda}msgs. hd\ msgs{\isacharparenright}\ {\isachardot}} \myhypertargetcolq{1} \\
\isa{\ \ \ \ \ \ \ \ (\ \ \,\ \ \ {\isasymlbrakk}{\isasymlambda}msgs. tl msgs{\isasymrbrakk}}\ %
\isa{call{\isacharparenleft}\qname{\isacharparenright}} \myhypertargetcolq{2} \\
\isa{\ \ \ \ \ \ \ \ \ \ \ \ {\isasymoplus}}\ 
\isa{receive{\isacharparenleft}{\isasymlambda}msg\ msgs{\isachardot}\ tl\ msgs\ 
{\isacharat}\ 
{\isacharbrackleft}msg{\isacharbrackright}{\isacharparenright}\ 
{\isachardot}}\ %
\isa{call{\isacharparenleft}\qname{\isacharparenright}{\isacharparenright}} 
& \qmsg{2} \\%
\isa{\ \ \ \ {\isasymoplus}}\ \isa{receive{\isacharparenleft}{\isasymlambda}msg\ msgs{\isachardot}\ msgs\ {\isacharat}\ {\isacharbrackleft}msg{\isacharbrackright}{\isacharparenright}\ {\isachardot}}\ %
\isa{call{\isacharparenleft}\qname{\isacharparenright}{\isacharparenright}{\isacharparenright}} & \qmsg{1} %
\end{tabular}} 
\caption{\ac{AWN}-specification of the queue process\label{fig:qmsg}.}
\end{figure}
Unlike the data state of the \isa{PToy} process, which mapped variable names 
to values, the data state \isa{msgs} of~\isa{Qmsg} is simply a list of messages.
The control term is always ready to receive a message 
(lines~\myhyperlinkq{0}, \myhyperlinkq{1}, and~\myhyperlinkq{2}), in which case it appends 
(\isa{\isacharat} concatenates lists) the received message onto the state.
When the state is not empty (line~\myhyperlinkq{0}), the first element can 
be sent (line~\myhyperlinkq{1}: \isa{hd} returns the head of a list), and, 
on doing so, removes it from the state (line~\myhyperlinkq{2}: \isa{tl} 
returns the tail of a list).
A \isa{receive} command must be repeated at each control location to 
ensure input enabledness.
Compared to the \isa{Qmsg} process in the original presentation of 
\ac{AWN}~\cite[Process 6]{FehnkerEtAl:AWN:2013}, there is an extra 
\isa{receive} at \myhyperlinkq{2}.
It is necessary due to the modelling of parameter passing by an assignment 
followed by a recursive call, which introduces a $\tau$-transition.
This is unfortunate, but eliminating parameter passing greatly simplifies 
the constructions presented in \refsec{proof-base}.
The corresponding automaton is instantiated with an initially empty list:
\begin{center}
\snippet{pqmsg_lhs} \isa{\isacharequal} \snippet{pqmsg_rhs}\puncgap,
\end{center}
\begin{figure}[t]
\tiny
    \begin{mathpar}
        \msnippet{parleft} \and
        \msnippet{parright} \and
        \msnippet{parboth}
    \end{mathpar}
\caption[]{\acworkaround{SOS} rules for parallel processes: 
\parpsos\label{fig:parp}.}
\end{figure}
\pagebreak[3]
The composition of the example protocol with the queue is expressed as
\begin{center}
\snippet{ptoy_qmsg_term}\puncgap.
\end{center}
This \emph{local parallel} operator is a function over automata:
\begin{center}
\snippet{par_comp}\puncgap.
\end{center}
This is an operator of type \isa{('s,\ 'a)\ automaton\ {\isasymRightarrow}\ 
('t,\ 'a)\ automaton\ {\isasymRightarrow}\ ('s\ {\isasymtimes}\ 't,\ 'a)\ 
automaton}.
The process (automaton) \isa{A\ {\isasymlangle}{\isasymlangle}\ B} is a 
parallel composition of \isa{A} and \isa{B},
running on the same network node. As formalized in~\reffig{parp}, an 
action \isa{receive\ m} of \isa{A}
synchronizes with an action  \isa{send\ m} of \isa{B} into an internal
action \isasymtau. The \isa{receive}~actions of \isa{A} and 
\isa{send}~actions of \isa{B} cannot occur separately. All other actions of \isa{A} and 
\isa{B}, including
\isa{send}~actions of \isa{A} and
\isa{receive}~actions of \isa{B}, occur interleaved in
\isa{A\ {\isasymlangle}{\isasymlangle}\ B}. 
A parallel process expression denotes a
parallel composition of sequential processes---each with states \isa{(\isasymxi, p)}---with information
flowing from right to left. The variables of different sequential
processes running on the same node are maintained separately, and thus
cannot be shared.

\subsection{Nodes} \label{sec:awn:node} 

At the node level, a  local (parallel) process~\isa{A} is wrapped in a layer 
that records its address~\isa{i} and tracks the set of neighbouring node 
addresses, initially~\isa{\Ri}.
We define a function from these two parameters and~\isa{A}, an arbitrary 
automaton, as
\begin{center}
\snippet{node_comp}\puncgap.
\end{center}
Node states are triples denoted \snippet{net_state_nodes}.
\reffig{node} presents the rules of \nodesos.
Output network synchronizations, like \isa{groupcast} or \isa{broadcast}, 
are filtered by the list of neighbours to become 
\isa{\isacharasterisk{}cast} actions.
So, an action 
\isa{R{\isacharcolon}{\isacharasterisk}cast{\isacharparenleft}m\isacharparenright} 
transmits a message~\isa{m}  that can
be received by the set~\isa{R} of network nodes.
A failed unicast attempt by the process \isa{A} is
modelled as an internal action {\isasymtau} of the node expression.

There is no rule for propagating \isa{send\ m} actions from sequential 
processes to the node level.
These actions may only occur locally when paired with a receive action; they 
then become $\tau$-transitions, which are propagated.
The \snippet{act_arrive} action---instantiated in \reffig{node} as 
\isa{{\isasymemptyset}{\isasymnot}{\isacharbraceleft}i{\isacharbraceright}{\isacharcolon}arrive{\isacharparenleft}m{\isacharparenright}}
and 
\isa{{\isacharbraceleft}i{\isacharbraceright}{\isasymnot}{\isasymemptyset}{\isacharcolon}arrive{\isacharparenleft}m{\isacharparenright}}%
---is used to model a message \isa{m} received simultaneously by nodes in~\isa{H} 
and not by those in~\isa{K}.
The rules for \isa{arrive\ m} in \reffig{node}
state that the arrival of a message at a node happens if and only if
the node receives it, whereas non-arrival can happen at any time.
This embodies the assumption that, at any time, any message that is
transmitted to a node within range of the sender is actually received
by that node \cite{ESOP12,FehnkerEtAl:AWN:2013}.

\begin{figure}[t]
\tiny
    \begin{mathpar}
\msnippet{node_ucast} \and
\msnippet{node_notucast} \and
\msnippet{node_bcast} \and
\msnippet{node_gcast} \and
\msnippet{node_receive} \and
\msnippet{node_deliver} \and
\msnippet{node_arrive} \and
\msnippet{node_tau} \and
\msnippet{node_connect1} \and
\msnippet{node_connect2} \and
\msnippet{node_disconnect1} \and
\msnippet{node_disconnect2} \and
\msnippet{node_connect_other} \and
\msnippet{node_disconnect_other}
    \end{mathpar}
\caption[]{\acworkaround{SOS} rules for nodes:
\nodesos.\label{fig:node}}
\end{figure}

Internal actions {\isasymtau} and the action 
\isa{{\isacharbraceleft}i{\isacharbraceright}{\isacharcolon}deliver(d)}
are simply inherited by node expressions from the processes that run on these
nodes. 
Finally, we allow actions 
\isa{connect{\isacharparenleft}i,\ i{\isacharprime}{\isacharparenright}}
and \isa{disconnect{\isacharparenleft}i,\ i{\isacharprime}{\isacharparenright}}
for nodes \isa{i} and \isa{i\isacharprime}. They
model changes in network topology. Each node must synchronize
with such an action. These actions can be thought of as occurring
nondeterministically or as actions instigated by the environment of
the modelled network protocol. In this formalization node  \isa{i\isacharprime} is in
the range of node  \isa{i}, meaning that  \isa{i\isacharprime} can receive
messages sent by  \isa{i}, if and only if  \isa{i} is in the
range of  \isa{i\isacharprime}. 

\subsection{Partial networks} \label{sec:awn:pnet} 

Partial networks are specified by values of type 
\snippet{net_tree}. A \snippet{net_tree} is either a 
node \snippet{pnet_node_term} with address \isa{i} and a set of initial 
neighbours \isa{\Ri}, or a composition of two \snippet{net_tree}s 
\snippet{pnet_par_term}.
Hence it denotes a network topology. The \snippet{net_tree}
\mbox{\snippet{eg1_nettree}}, for instance, puts the three nodes
\isa{1}, \isa{2}, and \isa{3} in a linear topology where \isa{2} is 
connected to \isa{1} and \isa{3}.
The name \snippet{net_tree} refers to the parse tree of its syntactic expression;
unlike in \cite{ESOP12,FehnkerEtAl:AWN:2013}, it is treated as a tree 
because we do not make use of the
associativity of the parallel composition.

The function \isa{pnet} maps such a value, together with the 
process~\isa{np\ i} to execute at each node~\isa{i}, here parameterized by 
an address, to an automaton:
\begin{center}
\begin{tabular}{lcl}
\snippet{pnet1_lhs} & \isa{\isacharequal} & \snippet{pnet1_rhs}\\
\snippet{pnet2_lhs} & \isa{\isacharequal} & \snippet{pnet2_rhs}\puncgap,
\end{tabular}
\end{center}
The states of such automata mirror the tree structure of the network term; 
we denote composed states by \snippet{net_state_subnets}.
This structure and the node addresses remain constant during an execution.

The preceding definitions for sequential processes, local parallel 
composition, nodes, and partial networks suffice to model an example
three-node network of toy processes:
\begin{center}
\snippet{eg1_pnet}\puncgap.
\end{center}

The function \isa{pnet} is not present in 
\cite{ESOP12,FehnkerEtAl:AWN:2013}, where a partial network is defined simply as a 
parallel composition of nodes,
where in principle a different process could be running on each node.
With \isa{pnet} we ensure that in fact the same process is running on each 
node, and that this process is specified separately from the network topology.

\begin{figure}[t]
\tiny
    \begin{mathpar}
\msnippet{pnet_cast1} \and
\msnippet{pnet_cast2} \and
\msnippet{pnet_arrive} \and
\msnippet{pnet_deliver1} \and
\msnippet{pnet_deliver2} \and
\msnippet{pnet_tau1} \and
\msnippet{pnet_tau2} \and
\msnippet{pnet_connect} \and
\msnippet{pnet_disconnect}
    \end{mathpar}
\caption[]{\acworkaround{SOS} rules for partial networks 
\pnetsos.\label{fig:pnet}}
\end{figure}

\reffig{pnet} presents the rules of \pnetsos.
An
\isa{R{\isacharcolon}{\isacharasterisk}cast{\isacharparenleft}m\isacharparenright}~action 
of one node
synchronizes with an action \isa{arrive\ m} of all other nodes, where this
\isa{arrive\ m} amalgamates the arrival of message \isa{m} at the nodes in
the transmission range \isa{R} of the \isa{\isacharasterisk cast\ m}, and 
the non-arrival at the
other nodes. 
The third rule of \reffig{pnet}, in combination with the rules
  for \isa{arrive} in Figure~\ref{fig:node} and the fact that \isa{qmsg} is 
  always ready to \snippet{act_receive}, ensures
that a partial network can always perform an 
\snippet{act_arrive} for any combination of \isa{H} and \isa{K} consistent 
with its node addresses. Yet pairing with an
\snippet{act_cast}, through the first two rules in Figure~\ref{fig:pnet},
is possible only for those \isa{H} and~\isa{K} that are consistent with the 
destinations in \isa{R}.

Internal actions {\isasymtau} and the action 
\isa{i{\isacharcolon}deliver(d)}
are interleaved in the parallel composition of nodes that makes up a 
network.
\newpage

\subsection{Complete networks} \label{sec:awn:cnet} 

The last layer closes a network to further interactions with an environment. 
It ensures that a message cannot be received unless it is sent within the 
network or it is a \isa{Newpkt}.
\begin{center}
\snippet{closed'} \isa{\isacharequal} \snippet{closed_term}\puncgap.
\end{center}
The rules for \cnetsos{} are straightforward and presented in  
\reffig{cnet}.

\begin{figure}[t]
\tiny
    \begin{mathpar}
\msnippet{cnet_connect} \and
\msnippet{cnet_disconnect} \and
\msnippet{cnet_cast} \and
\msnippet{cnet_tau} \and
\msnippet{cnet_deliver} \and
\msnippet{cnet_newpkt}
    \end{mathpar}
\caption[]{\acworkaround{SOS} rules for complete networks.\label{fig:cnet}}
\end{figure}

The \isa{closed}-operator passes through internal actions, as well as
the delivery of data to destination nodes, this being an interaction
with the outside world. The \isa{\isacharasterisk cast}~actions are 
declared internal at this level; they cannot be influenced by the outside 
world.
The connect and disconnect actions are passed through in
\reffig{cnet}, thereby placing them under the control of the
environment.
Actions \isa{arrive m} are simply blocked by the encapsulation---they
cannot occur without synchronizing with a \isa{\isacharasterisk cast\ 
m}---except for
\snippet{act_arrive_newpkt}.
This action
represents new data \isa{d} that is submitted by a
client of the modelled protocol to node \isa{i} for delivery at
destination \isa{dst}.

\section{Basic invariance}\label{sec:proof-base} 

This paper only considers proofs of invariance, that is, properties of 
reachable states and reachable transitions.
The basic definitions are classic~\cite[Part III]{Muller:PhD:1998}.

\begin{definition}[reachability]\label{def:reachable}
Given an automaton~\isa{A} and an assumption~\isa{I} over actions, 
\isa{reachable\ A\ I} is the smallest set defined by the rules:
\begin{mathpar}
\msnippet{reachable_init}
\and
\msnippet{reachable_step}\puncgap\mbox{.}
\end{mathpar}
\end{definition}

\noindent
As usual, all initial states are reachable, and so is any
  state that can be reached from a reachable state by a single
  \isa{a}-transition that satisfies property \isa{I}.

\begin{definition}[invariance]\label{def:invariant}
Given an automaton~\isa{A} and an assumption~\isa{I}, a predicate~\isa{P} is 
\emph{(state) invariant}, denoted \snippet{invariant_lhs}, iff 
\snippet{invariant_rhs}.
\end{definition}

\noindent
We define reachability relative to an assumption on (input) actions \isa{I}.
When \isa{I} is \mbox{\snippet{TT_rhs}}, we write simply 
\snippet{invariant_TT_lhs}.

Using this definition of invariance, we can state a basic property of an 
instance of the toy process:
\begin{equation}\label{eq:basicinv}
\mbox{\snippet{nhip_eq_ip}\puncgap.}
\end{equation}
This invariant states that between the lines labelled \isa{PToy-:2} and 
\isa{PToy-:8},
that is,
after the assignment of \isa{PToy-:1} until before the assignment of 
\isa{PToy-:8},
the values of \isa{nhid} and \isa{id} are equal. Here
\snippet{onl_lhs}, defined as \snippet{onl_rhs}, extracts labels from 
control states, thereby converting a predicate on data states and line numbers 
into one on data states and control terms.%
\footnote{Using labels in this way is standard, see, for 
instance,~\cite[Chap. 1]{MannaPnu:Safety}, or the `assertion networks' of
~\cite[\textsection 2.5.1]{deRoeverEtAl:ConcVer:2001}.}
Because a \isa{\isasymoplus}-control term is unlabelled, the
  function \isa{label} takes the labels of both of its arguments;
for this reason \isa{labels\ \isasymGamma~p} generally yields a set of
labels rather than a single label.
As a control state \isa{call(pn)} also is unlabelled, the function
\isa{label} associates labels with it by unwinding the recursion; to
enable this, \isa{label} takes the recursive specification
\isa{\isasymGamma} as an extra argument.

The statements of properties that are true of all reachable states (for 
example, \refeq{bigger_than_next}, given later) do not depend on the values 
of control states nor the associated labels, but their proofs will if they 
involve other invariants (like that of \refeq{basicinv}).
Technically, the labels then form an integral part of the process model.
While this is unfortunate, expressing invariants in terms of the underlying 
control states is simply impractical: the terms are unwieldy and susceptible 
to modification.

State invariants concentrate on single states only. It is, 
however, often useful to characterize properties describing possible changes 
of the state.

\begin{definition}[transition invariance]\label{def:step-invariant}
Given an automaton~\isa{A} and an assumption~\isa{I}, a predicate~\isa{P} is 
\emph{transition invariant}, denoted 
\snippet{step_invariant_lhs}, iff\\
\centerline{\snippet{step_invariant_rhs}\puncgap.}
\end{definition}

\noindent
An example for a transition invariant of our running example is 
that the value of~\isa{no} never decreases over time:
\begin{equation}\label{eq:stepinv}
\mbox{
\snippet{seq_nos_increases'}\puncgap.
}\end{equation}
Here, the assumption on (input) actions \isa{I} is \snippet{TT_rhs} and hence skipped.
In case we want to restrict the statement to specific line numbers, 
the mechanization provides a function that extracts labels from control 
states, similar to \isa{onl} for state invariance:
\begin{center}
\begin{tabular}{lcl}
\snippet{onll_lhs} & \isa{\isacharequal} & \snippet{onll_rhs}\puncgap.
\end{tabular}
\end{center}

Our invariance proofs follow the compositional strategy recommended by de 
Roever et al.\ %
in~\cite[\textsection 
1.6.2]{deRoeverEtAl:ConcVer:2001}.
That is, we show properties of sequential process automata using the 
induction principle of~\refdef{reachable}, and then apply 
generic proof rules to successively lift such properties over each of the 
other layers.
The inductive assertion method, as stated by Manna and Pnueli in rule 
\textsc{inv-b} of~\cite{MannaPnu:Safety}, requires a finite set of 
transition schemas, which, together with the obligation on initial states 
yields a set of sufficient verification conditions.
We develop this set in \refsec{cterms} and use it to derive the main proof 
rule presented in \refsec{showinv} together with some examples.

\subsection{Control terms}\label{sec:cterms} 

Given a specification \isa{\isasymGamma} over finitely many process names, 
we can generate a finite set of verification conditions because transitions 
from \isa{('s, 'p, 'l) seqp} terms always yield subterms of terms in 
\isa{\isasymGamma}.
But, rather than simply considering the set of all subterms, we prefer to 
define a subset of `control terms' that reduces the number of verification 
conditions, avoids tedious duplication in proofs, and corresponds with the 
obligations considered in pen-and-paper proofs.
The main idea is that the \isa{\isasymoplus} and \isa{call} operators serve 
only to combine process terms: they are, in a sense, executed recursively by 
\seqpsos{} (see \refsec{awn:seq}) to determine the actions that 
a term offers to its environment.
This is made precise by defining a relation between sequential process 
terms.

\begin{definition}[{\microstep{\isasymGamma}}]\label{def:microsteps}
For a (recursive) specification \isa{\isasymGamma}, let 
\microstep{\isasymGamma} be the smallest relation such that
\snippet{microstep_choiceI1},
\snippet{microstep_choiceI2}, and
\snippet{microstep_callI}.
\end{definition}

\noindent
We write \microsteprtcl{\isasymGamma} for its reflexive transitive closure.
We consider a specification to be \emph{well formed}, when the inverse of 
this relation is well founded:
\begin{center}
\snippet{wellformed}\puncgap.\footnote{A specification is well formed iff it can be converted into one that is
\emph{weakly guarded} in the sense of \cite{Mi90ccs}.}
\end{center}
Most of our lemmas apply only to well-formed specifications, since otherwise 
functions over the terms they contain cannot be guaranteed to terminate.
Neither of these two specifications is well formed:
\isa{\isasymGamma\isactrlisub{a}(1)\ \isacharequal\ p\ {\isasymoplus}\ 
call(1)};
\isa{\isasymGamma\isactrlisub{b}(n)\ \isacharequal\ call(n+1)}.

We will also need a set of `start terms' of a process---the subterms that 
can act directly.

\begin{definition}[sterms]\label{def:sterms}
Given a \snippet{wellformed_gamma} and a sequential process term \isa{p}, 
\snippet{sterms_other_lhs} is the set of maximal elements related to \isa{p} 
by the reflexive transitive closure of the \microstep{\isasymGamma} 
relation:\footnote{This characterization is equivalent to 
\snippet{sterms_maximal_microstep_rhs}.
Termination follows from \snippet{wellformed_gamma}, that is,
\snippet{sterms_termination} for all \isa{p}.}

\begin{tabular}{lcl}
  \snippet{sterms_choice_concl_lhs} &\isa{\isacharequal}& \snippet{sterms_choice_concl_rhs}\puncgap, \\
  \snippet{sterms_call_concl_lhs}   &\isa{\isacharequal}& \snippet{sterms_call_concl_rhs}\puncgap, and,\\
  \snippet{sterms_other_lhs} &\isa{\isacharequal}& 
  \snippet{sterms_other_rhs} otherwise\puncgap.
\end{tabular}
\end{definition}

\noindent
As an example, consider the \isa{sterms} of the \isa{\gammaqmsg\ \qname} 
process from \reffig{qmsg}.

\medskip
\noindent
\null\isa{\ sterms\ \gammaqmsg\ (\gammaqmsg\ \qname)} =\\[.5em]
\null\hfil\begin{tabular}{l@{}}
$\left\{\begin{array}{c}
\mbox{\isa{\qmsg{0}receive{\isacharparenleft}{\isasymlambda}msg\ 
msgs{\isachardot}\ msgs\ {\isacharat}\ 
{\isacharbrackleft}msg{\isacharbrackright}{\isacharparenright}\ 
{\isachardot}\ 
call{\isacharparenleft}\qname{\isacharparenright}}\puncgap,}
\\
\mbox{\isa{\qmsg{0}{\isasymlangle}{\isasymlambda}msgs{\isachardot}\ 
\textsf{if}\ msgs\ {\isasymnoteq}\ {\isacharbrackleft}\,{\isacharbrackright}\ 
\textsf{then}\ {\isacharbraceleft}msgs{\isacharbraceright}\ \textsf{else}\ 
{\isasymemptyset}{\isasymrangle}}}
\mbox{\isa{\ {\isacharparenleft}\qmsg{1}send{\isacharparenleft}{\isasymlambda}msgs. 
hd\ msgs{\isacharparenright}\ $\cdots${\isacharparenright}}}
\end{array}\right\}$,
\end{tabular}

\medskip
\noindent
which contains the two subterms from either side of the initial choice: one 
that receives and loops, and another that begins by testing the value of 
\isa{msgs}.
An execution of the \isa{\gammaqmsg\ \qname} process amounts to an execution 
of one of these two terms.

We also define `local start terms' by
\snippet{stermsl_choice} and otherwise 
\snippet{stermsl_other}
to permit the sufficient syntactic condition that
a specification \isa{\isasymGamma} is well formed if 
\snippet{wf_no_direct_calls_prem_1}.

Since \isa{sterms\ \gammaqmsg\ (\gammaqmsg\ 
\qname)\ \isacharequal\ stermsl\ (\gammaqmsg\ \qname)}, and \qname{} is the 
only process in \gammaqmsg{}, we can conclude that \gammaqmsg{} is well formed,

Similarly to the way that start terms act as direct sources of transitions, 
we define `derivative terms' giving possible `active' destinations of 
transitions.

\begin{definition}[dterms]\label{def:dterms}
Given a \snippet{wellformed_gamma} and a sequential process term \isa{p}, 
\isa{dterms p} is defined by:

\begin{tabular}{lcl}
\snippet{dterms_choice_concl_lhs}  &\isa{\isacharequal}& \snippet{dterms_choice_concl_rhs}\puncgap, \\
\snippet{dterms_call_concl_lhs}    &\isa{\isacharequal}& \snippet{dterms_call_concl_rhs}\puncgap,\\[1ex]
\snippet{dterms_other1_concl_lhs}  &\isa{\isacharequal}& \snippet{dterms_other1_concl_rhs}\puncgap, \\
\snippet{dterms_other2_concl_lhs}  &\isa{\isacharequal}& \snippet{dterms_other2_concl_rhs}\puncgap, \\
\snippet{dterms_unicast_concl_lhs} &\isa{\isacharequal}& \snippet{dterms_unicast_concl_rhs}\puncgap, \\
\snippet{dterms_broadcast_concl_lhs} &\isa{\isacharequal}& \snippet{dterms_broadcast_concl_rhs}\puncgap, \\
\snippet{dterms_groupcast_concl_lhs} &\isa{\isacharequal}& \snippet{dterms_groupcast_concl_rhs}\puncgap, \\
\snippet{dterms_send_concl_lhs} &\isa{\isacharequal}& \snippet{dterms_send_concl_rhs}\puncgap, \\
\snippet{dterms_deliver_concl_lhs} &\isa{\isacharequal}& \snippet{dterms_deliver_concl_rhs}\puncgap, and, \\
\snippet{dterms_receive_concl_lhs} &\isa{\isacharequal}& \snippet{dterms_receive_concl_rhs}\puncgap.
\end{tabular}
\end{definition}

\noindent
For \isa{\gammaqmsg\ \qname}, for example, we calculate
\isa{\ dterms\ \gammaqmsg\ (\gammaqmsg\ \qname)} =
\begin{center}
$\left\{\begin{array}{c}
\mbox{\isa{\qmsg{0}receive{\isacharparenleft}{\isasymlambda}msg\ 
msgs{\isachardot}\ msgs\ {\isacharat}\ 
{\isacharbrackleft}msg{\isacharbrackright}{\isacharparenright}\ 
{\isachardot}\ 
call{\isacharparenleft}\qname{\isacharparenright}}\puncgap,}
\\
\mbox{\isa{\qmsg{0}{\isasymlangle}{\isasymlambda}msgs{\isachardot}\ 
\textsf{if}\ msgs\ {\isasymnoteq}\ {\isacharbrackleft}\,{\isacharbrackright}\ 
\textsf{then}\ {\isacharbraceleft}msgs{\isacharbraceright}\ \textsf{else}\ 
{\isasymemptyset}{\isasymrangle}}}
\mbox{\isa{\ {\isacharparenleft}\qmsg{1}send{\isacharparenleft}{\isasymlambda}msgs. 
hd\ msgs{\isacharparenright}\ $\cdots${\isacharparenright}}\puncgap,}
\\
\mbox{\isa{\qmsg{1}send{\isacharparenleft}{\isasymlambda}msgs. hd\ msgs{\isacharparenright}\ {\isachardot}}\ 
\isa{(\qmsg{2}{\isasymlbrakk}{\isasymlambda}msgs. tl msgs{\isasymrbrakk}}\ 
\isa{call{\isacharparenleft}\qname{\isacharparenright}\ {\isasymoplus}\ $\cdots$)}\puncgap,}
\\
\mbox{\isa{\qmsg{1}receive{\isacharparenleft}{\isasymlambda}msg\ msgs{\isachardot}\ msgs\ {\isacharat}\ {\isacharbrackleft}msg{\isacharbrackright}{\isacharparenright}\ {\isachardot}}\ %
\isa{call{\isacharparenleft}\qname{\isacharparenright}}}
\end{array}\right\}$.
\end{center}

\noindent
These derivative terms overapproximate the set of
\isa{sterms} of processes that can be reached in exactly one 
transition, since they do not consider the truth of guards
(like \isa{msgs\ {\isasymnoteq}\ {\isacharbrackleft}\,{\isacharbrackright}})
nor the willingness of communication partners (like \isa{receive(...)}). 

These auxiliary definitions lead to a succinct definition of the set of 
control terms of a specification.

\begin{definition}[cterms]\label{def:cterms}
For a specification $\Gamma$, \isa{cterms} is the smallest set where:
\begin{mathpar}
\msnippet{cterms_SI}
\and
\msnippet{cterms_DI}
\and
\end{mathpar}
\end{definition}

\noindent
There are, for example, six control terms in \isa{cterms \gammaqmsg} =
\begin{center}
$\left\{\begin{array}{c}
\mbox{\isa{\qmsg{0}receive{\isacharparenleft}{\isasymlambda}msg\ 
msgs{\isachardot}\ msgs\ {\isacharat}\ 
{\isacharbrackleft}msg{\isacharbrackright}{\isacharparenright}\ 
{\isachardot}\ 
call{\isacharparenleft}\qname{\isacharparenright}}\puncgap,}
\\
\mbox{\isa{\qmsg{0}{\isasymlangle}{\isasymlambda}msgs{\isachardot}\ 
\textsf{if}\ msgs\ {\isasymnoteq}\ {\isacharbrackleft}\,{\isacharbrackright}\ 
\textsf{then}\ {\isacharbraceleft}msgs{\isacharbraceright}\ \textsf{else}\ 
{\isasymemptyset}{\isasymrangle}}}
\mbox{\isa{\ {\isacharparenleft}\qmsg{1}send{\isacharparenleft}{\isasymlambda}msgs. 
hd\ msgs{\isacharparenright}\ $\cdots${\isacharparenright}}\puncgap,}
\\
\mbox{\isa{\qmsg{1}send{\isacharparenleft}{\isasymlambda}msgs. hd\ msgs{\isacharparenright}\ {\isachardot}}\ 
\isa{(\qmsg{2}{\isasymlbrakk}{\isasymlambda}msgs. tl msgs{\isasymrbrakk}}\ 
\isa{call{\isacharparenleft}\qname{\isacharparenright}\ {\isasymoplus}\ 
$\cdots$)}\puncgap,}
\\
\mbox{\isa{\qmsg{2}{\isasymlbrakk}{\isasymlambda}msgs. tl msgs{\isasymrbrakk}}\ 
\isa{call{\isacharparenleft}\qname{\isacharparenright}}\puncgap,}
\\
\mbox{\isa{\qmsg{2}receive{\isacharparenleft}{\isasymlambda}msg\ msgs{\isachardot}\ tl\ msgs\ {\isacharat}\ {\isacharbrackleft}msg{\isacharbrackright}{\isacharparenright}\ {\isachardot}}\ %
\isa{call{\isacharparenleft}\qname{\isacharparenright}}\puncgap,}
\\
\mbox{\isa{\qmsg{1}receive{\isacharparenleft}{\isasymlambda}msg\ msgs{\isachardot}\ msgs\ {\isacharat}\ {\isacharbrackleft}msg{\isacharbrackright}{\isacharparenright}\ {\isachardot}}\ %
\isa{call{\isacharparenleft}\qname{\isacharparenright}}}
\end{array}\right\}$.
\end{center}
In terms of the main example, the set \isa{cterms\ \gammatoy} has
fourteen elements; exactly one for each printed line in Figure~\ref{fig:toy} or
each transition in Figure~\ref{fig:auto}.\footnote{Of all the control terms, only those beginning with \isa{unicast} may induce more than one transition.}

When proving state or transition invariants of the form \isa{onl\ 
{\isasymGamma}\ P} or \isa{onll\ {\isasymGamma}\ P}, these are
the only control states for which the conditions of
\refdefs{invariant}{step-invariant} need
be checked.

$\!$As for \isa{sterms}, it is useful to define a local version independent of 
any \mbox{specification}.
\pagebreak[3]

\begin{definition}[ctermsl]\label{def:ctermsl}
Let \isa{ctermsl} be the smallest set defined by:

\begin{tabular}{@{}lcl@{}}
\snippet{ctermsl_choice_lhs}    &\isa{\isacharequal}& \snippet{ctermsl_choice_rhs}, \\
\snippet{ctermsl_call_lhs}      &\isa{\isacharequal}& \snippet{ctermsl_call_rhs}, \\
\snippet{ctermsl_other1_lhs}    &\isa{\isacharequal}& \snippet{ctermsl_other1_rhs}\puncgap, \\
\snippet{ctermsl_other2_lhs}    &\isa{\isacharequal}& \snippet{ctermsl_other2_rhs}\puncgap, \\
\snippet{ctermsl_unicast_lhs}   &\isa{\isacharequal}& 
\isa{{\isacharbraceleft}\gray{{\isacharbraceleft}l{\isacharbraceright}}unicast{\isacharparenleft}\selip,\ 
\selmsg{\isacharparenright}\ {\isachardot}\ p\ {\isasymtriangleright}\ 
q{\isacharbraceright}}\\
& & \isa{{\isasymunion}\ {\isacharparenleft}ctermsl\ p\ {\isasymunion}\ 
ctermsl\ q{\isacharparenright}}\puncgap, \\
\snippet{ctermsl_broadcast_lhs} &\isa{\isacharequal}& 
\snippet{ctermsl_broadcast_rhs}\puncgap, \\
\snippet{ctermsl_groupcast_lhs} &\isa{\isacharequal}& 
\snippet{ctermsl_groupcast_rhs}\puncgap, \\
\snippet{ctermsl_send_lhs}      &\isa{\isacharequal}& \snippet{ctermsl_send_rhs}\puncgap, \\
\snippet{ctermsl_deliver_lhs}   &\isa{\isacharequal}& \snippet{ctermsl_deliver_rhs}\puncgap, and, \\
\snippet{ctermsl_receive_lhs}   &\isa{\isacharequal}& \snippet{ctermsl_receive_rhs}\puncgap.
\end{tabular}
\end{definition}

\noindent
For our running example we have \isa{
ctermsl\ (\gammaqmsg{}\ Qmsg)\ \isacharequal\ cterms\ \gammaqmsg\ \ 
\isasymunion\ \  {\isacharbraceleft}\isa{call(Qmsg)}{\isacharbraceright}}.
\noindent
Including \isa{call} terms ensures that \snippet{stermsl_ctermsl_prem}
implies \snippet{stermsl_ctermsl_concl}, which facilitates proofs.
For \snippet{wellformed_gamma}$\!$, \isa{ctermsl} allows an alternative 
definition of \isa{cterms},
\begin{equation}\label{eq:cterms_def'}
\mbox{\snippet{cterms_def'_concl}\puncgap.}
\end{equation}
While the original definition is convenient for developing the meta-theory, 
due to the accompanying induction principle, this one is more useful for 
systematically generating the set of control terms of a specification, and 
thus, as we will see, sets of verification conditions.
And, for \snippet{wellformed_gamma}, we have as a corollary that
\begin{equation}\label{eq:cterms_subterms}
\mbox{\snippet{cterms_subterms_concl}\puncgap,}
\end{equation}
where \isa{subterms}, \isa{not-call}, and \isa{not-choice} are defined in 
the obvious way.

Our example already indicates that  \isa{cterms}
  over-approximates the set of start terms of reachable control states. 
  Formally we have the following theorem.

\begin{lemma}\label{thm:seq_reachable_in_cterms}
For
\snippet{seq_reachable_in_cterms_prem_1} and
automaton~\isa{A}
where
\snippet{seq_reachable_in_cterms_prem_2}
and
\snippet{seq_reachable_in_cterms_prem_3},
if \snippet{seq_reachable_in_cterms_prem_4} and 
\snippet{seq_reachable_in_cterms_prem_5} then
\snippet{seq_reachable_in_cterms_concl}.
\end{lemma}

\noindent
The predicate \snippet{control_within} serves to state that the initial 
control state is within the specification.

\subsection{Basic proof rule and invariants}\label{sec:showinv} 

State invariants such as \refeq{basicinv} are solved using a procedure whose 
soundness is justified as a theorem.
The proof exploits \refeq{cterms_def'} and \reflem{seq_reachable_in_cterms}.

\begin{theorem}\label{thm:seq_invariant_ctermsI} 
To prove \snippet{seq_invariant_ctermsI_concl}, where
\snippet{seq_invariant_ctermsI_prem_1},
\snippet{seq_invariant_ctermsI_prem_3},
\snippet{seq_invariant_ctermsI_prem_2}, and
\snippet{seq_invariant_ctermsI_prem_4}, it suffices
\begin{description}[leftmargin=4.1em,style=nextline]

\item[(init)]
for arbitrary
\snippet{seq_invariant_ctermsI_init_1} and
\snippet{seq_invariant_ctermsI_init_2}, to show
\snippet{seq_invariant_ctermsI_init_3}, and,

\item[(trans)]
for arbitrary
\snippet{seq_invariant_ctermsI_trans_1}, but 
\snippet{seq_invariant_ctermsI_trans_2}, and
\snippet{seq_invariant_ctermsI_trans_3},
given that
\snippet{seq_invariant_ctermsI_trans_8} for some
\snippet{seq_invariant_ctermsI_trans_7},
to assume
\snippet{seq_invariant_ctermsI_trans_4},
and then for any \isa{a} with \snippet{seq_invariant_ctermsI_trans_9} and any \isa{(\isasymxi\isacharprime,\ q)} such that
\snippet{seq_invariant_ctermsI_trans_5}
and
\snippet{seq_invariant_ctermsI_trans_6},
to show
\snippet{seq_invariant_ctermsI_trans_10}.
\end{description}
\end{theorem} 

\noindent
Here, \snippet{simple_labels}: each subterm must have exactly one label, 
that is, \isa{\isasymoplus} terms must be labelled consistently.
The specification \snippet{not_simple_labels_eg}, for updates \isa{f} and 
\isa{g}, does not satisfy \isa{simple-labels}.
Overlooking the technicalities, \refthm{seq_invariant_ctermsI} defines 
the expected set of verification conditions: we must show that a property 
\isa{P} holds of all initial states and that it is preserved by all 
transitions from control terms in a specification \isa{\isasymGamma}.

We incorporate this theorem into a generic tactic that \begin{inparaenum}

\item
applies it as an introduction rule,

\item
replaces \snippet{seq_invariant_ctermsI_trans_1} by a disjunction over the 
values of \isa{pn},

\item
applies \refdef{ctermsl} and repeated simplifications of \isa{\isasymGamma}s 
and eliminations on disjunctions to generate one subgoal (verification 
condition) for each control term,

\item \label{step:elimder}
replaces control term derivatives, the subterms in \refdef{dterms}, by fresh 
variables, and, finally,

\item
tries to solve each subgoal by simplification.

\end{inparaenum}
\refstep{elimder} replaces potentially large control terms by their 
(labelled) heads, which is important for readability and prover performance.
The tactic takes as arguments a list of existing invariants to include after 
having applied the introduction rule and a list of lemmas for trying to 
solve any subgoals that survive the final simplification.
There are no schematic variables in the subgoals and we benefit greatly from
Isabelle's \textsc{parallel\_goals} tactical~\cite{Wenzel:ParITP:2013}.

In practice, one states an invariant, applies the tactic, and examines the 
resulting goals.
One may need new lemmas for functions over the data state or explicit proofs 
for difficult goals.
That said, we find that the tactic generally dispatches the 
uninteresting goals, and the remaining ones typically correspond with the 
cases treated explicitly in the
pen-and-paper proofs~\cite{ATVA14}.

Using the generic tactic, the verification of \refeq{basicinv} 
is fully automatic. Isabelle rapidly dispatches the fourteen cases; one for 
each element of \isa{ctermsl\ \gammatoy}.

For transition invariants, we show a counterpart to 
\refthm{seq_invariant_ctermsI}, and declare it to the tactic described 
above.

\begin{theorem}\label{thm:seq_step_invariant_ctermsI} 
To prove \snippet{seq_step_invariant_ctermsI_concl}, where
\snippet{seq_step_invariant_ctermsI_prem_1},
\snippet{seq_step_invariant_ctermsI_prem_3},
\snippet{seq_step_invariant_ctermsI_prem_2}, and
\snippet{seq_step_invariant_ctermsI_prem_4}\!, it suffices
for arbitrary
\mbox{\snippet{seq_step_invariant_ctermsI_trans_1}}, but 
\snippet{seq_step_invariant_ctermsI_trans_2}, and
\snippet{seq_step_invariant_ctermsI_trans_3},
given that
\snippet{seq_step_invariant_ctermsI_trans_7} for some
\mbox{\snippet{seq_step_invariant_ctermsI_trans_6}},
for any \isa{a} with \snippet{seq_step_invariant_ctermsI_trans_8}, and for 
any \isa{(\isasymxi\isacharprime,\ q)} such that
\snippet{seq_step_invariant_ctermsI_trans_4}
and
\snippet{seq_step_invariant_ctermsI_trans_5},
to show
\snippet{seq_step_invariant_ctermsI_trans_9}.
\end{theorem} 

Again, stripped of its technicalities, this theorem simply requires checking 
a predicate \isa{P} across all transitions from all control terms in a 
specification \isa{\isasymGamma}.

Using \refthm{seq_step_invariant_ctermsI} we can prove that, within our 
toy-protocol, the value of \isa{no} never decreases 
(Equation~\refeq{stepinv}). Isabelle dispatches all cases but one, leaving 
the goal
\isa{no\ \isasymxi\ {\isasymle}\ no\ 
{\isasymxi}{\isacharprime}}
to be shown after the update 
\isa{{\isasymlbrakk}{\isasymlambda}{\isasymxi}{\isachardot}\ {\isasymxi}\ 
{\isasymlparr}nhid\ {\isacharcolon}{\isacharequal}\ sid\ 
{\isasymxi}{\isasymrparr}{\isasymrbrakk}} at line~\myhyperlink{7}. In fact, 
Isabelle determines that \isa{no\ \isasymxi{\isacharprime}\ {\isacharequal}\ 
num\ \isasymxi}, and hence it suffices to prove \mbox{\isa{no\ \isasymxi\ 
{\isasymle}\ num\ {\isasymxi}}} before the update. A manual inspection shows 
that neither \mbox{\isa{{no\ \isasymxi}}} nor \isa{num\ \isasymxi} change 
after the guard is evaluated and hence that the statement must be true. 
However, Isabelle cannot `inspect' the specification and we must introduce 
an auxiliary invariant:
\begin{center}
\snippet{seq_no_leq_num_restricted}\puncgap.
\end{center}
This state invariant is proven by Isabelle immediately, using our tactic; 
afterwards the transition invariant \snippet{seq_nos_increases'} passes 
without difficulty.

\section{Open invariance}\label{sec:proof-comp} 

The analysis of network protocols often requires `inter-node' invariants, 
like
\begin{multline}\label{eq:bigger_than_next}
\snippet{bigger_than_next}\puncgap,
\end{multline}

\noindent%
which states that, for any network topology, specified as a \isa{net-tree} 
with disjoint node addresses (\isa{wf-net-tree\ \isasymPsi}), the value of 
\isa{no} at a node is never greater than its value at the `next hop'---the 
address in \isa{nhid}.
This is a property of a global state~\isa{\isasymsigma} mapping addresses to 
corresponding local data states \isa{\isasymxi}.

We build a global state in two steps.
The first step maps a tree of states to a partial function from addresses to 
the data states of node processes:
\begin{center}
\begin{tabular}{lcl}
\snippet{netlift1_lhs} & \isa{\isacharequal} & 
\snippet{netlift1_rhs}\puncgap,\\
\snippet{netlift2_lhs} & \isa{\isacharequal} & 
\snippet{netlift2_rhs}\puncgap.
\end{tabular}
\end{center}
The \isa{netlift} function is parameterized by a `process selection' function~\isa{ps} that is 
applied to the state of a node process---that is, a state of the~\isa{np\ i} 
of \refsec{awn:pnet}.
In typical applications, such a state is
the local parallel composition of a protocol process and a message queue (see~\refsec{awn:par}).
In such a case, \isa{ps} selects just the protocol process, while 
abstracting
from the queue.
The first \isa{netlift} rule associates the node address~\isa{i} with the 
process data state.
The \isa{fst} elides the local component of the process state.
The second rule concatenates the partial maps generated for each branch of 
the state tree.
The assumption of disjoint node addresses is critical for reasoning about 
the resulting map.

The idea is to treat all (local) data states~\isa{\isasymxi}
as a single global state~\isa{\isasymsigma} and to abstract from local 
details like the process control state and queue.
The local details are important for stating and showing intermediate lemmas, 
but their inclusion in global invariants would be an unnecessary 
complication.

The second step in building the global state is to add default 
elements~\isa{df} for undefined addresses~\isa{i}. We first define the 
auxiliary function
\begin{center}
\snippet{default}\puncgap,
\end{center}
and then apply it to the result of \isa{netlift} in the definition of 
\isa{netglobal}. For our example we set
\begin{center}
\snippet{netglobal}\puncgap.
\end{center}
Basically, we associate a state with every node address by setting the state 
at non-existent addresses to the initial state (here \isa{toy-init}).
The advantage is that invariants and associated proofs need not consider the 
possibility of an undefined state or, in other words, that 
\isa{\isasymsigma\ i} could be \isa{None}.
In \refeq{bigger_than_next}, for example, this convention avoids three 
guards on address definedness.
One must decide, however, whether this convention is appropriate for a given 
property.

While we can readily state inter-node invariants of a complete model, 
showing them compositionally is another issue.
\refsecs{omodel}{oinv} present a way to state and prove such invariants at 
the level of sequential processes---in our example that is, with 
only~\isa{ptoy\ i} left of the turnstile.
\refsecs{lift}{transfer} present, respectively, rules for lifting such 
results to network models and for recovering invariants 
like~\refeq{bigger_than_next}.

\subsection{The open model}\label{sec:omodel} 

In (the standard model of) \ac{AWN} as presented in Section~\ref{sec:awn}, a 
network state is a \isa{closed} parallel composition of the states of the 
nodes in the network, arranged in a tree structure.
A state of a node is, in turn, a wrapper around a local parallel composition 
of states of sequential processes, each consisting of a local data 
state~\isa{\isasymxi} and a control term~\isa{p}.\linebreak[2]
To reason compositionally about the relations between these local data 
states, we introduce
the \emph{open model} of \ac{AWN}. This model collects relevant information from the 
individual local states \isa{\isasymxi}
into a single global state \isa{\isasymsigma}.
For our applications so far, we
have not needed to include \emph{all} local state elements in this global
data state; in fact we need only one local
data state per node, namely the one stemming from the leftmost component of 
the (non-commutative) local parallel composition of processes running on 
that node.
The leftmost component is by convention the main protocol process.
This type of global state is not only sufficient for our
purposes, but also easier to manipulate in Isabelle.

Recall that the data type \isa{ip} contains identifiers for all nodes that 
could occur in a network.
Since the leftmost parallel process running on a node has a local 
data state of type \isa{'k}, our global data state 
is of type \isa{ip {\isasymRightarrow} 'k}.
As described previously, identifiers that are not in a given network are 
mapped to default values.

\begin{figure}[t]\tiny
    \begin{mathpar}
\msnippet{oassignT} \and
\msnippet{oguardT} \and
\msnippet{ounicastT} \and
\msnippet{onotunicastT} \and
\msnippet{obroadcastT} \and
\msnippet{ogroupcastT} \and
\msnippet{osendT} \and
\msnippet{oreceiveT} \and
\msnippet{odeliverT} \and
\msnippet{ochoiceT1} \and
\msnippet{ochoiceT2} \and
\msnippet{ocallT}
    \end{mathpar}
\caption[]{Sequential processes: \oseqpsos.\label{fig:oseqp:sos}}
\end{figure}

In the open model, a state of a network is described as a pair 
\isa{(\isasymsigma, s)} of such a global state and a \isa{closed} parallel 
composition \isa{s} of the \emph{control} states of the nodes in the 
network.
The control state of a node is a wrapper around a local parallel composition 
of states of sequential processes, where we take only the control 
term~\isa{p} from the state~\isa{(\isasymxi, p)} of the leftmost parallel 
process running on the node, and the entire state from all other components.
As a result, a state in the open model contains
exactly the same information as a state in the default model, 
even if it is arranged differently.

\reffigr{oseqp:sos}{cpnet:sos} present the SOS rules for the open model.
Many of them are similar to the rules presented in \refsec{awn}; 
for the sake of completeness we list them nevertheless.

\subsubsection{Sequential Processes} 

The rules for the sequential control terms in the open model, \oseqpsos{}, 
are presented in \reffig{oseqp:sos}.
They are nearly identical to the ones in the original model, but have to be 
parameterized by an address \isa{i} and constrain only that entry of the 
global state, either to say how it changes (\snippet{oassignT_prem_1}) or 
that it does not change
\mbox{(\snippet{ounicastT_prem_1})}.
These rules do not restrict changes in the data state of any other 
node~\isa{j} (\isa{j{\isasymnoteq}i}).
In principle, the data states of these nodes can change
arbitrarily, so that any state \mbox{\isa{(\isasymsigma, p)}} has
infinitely many outgoing transitions.
However, the composition with other nodes, introduced in a higher layer of 
the process algebra, will limit the set of outgoing transitions by combining 
the restrictions imposed by each of the nodes.

\subsubsection{Local parallel composition} 

The states \isa{(\isasymsigma, s)} in an automaton of the open model are of type
\isa{(ip\ {\isasymRightarrow}\ 'k)\ \isasymtimes\ 's} with \mbox{\isa{ip\ 
{\isasymRightarrow}\ 'k}} the type of global data
states and \isa{'s} the type of control states.
Hence, such an automaton has type
\isa{((ip\ {\isasymRightarrow}\ 'k)\ \isasymtimes\ 's,\ 'a)\ automaton}.
The local parallel composition of the open model pairs an open automaton 
with a standard one, and thus has type\linebreak
\isa{((ip\ {\isasymRightarrow}\ 'k)\ \isasymtimes\ 's,\ 'a)\ automaton\ 
{\isasymRightarrow}\ ('t,\ 'a)\ automaton\ {\isasymRightarrow}\ ((ip\ 
{\isasymRightarrow}\ 'k)\ \isasymtimes\ ('s\ \isasymtimes\ 't),\ 'a)\ 
automaton}.

\begin{figure}[t]\tiny
\begin{mathpar}
\msnippet{oparleft} \and
\msnippet{oparright} \and
\msnippet{oparboth}
    \end{mathpar}
\caption[]{Parallel processes: \oparpsos.\label{fig:oparp:sos}}
\end{figure}

\begin{figure}[tb]\tiny
    \begin{mathpar}
\msnippet{onode_ucast} \and
\msnippet{onode_notucast} \and
\msnippet{onode_bcast} \and
\msnippet{onode_gcast} \and
\msnippet{onode_receive} \and
\msnippet{onode_deliver} \and
\msnippet{onode_tau} \and
\msnippet{onode_arrive} \and
\msnippet{onode_connect1} \and
\msnippet{onode_connect2} \and
\msnippet{onode_disconnect1} \and
\msnippet{onode_disconnect2} \and
\msnippet{onode_connect_other} \and
\msnippet{onode_disconnect_other}
    \end{mathpar}
\caption[]{Nodes: \onodesos.\label{fig:onode:sos}}
\end{figure}

The rules for \oparpsos{}, depicted in \reffig{oparp:sos},
only allow the first sub-process to constrain 
\isa{\isasymsigma}:
the global data state that appears in the
parallel composition is simply taken from its first component.
This choice precludes comparing the states of \isa{qmsg}s 
(and any other local filters) across a network, but it also simplifies the 
mechanics and use of this layer of the framework.
Since our mechanization aims at the verification of (routing) 
protocols~\cite{ATVA14}, which nearly always implement a queue, simplifying 
the
mechanization in this way seems reasonable.
The treatment of the other layers is independent of this choice.
So, if our work were to be applied in another setting where 
queues are not used, or where data states of 
more than one parallel control term need to be lifted to a control state, 
only this layer need be adapted.
\newpage

\subsubsection{Nodes and partial networks}\label{sec:onodes} 

The sets \onodesos{} (\reffig{onode:sos}) and \opnetsos{} 
(\reffig{opnet:sos}) need not be parameterized by an address since they are 
generated inductively from lower layers.
Together they constrain parts of \isa{\isasymsigma}.
This occurs naturally for rules like those for \isa{arrive} and 
\isa{\isacharasterisk cast}, where the synchronous communication serves as a 
conjunction of constraints on different parts of \isa{\isasymsigma}. But for 
others that normally only constrain a single element, like those for 
\isa{\isasymtau}, assumptions (\snippet{onode_tau_prem_2}) are introduced 
here and dispatched later (\refsec{transfer}).
Such assumptions aid later proofs, but they must be justified when transferring results to closed  systems.

\begin{figure}[tb]\tiny
    \begin{mathpar}
\msnippet{opnet_cast1} \and
\msnippet{opnet_cast2} \and
\msnippet{opnet_arrive} \and
\msnippet{opnet_deliver1} \and
\msnippet{opnet_deliver2} \and
\msnippet{opnet_tau1} \and
\msnippet{opnet_tau2} \and
\msnippet{opnet_connect} \and
\msnippet{opnet_disconnect}
    \end{mathpar}
\caption[]{Partial networks: \opnetsos.}\label{fig:opnet:sos}
\end{figure}

\subsubsection{Complete networks}\label{sec:oclosed} 
The rules for \ocnetsos{} are shown in \reffig{cpnet:sos}.
Each rule includes a precondition that ensures that elements not 
addressed within a model do not change:
\isa{net-ips} gives the set of node addresses in a 
state of a partial network.

\begin{figure}[tb]
\tiny
    \begin{mathpar}
\msnippet{ocnet_connect} \and
\msnippet{ocnet_disconnect} \and
\msnippet{ocnet_cast} \and
\msnippet{ocnet_tau} \and
\msnippet{ocnet_deliver} \and
\msnippet{ocnet_newpkt}
    \end{mathpar}
\caption[]{Complete networks: \ocnetsos.\label{fig:cpnet:sos}}
\end{figure}

\subsubsection{Application} 

We now show how to construct an open model.
For the running example, a sequential instance of the toy protocol is 
defined as
\begin{center}
\snippet{optoy_lhs} \isa{\isacharequal} \snippet{optoy_rhs}\puncgap,
\end{center}
combined with the standard \isa{qmsg} process into
\begin{center}
\isa{onp i} = \snippet{optoy_qmsg_term}\puncgap,
\end{center}
\noindent
using the operator
\begin{align*}
\snippet{opar_comp'}\puncgap,
\end{align*}

\noindent and lifted to the node level via the open node constructor
\begin{center}
\snippet{onode_comp}\puncgap.
\end{center}
Similarly, to map a \snippet{net_tree} term to an open model we define:

\smallskip
\begin{center}
\begin{tabular}{@{\hspace{-5.1mm}}l@{\,}c@{\,}l}
\centering
\snippet{opnet1_lhs} & \isa{\isacharequal} & \snippet{opnet1_rhs} \\
\snippet{opnet2_lhs} & \isa{\isacharequal} & \snippet{opnet2_rhs}\puncgap.
\end{tabular}
\end{center}

\noindent
The third requirement on initial states makes the open model non-empty
only for \snippet{net_tree}s with disjoint node addresses.
Including such a constraint within the open model, rather than as a separate 
assumption like
the \isa{wf-net-tree\ n} in \refeq{bigger_than_next},
eliminates an annoying technicality from the inductions described in 
\refsec{lift}.
As with the extra premises in the open \ac{SOS} rules, we can freely adjust 
the open model to facilitate proofs, but each `encoded assumption' 
becomes an obligation that must be discharged in the transfer 
lemma of \refsec{transfer}.

Of course, the above constructs apply to any function \isa{onp} from
addresses to automata in the open model, that is, any \isa{onp} of type
\isa{ip $\Rightarrow$ ((ip $\Rightarrow$ 'k) $\times$ 's,~'a) automaton}.

An operator for adding the last layer is also readily defined by
\begin{center}
\snippet{oclosed'} \isa{\isacharequal} \snippet{oclosed_term}\puncgap,
\end{center}
giving all the definitions necessary to turn a standard model into an open 
one.

\subsection{Open invariants}\label{sec:oinv} 

The basic definitions of reachability, invariance, and 
transition invariance, 
Definitions~\ref{def:reachable}--\ref{def:step-invariant}, apply to open 
models since they are given for generic automata, but constructing a 
compositional proof requires considering the effects of both synchronized 
and interleaved actions of possible en\-vir\-on\-ments.
Our automaton \isa{A} could, for instance, be a partial network, 
consisting of several nodes, and the environment could be another partial 
network running in parallel.
An action performed by the environment and the automaton together, or 
indeed, since the distinction is unimportant here, by the automaton alone, 
is termed \emph{synchronized} and an action made by the environment without 
the participation of the automaton is termed \emph{interleaved}.
We identify the nature of a synchronized action by the environment through 
the action of \isa{A} that synchronizes with it.
We focus first of all on the case where \isa{A} is a single node \isa{i}.

The proper analysis of properties of \isa{A}, such as 
(\ref{eq:bigger_than_next}), often requires assumptions on the 
behaviour of the environment.
We consider assumptions on both synchronized and interleaved actions.

A typical example for an assumption on synchronized actions is
\begin{equation}\label{eq:syncenv}
\mbox{\isa{{\isacharparenleft}{\isasymforall}j{\isachardot}\ j\ {\isasymnoteq}\ i\ {\isasymlongrightarrow}\ no\ {\isacharparenleft}{\isasymsigma}\ j{\isacharparenright}\ \isasymle\ no\ {\isacharparenleft}{\isasymsigma}{\isacharprime}\ j{\isacharparenright}{\isacharparenright}\ {\isasymand}\ orecvmsg\ msg{\isacharunderscore}ok\ {\isasymsigma}\ a}\puncgap,}
\end{equation}
where \isa{orecvmsg} applies a predicate (here
\isa{msg{\isacharunderscore}ok}) to \isa{receive}~actions and is otherwise 
true:
\snippet{msg_num_ok_pkt} and \snippet{msg_num_ok_newpkt2}.
So, the assumption manifests two properties of the environment (nodes that are not equal to \isa{i}) 
(1) it guarantees that all nodes different from \isa{i} preserve the 
property that the value of \isa{no} cannot be decreased by the protocol; (2) 
whenever a \isa{Pkt} message is sent, the value 
\isa{d} stored in the message is smaller than or equal to the current value 
of \isa{no}, stored at the sender of the message \isa{src}.
The synchronization occurs via the exchange of messages.

A typical example for an interleaved (un-synchronized) action \emph{from the environment} is
\begin{equation}\label{eq:interenv}
\mbox{
\isa{{\isacharparenleft}{\isasymforall}j{\isachardot}\ j\ {\isasymnoteq}\ i\ {\isasymlongrightarrow}\ no\ {\isacharparenleft}{\isasymsigma}\ j{\isacharparenright}\ \isasymle\ no\ {\isacharparenleft}{\isasymsigma}{\isacharprime}\ j{\isacharparenright}{\isacharparenright}\ 
{\isasymand}\ {\isasymsigma}{\isacharprime}\ i\ {\isacharequal}\ {\isasymsigma}\ i\puncgap,
 }}
 \end{equation}
This assumption states that 
(1) nodes that are not equal to \isa{i}  do not decrease the value of 
\isa{no}---as before---and (2) that the data state at node \isa{i} does not 
change. So transitions of the environment may interleave 
with actions performed by node \isa{i}, as long as they are
`well-behaved' and do not interfere with the state of \isa{i}.

\begin{definition}[open reachability]\label{def:oreachable}
Given an automaton~\isa{A} and assumptions~\isa{S} and~\isa{U} over, 
respectively, synchronized and interleaved actions of the environment, \isa{oreachable\ A\ S\ 
U} is the smallest set defined by the rules:
\begin{mathpar}
\msnippet{oreachable_init}
\and
\msnippet{oreachable_other'}
\\\vspace{-3mm}
\msnippet{oreachable_local'}\mbox{.}
\end{mathpar}
\end{definition}

\begin{figure}[t]
\centering
\begin{tikzpicture}[
  -latex,
  semithick,
  start chain=going right,
  node distance=2.5cm,
  every state/.style={on chain,text width=7mm,align=center}]

   \node[state] (state 1) {\isa{\isasymsigma, s}};
   \node[state] (state 2) {\isa{\isasymsigma{\isacharprime}, s{\isacharprime}}};
   \node[state] (state 3) {\isa{\isasymsigma{\isacharprime}{\isacharprime},\,s{\isacharprime}}};

   \path (state 1) edge [bend left=10] node [above] {\isa{a}} (state 2)
         (state 1.50) edge [bend left=10,dashed] (state 2.130)
         (state 2) edge [bend left=10,dashed] (state 3)
     ;

   \path[every node/.style={yshift=-.5cm}]
     (state 1) -- node [midway] (local label) {\isa{S\ \isasymsigma\ \isasymsigma'\ a}} (state 2)
     (state 2) -- node [midway] (other label) {\isa{U\ \isasymsigma'\ \isasymsigma''}} (state 3);

\end{tikzpicture}
\caption{Open reachability with assumptions on synchronized and interleaved 
actions.}
\label{fig:oreach}
\end{figure}
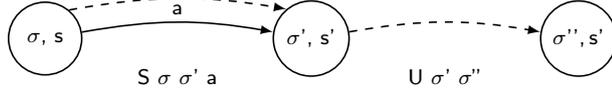

\noindent
The first rule declares all initial states reachable. The second declares as 
reachable all states that result from an interleaving transition of the 
environment that satisfies \isa{U} and where the process \isa{s} does not 
perform any action.
In the third rule process \isa{s} performs an action that yields a reachable 
state if \isa{s} in combination with the
global data state was reachable and 
if the assumption \isa{S} is respected by the action and the environment.
\reffig{oreach} illustrates the main idea of synchronized and interleaved 
actions---the solid arrow
represents an action~\isa{a} performed by state~(\mbox{\isa{\isasymsigma, 
s}}), the dashed arrows indicate
transitions taken by other nodes (on the left in synchrony with action \isa{a}).

In practice, we use restricted forms \isa{otherwith E N I} and \isa{other E 
N} of the assumptions~\isa{S} and~\isa{U}, respectively:
\begin{align}
\mbox{\snippet{otherwith_def_lhs}} &= 
\mbox{\snippet{otherwith_def_rhs}}\puncgap\mbox{, and}\label{eq:otherwith}\\
\mbox{\snippet{other_def_lhs}} &= 
\mbox{\snippet{other_def_rhs}}\puncgap.\label{eq:other}
\end{align}
The requirements \refeq{syncenv} and \refeq{interenv}, presented above, 
have exactly these forms. 

The assumptions \isa{otherwith} and \isa{other} are parameterized with a set 
\isa{N} of type
\isa{ip\ set} of \emph{scoped} nodes---those that occur in the control 
states of the automaton.
They both restrict the environments under consideration
by applying a predicate~\isa{E} of type \isa{{\isacharprime}s\ 
{\isasymRightarrow}\ {\isacharprime}s\ {\isasymRightarrow}\ bool} to 
possible changes of (local) data states of nodes~\isa{i} of the environment. 
In addition, \isa{otherwith} permits constraints on the information~\isa{I}
from shared actions,
like \isa{broadcast} or \isa{receive}. These constraints refer to the action 
\isa{a} and the global data state \isa{\isasymsigma}.

In contrast to \refeq{otherwith}, Equation\ \refeq{other} excludes changes 
in scoped nodes (\isa{\isasymsigma'\ i\ \isacharequal\ \isasymsigma\ i}).

\begin{definition}[open invariance]\label{def:oinvariant}
Given an automaton~\isa{A} and assumptions~\isa{S} and~\isa{U} over 
synchronized and interleaved actions, respectively, a predicate~\isa{P} is 
an \emph{open invariant}, denoted \snippet{oinvariant_lhs}, iff 
\snippet{oinvariant_rhs}.
\end{definition}

\noindent
It follows easily that existing invariants can be made open. In 
practice, this means that most invariants can be shown in the basic context 
but still exploited in the more complicated one.

\begin{lemma}\label{thm:open_seq_invariant}
Given an invariant
\snippet{open_seq_invariant_prem_1}
where \snippet{open_seq_invariant_prem_4},
and any predicate~\isa{F},
there is an open invariant
\snippet{open_seq_invariant_concl'}
where \snippet{open_seq_invariant_prem_3},
provided that
\snippet{open_seq_invariant_prem_2'}.
\end{lemma}

\noindent
Open transition invariance and a similar transfer lemma are 
defined similarly.
The meta theory for basic invariants is also readily adapted, in particular,

\begin{theorem}\label{thm:oseq_invariant_ctermsI} 
To show \snippet{oseq_invariant_ctermsI_concl}, in addition to the 
conditions and the obligations (init)  and (trans) of 
\refthm{seq_invariant_ctermsI}, suitably adjusted, it suffices,
\begin{description}[leftmargin=4.1em,style=nextline]

\item[(env)]
for arbitrary \snippet{oseq_invariant_ctermsI_trans_1}
and \snippet{oseq_invariant_ctermsI_trans_2}\puncgap,\\
to assume both
\snippet{oseq_invariant_ctermsI_trans_3} and
\snippet{oseq_invariant_ctermsI_trans_4},
and then to show
\snippet{oseq_invariant_ctermsI_trans_5}\puncgap.
\end{description}
\end{theorem} 

\noindent
This theorem (together the counterpart of 
\refthm{seq_step_invariant_ctermsI} for open transition
invariance) is declared to the tactic described in \refsec{showinv} and 
proofs proceed as before, but with the new obligation to show invariance 
over interleaved transitions.

We finally have sufficient machinery to state and prove Invariant 
    \refeq{bigger_than_next} at the level of a sequential process:
\begin{equation}\label{eq:oseq_bigger_than_next}
\begin{tabular}{l}
\snippet{oseq_bigger_than_next}\puncgap,
\end{tabular}
\end{equation}
where \snippet{nos_increase},
So, given that the variables \isa{no} in the environment never decrease and 
that incoming \isa{Pkt}s reflect the state of the sender, there is a 
relation between the local node and the next hop.
Similar invariants occur in proofs of realistic protocols~\cite{ATVA14}.

\subsection{Lifting open invariants}\label{sec:lift} 

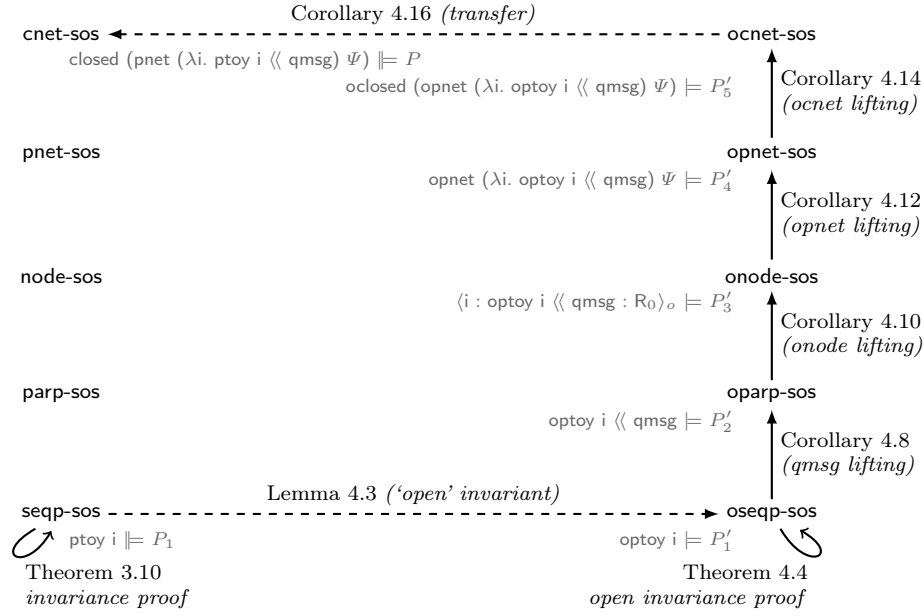
\begin{figure}[t]

  \tikzstyle{lift}=[thick,-latex,
                    every node/.style={right,align=left}]
  \tikzstyle{transfer}=[lift,dashed]
  \tikzstyle{example}=[gray]

  \newcommand{\cnetprop}{%
    \isasmall{closed\ {\isacharparenleft}pnet\ 
    {\isacharparenleft}{\isasymlambda}i{\isachardot}\ ptoy\ i\ 
    {\isasymlangle}\!{\isasymlangle}\ qmsg{\isacharparenright}\ 
    {\isasymPsi}{\isacharparenright}\ {\isasymTTurnstile}\ $P$}}
  \newcommand{\ocnetprop}{%
    \isasmall{oclosed\ {\isacharparenleft}opnet\ 
    {\isacharparenleft}{\isasymlambda}i{\isachardot}\ optoy\ i\ 
    {\isasymlangle}\!{\isasymlangle}\ qmsg{\isacharparenright}\ 
    {\isasymPsi}{\isacharparenright}\ {\isasymTurnstile}\ $P'_5$}}
  \newcommand{\opnetprop}{%
    \isasmall{opnet\ {\isacharparenleft}{\isasymlambda}i{\isachardot}\ 
    optoy\ i\ {\isasymlangle}\!{\isasymlangle}\ qmsg{\isacharparenright}\ 
    {\isasymPsi}\ {\isasymTurnstile}\ $P'_4$}}
  \newcommand{\onodeprop}{%
    \isasmall{{\isasymlangle}i\ :\ optoy\ i\ 
    {\isasymlangle}\!{\isasymlangle}\ qmsg\ :\ \Ri{\isasymrangle}\isactrlsub 
    o
    {\isasymTurnstile}\ $P'_3$}}
  \newcommand{\oparpprop}{%
    \isasmall{optoy\ i\ {\isasymlangle}\!{\isasymlangle}\ qmsg\ 
    {\isasymTurnstile}\ $P'_2$}}
  \newcommand{\oseqpprop}{%
    \isasmall{optoy\ i\ {\isasymTurnstile}\ $P'_1$}}
  \newcommand{\seqpprop}{%
    \isasmall{ptoy\ i\ {\isasymTTurnstile}\ $P_1$}}
  \hspace{-10pt}
  \begin{tikzpicture}
    \matrix (m)
      [column sep=8cm, row sep=1.2cm,
       matrix of nodes,
       nodes in empty cells,
       ampersand replacement=\&] at (0,-.3) {
        \cnetsos \& \ocnetsos \\
        \pnetsos \& \opnetsos \\
        \nodesos \& \onodesos \\
        \parpsos \& \oparpsos \\
        \seqpsos \& \oseqpsos \\
    };

    \node[shift={(  0,-.35)},below,right,example] at (m-1-1) {\cnetprop};
    \node[shift={(  0,-.35)},below,right,example] at (m-5-1) {\seqpprop};
    \node[shift={(-.4,-.35)},below,left,example] at (m-5-2)  {\oseqpprop};
    \node[shift={(-.4,-.35)},below,left,example] at (m-4-2)  {\oparpprop};
    \node[shift={(-.4,-.35)},below,left,example] at (m-3-2)  {\onodeprop};
    \node[shift={(-.4,-.35)},below,left,example] at (m-2-2)  {\opnetprop};
    \node[shift={(-.4,-.7)},below,left,example] at (m-1-2)  {\ocnetprop};

    \draw[lift] (m-5-2) --
      node {\refcoro{par_qmsg_lift}\\\emph{(qmsg lifting)}} (m-4-2);

    \draw[lift] (m-4-2) --
      node {\refcoro{node_proc_lift}\\\emph{(onode lifting)}} (m-3-2);

    \draw[lift] (m-3-2) --
      node {\refcoro{opnet_lifting}\\\emph{(opnet lifting)}} (m-2-2);

    \draw[lift] (m-2-2) --
     node {\refcoro{ocnet_lifting}\\\emph{(ocnet lifting)}} (m-1-2);

    \draw[transfer] (m-1-2) --
      node[above] {\refcoro{close_lift} \emph{(transfer)}} (m-1-1);

    \draw[transfer] (m-5-1) -- node[above]
      {\reflem{open_seq_invariant} \emph{(`open' invariant)}} (m-5-2);

    \draw[lift] (m-5-1) edge [in=240,out=210,loop]
      node[below right] {\refthm{seq_invariant_ctermsI}\\
                         \emph{invariance proof}} (m-5-1);

    \draw[lift] (m-5-2) edge [in=330,out=300,loop]
      node[below left,align=right]
        {\refthm{oseq_invariant_ctermsI}\\
         \emph{open invariance proof}} (m-5-2);
  \end{tikzpicture} 
\caption{Schema of the overall proof structure.}
\label{fig:oschema}
\end{figure}

The preceding two sections provide enough machinery to state and show global 
invariants at the level of sequential processes, that is, over automata like 
\isa{optoy\ i} in Invariant \refeq{oseq_bigger_than_next}.
It still remains to extend such results to models of entire networks, and 
ultimately to re-establish them in the original model of~\refsec{awn}.

Our approach is sketched in~\reffig{oschema}.
We prove as many invariants as possible in the closed sequential model 
(\seqpsos) as described in~\refsec{showinv}.
These invariants are extended to the open sequential model (\oseqpsos)
using~\reflem{open_seq_invariant}, where they support proofs of the forms of 
global invariants described in \refsec{oinv}.
Invariants that cannot be stated in \seqpsos,
because they interrelate the states of multiple nodes,
are proved directly in \oseqpsos\ using \refthm{oseq_invariant_ctermsI}
and its counter part for open transition invariance. Once 
established in \oseqpsos, global invariants can be lifted successively over 
the composition operators of the open model (\oparpsos, \onodesos, 
\opnetsos, \ocnetsos), using the lemmas described in this section, and then 
transferred into the closed complete model (\cnetsos), using the lemma 
described in the next section.
\reffig{oschema} shows, in grey, examples of the forms of invariants at each 
stage.
The goal is to show a property~$P$ over an entire arbitrary network in the 
closed model (at top-left).
The property $P$ is proven via a succession of intermediate
invariants, starting with~$P_1$, which is expressed relative to a
single node, possibly in relation to the rest of the network~($P'_1$).
At each step its form changes slightly ($P'_2$, $P'_3$, $P'_4$, and~$P'_5$) 
to hide technical details introduced at each layer and as its range extends 
to multiple nodes.

The first lifting rule (\refcoro{par_qmsg_lift}) treats composition with the 
\isa{qmsg} process.
It mixes \isa{oreachable} and \isa{reachable} predicates: the former for the 
automaton being lifted, the latter for properties of \isa{qmsg}.
Two main properties of \isa{qmsg} are required: only received messages are 
added to the queue and sent messages come from the queue. They are shown 
using the techniques of \refsec{proof-base}.

\begin{lemma}
\isa{%
qmsg\ \stepinv\ 
{\isacharparenleft}{\isasymlambda}{\isacharparenleft}{\isacharparenleft}msgs,\ 
q{\isacharparenright},\ a,\ {\isacharparenleft}msgs{\isacharprime},\ 
q{\isacharprime}{\isacharparenright}{\isacharparenright}{\isachardot}} \\
\null\hspace{4.8cm}
\isa{\textsf{case}\ a\ \textsf{of}\ receive\ m\ {\isasymRightarrow}\ 
set\ msgs{\isacharprime}\ {\isasymsubseteq}\ set\ (msgs\ {\isacharat}\ 
{\isacharbrackleft}m{\isacharbrackright})}
\\
\null\hspace{5.8cm}
\isa{{\isacharbar}\ 
{\isacharunderscore}}\hspace{1.2cm}\isa{{\isasymRightarrow}\ set\ 
msgs{\isacharprime}\ {\isasymsubseteq}\ set\ 
msgs{\isacharparenright}}\puncgap.%
\end{lemma}

\begin{lemma}
\snippet{qmsg_send_from_queue}\puncgap.
\end{lemma}

These two properties of \isa{qmsg} are used to prove a lemma that 
decomposes open reachability of \isa{A\ 
{\isasymlangle}{\isasymlangle}\isactrlbsub i\isactrlesub \ qmsg} into open 
reachability of \isa{A} and reachability of \isa{qmsg}.

\begin{lemma}[qmsg reachability]\label{lem:par_qmsg_lift} 
\newline
Given
\snippet{par_qmsg_oreachable_prem_1},
with assumptions on synchronizing and interleaved transitions
\snippet{par_qmsg_oreachable_L} and
\snippet{par_qmsg_oreachable_E}, and provided
\begin{enumerate}
\item
\isa{F} is reflexive,\label{eq:qmsglift:Frefl}
\item
for all
\isa{\isasymxi,\ \isasymxi\isacharprime},
\snippet{par_qmsg_oreachable_Exixi'} implies
\snippet{par_qmsg_oreachable_Fxixi'},\label{eq:qmsglift:EF}
\item
\snippet{par_qmsg_oreachable_prem_2}, and,\label{eq:qmsglift:F}
\item
for all
\isa{\isasymsigma,\ \isasymsigma',\ m},
\snippet{par_qmsg_oreachable_Fsigmaj}
and
\snippet{par_qmsg_oreachable_Rsigmam}
imply
\snippet{par_qmsg_oreachable_Rsigma'm},\label{eq:qmsglift:RR}
\end{enumerate}
then
\snippet{par_qmsg_oreachable_conj1} and
\snippet{par_qmsg_oreachable_conj2}, and furthermore
\snippet{par_qmsg_oreachable_conj3}.
\end{lemma} 

\noindent
In the \isa{qmsg} part of the local state \isa{(msgs, q)}, \isa{msgs} 
is a list of messages and \isa{q} is the control state of the queue.
We write \isa{set\ msgs} to generate a set of messages from the list of 
messages.
The key intuition behind the four clauses is that every message~\isa{m} 
received, queued, and sent by \isa{qmsg} must satisfy \isa{M\ \isasymsigma\ 
m}.
That is, the properties of messages received into the queue continue to 
hold---even as the environment and thus the original senders act---until 
those messages are transmitted from the queue to the automaton~\isa{A}.
The proof is by induction over \isa{oreachable}.
The \isa{M}'s are preserved when the external environment acts independently 
\refeqs{qmsglift:Frefl}{qmsglift:RR}, when it acts synchronously 
\refeqs{qmsglift:EF}{qmsglift:RR}, and when the local process acts 
\refeqs{qmsglift:F}{qmsglift:RR}.

The preceding lemma allows open reachability of the parallel composition to 
be decomposed into (open) reachability of its components.
It follows easily that an invariant of the principle component (\isa{A}) is 
also an invariant of the composition.

\newpage
\begin{corollary}[qmsg lifting]\label{thm:par_qmsg_lift} 
Given
\snippet{par_qmsg_lift_prem_1},
with the predicates
\snippet{par_qmsg_oreachable_L} and
\snippet{par_qmsg_oreachable_E}, and provided
\begin{enumerate}
\item
\isa{F} is reflexive,
\item
for all
\isa{\isasymxi,\ \isasymxi\isacharprime},
\snippet{par_qmsg_oreachable_Exixi'} implies
\snippet{par_qmsg_oreachable_Fxixi'},
\item
\snippet{par_qmsg_oreachable_prem_2}\puncgap, and,
\item
for all
\isa{\isasymsigma,\ \isasymsigma',\ m},
\snippet{par_qmsg_oreachable_Fsigmaj}
and
\snippet{par_qmsg_oreachable_Rsigmam}
imply
\snippet{par_qmsg_oreachable_Rsigma'm},
\end{enumerate}
then
\snippet{par_qmsg_lift_concl}.
\end{corollary} 

\noindent
This lifting rule is specific to the queue model presented in \reffig{qmsg}.
Changes to the model may necessitate changes to the rule or its proof, or 
indeed, a different parallel composition would require a new rule and proof.
No assumptions are made, however, on the structure of \isa{A}, the automaton 
being lifted.

The rule for lifting to the node level is straightforward since a node 
simply encapsulates the state of the underlying model.
It is necessary, though, to adapt assumptions on \isa{receive} actions 
(\isa{orecvmsg}) to \isa{arrive} actions (\isa{oarrivemsg}), but this is 
essentially a minor technicality.

\begin{lemma}[onode reachability]\label{thm:node_proc_reachable} 
If, for all \isa{\isasymxi} and \isa{\isasymxi\isacharprime},
\snippet{node_proc_reachable_prem_2_prem}
implies
\snippet{node_proc_reachable_prem_2_concl},
then given
\snippet{node_proc_reachable_prem_1}
it follows that
\snippet{node_proc_reachable_concl}.
\end{lemma} 

\noindent
The sole condition \snippet{node_proc_reachable_prem_2_prem}$\ \Rightarrow$
\snippet{node_proc_reachable_prem_2_concl} is needed because certain 
node-level actions---namely \isa{connect}, \isa{disconnect}, and 
\snippet{act_not_arrive}---synchronize with the environment (giving \isa{E\ 
\isasymxi\ \isasymxi'}) but appear to `stutter' (requiring \isa{F\ 
\isasymxi\ \isasymxi'}) relative to the underlying process.
That is, showing \isa{oreachable} by induction involves the three cases 
in \refdef{oreachable}.
The first two, initial states and interleaved actions, follow directly from 
the induction hypothesis at the node level.
For the third, synchronized actions where \isa{A} participates also follow 
directly, but when the node layer acts alone we only know that the state 
component of \isa{A} is unchanged (\isa{\isasymsigma'\ i\ =\ \isasymsigma\ 
i}) and that changes in the environment components (\isa{\isasymsigma'\ j} 
for all \isa{j{\isasymnoteq}i}) satisfy the synchronizing assumption 
(\isa{E}).
The implication between \isa{E} and \isa{F} gives \isa{other\ F\ 
{\{}{i}{\}}} and thus open reachability in \isa{A} is preserved by applying 
the rule for an interleaved transition.

\begin{corollary}[onode lifting]\label{thm:node_proc_lift} 
If, for all \isa{\isasymxi} and \isa{\isasymxi\isacharprime},
\snippet{node_proc_reachable_prem_2_prem}
implies
\snippet{node_proc_reachable_prem_2_concl},
then given
\snippet{node_proc_lift_prem_1}
it follows that
\begin{center}
\snippet{node_proc_lift_concl}.
\end{center}
\end{corollary} 

The lifting rule for partial networks is the most demanding to state and 
prove.
We require the function \snippet{net_tree_ips}, which gives the set of 
addresses in a \snippet{net_tree}.
It is defined in the obvious way:
\begin{center}
\begin{tabular}{lcl}
\snippet{net_tree_ips_node_lhs} & \isa{\isacharequal}
    & \snippet{net_tree_ips_node_rhs}\puncgap, and\\
\snippet{net_tree_ips_par_lhs} & \isa{\isacharequal}
    & \snippet{net_tree_ips_par_rhs}\puncgap.
\end{tabular}
\end{center}
\noindent%
This function is important for bookkeeping in inductions over 
\isa{net-tree}s since it allows the identification of the nodes on either 
side of a partial network composition.\footnote{Recall that the 
\isa{wf-net-tree} condition on the disjointness of such sets is encoded in 
\isa{opnet}.}
In turn, this identification determines which nodes are scoped to each side 
of the composition, and whose properties are assured by the induction 
hypothesis, and which are in the environment, and whose properties are thus 
assumed by the induction hypothesis.

Similarly to the treatment of parallel composition with \isa{qmsg}, it is 
necessary to break the open reachability of a composition of partial 
networks into open reachability of both components.
For this, we require transition invariants 
guaranteeing that the messages sent by nodes in one partial network 
satisfy the assumptions made by nodes in the other partial 
network on messages arriving from their environment.
We therefore introduce the \isa{castmsg} predicate:
\snippet{castmsg_cast} iff~\mbox{\isa{M\ \isasymsigma\ m}}, while
\snippet{castmsg_other} is true for all other \isa{a}.

\begin{lemma}[opnet reachability]\label{thm:subnet_oreachable} 
If $\!$\snippet{subnet_oreachable_prem_1},
with
\snippet{subnet_oreachable_prem_1_L},
\snippet{subnet_oreachable_prem_1_E},
and
\isa{E} and \isa{F} reflexive,
and given
\begin{enumerate}
\item\label{eq:opnetr:cast} 
\isa{{\isasymlangle}i\ {\isacharcolon}\ onp\ i\ {\isacharcolon}\ 
\Ri{\isasymrangle}\isactrlsub o\ \ostepinv\ 
{\isacharparenleft}{\isasymlambda}{\isasymsigma}\ 
{\isacharunderscore}{\isachardot}\ oarrivemsg\ M\ {\isasymsigma},\ other\ F\ 
{\isacharbraceleft}i{\isacharbraceright}\ 
{\isasymrightarrow}{\isacharparenright}}\\
\null\hspace{1.33cm}%
\isa{{\isacharparenleft}{\isasymlambda}{\isacharparenleft}{\isacharparenleft}{\isasymsigma},\ 
{\isacharunderscore}{\isacharparenright},\ a,\ 
({\isasymsigma}{\isacharprime},\ 
{\isacharunderscore}){\isacharparenright}{\isachardot}\ castmsg\ M\ 
{\isasymsigma}\ a{\isacharparenright}}\puncgap,
\item\label{eq:opnetr:E} 
\isa{{\isasymlangle}i\ {\isacharcolon}\ onp\ i\ {\isacharcolon}\ 
\Ri{\isasymrangle}\isactrlsub o\ \ostepinv\ 
{\isacharparenleft}{\isasymlambda}{\isasymsigma}\ 
{\isacharunderscore}{\isachardot}\ oarrivemsg\ M\ {\isasymsigma},\ other\ F\ 
{\isacharbraceleft}i{\isacharbraceright}\ 
{\isasymrightarrow}{\isacharparenright}}\\
\null\hspace{1.33cm}%
\isa{{\isacharparenleft}{\isasymlambda}{\isacharparenleft}{\isacharparenleft}{\isasymsigma},\ 
{\isacharunderscore}{\isacharparenright},\ a,\ 
({\isasymsigma}{\isacharprime},\ 
{\isacharunderscore}){\isacharparenright}{\isachardot}\ a\ {\isasymnoteq}\ 
{\isasymtau}\ {\isasymand}\ 
{\isacharparenleft}{\isasymforall}d{\isachardot}\ a\ {\isasymnoteq}\ 
i{\isacharcolon}deliver{\isacharparenleft}d{\isacharparenright}{\isacharparenright}\ 
{\isasymlongrightarrow}\ E\ {\isacharparenleft}{\isasymsigma}\ 
i{\isacharparenright}\ {\isacharparenleft}{\isasymsigma}{\isacharprime}\ 
i{\isacharparenright}{\isacharparenright}}\puncgap, and
\item\label{eq:opnetr:F} 
\isa{{\isasymlangle}i\ {\isacharcolon}\ onp\ i\ {\isacharcolon}\ 
\Ri{\isasymrangle}\isactrlsub o\ \ostepinv\ 
{\isacharparenleft}{\isasymlambda}{\isasymsigma}\ 
{\isacharunderscore}{\isachardot}\ oarrivemsg\ M\ {\isasymsigma},\ other\ F\ 
{\isacharbraceleft}i{\isacharbraceright}\ 
{\isasymrightarrow}{\isacharparenright}}\\
\null\hspace{1.33cm}%
\isa{{\isacharparenleft}{\isasymlambda}{\isacharparenleft}{\isacharparenleft}{\isasymsigma},\ 
{\isacharunderscore}{\isacharparenright},\ a,\ 
({\isasymsigma}{\isacharprime},\ 
{\isacharunderscore}){\isacharparenright}{\isachardot}\ a\ {\isacharequal}\ 
{\isasymtau}\ {\isasymor}\ {\isacharparenleft}{\isasymexists}d{\isachardot}\ 
a\ {\isacharequal}\ 
i{\isacharcolon}deliver{\isacharparenleft}d{\isacharparenright}{\isacharparenright}\ 
{\isasymlongrightarrow}\ F\ {\isacharparenleft}{\isasymsigma}\ 
i{\isacharparenright}\ {\isacharparenleft}{\isasymsigma}{\isacharprime}\ 
i{\isacharparenright}{\isacharparenright}}\puncgap,
\end{enumerate}
then it follows that both
\begin{enumerate}
\item
\snippet{subnet_oreachable_concl_1}\puncgap and
\item
\snippet{subnet_oreachable_concl_2}\puncgap,
\end{enumerate}
where
\snippet{subnet_oreachable_prem_9}\puncgap,
\snippet{subnet_oreachable_prem_10}\puncgap,\\
\snippet{subnet_oreachable_prem_11}\puncgap, and
\snippet{subnet_oreachable_prem_12}\puncgap.
\end{lemma} 

\noindent
The proof is by induction over \isa{oreachable}.
The initial and interleaved cases are trivial.
For the local case, given open reachability of \isa{(\isasymsigma, s)} and 
\isa{(\isasymsigma, t)} for \isa{\isasymPsi\isactrlsub 1}
and~\isa{\isasymPsi\isactrlsub 2}, respectively, and 
\snippet{subnet_oreachable_proof_step}, we must show open reachability of
\isa{(\isasymsigma\isacharprime, s\isacharprime)} and 
\isa{(\isasymsigma\isacharprime, t\isacharprime)}.
The proof proceeds by a case distinction on actions~\isa{a}.
The key step is to have stated the lemma without introducing cyclic 
dependencies between (synchronizing) assumptions and (transition-invariant) 
guarantees: that is, each partial network~\isa{\isasymPsi\isactrlsub 
i} assumes that the other partial network~\isa{\isasymPsi\isactrlsub j} 
satisfies
\isa{S\isactrlsub {\isadigit{j}}} and \isa{U\isactrlsub {\isadigit{j}}}, 
while itself guaranteeing
\isa{S\isactrlsub {\isadigit{i}}} and \isa{U\isactrlsub {\isadigit{i}}} 
thanks to the lifting of Conditions~\refeqsss{opnetr:E}{opnetr:F}.
For a synchronizing action like \isa{arrive}, \refdef{oreachable} requires 
satisfaction of~\snippet{S1} in order to advance
in~\isa{\isasymPsi\isactrlsub 1} and of \snippet{S2} to advance in 
\isa{\isasymPsi\isactrlsub 2}, but the assumption \isa{S} only guarantees 
that~\isa{E} holds for addresses \snippet{subnet_oreachable_proof_jL}---the 
gap is filled by Assumption~\refeq{opnetr:E}.
This is why the transition invariants required of nodes 
(Conditions~\refeqss{opnetr:cast}{opnetr:F}) may not assume
\isa{otherwith\ E\ \{i\}}.
This is not unduly restrictive, since the transition invariants provide 
guarantees for individual local state elements and not between network 
nodes.
The assumption \snippet{subnet_oreachable_proof_oarrivemsg} is never cyclic: 
it is either assumed of the environment for paired \isa{arrive}s, or 
trivially satisfied for the side that \isa{\isacharasterisk cast}s.

The transition invariants are lifted from nodes to networks by induction 
over \snippet{net_tree}s, using the above
decomposition of open reachability.
For non-synchronizing actions, we exploit the extra guarantees built into 
the open \ac{SOS} rules.

\newpage
\begin{corollary}[opnet lifting]\label{thm:opnet_lifting} 
Given
\begin{center}
\isa{{\isasymlangle}i\ {\isacharcolon}\ onp\ i\ {\isacharcolon}\ 
\Ri{\isasymrangle}\isactrlsub o\ {\isasymTurnstile}\ 
{\isacharparenleft}otherwith\ E\ {\isacharbraceleft}i{\isacharbraceright}\ 
{\isacharparenleft}oarrivemsg\ M{\isacharparenright},\ other\ F\ 
{\isacharbraceleft}i{\isacharbraceright}\ 
{\isasymrightarrow}{\isacharparenright}\ 
{\isacharparenleft}{\isasymlambda}{\isacharparenleft}{\isasymsigma},\ 
{\isacharunderscore}{\isacharparenright}{\isachardot}\ P\ i\ 
{\isasymsigma}{\isacharparenright}}
\end{center}
and provided that,
\begin{enumerate}
\item 
\isa{{\isasymlangle}i\ {\isacharcolon}\ onp\ i\ {\isacharcolon}\ 
\Ri{\isasymrangle}\isactrlsub o\ \ostepinv\
{\isacharparenleft}{\isasymlambda}{\isacharparenleft}{\isasymsigma},\ 
{\isacharunderscore}{\isacharparenright}
{\isachardot}\ oarrivemsg\ M\ {\isasymsigma},\ other\ F\ 
{\isacharbraceleft}i{\isacharbraceright}\ 
{\isasymrightarrow}{\isacharparenright}}\\
\null\hspace{1.435cm}\isa{{\isacharparenleft}{\isasymlambda}{\isacharparenleft}{\isacharparenleft}{\isasymsigma},\ 
{\isacharunderscore}{\isacharparenright},\ a,\ 
{\isacharparenleft}{\isasymsigma}\isacharprime,\ 
{\isacharunderscore}{\isacharparenright}{\isacharparenright}{\isachardot}\ 
castmsg\ M\ {\isasymsigma}\ a{\isacharparenright}}\puncgap,
\item 
\isa{{\isasymlangle}i\ {\isacharcolon}\ onp\ i\ {\isacharcolon}\
\Ri{\isasymrangle}\isactrlsub o\ \ostepinv\ 
{\isacharparenleft}{\isasymlambda}{\isacharparenleft}{\isasymsigma},\ 
{\isacharunderscore}{\isacharparenright}
{\isachardot}\ oarrivemsg\ M\ {\isasymsigma},\ other\ F\ 
{\isacharbraceleft}i{\isacharbraceright}\
{\isasymrightarrow}{\isacharparenright}}\\
\null\hfil\hfil\hfil\isa{{\isacharparenleft}{\isasymlambda}{\isacharparenleft}{\isacharparenleft}{\isasymsigma},\ 
{\isacharunderscore}{\isacharparenright},\ a,\ 
\isacharparenleft{\isasymsigma}{\isacharprime},\ 
{\isacharunderscore}\isacharparenright{\isacharparenright}{\isachardot}\ a\ 
{\isasymnoteq}\ {\isasymtau}\ {\isasymand}\ 
{\isacharparenleft}{\isasymforall}d{\isachardot}\ a\ {\isasymnoteq}\ 
i{\isacharcolon}deliver{\isacharparenleft}d{\isacharparenright}{\isacharparenright}\ 
{\isasymlongrightarrow}\ E\ {\isacharparenleft}{\isasymsigma}\ 
i{\isacharparenright}\ {\isacharparenleft}{\isasymsigma}{\isacharprime}\ 
i{\isacharparenright}{\isacharparenright}}\puncgap, and
\item 
\isa{{\isasymlangle}i\ {\isacharcolon}\ onp\ i\ {\isacharcolon}\ 
\Ri{\isasymrangle}\isactrlsub o\ \ostepinv\ 
{\isacharparenleft}{\isasymlambda}{\isacharparenleft}{\isasymsigma},\ 
{\isacharunderscore}{\isacharparenright}
{\isachardot}\ oarrivemsg\ M\ {\isasymsigma},\ other\ F\ 
{\isacharbraceleft}i{\isacharbraceright}\ 
{\isasymrightarrow}{\isacharparenright}}\\ 
\null\hspace{1.43cm}%
\isa{{\isacharparenleft}{\isasymlambda}{\isacharparenleft}{\isacharparenleft}{\isasymsigma},\ 
{\isacharunderscore}{\isacharparenright},\ a,\ 
\isacharparenleft{\isasymsigma}{\isacharprime},\ 
{\isacharunderscore}\isacharparenright{\isacharparenright}{\isachardot}\ a\ 
{\isacharequal}\ {\isasymtau}\ {\isasymor}\ 
{\isacharparenleft}{\isasymexists}d{\isachardot}\ a\ {\isacharequal}\ 
i{\isacharcolon}deliver{\isacharparenleft}d{\isacharparenright}{\isacharparenright}\ 
{\isasymlongrightarrow}\ F\ {\isacharparenleft}{\isasymsigma}\ 
i{\isacharparenright}\ {\isacharparenleft}{\isasymsigma}{\isacharprime}\ 
i{\isacharparenright}{\isacharparenright}}\puncgap,
\end{enumerate}
for all \isa{i} and \isa{\Ri},
with \isa{E} and \isa{F} reflexive, then\\
\null\isa{\ \ opnet\ onp\ {\isasymPsi}\ {\isasymTurnstile}\ 
{\isacharparenleft}otherwith\ E\ 
{\isacharparenleft}net{\isacharunderscore}tree{\isacharunderscore}ips\ 
{\isasymPsi}{\isacharparenright}\ {\isacharparenleft}oarrivemsg\ 
M{\isacharparenright},\ other\ F\ 
{\isacharparenleft}net{\isacharunderscore}tree{\isacharunderscore}ips\ 
{\isasymPsi}{\isacharparenright}\ {\isasymrightarrow}{\isacharparenright}}\\ 
\null\hfil\hfil\hfil\hfil\hfil\hfil\hfil\hfil\hfil\hfil\hfil\hfil%
\isa{{\isacharparenleft}{\isasymlambda}{\isacharparenleft}{\isasymsigma},\ 
{\isacharunderscore}{\isacharparenright}{\isachardot}\ 
{\isasymforall}i{\isasymin}net{\isacharunderscore}tree{\isacharunderscore}ips\ 
{\isasymPsi}{\isachardot}\ P\ i\ 
{\isasymsigma}{\isacharparenright}}\puncgap.
\end{corollary} 

For transition invariants, we obtain results
similar to \refcoross{par_qmsg_lift}{node_proc_lift}{opnet_lifting}.
They are essential for discharging the three conditions of 
\refcoro{opnet_lifting}.

The rule for closed networks is similar to the others.
Its important function is to eliminate the synchronizing assumption (\isa{S} 
in the lemmas above), since messages no longer arrive from the environment.
The conclusion of this rule has the form required by the transfer 
lemma of the next section.

\begin{lemma}[ocnet reachability]\label{thm:ocnet_oreachable} 
\\From \snippet{oclosed_oreachable_inclosed_1},
it follows that\\
\snippet{oclosed_oreachable_inclosed_concl}\puncgap,\footnotemark{}
where
\snippet{inoclosed_arrive_newpkt},
but
\snippet{inoclosed_arrive_other},
for all other \isa{m},
\snippet{inoclosed_newpkt}, and otherwise, for all other \isa{a},
\snippet{inoclosed_other}.
\end{lemma} 
\footnotetext{The predicate \isa{(op\ =)} simply compares its two arguments for equality.}
That is, reachability in \isa{opnet\ onp\ p} prior to closing need not 
consider transitions with the action \isa{arrive} for any message other than 
a \isa{Newpkt}, nor with the action~\isa{newpkt}.

\begin{corollary}[ocnet lifting]\label{thm:ocnet_lifting} 
\\From \snippet{inclosed_closed_1}, it follows that
\snippet{inclosed_closed_concl}.
\end{corollary} 

\subsection{Transferring open invariants}\label{sec:transfer} 

The rules in the last section extend invariants over sequential processes, 
like \refeq{oseq_bigger_than_next}, for example, to 
arbitrary, open network models.
All that remains is to transfer the extended invariants to the standard 
model.
Our approach is to define a relation between a standard automaton and an 
open automaton, for instance at the level of local parallel processes, and 
then to show that this relation implies the desired transfer property 
between the respective network models at the closed level.

\begin{figure}[t]
\centering
  \begin{tikzpicture}
    \matrix (m)
      [column sep=2cm, row sep=0cm,
       matrix of nodes,
       nodes in empty cells,
       ampersand replacement=\&] at (0,-.3) {
        \isa{\phantom{'}\isasymsigma\ i\ =\ fst\ (sr\ s)} \& \& \\
        \isa{\isasymand} \& \& \\
        \isa{s} \&
        \& \isa{\phantom{'}(\isasymsigma,\phantom{'}\ snd(sr\ s))} \\
        \& \& \\
        \& \rlap{\isa{\isasymLongrightarrow}} \& \\
        \& \& \\
        \isa{s\rlap{'}} \&
        \& \isa{\phantom{'}(\isasymsigma',\ snd(sr\ s'))} \\
        \isa{\isasymand} \& \& \\
        \isa{\isasymsigma'\ i\ =\ fst\ (sr s')} \& \& \\
    };
    \draw[-latex]
        (m-3-1) --
        node[left] {\isa{a}}
        node[right] {\hspace{1em}\isa{\isasymin\ trans\ (np\ i)}}
        (m-7-1);
    \draw[-latex]
        (m-3-3) --
        node[left] {\isa{a}}
        node[right] {\hspace{1em}\isa{\isasymin\ trans\ (onp\ i)}}
        (m-7-3);
  \end{tikzpicture} 
\caption{Schema of the \snippet{pnet_reachable_transfer_prem_1}
relation.}
\label{fig:opnetschema}
\end{figure}

We construct our proofs using Isabelle's \emph{locale} 
feature~\cite{KammullerWenPau:Locales:1999}, which allows one to fix a set 
of constants and their properties, and then to derive lemmas about them.
The constants can later be instantiated with any terms that satisfy the 
assumed properties and the system automatically specializes the associated 
lemmas.
Specifically, we define the locale \snippet{pnet_reachable_transfer_prem_1}, 
which relates the three constants
\begin{enumerate}
\item \isa{np} of type \snippet{openproc_np_type}\puncgap,
\item \isa{onp} of type \snippet{openproc_onp_type}\puncgap, and
\item \isa{sr} of type \snippet{openproc_sr_type}\puncgap,
\end{enumerate}
where \isa{proc-action} stands for the actions on transitions in 
\reffigs{seqp:sos}{parp}, and where~\isa{sr} is a simulation relation that 
effectively divides the states of~\isa{np i} into data and control states.
Unlike for the process selection function \isa{ps} described in 
\refsec{proof-comp}, we cannot simply discard control state elements because 
they are critical to formalizing and reasoning about the relationship 
between the two automata.
The three constants must satisfy two technical conditions that guarantee 
that the initial states of \isa{np\ i} are correctly `embedded' into the 
(global) initial states of \mbox{\isa{onp\ i}}, which we do not detail here, 
and a condition relating transitions across the two models.
The condition on transitions is illustrated in \reffig{opnetschema}: for 
every transition \mbox{\snippet{openproc_trans_prem_4}}, and given
\snippet{openproc_trans_prem_2} and
\snippet{openproc_trans_prem_3}, it must be the case that
\snippet{openproc_trans_concl}.
In other words, \snippet{pnet_reachable_transfer_prem_1} holds if \isa{onp} 
simulates \isa{np} for each component~\isa{i} of \isa{\isasymsigma}.

The simulation requirement ensures that any step of the standard model 
is taken into account by the corresponding open model.
Indeed, for any state reachable in the standard model, a corresponding state 
is reachable in the open model.%

\begin{lemma}[transfer reachability]\label{thm:close_opnet}
Given \isa{np}, \isa{onp}, and \isa{sr} such that 
\snippet{pnet_reachable_transfer_prem_1}, then for any 
\snippet{pnet_reachable_transfer_prem_2}$\!$ and 
\snippet{pnet_reachable_transfer_prem_3}, it follows that\\
\snippet{pnet_reachable_transfer_concl}.
\end{lemma}
This lemma uses two \isa{openproc} constants:
\snippet{someinit_concl_lhs} chooses an arbitrary initial data state from 
\isa{np\ i},\footnote{\snippet{someinit_concl_rhs}, where the `\isa{`}' 
is the image operator.} with which \isa{default} completes missing state 
elements, and \isa{netliftc} lifts the control part of a process state to 
nodes and partial networks:
\begin{center}
\begin{tabular}{lcl}
\snippet{netliftl1_lhs} & \isa{\isacharequal} & \snippet{netliftl1_rhs} \\
\snippet{netliftl2_lhs} & \isa{\isacharequal} & 
\snippet{netliftl2_rhs}\puncgap. \\[.7em]
\end{tabular}
\end{center}
\reflem{close_opnet} is shown by lifting the simulation relation to nodes 
and partial networks by an induction on \isa{\isasymPsi}.
A separate subproof is required for each type of action and the assumptions 
incorporated into the corresponding open \ac{SOS} rules (see 
\refsecs{onodes}{oclosed}) are `discharged' using contextual information 
from the transition in the standard model.
The result is that every transition in the standard model (\isa{closed\ 
(pnet\ np\ \isasymPsi)}) is simulated by a transition in the open model 
(\isa{oclosed\ (opnet\ onp\ \isasymPsi)}).
An implication from an open invariant on an open model to an invariant on 
the corresponding standard model follows directly.

\begin{corollary}[transfer]\label{thm:close_lift}
Given \isa{np}, \isa{onp}, and \isa{sr} such that 
\snippet{close_opnet_prem_1}, and provided \snippet{close_opnet_prem_2}, 
then from \mbox{\snippet{close_opnet_prem_3}}, it follows that 
\snippet{close_opnet_concl}.
\end{corollary}

In terms of our running example, we first show \isa{openproc\ ptoy\ 
optoy\ id}.
We then apply a generic sublocale relation for parallel composition with 
\isa{qmsg} to obtain
\isa{openproc
     \ (\isasymlambda i. ptoy\ i\ {\isasymlangle}{\isasymlangle}\ qmsg)
     \ (\isasymlambda i. optoy\ i\ 
     {\isasymlangle}{\isasymlangle}\isactrlbsub i\isactrlesub \ qmsg)\ 
     (\isasymlambda((\isasymxi, p),\ q).\ (\isasymxi,\ (p,\ q)))},
to which we can apply \refcoro{close_lift} to obtain an appropriate transfer 
lemma. Compared to the \isa{netglobal} constant in 
Invariant~\refeq{bigger_than_next}, the one in \refcoro{close_lift} is 
defined generically within the \isa{openproc} locale and is therefore 
parameterized by \isa{np} and \isa{sr}. The former is obtained from the 
latter by a simple instantiation.

\paragraph{Summary.}

The technicalities of the lemmas in this and the preceding section are 
essential for the underlying proofs to succeed.
The key idea is that through an open version of \ac{AWN} where automaton 
states are segregated into data and control 
components, one can reason locally about global properties, but still, using 
the transfer and lifting results, obtain a result over the original model 
(c.f. \reffig{oschema}).

\section{Concluding remarks} 

We present a mechanization of \ac{AWN}, a modelling language for \ac{MANET} 
and \ac{WMN} protocols, including a streamlined adaptation of standard 
theory for showing invariants of individual reactive processes, and a novel 
and compositional framework for lifting such results to network models.
The framework allows the statement and proof of inter-node properties.
We think that many elements of our approach would apply to similarly 
structured models in other formalisms.

It is reasonable to ask whether the basic model presented in 
\refsec{awn} could not simply be abandoned in favour of the open model of 
\refsec{omodel}.
We believe, however, that the basic model is the most 
natural way of describing what \ac{AWN} means, proving semantic properties 
of the language, showing `node-only' invariants, and, potentially, for 
showing refinement relations.
Having such a reference model allows us to freely incorporate assumptions 
into the open \ac{SOS} rules, knowing that their soundness will
later be justified.

\paragraph{The \ac{AODV} case study.}
The framework we present in this paper was successfully applied 
in the mechanization of a proof of loop freedom~\cite[\textsection 
7]{FehnkerEtAl:AWN:2013} of the \ac{AODV} protocol~\cite{RFC3561},
a widely-used routing protocol designed for \acp{MANET}, and one of the four 
protocols currently standardized by the IETF \ac{MANET} working group.
The model has about~$100$ control locations across~$6$ different processes, 
and
uses about~$40$ functions to 
manipulate the data state.
The main property (loop freedom) roughly states that `a data packet is never 
sent round in circles without being delivered'.
To establish this property, we proved around 
$400$ lemmas.
Due to the complexity of the protocol logic and the length of the 
proof, we present the 
details elsewhere~\cite{ATVA14}.
The case study shows that the presented framework can be applied to 
verification tasks of industrial relevance.

\paragraph{Verifying implementations.}
$\!\!$We argue in the introduction that \ac{AWN} is well-adapted for 
modelling \ac{MANET} and \ac{WMN} protocols due to its support for their 
data structures and specialized communication primitives, and also because 
of its operational style.
In the rest of the paper, we present techniques for the machine-assisted and 
compositional verification of safety properties of networks of cooperating 
nodes; and we claim that the \ac{AODV} case study is testament to the 
effectiveness of this approach.
An important question remains: \emph{are \ac{AWN} models suitable 
specifications for protocol implementations?}
For instance, is it feasible to prove that a program written in C or a 
similar programming language correctly implements a sequential \ac{AWN} 
process?
Would it be better to try to refine or transform an \ac{AWN} process into an 
executable form?
Or simply to analyse network traces against an instantiation of the 
model~\cite{BishopEtAl:TCPinHOL:2006}?
In any case, all of these challenges require precise, and ideally 
mechanized, protocol models, and proofs that they satisfy given properties.

\begin{acknowledgements} 
We thank G.~Klein and M.~Pouzet for their support and complaisance, M.~Daum 
for his participation in
early discussions, and T.\ Sewell for sharing his talent with Isabelle.
The tools\ 
Isabelle/jEdit~\cite{Wenzel:jEdit:2012}, 
Sledge\-hammer~\cite{BlanchetteBohPau:Sledgehammer:2011}, parallel processing~\cite{Wenzel:ParITP:2013}, and the TPTP 
project~\cite{Sutcliffe:TPTP} were invaluable.
\end{acknowledgements}

\bibliographystyle{spmpsci}
\bibliography{paper}
\end{document}

%% file: snippets.tex
\snip{toy_init}{0}{0}{\isa{{\isasymlparr}id\ {\isacharequal}\ i,\ no\ {\isacharequal}\ {\isadigit{0}},\ nhid\ {\isacharequal}\ i,\ msg\ {\isacharequal}\ SOME\ x{\isachardot}\ True,\ num\ {\isacharequal}\ SOME\ x{\isachardot}\ True,\ sid\ {\isacharequal}\ SOME\ x{\isachardot}\ True{\isasymrparr}}}
\snip{ptoy_lhs}{0}{0}{\isa{ptoy\ i}}
\snip{ptoy_rhs}{0}{0}{\isa{{\isasymlparr}init\ {\isacharequal}\ {\isacharbraceleft}{\isacharparenleft}toy{\isacharunderscore}init\ i,\ \gammatoy\ PToy{\isacharparenright}{\isacharbraceright},\ trans\ {\isacharequal}\ \seqpsos\ \gammatoy{\isasymrparr}}}

\snip{optoy_lhs}{0}{0}{\isa{optoy\ i}}
\snip{optoy_rhs}{0}{0}{\isa{{\isasymlparr}init\ {\isacharequal}\ {\isacharbraceleft}{\isacharparenleft}toy{\isacharunderscore}init,\ \gammatoy\ PToy{\isacharparenright}{\isacharbraceright},\ trans\ {\isacharequal}\ \oseqpsos\ \gammatoy\ i{\isasymrparr}}}

\snip{pqmsg_lhs}{0}{0}{\isa{qmsg}}
\snip{pqmsg_rhs}{0}{0}{\isa{{\isasymlparr}init\ {\isacharequal}\ {\isacharbraceleft}{\isacharparenleft}{\isacharbrackleft}\,{\isacharbrackright},\ \gammaqmsg\ Qmsg{\isacharparenright}{\isacharbraceright},\ trans\ {\isacharequal}\ \seqpsos\ \gammaqmsg{\isasymrparr}}}

\snip{pkt}{0}{0}{\isa{pkt\ {\isacharparenleft}d,\ src{\isacharparenright}}}
\snip{newpkt}{0}{0}{\isa{Newpkt\ d\ dst}}
\snip{is_newpkt_def}{0}{0}{\isa{is{\isacharunderscore}newpkt\ {\isasymxi}\ {\isacharequal}\ \textsf{case}\ msg\ {\isasymxi}\ \textsf{of}\ Pkt\ nat{\isadigit{1}}\ nat{\isadigit{2}}\ {\isasymRightarrow}\ {\isasymemptyset}\ {\isacharbar}\ Newpkt\ data\ dip\ {\isasymRightarrow}\ {\isacharbraceleft}{\isasymxi}{\isasymlparr}num\ {\isacharcolon}{\isacharequal}\ data{\isasymrparr}{\isacharbraceright}}}

\snip{automaton_type}{0}{0}{\isa{{\isacharparenleft}{\isacharprime}s,\ {\isacharprime}a{\isacharparenright}\ automaton}}

\snip{seqp_choice}{0}{0}{\isa{p\ {\isasymoplus}\ q}}
\snip{seqp_call}{0}{0}{\isa{call{\isacharparenleft}pn{\isacharparenright}}}

\snip{lseqp_guard}{0}{0}{\isa{\gray{{\isacharbraceleft}l{\isacharbraceright}}{\isasymlangle}g{\isasymrangle}\ p}}
\snip{lseqp_assign}{0}{0}{\isa{\gray{{\isacharbraceleft}l{\isacharbraceright}}{\isasymlbrakk}u{\isasymrbrakk}\ p}}
\snip{lseqp_ucast}{0}{0}{\isa{\gray{{\isacharbraceleft}l{\isacharbraceright}}unicast{\isacharparenleft}\selip,\ \selmsg{\isacharparenright}\ {\isachardot}\ p\ {\isasymtriangleright}\ q}}
\snip{lseqp_bcast}{0}{0}{\isa{\gray{{\isacharbraceleft}l{\isacharbraceright}}broadcast{\isacharparenleft}\selmsg{\isacharparenright}\ {\isachardot}\ p}}
\snip{lseqp_gcast}{0}{0}{\isa{\gray{{\isacharbraceleft}l{\isacharbraceright}}groupcast{\isacharparenleft}\selips,\ \selmsg{\isacharparenright}\ {\isachardot}\ p}}
\snip{lseqp_send}{0}{0}{\isa{\gray{{\isacharbraceleft}l{\isacharbraceright}}send{\isacharparenleft}\selmsg{\isacharparenright}\ {\isachardot}\ p}}
\snip{lseqp_deliver}{0}{0}{\isa{\gray{{\isacharbraceleft}l{\isacharbraceright}}deliver{\isacharparenleft}\seldata{\isacharparenright}\ {\isachardot}\ p}}
\snip{lseqp_receive}{0}{0}{\isa{\gray{{\isacharbraceleft}l{\isacharbraceright}}receive{\isacharparenleft}\updmsg{\isacharparenright}\ {\isachardot}\ p}}

\snip{seqp_guard}{0}{0}{\isa{{\isasymlangle}g{\isasymrangle}\ p}}
\snip{seqp_assign}{0}{0}{\isa{{\isasymlbrakk}u{\isasymrbrakk}\ p}}
\snip{seqp_ucast}{0}{0}{\isa{unicast{\isacharparenleft}\selip,\ \selmsg{\isacharparenright}\ {\isachardot}\ p\ {\isasymtriangleright}\ q}}
\snip{seqp_bcast}{0}{0}{\isa{broadcast{\isacharparenleft}\selmsg{\isacharparenright}\ {\isachardot}\ p}}
\snip{seqp_gcast}{0}{0}{\isa{groupcast{\isacharparenleft}\selips,\ \selmsg{\isacharparenright}\ {\isachardot}\ p}}
\snip{seqp_send}{0}{0}{\isa{send{\isacharparenleft}\selmsg{\isacharparenright}\ {\isachardot}\ p}}
\snip{seqp_deliver}{0}{0}{\isa{deliver{\isacharparenleft}\seldata{\isacharparenright}\ {\isachardot}\ p}}
\snip{seqp_receive}{0}{0}{\isa{receive{\isacharparenleft}\updmsg{\isacharparenright}\ {\isachardot}\ p}}

\snip{seqp_guard_map}{0}{0}{\isa{{\isacharprime}k\ {\isasymRightarrow}\ {\isacharprime}k\ set}}
\snip{seqp_assign_map}{0}{0}{\isa{{\isacharprime}k\ {\isasymRightarrow}\ {\isacharprime}k}}
\snip{seqp_cast_ip_map}{0}{0}{\isa{{\isacharprime}k\ {\isasymRightarrow}\ ip}}
\snip{seqp_cast_msg_map}{0}{0}{\isa{{\isacharprime}k\ {\isasymRightarrow}\ {\isacharprime}m}}
\snip{seqp_cast_ip_set_map}{0}{0}{\isa{{\isacharprime}k\ {\isasymRightarrow}\ ip\ set}}

\snip{gamma_fun}{0}{0}{\isa{{\isasymGamma}{\isasymColon}{\isacharprime}p\ {\isasymRightarrow}\ {\isacharparenleft}{\isacharprime}k,\ {\isacharprime}m,\ {\isacharprime}p,\ {\isacharprime}l{\isacharparenright}\ seqp}}
\snip{wellformed_gamma}{0}{0}{\isa{wellformed\ {\isasymGamma}}}

\snip{broadcastT}{0}{0}{\isa{{\isacharparenleft}{\isacharparenleft}{\isasymxi},\ \gray{{\isacharbraceleft}l{\isacharbraceright}}broadcast{\isacharparenleft}\selmsg{\isacharparenright}\ {\isachardot}\ p{\isacharparenright},\ broadcast\ {\isacharparenleft}\selmsg\ {\isasymxi}{\isacharparenright},\ {\isacharparenleft}{\isasymxi},\ p{\isacharparenright}{\isacharparenright}{\isasymin}\seqpsos\ {\isasymGamma}}}
\snip{groupcastT}{0}{0}{\isa{{\isacharparenleft}{\isacharparenleft}{\isasymxi},\ \gray{{\isacharbraceleft}l{\isacharbraceright}}groupcast{\isacharparenleft}\selips,\ \selmsg{\isacharparenright}\ {\isachardot}\ p{\isacharparenright},\ groupcast\ {\isacharparenleft}\selips\ {\isasymxi}{\isacharparenright}\ {\isacharparenleft}\selmsg\ {\isasymxi}{\isacharparenright},\ {\isacharparenleft}{\isasymxi},\ p{\isacharparenright}{\isacharparenright}{\isasymin}\seqpsos\ {\isasymGamma}}}
\snip{unicastT}{0}{0}{\isa{{\isacharparenleft}{\isacharparenleft}{\isasymxi},\ \gray{{\isacharbraceleft}l{\isacharbraceright}}unicast{\isacharparenleft}\selip,\ \selmsg{\isacharparenright}\ {\isachardot}\ p\ {\isasymtriangleright}\ q{\isacharparenright},\ unicast\ {\isacharparenleft}\selip\ {\isasymxi}{\isacharparenright}\ {\isacharparenleft}\selmsg\ {\isasymxi}{\isacharparenright},\ {\isacharparenleft}{\isasymxi},\ p{\isacharparenright}{\isacharparenright}{\isasymin}\seqpsos\ {\isasymGamma}}}
\snip{notunicastT}{0}{0}{\isa{{\isacharparenleft}{\isacharparenleft}{\isasymxi},\ \gray{{\isacharbraceleft}l{\isacharbraceright}}unicast{\isacharparenleft}\selip,\ \selmsg{\isacharparenright}\ {\isachardot}\ p\ {\isasymtriangleright}\ q{\isacharparenright},\ {\isasymnot}unicast\ {\isacharparenleft}\selip\ {\isasymxi}{\isacharparenright},\ {\isacharparenleft}{\isasymxi},\ q{\isacharparenright}{\isacharparenright}{\isasymin}\seqpsos\ {\isasymGamma}}}
\snip{sendT}{0}{0}{\isa{{\isacharparenleft}{\isacharparenleft}{\isasymxi},\ \gray{{\isacharbraceleft}l{\isacharbraceright}}send{\isacharparenleft}\selmsg{\isacharparenright}\ {\isachardot}\ p{\isacharparenright},\ send\ {\isacharparenleft}\selmsg\ {\isasymxi}{\isacharparenright},\ {\isacharparenleft}{\isasymxi},\ p{\isacharparenright}{\isacharparenright}{\isasymin}\seqpsos\ {\isasymGamma}}}
\snip{deliverT}{0}{0}{\isa{{\isacharparenleft}{\isacharparenleft}{\isasymxi},\ \gray{{\isacharbraceleft}l{\isacharbraceright}}deliver{\isacharparenleft}\seldata{\isacharparenright}\ {\isachardot}\ p{\isacharparenright},\ deliver\ {\isacharparenleft}\seldata\ {\isasymxi}{\isacharparenright},\ {\isacharparenleft}{\isasymxi},\ p{\isacharparenright}{\isacharparenright}{\isasymin}\seqpsos\ {\isasymGamma}}}
\snip{receiveT}{0}{0}{\isa{{\isacharparenleft}{\isacharparenleft}{\isasymxi},\ \gray{{\isacharbraceleft}l{\isacharbraceright}}receive{\isacharparenleft}\updmsg{\isacharparenright}\ {\isachardot}\ p{\isacharparenright},\ receive\ msg,\ {\isacharparenleft}\updmsg\ msg\ {\isasymxi},\ p{\isacharparenright}{\isacharparenright}{\isasymin}\seqpsos\ {\isasymGamma}}}
\snip{assignT}{0}{0}{\isa{{\isacharparenleft}{\isacharparenleft}{\isasymxi},\ \gray{{\isacharbraceleft}l{\isacharbraceright}}{\isasymlbrakk}u{\isasymrbrakk}\ p{\isacharparenright},\ {\isasymtau},\ {\isacharparenleft}u\ {\isasymxi},\ p{\isacharparenright}{\isacharparenright}{\isasymin}\seqpsos\ {\isasymGamma}}}
\snip{assignT'}{0}{0}{\isa{\mbox{}\inferrule{\mbox{{\isasymxi}{\isacharprime}\ {\isacharequal}\ u\ {\isasymxi}}}{\mbox{{\isacharparenleft}{\isacharparenleft}{\isasymxi},\ \gray{{\isacharbraceleft}l{\isacharbraceright}}{\isasymlbrakk}u{\isasymrbrakk}\ p{\isacharparenright},\ {\isasymtau},\ {\isacharparenleft}{\isasymxi}{\isacharprime},\ p{\isacharparenright}{\isacharparenright}{\isasymin}\seqpsos\ {\isasymGamma}}}}}
\snip{callT}{0}{0}{\isa{\mbox{}\inferrule{\mbox{{\isacharparenleft}{\isacharparenleft}{\isasymxi},\ {\isasymGamma}\ pn{\isacharparenright},\ a,\ {\isacharparenleft}{\isasymxi}{\isacharprime},\ p{\isacharprime}{\isacharparenright}{\isacharparenright}{\isasymin}\seqpsos\ {\isasymGamma}}}{\mbox{{\isacharparenleft}{\isacharparenleft}{\isasymxi},\ call{\isacharparenleft}pn{\isacharparenright}{\isacharparenright},\ a,\ {\isacharparenleft}{\isasymxi}{\isacharprime},\ p{\isacharprime}{\isacharparenright}{\isacharparenright}{\isasymin}\seqpsos\ {\isasymGamma}}}}}
\snip{choiceT1}{0}{0}{\isa{\mbox{}\inferrule{\mbox{{\isacharparenleft}{\isacharparenleft}{\isasymxi},\ p{\isacharparenright},\ a,\ {\isacharparenleft}{\isasymxi}{\isacharprime},\ p{\isacharprime}{\isacharparenright}{\isacharparenright}{\isasymin}\seqpsos\ {\isasymGamma}}}{\mbox{{\isacharparenleft}{\isacharparenleft}{\isasymxi},\ p\ {\isasymoplus}\ q{\isacharparenright},\ a,\ {\isacharparenleft}{\isasymxi}{\isacharprime},\ p{\isacharprime}{\isacharparenright}{\isacharparenright}{\isasymin}\seqpsos\ {\isasymGamma}}}}}
\snip{choiceT2}{0}{0}{\isa{\mbox{}\inferrule{\mbox{{\isacharparenleft}{\isacharparenleft}{\isasymxi},\ q{\isacharparenright},\ a,\ {\isacharparenleft}{\isasymxi}{\isacharprime},\ q{\isacharprime}{\isacharparenright}{\isacharparenright}{\isasymin}\seqpsos\ {\isasymGamma}}}{\mbox{{\isacharparenleft}{\isacharparenleft}{\isasymxi},\ p\ {\isasymoplus}\ q{\isacharparenright},\ a,\ {\isacharparenleft}{\isasymxi}{\isacharprime},\ q{\isacharprime}{\isacharparenright}{\isacharparenright}{\isasymin}\seqpsos\ {\isasymGamma}}}}}
\snip{guardT}{0}{0}{\isa{\mbox{}\inferrule{\mbox{{\isasymxi}{\isacharprime}{\isasymin}g\ {\isasymxi}}}{\mbox{{\isacharparenleft}{\isacharparenleft}{\isasymxi},\ \gray{{\isacharbraceleft}l{\isacharbraceright}}{\isasymlangle}g{\isasymrangle}\ p{\isacharparenright},\ {\isasymtau},\ {\isacharparenleft}{\isasymxi}{\isacharprime},\ p{\isacharparenright}{\isacharparenright}{\isasymin}\seqpsos\ {\isasymGamma}}}}}

\snip{parleft}{0}{0}{\isa{\mbox{}\inferrule{\mbox{{\isacharparenleft}s,\ a,\ s{\isacharprime}{\isacharparenright}{\isasymin}T\isactrlsub A}\\\ \mbox{{\isasymAnd}m{\isachardot}\ a\ {\isasymnoteq}\ receive\ m}}{\mbox{{\isacharparenleft}{\isacharparenleft}s,\ t{\isacharparenright},\ a,\ {\isacharparenleft}s{\isacharprime},\ t{\isacharparenright}{\isacharparenright}{\isasymin}\parpsos\ T\isactrlsub A\ T\isactrlsub B}}}}
\snip{parright}{0}{0}{\isa{\mbox{}\inferrule{\mbox{{\isacharparenleft}t,\ a,\ t{\isacharprime}{\isacharparenright}{\isasymin}T\isactrlsub B}\\\ \mbox{{\isasymAnd}m{\isachardot}\ a\ {\isasymnoteq}\ send\ m}}{\mbox{{\isacharparenleft}{\isacharparenleft}s,\ t{\isacharparenright},\ a,\ {\isacharparenleft}s,\ t{\isacharprime}{\isacharparenright}{\isacharparenright}{\isasymin}\parpsos\ T\isactrlsub A\ T\isactrlsub B}}}}
\snip{parboth}{0}{0}{\isa{\mbox{}\inferrule{\mbox{{\isacharparenleft}s,\ receive\ m,\ s{\isacharprime}{\isacharparenright}{\isasymin}T\isactrlsub A}\\\ \mbox{{\isacharparenleft}t,\ send\ m,\ t{\isacharprime}{\isacharparenright}{\isasymin}T\isactrlsub B}}{\mbox{{\isacharparenleft}{\isacharparenleft}s,\ t{\isacharparenright},\ {\isasymtau},\ {\isacharparenleft}s{\isacharprime},\ t{\isacharprime}{\isacharparenright}{\isacharparenright}{\isasymin}\parpsos\ T\isactrlsub A\ T\isactrlsub B}}}}

\snip{par_comp}{0}{0}{\isa{A\ {\isasymlangle}{\isasymlangle}\ B\ {\isacharequal}\ {\isasymlparr}init\ {\isacharequal}\ init\ A\ {\isasymtimes}\ init\ B,\ trans\ {\isacharequal}\ \parpsos\ {\isacharparenleft}trans\ A{\isacharparenright}\ {\isacharparenleft}trans\ B{\isacharparenright}{\isasymrparr}}}
\snip{par_comp_term}{0}{0}{\isa{A\ {\isasymlangle}{\isasymlangle}\ B}}
\snip{par_comp_type}{0}{0}{\isa{{\isacharparenleft}{\isacharprime}s,\ {\isacharprime}m\ seq{\isacharunderscore}action{\isacharparenright}\ automaton\ {\isasymRightarrow}\ {\isacharparenleft}{\isacharprime}p,\ {\isacharprime}m\ seq{\isacharunderscore}action{\isacharparenright}\ automaton\ {\isasymRightarrow}\ {\isacharparenleft}{\isacharprime}s\ {\isasymtimes}\ {\isacharprime}p,\ {\isacharprime}m\ seq{\isacharunderscore}action{\isacharparenright}\ automaton}}

\snip{node_bcast}{0}{0}{\isa{\mbox{}\inferrule{\mbox{{\isacharparenleft}s,\ broadcast\ m,\ s{\isacharprime}{\isacharparenright}{\isasymin}T\isactrlsub A}}{\mbox{{\isacharparenleft}\NodeS\ i\ s\ R,\ R{\isacharcolon}{\isacharasterisk}cast{\isacharparenleft}m{\isacharparenright},\ \NodeS\ i\ s{\isacharprime}\ R{\isacharparenright}{\isasymin}\nodesos\ T\isactrlsub A}}}}
\snip{node_gcast}{0}{0}{\isa{\mbox{}\inferrule{\mbox{{\isacharparenleft}s,\ groupcast\ D\ m,\ s{\isacharprime}{\isacharparenright}{\isasymin}T\isactrlsub A}}{\mbox{{\isacharparenleft}\NodeS\ i\ s\ R,\ {\isacharparenleft}R\ {\isasyminter}\ D{\isacharparenright}{\isacharcolon}{\isacharasterisk}cast{\isacharparenleft}m{\isacharparenright},\ \NodeS\ i\ s{\isacharprime}\ R{\isacharparenright}{\isasymin}\nodesos\ T\isactrlsub A}}}}
\snip{node_ucast}{0}{0}{\isa{\mbox{}\inferrule{\mbox{{\isacharparenleft}s,\ unicast\ dst\ m,\ s{\isacharprime}{\isacharparenright}{\isasymin}T\isactrlsub A}\\\ \mbox{dst{\isasymin}R}}{\mbox{{\isacharparenleft}\NodeS\ i\ s\ R,\ {\isacharbraceleft}dst{\isacharbraceright}{\isacharcolon}{\isacharasterisk}cast{\isacharparenleft}m{\isacharparenright},\ \NodeS\ i\ s{\isacharprime}\ R{\isacharparenright}{\isasymin}\nodesos\ T\isactrlsub A}}}}
\snip{node_notucast}{0}{0}{\isa{\mbox{}\inferrule{\mbox{{\isacharparenleft}s,\ {\isasymnot}unicast\ dst,\ s{\isacharprime}{\isacharparenright}{\isasymin}T\isactrlsub A}\\\ \mbox{dst\ {\isasymnotin}\ R}}{\mbox{{\isacharparenleft}\NodeS\ i\ s\ R,\ {\isasymtau},\ \NodeS\ i\ s{\isacharprime}\ R{\isacharparenright}{\isasymin}\nodesos\ T\isactrlsub A}}}}
\snip{node_deliver}{0}{0}{\isa{\mbox{}\inferrule{\mbox{{\isacharparenleft}s,\ deliver\ d,\ s{\isacharprime}{\isacharparenright}{\isasymin}T\isactrlsub A}}{\mbox{{\isacharparenleft}\NodeS\ i\ s\ R,\ i{\isacharcolon}deliver{\isacharparenleft}d{\isacharparenright},\ \NodeS\ i\ s{\isacharprime}\ R{\isacharparenright}{\isasymin}\nodesos\ T\isactrlsub A}}}}
\snip{node_receive}{0}{0}{\isa{\mbox{}\inferrule{\mbox{{\isacharparenleft}s,\ receive\ m,\ s{\isacharprime}{\isacharparenright}{\isasymin}T\isactrlsub A}}{\mbox{{\isacharparenleft}\NodeS\ i\ s\ R,\ {\isacharbraceleft}i{\isacharbraceright}{\isasymnot}{\isasymemptyset}{\isacharcolon}arrive{\isacharparenleft}m{\isacharparenright},\ \NodeS\ i\ s{\isacharprime}\ R{\isacharparenright}{\isasymin}\nodesos\ T\isactrlsub A}}}}
\snip{node_tau}{0}{0}{\isa{\mbox{}\inferrule{\mbox{{\isacharparenleft}s,\ {\isasymtau},\ s{\isacharprime}{\isacharparenright}{\isasymin}T\isactrlsub A}}{\mbox{{\isacharparenleft}\NodeS\ i\ s\ R,\ {\isasymtau},\ \NodeS\ i\ s{\isacharprime}\ R{\isacharparenright}{\isasymin}\nodesos\ T\isactrlsub A}}}}
\snip{node_arrive}{0}{0}{\isa{{\isacharparenleft}\NodeS\ i\ s\ R,\ {\isasymemptyset}{\isasymnot}{\isacharbraceleft}i{\isacharbraceright}{\isacharcolon}arrive{\isacharparenleft}m{\isacharparenright},\ \NodeS\ i\ s\ R{\isacharparenright}{\isasymin}\nodesos\ T\isactrlsub A}}
\snip{node_connect1}{0}{0}{\isa{{\isacharparenleft}\NodeS\ i\ s\ R,\ connect{\isacharparenleft}i,\ i{\isacharprime}{\isacharparenright},\ \NodeS\ i\ s\ {R\ {\isasymunion}\ {\isacharbraceleft}i{\isacharprime}{\isacharbraceright}}{\isacharparenright}{\isasymin}\nodesos\ T\isactrlsub A}}
\snip{node_connect2}{0}{0}{\isa{{\isacharparenleft}\NodeS\ i\ s\ R,\ connect{\isacharparenleft}i{\isacharprime},\ i{\isacharparenright},\ \NodeS\ i\ s\ {R\ {\isasymunion}\ {\isacharbraceleft}i{\isacharprime}{\isacharbraceright}}{\isacharparenright}{\isasymin}\nodesos\ T\isactrlsub A}}
\snip{node_disconnect1}{0}{0}{\isa{{\isacharparenleft}\NodeS\ i\ s\ R,\ disconnect{\isacharparenleft}i,\ i{\isacharprime}{\isacharparenright},\ \NodeS\ i\ s\ {R\ {\isacharminus}\ {\isacharbraceleft}i{\isacharprime}{\isacharbraceright}}{\isacharparenright}{\isasymin}\nodesos\ T\isactrlsub A}}
\snip{node_disconnect2}{0}{0}{\isa{{\isacharparenleft}\NodeS\ i\ s\ R,\ disconnect{\isacharparenleft}i{\isacharprime},\ i{\isacharparenright},\ \NodeS\ i\ s\ {R\ {\isacharminus}\ {\isacharbraceleft}i{\isacharprime}{\isacharbraceright}}{\isacharparenright}{\isasymin}\nodesos\ T\isactrlsub A}}
\snip{node_connect_other}{0}{0}{\isa{\mbox{}\inferrule{\mbox{i\ {\isasymnoteq}\ i{\isacharprime}}\\\ \mbox{i\ {\isasymnoteq}\ i{\isacharprime}{\isacharprime}}}{\mbox{{\isacharparenleft}\NodeS\ i\ s\ R,\ connect{\isacharparenleft}i{\isacharprime},\ i{\isacharprime}{\isacharprime}{\isacharparenright},\ \NodeS\ i\ s\ R{\isacharparenright}{\isasymin}\nodesos\ T\isactrlsub A}}}}
\snip{node_disconnect_other}{0}{0}{\isa{\mbox{}\inferrule{\mbox{i\ {\isasymnoteq}\ i{\isacharprime}}\\\ \mbox{i\ {\isasymnoteq}\ i{\isacharprime}{\isacharprime}}}{\mbox{{\isacharparenleft}\NodeS\ i\ s\ R,\ disconnect{\isacharparenleft}i{\isacharprime},\ i{\isacharprime}{\isacharprime}{\isacharparenright},\ \NodeS\ i\ s\ R{\isacharparenright}{\isasymin}\nodesos\ T\isactrlsub A}}}}

\snip{node_comp}{0}{0}{\isa{{\isasymlangle}i\ {\isacharcolon}\ A\ {\isacharcolon}\ \Ri{\isasymrangle}\ {\isacharequal}\ {\isasymlparr}init\ {\isacharequal}\ {\isacharbraceleft}\NodeS\ i\ s\ \Ri\ {\isacharbar}\ s{\isasymin}init\ A{\isacharbraceright},\ trans\ {\isacharequal}\ \nodesos\ {\isacharparenleft}trans\ A{\isacharparenright}{\isasymrparr}}}
\snip{node_comp_term}{0}{0}{\isa{{\isasymlangle}i\ {\isacharcolon}\ S\ {\isacharcolon}\ \Ri{\isasymrangle}}}
\snip{node_comp_type}{0}{0}{\isa{{\isacharparenleft}{\isacharprime}a\ net{\isacharunderscore}state,\ {\isacharprime}b\ node{\isacharunderscore}action{\isacharparenright}\ automaton}}

\snip{pnet_cast1}{0}{0}{\isa{\mbox{}\inferrule{\mbox{{\isacharparenleft}s,\ R{\isacharcolon}{\isacharasterisk}cast{\isacharparenleft}m{\isacharparenright},\ s{\isacharprime}{\isacharparenright}{\isasymin}T\isactrlsub A}\\\ \mbox{{\isacharparenleft}t,\ H{\isasymnot}K{\isacharcolon}arrive{\isacharparenleft}m{\isacharparenright},\ t{\isacharprime}{\isacharparenright}{\isasymin}T\isactrlsub B}\\\ \mbox{H\ {\isasymsubseteq}\ R}\\\ \mbox{K\ {\isasyminter}\ R\ {\isacharequal}\ {\isasymemptyset}}}{\mbox{{\isacharparenleft}\subnets{s}{t},\ R{\isacharcolon}{\isacharasterisk}cast{\isacharparenleft}m{\isacharparenright},\ \subnetsadj{s{\isacharprime}}{t{\isacharprime}}{\isacharparenright}{\isasymin}\pnetsos\ T\isactrlsub A\ T\isactrlsub B}}}}
\snip{pnet_cast2}{0}{0}{\isa{\mbox{}\inferrule{\mbox{{\isacharparenleft}s,\ H{\isasymnot}K{\isacharcolon}arrive{\isacharparenleft}m{\isacharparenright},\ s{\isacharprime}{\isacharparenright}{\isasymin}T\isactrlsub A}\\\ \mbox{{\isacharparenleft}t,\ R{\isacharcolon}{\isacharasterisk}cast{\isacharparenleft}m{\isacharparenright},\ t{\isacharprime}{\isacharparenright}{\isasymin}T\isactrlsub B}\\\ \mbox{H\ {\isasymsubseteq}\ R}\\\ \mbox{K\ {\isasyminter}\ R\ {\isacharequal}\ {\isasymemptyset}}}{\mbox{{\isacharparenleft}\subnets{s}{t},\ R{\isacharcolon}{\isacharasterisk}cast{\isacharparenleft}m{\isacharparenright},\ \subnetsadj{s{\isacharprime}}{t{\isacharprime}}{\isacharparenright}{\isasymin}\pnetsos\ T\isactrlsub A\ T\isactrlsub B}}}}
\snip{pnet_arrive}{0}{0}{\isa{\mbox{}\inferrule{\mbox{{\isacharparenleft}s,\ H{\isasymnot}K{\isacharcolon}arrive{\isacharparenleft}m{\isacharparenright},\ s{\isacharprime}{\isacharparenright}{\isasymin}T\isactrlsub A}\\\ \mbox{{\isacharparenleft}t,\ H{\isacharprime}{\isasymnot}K{\isacharprime}{\isacharcolon}arrive{\isacharparenleft}m{\isacharparenright},\ t{\isacharprime}{\isacharparenright}{\isasymin}T\isactrlsub B}}{\mbox{{\isacharparenleft}\subnets{s}{t},\ {\isacharparenleft}H\ {\isasymunion}\ H{\isacharprime}{\isacharparenright}{\isasymnot}{\isacharparenleft}K\ {\isasymunion}\ K{\isacharprime}{\isacharparenright}{\isacharcolon}arrive{\isacharparenleft}m{\isacharparenright},\ \subnetsadj{s{\isacharprime}}{t{\isacharprime}}{\isacharparenright}{\isasymin}\pnetsos\ T\isactrlsub A\ T\isactrlsub B}}}}
\snip{pnet_deliver1}{0}{0}{\isa{\mbox{}\inferrule{\mbox{{\isacharparenleft}s,\ i{\isacharcolon}deliver{\isacharparenleft}d{\isacharparenright},\ s{\isacharprime}{\isacharparenright}{\isasymin}T\isactrlsub A}}{\mbox{{\isacharparenleft}\subnets{s}{t},\ i{\isacharcolon}deliver{\isacharparenleft}d{\isacharparenright},\ \subnetsadj{s{\isacharprime}}{t}{\isacharparenright}{\isasymin}\pnetsos\ T\isactrlsub A\ T\isactrlsub B}}}}
\snip{pnet_deliver2}{0}{0}{\isa{\mbox{}\inferrule{\mbox{{\isacharparenleft}t,\ i{\isacharcolon}deliver{\isacharparenleft}d{\isacharparenright},\ t{\isacharprime}{\isacharparenright}{\isasymin}T\isactrlsub B}}{\mbox{{\isacharparenleft}\subnets{s}{t},\ i{\isacharcolon}deliver{\isacharparenleft}d{\isacharparenright},\ \subnets{s}{t{\isacharprime}}{\isacharparenright}{\isasymin}\pnetsos\ T\isactrlsub A\ T\isactrlsub B}}}}
\snip{pnet_tau1}{0}{0}{\isa{\mbox{}\inferrule{\mbox{{\isacharparenleft}s,\ {\isasymtau},\ s{\isacharprime}{\isacharparenright}{\isasymin}T\isactrlsub A}}{\mbox{{\isacharparenleft}\subnets{s}{t},\ {\isasymtau},\ \subnetsadj{s{\isacharprime}}{t}{\isacharparenright}{\isasymin}\pnetsos\ T\isactrlsub A\ T\isactrlsub B}}}}
\snip{pnet_tau2}{0}{0}{\isa{\mbox{}\inferrule{\mbox{{\isacharparenleft}t,\ {\isasymtau},\ t{\isacharprime}{\isacharparenright}{\isasymin}T\isactrlsub B}}{\mbox{{\isacharparenleft}\subnets{s}{t},\ {\isasymtau},\ \subnets{s}{t{\isacharprime}}{\isacharparenright}{\isasymin}\pnetsos\ T\isactrlsub A\ T\isactrlsub B}}}}
\snip{pnet_connect}{0}{0}{\isa{\mbox{}\inferrule{\mbox{{\isacharparenleft}s,\ connect{\isacharparenleft}i,\ i{\isacharprime}{\isacharparenright},\ s{\isacharprime}{\isacharparenright}{\isasymin}T\isactrlsub A}\\\ \mbox{{\isacharparenleft}t,\ connect{\isacharparenleft}i,\ i{\isacharprime}{\isacharparenright},\ t{\isacharprime}{\isacharparenright}{\isasymin}T\isactrlsub B}}{\mbox{{\isacharparenleft}\subnets{s}{t},\ connect{\isacharparenleft}i,\ i{\isacharprime}{\isacharparenright},\ \subnetsadj{s{\isacharprime}}{t{\isacharprime}}{\isacharparenright}{\isasymin}\pnetsos\ T\isactrlsub A\ T\isactrlsub B}}}}
\snip{pnet_disconnect}{0}{0}{\isa{\mbox{}\inferrule{\mbox{{\isacharparenleft}s,\ disconnect{\isacharparenleft}i,\ i{\isacharprime}{\isacharparenright},\ s{\isacharprime}{\isacharparenright}{\isasymin}T\isactrlsub A}\\\ \mbox{{\isacharparenleft}t,\ disconnect{\isacharparenleft}i,\ i{\isacharprime}{\isacharparenright},\ t{\isacharprime}{\isacharparenright}{\isasymin}T\isactrlsub B}}{\mbox{{\isacharparenleft}\subnets{s}{t},\ disconnect{\isacharparenleft}i,\ i{\isacharprime}{\isacharparenright},\ \subnetsadj{s{\isacharprime}}{t{\isacharprime}}{\isacharparenright}{\isasymin}\pnetsos\ T\isactrlsub A\ T\isactrlsub B}}}}

\snip{pnet}{0}{0}{\isa{pnet}}
\snip{pnet_node_term}{0}{0}{\isa{{\isasymlangle}i{\isacharsemicolon}\ \Ri{\isasymrangle}}}
\snip{pnet_node_type}{0}{0}{\isa{net{\isacharunderscore}tree}}
\snip{pnet_par_term}{0}{0}{\isa{{\isasymPsi}\isactrlsub {\isadigit{1}}\!\parallelcomp{}{\isasymPsi}\isactrlsub {\isadigit{2}}}}
\snip{pnet_par_type}{0}{0}{\isa{net{\isacharunderscore}tree}}
\snip{net_tree}{0}{0}{\isa{net{\isacharunderscore}tree}}
\snip{net_tree_ips}{0}{0}{\isa{net{\isacharunderscore}tree{\isacharunderscore}ips}}

\snip{act_arrive}{0}{0}{\isa{H{\isasymnot}K{\isacharcolon}arrive{\isacharparenleft}m{\isacharparenright}}}
\snip{act_not_arrive}{0}{0}{\isa{{\isasymemptyset}{\isasymnot}{\isacharbraceleft}i{\isacharbraceright}{\isacharcolon}arrive{\isacharparenleft}m{\isacharparenright}}}
\snip{act_connect}{0}{0}{\isa{connect{\isacharparenleft}i,\ i{\isacharprime}{\isacharparenright}}}
\snip{act_disconnect}{0}{0}{\isa{disconnect{\isacharparenleft}i,\ i{\isacharprime}{\isacharparenright}}}
\snip{act_receive}{0}{0}{\isa{receive\ m}}
\snip{act_cast}{0}{0}{\isa{R{\isacharcolon}{\isacharasterisk}cast{\isacharparenleft}m{\isacharparenright}}}

\snip{act_arrive_newpkt}{0}{0}{\isa{{\isacharbraceleft}i{\isacharbraceright}{\isasymnot}K{\isacharcolon}arrive{\isacharparenleft}Newpkt\ d\ dst{\isacharparenright}}}

\snip{net_state}{0}{0}{\isa{{\isacharprime}s\ net{\isacharunderscore}state}}
\snip{net_state_nodes}{0}{0}{\isa{\NodeS\ i\ s\ R}}
\snip{net_state_subnets}{0}{0}{\isa{\subnetsadj{s\isactrlsub {\isadigit{1}}}{s\isactrlsub {\isadigit{2}}}}}

\snip{wf_net_tree}{0}{0}{\isa{wf{\isacharunderscore}net{\isacharunderscore}tree\ {\isacharparenleft}{\isasymPsi}\isactrlsub {\isadigit{1}}\parallelcomp{}{\isasymPsi}\isactrlsub {\isadigit{2}}{\isacharparenright}\ {\isacharequal}\ {\isacharparenleft}net{\isacharunderscore}tree{\isacharunderscore}ips\ {\isasymPsi}\isactrlsub {\isadigit{1}}\ {\isasyminter}\ net{\isacharunderscore}tree{\isacharunderscore}ips\ {\isasymPsi}\isactrlsub {\isadigit{2}}\ {\isacharequal}\ {\isasymemptyset}\ {\isasymand}\ wf{\isacharunderscore}net{\isacharunderscore}tree\ {\isasymPsi}\isactrlsub {\isadigit{1}}\ {\isasymand}\ wf{\isacharunderscore}net{\isacharunderscore}tree\ {\isasymPsi}\isactrlsub {\isadigit{2}}{\isacharparenright}\isasep\isanewline%
wf{\isacharunderscore}net{\isacharunderscore}tree\ {\isasymlangle}i{\isacharsemicolon}\ R{\isasymrangle}\ {\isacharequal}\ True}}
\snip{pnet_rules}{0}{0}{\isa{pnet\ np\ {\isasymlangle}i{\isacharsemicolon}\ \Ri{\isasymrangle}\ {\isacharequal}\ {\isasymlangle}i\ {\isacharcolon}\ np\ i\ {\isacharcolon}\ \Ri{\isasymrangle}\isasep\isanewline%
pnet\ np\ {\isacharparenleft}p\isactrlsub {\isadigit{1}}\parallelcomp{}p\isactrlsub {\isadigit{2}}{\isacharparenright}\ {\isacharequal}\ {\isasymlparr}init\ {\isacharequal}\ {\isacharbraceleft}\subnets{s\isactrlsub {\isadigit{1}}}{s\isactrlsub {\isadigit{2}}}\ {\isacharbar}\ s\isactrlsub {\isadigit{1}}{\isasymin}init\ {\isacharparenleft}pnet\ np\ p\isactrlsub {\isadigit{1}}{\isacharparenright}\ {\isasymand}\ s\isactrlsub {\isadigit{2}}{\isasymin}init\ {\isacharparenleft}pnet\ np\ p\isactrlsub {\isadigit{2}}{\isacharparenright}{\isacharbraceright},\ trans\ {\isacharequal}\ \pnetsos\ {\isacharparenleft}trans\ {\isacharparenleft}pnet\ np\ p\isactrlsub {\isadigit{1}}{\isacharparenright}{\isacharparenright}\ {\isacharparenleft}trans\ {\isacharparenleft}pnet\ np\ p\isactrlsub {\isadigit{2}}{\isacharparenright}{\isacharparenright}{\isasymrparr}}}

\snip{pnet1_lhs}{0}{0}{\isa{pnet\ np\ {\isasymlangle}i{\isacharsemicolon}\ \Ri{\isasymrangle}}}
\snip{pnet1_rhs}{0}{0}{\isa{{\isasymlangle}i\ {\isacharcolon}\ np\ i\ {\isacharcolon}\ \Ri{\isasymrangle}}}
\snip{pnet2_lhs}{0}{0}{\isa{pnet\ np\ {\isacharparenleft}{\isasymPsi}\isactrlsub {\isadigit{1}}\parallelcomp{}{\isasymPsi}\isactrlsub {\isadigit{2}}{\isacharparenright}}}
\snip{pnet2_rhs}{0}{0}{\isa{{\isasymlparr}init\ {\isacharequal}\ {\isacharbraceleft}\subnets{s\isactrlsub {\isadigit{1}}}{s\isactrlsub {\isadigit{2}}}\ {\isacharbar}\ s\isactrlsub {\isadigit{1}}{\isasymin}init\ {\isacharparenleft}pnet\ np\ {\isasymPsi}\isactrlsub {\isadigit{1}}{\isacharparenright}\ {\isasymand}\ s\isactrlsub {\isadigit{2}}{\isasymin}init\ {\isacharparenleft}pnet\ np\ {\isasymPsi}\isactrlsub {\isadigit{2}}{\isacharparenright}{\isacharbraceright},}\\ && \isa{\phantom{\isasymlparr}trans {\isacharequal}\ \pnetsos\ {\isacharparenleft}trans\ {\isacharparenleft}pnet\ np\ {\isasymPsi}\isactrlsub {\isadigit{1}}{\isacharparenright}{\isacharparenright}\ {\isacharparenleft}trans\ {\isacharparenleft}pnet\ np\ {\isasymPsi}\isactrlsub {\isadigit{2}}{\isacharparenright}{\isacharparenright}{\isasymrparr}}}

\snip{opnet1_lhs}{0}{0}{\isa{opnet\ onp\ {\isasymlangle}i{\isacharsemicolon}\ \Ri{\isasymrangle}}}
\snip{opnet1_rhs}{0}{0}{\isa{{\isasymlangle}i\ {\isacharcolon}\ onp\ i\ {\isacharcolon}\ \Ri{\isasymrangle}\isactrlsub o}}
\snip{opnet2_lhs}{0}{0}{\isa{opnet\ onp\ {\isacharparenleft}{\isasymPsi}\isactrlsub {\isadigit{1}}\!\parallelcomp{}{\isasymPsi}\isactrlsub {\isadigit{2}}{\isacharparenright}}}
\snip{opnet2_rhs}{0}{0}{\isa{{\isasymlparr}init\ {\isacharequal}\ {\isacharbraceleft}{\isacharparenleft}{\isasymsigma},\ \subnetsadj{s\isactrlsub {\isadigit{1}}}{s\isactrlsub {\isadigit{2}}}{\isacharparenright}\ {\isacharbar}\ {\isacharparenleft}{\isasymsigma},\ s\isactrlsub {\isadigit{1}}{\isacharparenright}{\isasymin}init\ {\isacharparenleft}opnet\ onp\ {\isasymPsi}\isactrlsub {\isadigit{1}}{\isacharparenright}}\\ && \isa{\phantom{\isasymlparr init\ = \ {\isacharbraceleft}{\isacharparenleft}{\isasymsigma}{\isacharcomma}\ \subnets{s\isactrlsub {\isadigit{1}}}{s\isactrlsub {\isadigit{2}}}{\isacharparenright}\ {\isacharbar}}{\isasymand}\ {\isacharparenleft}{\isasymsigma},\ s\isactrlsub {\isadigit{2}}{\isacharparenright}{\isasymin}init\ {\isacharparenleft}opnet\ onp\ {\isasymPsi}\isactrlsub {\isadigit{2}}{\isacharparenright}}\\ && \isa{\phantom{\isasymlparr init\ = \ {\isacharbraceleft}{\isacharparenleft}{\isasymsigma}{\isacharcomma}\ \subnets{s\isactrlsub {\isadigit{1}}}{s\isactrlsub {\isadigit{2}}}{\isacharparenright}\ {\isacharbar}}{\isasymand}\ net{\isacharunderscore}ips\ s\isactrlsub {\isadigit{1}}\ {\isasyminter}\ net{\isacharunderscore}ips\ s\isactrlsub {\isadigit{2}}\ {\isacharequal}\ {\isasymemptyset}{\isacharbraceright},}\\ && \isa{\phantom{\isasymlparr}trans {\isacharequal}\ \opnetsos\ {\isacharparenleft}trans\ {\isacharparenleft}opnet\ onp\ {\isasymPsi}\isactrlsub {\isadigit{1}}{\isacharparenright}{\isacharparenright}\ {\isacharparenleft}trans\ {\isacharparenleft}opnet\ onp\ {\isasymPsi}\isactrlsub {\isadigit{2}}{\isacharparenright}{\isacharparenright}{\isasymrparr}}}

\snip{oclosed'}{0}{0}{\isa{oclosed\ A}}
\snip{oclosed_term}{0}{0}{\isa{A{\isasymlparr}trans\ {\isacharcolon}{\isacharequal}\ \ocnetsos\ {\isacharparenleft}trans\ A{\isacharparenright}{\isasymrparr}}}

\snip{netglobal}{0}{0}{\isa{netglobal\ P\ {\isacharequal}\ {\isasymlambda}s{\isachardot}\ P\ {\isacharparenleft}default\ toy{\isacharunderscore}init\ {\isacharparenleft}netlift\ fst\ s{\isacharparenright}{\isacharparenright}}}

\snip{default}{0}{0}{\isa{default\ df\ f\ {\isacharequal}\ {\isacharparenleft}{\isasymlambda}i{\isachardot}\ \textsf{case}\ f\ i\ \textsf{of}\ None\ {\isasymRightarrow}\ df\ i\ {\isacharbar}\ Some\ s\ {\isasymRightarrow}\ s{\isacharparenright}}}
\snip{default_lhs}{0}{0}{\isa{default\ df\ f}}
\snip{default_rhs}{0}{0}{\isa{{\isasymlambda}i{\isachardot}\ \textsf{case}\ f\ i\ \textsf{of}\ None\ {\isasymRightarrow}\ df\ i\ {\isacharbar}\ Some\ s\ {\isasymRightarrow}\ s}}

\snip{netlift1_lhs}{0}{0}{\isa{netlift\ ps\ {\isacharparenleft}\NodeS\ i\ s\ R{\isacharparenright}}}
\snip{netlift1_rhs}{0}{0}{\isa{{\isacharbrackleft}i\ {\isasymmapsto}\ fst\ {\isacharparenleft}ps\ s{\isacharparenright}{\isacharbrackright}}}
\snip{netlift2_lhs}{0}{0}{\isa{netlift\ ps\ {\isacharparenleft}\subnets{s}{t}{\isacharparenright}}}
\snip{netlift2_rhs}{0}{0}{\isa{netlift\ ps\ s\ \listconcat\ netlift\ ps\ t}}

\snip{netliftl1_lhs}{0}{0}{\isa{netliftc\ sr\ {\isacharparenleft}\NodeS\ i\ s\ R{\isacharparenright}}}
\snip{netliftl1_rhs}{0}{0}{\isa{\NodeS\ i\ {{\isacharparenleft}snd\ {\isacharparenleft}sr\ s{\isacharparenright}{\isacharparenright}}\ R}}
\snip{netliftl2_lhs}{0}{0}{\isa{netliftc\ sr\ {\isacharparenleft}\subnets{s}{t}{\isacharparenright}}}
\snip{netliftl2_rhs}{0}{0}{\isa{\subnets{{\isacharparenleft}netliftc\ sr\ s{\isacharparenright}}{{\isacharparenleft}netliftc\ sr\ t{\isacharparenright}}}}

\snip{netgmap1_lhs}{0}{0}{\isa{netgmap\ sr\ {\isacharparenleft}\NodeS\ i\ s\ R{\isacharparenright}}}
\snip{netgmap1_rhs}{0}{0}{\isa{{\isacharparenleft}{\isacharbrackleft}i\ {\isasymmapsto}\ fst\ {\isacharparenleft}sr\ s{\isacharparenright}{\isacharbrackright},\ \NodeS\ i\ {{\isacharparenleft}snd\ {\isacharparenleft}sr\ s{\isacharparenright}{\isacharparenright}}\ R{\isacharparenright}}}
\snip{netgmap2_lhs}{0}{0}{\isa{netgmap\ sr\ {\isacharparenleft}\subnets{s\isactrlsub {\isadigit{1}}}{s\isactrlsub {\isadigit{2}}}{\isacharparenright}}}
\snip{netgmap2_rhs}{0}{0}{\isa{\textsf{let}\ {\isacharparenleft}{\isasymsigma}\isactrlsub {\isadigit{1}},\ ss{\isacharparenright}\ {\isacharequal}\ netgmap\ sr\ s\isactrlsub {\isadigit{1}}{\isacharsemicolon}}\\&&\isa{\phantom{let\ }{\isacharparenleft}{\isasymsigma}\isactrlsub {\isadigit{2}},\ tt{\isacharparenright}\ {\isacharequal}\ netgmap\ sr\ s\isactrlsub {\isadigit{2}}\ }\\&&\isa{\textsf{in}\ {\isacharparenleft}{\isasymsigma}\isactrlsub {\isadigit{1}}\ \listconcat\ {\isasymsigma}\isactrlsub {\isadigit{2}},\ \subnets{ss}{tt}{\isacharparenright}}}
\snip{initmissing_def}{0}{0}{\isa{initmissing\ {\isasymsigma}\ {\isacharequal}\ {\isacharparenleft}{\isasymlambda}i{\isachardot}\ \textsf{case}\ fst\ {\isasymsigma}\ i\ \textsf{of}\ None\ {\isasymRightarrow}\ toy{\isacharunderscore}init\ i\ {\isacharbar}\ Some\ s\ {\isasymRightarrow}\ s,\ snd\ {\isasymsigma}{\isacharparenright}}}
\snip{initmissing_lhs}{0}{0}{\isa{initmissing\ {\isasymsigma}}}
\snip{initmissing_rhs}{0}{0}{\isa{{\isacharparenleft}{\isasymlambda}i{\isachardot}\ \textsf{case}\ fst\ {\isasymsigma}\ i\ \textsf{of}\ None\ {\isasymRightarrow}\ toy{\isacharunderscore}init\ i\ {\isacharbar}\ Some\ s\ {\isasymRightarrow}\ s,\ snd\ {\isasymsigma}{\isacharparenright}}}

\snip{onl}{0}{0}{\isa{onl\ {\isasymGamma}\ P\ {\isacharequal}\ {\isasymlambda}{\isacharparenleft}{\isasymxi},\ p{\isacharparenright}{\isachardot}\ {\isasymforall}l{\isasymin}labels\ {\isasymGamma}\ p{\isachardot}\ P\ {\isacharparenleft}{\isasymxi},\ l{\isacharparenright}}}
\snip{onl_lhs}{0}{0}{\isa{onl\ {\isasymGamma}\ P}}
\snip{onl_rhs}{0}{0}{\isa{{\isasymlambda}{\isacharparenleft}{\isasymxi},\ p{\isacharparenright}{\isachardot}\ {\isasymforall}l{\isasymin}labels\ {\isasymGamma}\ p{\isachardot}\ P\ {\isacharparenleft}{\isasymxi},\ l{\isacharparenright}}}

\snip{onll}{0}{0}{\isa{onll\ {\isasymGamma}\ P\ {\isacharequal}\ {\isasymlambda}{\isacharparenleft}{\isacharparenleft}{\isasymxi},\ p{\isacharparenright},\ a,\ {\isasymxi}{\isacharprime},\ p{\isacharprime}{\isacharparenright}{\isachardot}\ {\isasymforall}l{\isasymin}labels\ {\isasymGamma}\ p{\isachardot}\ {\isasymforall}l{\isacharprime}{\isasymin}labels\ {\isasymGamma}\ p{\isacharprime}{\isachardot}\ P\ {\isacharparenleft}{\isacharparenleft}{\isasymxi},\ l{\isacharparenright},\ a,\ {\isasymxi}{\isacharprime},\ l{\isacharprime}{\isacharparenright}}}
\snip{onll_lhs}{0}{0}{\isa{onll\ {\isasymGamma}\ P}}
\snip{onll_rhs}{0}{0}{\isa{{\isasymlambda}{\isacharparenleft}{\isacharparenleft}{\isasymxi},\ p{\isacharparenright},\ a,\ {\isacharparenleft}{\isasymxi}{\isacharprime},\ p{\isacharprime}{\isacharparenright}{\isacharparenright}{\isachardot}\ {\isasymforall}l{\isasymin}labels\ {\isasymGamma}\ p{\isachardot}\ {\isasymforall}l{\isacharprime}{\isasymin}labels\ {\isasymGamma}\ p{\isacharprime}{\isachardot}\ P\ {\isacharparenleft}{\isacharparenleft}{\isasymxi},\ l{\isacharparenright},\ a,\ {\isacharparenleft}{\isasymxi}{\isacharprime},\ l{\isacharprime}{\isacharparenright}{\isacharparenright}}}

\snip{cnet_connect}{0}{0}{\isa{\mbox{}\inferrule{\mbox{{\isacharparenleft}s,\ connect{\isacharparenleft}i,\ i{\isacharprime}{\isacharparenright},\ s{\isacharprime}{\isacharparenright}{\isasymin}T\isactrlsub A}}{\mbox{{\isacharparenleft}s,\ connect{\isacharparenleft}i,\ i{\isacharprime}{\isacharparenright},\ s{\isacharprime}{\isacharparenright}{\isasymin}\cnetsos\ T\isactrlsub A}}}}
\snip{cnet_disconnect}{0}{0}{\isa{\mbox{}\inferrule{\mbox{{\isacharparenleft}s,\ disconnect{\isacharparenleft}i,\ i{\isacharprime}{\isacharparenright},\ s{\isacharprime}{\isacharparenright}{\isasymin}T\isactrlsub A}}{\mbox{{\isacharparenleft}s,\ disconnect{\isacharparenleft}i,\ i{\isacharprime}{\isacharparenright},\ s{\isacharprime}{\isacharparenright}{\isasymin}\cnetsos\ T\isactrlsub A}}}}
\snip{cnet_cast}{0}{0}{\isa{\mbox{}\inferrule{\mbox{{\isacharparenleft}s,\ R{\isacharcolon}{\isacharasterisk}cast{\isacharparenleft}m{\isacharparenright},\ s{\isacharprime}{\isacharparenright}{\isasymin}T\isactrlsub A}}{\mbox{{\isacharparenleft}s,\ {\isasymtau},\ s{\isacharprime}{\isacharparenright}{\isasymin}\cnetsos\ T\isactrlsub A}}}}
\snip{cnet_tau}{0}{0}{\isa{\mbox{}\inferrule{\mbox{{\isacharparenleft}s,\ {\isasymtau},\ s{\isacharprime}{\isacharparenright}{\isasymin}T\isactrlsub A}}{\mbox{{\isacharparenleft}s,\ {\isasymtau},\ s{\isacharprime}{\isacharparenright}{\isasymin}\cnetsos\ T\isactrlsub A}}}}
\snip{cnet_deliver}{0}{0}{\isa{\mbox{}\inferrule{\mbox{{\isacharparenleft}s,\ i{\isacharcolon}deliver{\isacharparenleft}d{\isacharparenright},\ s{\isacharprime}{\isacharparenright}{\isasymin}T\isactrlsub A}}{\mbox{{\isacharparenleft}s,\ i{\isacharcolon}deliver{\isacharparenleft}d{\isacharparenright},\ s{\isacharprime}{\isacharparenright}{\isasymin}\cnetsos\ T\isactrlsub A}}}}
\snip{cnet_newpkt}{0}{0}{\isa{\mbox{}\inferrule{\mbox{{\isacharparenleft}s,\ {\isacharbraceleft}i{\isacharbraceright}{\isasymnot}K{\isacharcolon}arrive{\isacharparenleft}Newpkt\ d\ dst{\isacharparenright},\ s{\isacharprime}{\isacharparenright}{\isasymin}T\isactrlsub A}}{\mbox{{\isacharparenleft}s,\ i{\isacharcolon}newpkt{\isacharparenleft}d,\ dst{\isacharparenright},\ s{\isacharprime}{\isacharparenright}{\isasymin}\cnetsos\ T\isactrlsub A}}}}

\snip{closed}{0}{0}{\isa{closed\ {\isacharparenleft}pnet\ {\isacharparenleft}{\isasymlambda}i{\isachardot}\ ptoy\ i\ {\isasymlangle}{\isasymlangle}\ qmsg{\isacharparenright}\ n{\isacharparenright}}}
\snip{closed'}{0}{0}{\isa{closed\ A}}
\snip{closed_term}{0}{0}{\isa{A{\isasymlparr}trans\ {\isacharcolon}{\isacharequal}\ \cnetsos\ {\isacharparenleft}trans\ A{\isacharparenright}{\isasymrparr}}}

\snip{closed_type}{0}{0}{\isa{{\isacharparenleft}{\isacharparenleft}{\isacharparenleft}state\ {\isasymtimes}\ {\isacharparenleft}state,\ msg,\ pseqp,\ pseqp\ label{\isacharparenright}\ seqp{\isacharparenright}\ {\isasymtimes}\ msg\ list\ {\isasymtimes}\ {\isacharparenleft}msg\ list,\ msg,\ unit,\ unit\ label{\isacharparenright}\ seqp{\isacharparenright}\ net{\isacharunderscore}state,\ msg\ node{\isacharunderscore}action{\isacharparenright}\ automaton}}
\snip{state_type}{0}{0}{\isa{state}}
\snip{sigma_type}{0}{0}{\isa{ip\ {\isasymRightarrow}\ state}}

\snip{obroadcastT}{0}{0}{\isa{\mbox{}\inferrule{\mbox{{\isasymsigma}{\isacharprime}\ i\ {\isacharequal}\ {\isasymsigma}\ i}}{\mbox{{\isacharparenleft}{\isacharparenleft}{\isasymsigma},\ \gray{{\isacharbraceleft}l{\isacharbraceright}}broadcast{\isacharparenleft}\selmsg{\isacharparenright}\ {\isachardot}\ p{\isacharparenright},\ broadcast\ {\isacharparenleft}\selmsg\ {\isacharparenleft}{\isasymsigma}\ i{\isacharparenright}{\isacharparenright},\ {\isacharparenleft}{\isasymsigma}{\isacharprime},\ p{\isacharparenright}{\isacharparenright}{\isasymin}\oseqpsos\ {\isasymGamma}\ i}}}}
\snip{ogroupcastT}{0}{0}{\isa{\mbox{}\inferrule{\mbox{{\isasymsigma}{\isacharprime}\ i\ {\isacharequal}\ {\isasymsigma}\ i}}{\mbox{{\isacharparenleft}{\isacharparenleft}{\isasymsigma},\ \gray{{\isacharbraceleft}l{\isacharbraceright}}groupcast{\isacharparenleft}\selips,\ \selmsg{\isacharparenright}\ {\isachardot}\ p{\isacharparenright},\ groupcast\ {\isacharparenleft}\selips\ {\isacharparenleft}{\isasymsigma}\ i{\isacharparenright}{\isacharparenright}\ {\isacharparenleft}\selmsg\ {\isacharparenleft}{\isasymsigma}\ i{\isacharparenright}{\isacharparenright},\ {\isacharparenleft}{\isasymsigma}{\isacharprime},\ p{\isacharparenright}{\isacharparenright}{\isasymin}\oseqpsos\ {\isasymGamma}\ i}}}}
\snip{ounicastT}{0}{0}{\isa{\mbox{}\inferrule{\mbox{{\isasymsigma}{\isacharprime}\ i\ {\isacharequal}\ {\isasymsigma}\ i}}{\mbox{{\isacharparenleft}{\isacharparenleft}{\isasymsigma},\ \gray{{\isacharbraceleft}l{\isacharbraceright}}unicast{\isacharparenleft}\selip,\ \selmsg{\isacharparenright}\ {\isachardot}\ p\ {\isasymtriangleright}\ q{\isacharparenright},\ unicast\ {\isacharparenleft}\selip\ {\isacharparenleft}{\isasymsigma}\ i{\isacharparenright}{\isacharparenright}\ {\isacharparenleft}\selmsg\ {\isacharparenleft}{\isasymsigma}\ i{\isacharparenright}{\isacharparenright},\ {\isacharparenleft}{\isasymsigma}{\isacharprime},\ p{\isacharparenright}{\isacharparenright}{\isasymin}\oseqpsos\ {\isasymGamma}\ i}}}}
\snip{onotunicastT}{0}{0}{\isa{\mbox{}\inferrule{\mbox{{\isasymsigma}{\isacharprime}\ i\ {\isacharequal}\ {\isasymsigma}\ i}}{\mbox{{\isacharparenleft}{\isacharparenleft}{\isasymsigma},\ \gray{{\isacharbraceleft}l{\isacharbraceright}}unicast{\isacharparenleft}\selip,\ \selmsg{\isacharparenright}\ {\isachardot}\ p\ {\isasymtriangleright}\ q{\isacharparenright},\ {\isasymnot}unicast\ {\isacharparenleft}\selip\ {\isacharparenleft}{\isasymsigma}\ i{\isacharparenright}{\isacharparenright},\ {\isacharparenleft}{\isasymsigma}{\isacharprime},\ q{\isacharparenright}{\isacharparenright}{\isasymin}\oseqpsos\ {\isasymGamma}\ i}}}}
\snip{osendT}{0}{0}{\isa{\mbox{}\inferrule{\mbox{{\isasymsigma}{\isacharprime}\ i\ {\isacharequal}\ {\isasymsigma}\ i}}{\mbox{{\isacharparenleft}{\isacharparenleft}{\isasymsigma},\ \gray{{\isacharbraceleft}l{\isacharbraceright}}send{\isacharparenleft}\selmsg{\isacharparenright}\ {\isachardot}\ p{\isacharparenright},\ send\ {\isacharparenleft}\selmsg\ {\isacharparenleft}{\isasymsigma}\ i{\isacharparenright}{\isacharparenright},\ {\isacharparenleft}{\isasymsigma}{\isacharprime},\ p{\isacharparenright}{\isacharparenright}{\isasymin}\oseqpsos\ {\isasymGamma}\ i}}}}
\snip{odeliverT}{0}{0}{\isa{\mbox{}\inferrule{\mbox{{\isasymsigma}{\isacharprime}\ i\ {\isacharequal}\ {\isasymsigma}\ i}}{\mbox{{\isacharparenleft}{\isacharparenleft}{\isasymsigma},\ \gray{{\isacharbraceleft}l{\isacharbraceright}}deliver{\isacharparenleft}\seldata{\isacharparenright}\ {\isachardot}\ p{\isacharparenright},\ deliver\ {\isacharparenleft}\seldata\ {\isacharparenleft}{\isasymsigma}\ i{\isacharparenright}{\isacharparenright},\ {\isacharparenleft}{\isasymsigma}{\isacharprime},\ p{\isacharparenright}{\isacharparenright}{\isasymin}\oseqpsos\ {\isasymGamma}\ i}}}}
\snip{oreceiveT}{0}{0}{\isa{\mbox{}\inferrule{\mbox{{\isasymsigma}{\isacharprime}\ i\ {\isacharequal}\ \updmsg\ msg\ {\isacharparenleft}{\isasymsigma}\ i{\isacharparenright}}}{\mbox{{\isacharparenleft}{\isacharparenleft}{\isasymsigma},\ \gray{{\isacharbraceleft}l{\isacharbraceright}}receive{\isacharparenleft}\updmsg{\isacharparenright}\ {\isachardot}\ p{\isacharparenright},\ receive\ msg,\ {\isacharparenleft}{\isasymsigma}{\isacharprime},\ p{\isacharparenright}{\isacharparenright}{\isasymin}\oseqpsos\ {\isasymGamma}\ i}}}}
\snip{oassignT}{0}{0}{\isa{\mbox{}\inferrule{\mbox{{\isasymsigma}{\isacharprime}\ i\ {\isacharequal}\ u\ {\isacharparenleft}{\isasymsigma}\ i{\isacharparenright}}}{\mbox{{\isacharparenleft}{\isacharparenleft}{\isasymsigma},\ \gray{{\isacharbraceleft}l{\isacharbraceright}}{\isasymlbrakk}u{\isasymrbrakk}\ p{\isacharparenright},\ {\isasymtau},\ {\isacharparenleft}{\isasymsigma}{\isacharprime},\ p{\isacharparenright}{\isacharparenright}{\isasymin}\oseqpsos\ {\isasymGamma}\ i}}}}
\snip{ocallT}{0}{0}{\isa{\mbox{}\inferrule{\mbox{{\isacharparenleft}{\isacharparenleft}{\isasymsigma},\ {\isasymGamma}\ pn{\isacharparenright},\ a,\ {\isacharparenleft}{\isasymsigma}{\isacharprime},\ p{\isacharprime}{\isacharparenright}{\isacharparenright}{\isasymin}\oseqpsos\ {\isasymGamma}\ i}}{\mbox{{\isacharparenleft}{\isacharparenleft}{\isasymsigma},\ call{\isacharparenleft}pn{\isacharparenright}{\isacharparenright},\ a,\ {\isacharparenleft}{\isasymsigma}{\isacharprime},\ p{\isacharprime}{\isacharparenright}{\isacharparenright}{\isasymin}\oseqpsos\ {\isasymGamma}\ i}}}}
\snip{ochoiceT1}{0}{0}{\isa{\mbox{}\inferrule{\mbox{{\isacharparenleft}{\isacharparenleft}{\isasymsigma},\ p{\isacharparenright},\ a,\ {\isacharparenleft}{\isasymsigma}{\isacharprime},\ p{\isacharprime}{\isacharparenright}{\isacharparenright}{\isasymin}\oseqpsos\ {\isasymGamma}\ i}}{\mbox{{\isacharparenleft}{\isacharparenleft}{\isasymsigma},\ p\ {\isasymoplus}\ q{\isacharparenright},\ a,\ {\isacharparenleft}{\isasymsigma}{\isacharprime},\ p{\isacharprime}{\isacharparenright}{\isacharparenright}{\isasymin}\oseqpsos\ {\isasymGamma}\ i}}}}
\snip{ochoiceT2}{0}{0}{\isa{\mbox{}\inferrule{\mbox{{\isacharparenleft}{\isacharparenleft}{\isasymsigma},\ q{\isacharparenright},\ a,\ {\isacharparenleft}{\isasymsigma}{\isacharprime},\ q{\isacharprime}{\isacharparenright}{\isacharparenright}{\isasymin}\oseqpsos\ {\isasymGamma}\ i}}{\mbox{{\isacharparenleft}{\isacharparenleft}{\isasymsigma},\ p\ {\isasymoplus}\ q{\isacharparenright},\ a,\ {\isacharparenleft}{\isasymsigma}{\isacharprime},\ q{\isacharprime}{\isacharparenright}{\isacharparenright}{\isasymin}\oseqpsos\ {\isasymGamma}\ i}}}}
\snip{oguardT}{0}{0}{\isa{\mbox{}\inferrule{\mbox{{\isasymsigma}{\isacharprime}\ i{\isasymin}g\ {\isacharparenleft}{\isasymsigma}\ i{\isacharparenright}}}{\mbox{{\isacharparenleft}{\isacharparenleft}{\isasymsigma},\ \gray{{\isacharbraceleft}l{\isacharbraceright}}{\isasymlangle}g{\isasymrangle}\ p{\isacharparenright},\ {\isasymtau},\ {\isacharparenleft}{\isasymsigma}{\isacharprime},\ p{\isacharparenright}{\isacharparenright}{\isasymin}\oseqpsos\ {\isasymGamma}\ i}}}}

\snip{oassignT_prem_1}{0}{0}{\isa{{\isasymsigma}{\isacharprime}\ i\ {\isacharequal}\ u\ {\isacharparenleft}{\isasymsigma}\ i{\isacharparenright}}}
\snip{ounicastT_prem_1}{0}{0}{\isa{{\isasymsigma}{\isacharprime}\ i\ {\isacharequal}\ {\isasymsigma}\ i}}

\snip{oparleft}{0}{0}{\isa{\mbox{}\inferrule{\mbox{{\isacharparenleft}{\isacharparenleft}{\isasymsigma},\ s{\isacharparenright},\ a,\ {\isacharparenleft}{\isasymsigma}{\isacharprime},\ s{\isacharprime}{\isacharparenright}{\isacharparenright}{\isasymin}T\isactrlsub A}\\\ \mbox{{\isasymAnd}m{\isachardot}\ a\ {\isasymnoteq}\ receive\ m}}{\mbox{{\isacharparenleft}{\isacharparenleft}{\isasymsigma},\ {\isacharparenleft}s,\ t{\isacharparenright}{\isacharparenright},\ a,\ {\isacharparenleft}{\isasymsigma}{\isacharprime}, {\isacharparenleft}s{\isacharprime},\ t{\isacharparenright}{\isacharparenright}{\isacharparenright}{\isasymin}\oparpsos\ i\ T\isactrlsub A\ T\isactrlsub B}}}}
\snip{oparright}{0}{0}{\isa{\mbox{}\inferrule{\mbox{{\isacharparenleft}t,\ a,\ t{\isacharprime}{\isacharparenright}{\isasymin}T\isactrlsub B}\\\ \mbox{{\isasymAnd}m{\isachardot}\ a\ {\isasymnoteq}\ send\ m}\\\ \mbox{{\isasymsigma}{\isacharprime}\ i\ {\isacharequal}\ {\isasymsigma}\ i}}{\mbox{{\isacharparenleft}{\isacharparenleft}{\isasymsigma},\ {\isacharparenleft}s,\ t{\isacharparenright}{\isacharparenright},\ a,\ {\isacharparenleft}{\isasymsigma}{\isacharprime}, {\isacharparenleft}s,\ t{\isacharprime}{\isacharparenright}{\isacharparenright}{\isacharparenright}{\isasymin}\oparpsos\ i\ T\isactrlsub A\ T\isactrlsub B}}}}
\snip{oparboth}{0}{0}{\isa{\mbox{}\inferrule{\mbox{{\isacharparenleft}{\isacharparenleft}{\isasymsigma},\ s{\isacharparenright},\ receive\ m,\ {\isacharparenleft}{\isasymsigma}{\isacharprime},\ s{\isacharprime}{\isacharparenright}{\isacharparenright}{\isasymin}T\isactrlsub A}\\\ \mbox{{\isacharparenleft}t,\ send\ m,\ t{\isacharprime}{\isacharparenright}{\isasymin}T\isactrlsub B}}{\mbox{{\isacharparenleft}{\isacharparenleft}{\isasymsigma},\ {\isacharparenleft}s,\ t{\isacharparenright}{\isacharparenright},\ {\isasymtau},\ {\isacharparenleft}{\isasymsigma}{\isacharprime},\ {\isacharparenleft}s{\isacharprime},\ t{\isacharprime}{\isacharparenright}{\isacharparenright}{\isacharparenright}{\isasymin}\oparpsos\ i\ T\isactrlsub A\ T\isactrlsub B}}}}

\snip{opar_comp}{0}{0}{\mbox{\isa{s\ {\isasymlangle}{\isasymlangle}\isactrlbsub i\isactrlesub \ t\ {\isacharequal}\ {\isasymlparr}}} & \mbox{\isa{init\ {\isacharequal}\ extg\ {\isacharbackquote}\ {\isacharparenleft}init\ s\ {\isasymtimes}\ init\ t{\isacharparenright},}}\\ & \mbox{\isa{trans\ {\isacharequal}\ \oparpsos\ i\ {\isacharparenleft}trans\ s{\isacharparenright}\ {\isacharparenleft}trans\ t{\isacharparenright}{\isasymrparr}}}}
\snip{opar_comp'}{0}{0}{\mbox{\isa{A\ {\isasymlangle}{\isasymlangle}\isactrlbsub i\isactrlesub \ B\ {\isacharequal}\ {\isasymlparr}}} & \mbox{\isa{init\ {\isacharequal}\ {\isacharbraceleft}{\isacharparenleft}{\isasymsigma},\ (s,\ t){\isacharparenright}\ {\isacharbar}\ {\isacharparenleft}{\isasymsigma},\ s{\isacharparenright}{\isasymin}init\ A\ {\isasymand}\ t{\isasymin}init\ B{\isacharbraceright},}}\\ & \mbox{\isa{trans\ {\isacharequal}\ \oparpsos\ i\ {\isacharparenleft}trans\ A{\isacharparenright}\ {\isacharparenleft}trans\ B{\isacharparenright}{\isasymrparr}}}}
\snip{opar_comp_term}{0}{0}{\mbox{\isa{s\ {\isasymlangle}{\isasymlangle}\isactrlbsub i\isactrlesub \ t{\isasymColon}{\isacharparenleft}{\isacharparenleft}nat\ {\isasymRightarrow}\ {\isacharprime}a{\isacharparenright}\ {\isasymtimes}\ {\isacharprime}b\ {\isasymtimes}\ {\isacharprime}d,\ {\isacharprime}c\ seq{\isacharunderscore}action{\isacharparenright}\ automaton}}}

\snip{onode_bcast}{0}{0}{\isa{\mbox{}\inferrule{\mbox{{\isacharparenleft}{\isacharparenleft}{\isasymsigma},\ s{\isacharparenright},\ broadcast\ m,\ {\isacharparenleft}{\isasymsigma}{\isacharprime},\ s{\isacharprime}{\isacharparenright}{\isacharparenright}{\isasymin}T\isactrlsub A}}{\mbox{{\isacharparenleft}{\isacharparenleft}{\isasymsigma},\ \NodeS\ i\ s\ R{\isacharparenright},\ R{\isacharcolon}{\isacharasterisk}cast{\isacharparenleft}m{\isacharparenright},\ {\isacharparenleft}{\isasymsigma}{\isacharprime},\ \NodeS\ i\ s{\isacharprime}\ R{\isacharparenright}{\isacharparenright}{\isasymin}\onodesos\ T\isactrlsub A}}}}
\snip{onode_gcast}{0}{0}{\isa{\mbox{}\inferrule{\mbox{{\isacharparenleft}{\isacharparenleft}{\isasymsigma},\ s{\isacharparenright},\ groupcast\ D\ m,\ {\isacharparenleft}{\isasymsigma}{\isacharprime},\ s{\isacharprime}{\isacharparenright}{\isacharparenright}{\isasymin}T\isactrlsub A}}{\mbox{{\isacharparenleft}{\isacharparenleft}{\isasymsigma},\ \NodeS\ i\ s\ R{\isacharparenright},\ {\isacharparenleft}R\ {\isasyminter}\ D{\isacharparenright}{\isacharcolon}{\isacharasterisk}cast{\isacharparenleft}m{\isacharparenright},\ {\isacharparenleft}{\isasymsigma}{\isacharprime},\ \NodeS\ i\ s{\isacharprime}\ R{\isacharparenright}{\isacharparenright}{\isasymin}\onodesos\ T\isactrlsub A}}}}
\snip{onode_ucast}{0}{0}{\isa{\mbox{}\inferrule{\mbox{{\isacharparenleft}{\isacharparenleft}{\isasymsigma},\ s{\isacharparenright},\ unicast\ dst\ m,\ {\isacharparenleft}{\isasymsigma}{\isacharprime},\ s{\isacharprime}{\isacharparenright}{\isacharparenright}{\isasymin}T\isactrlsub A}\\\ \mbox{dst{\isasymin}R}}{\mbox{{\isacharparenleft}{\isacharparenleft}{\isasymsigma},\ \NodeS\ i\ s\ R{\isacharparenright},\ {\isacharbraceleft}dst{\isacharbraceright}{\isacharcolon}{\isacharasterisk}cast{\isacharparenleft}m{\isacharparenright},\ {\isacharparenleft}{\isasymsigma}{\isacharprime},\ \NodeS\ i\ s{\isacharprime}\ R{\isacharparenright}{\isacharparenright}{\isasymin}\onodesos\ T\isactrlsub A}}}}
\snip{onode_notucast}{0}{0}{\isa{\mbox{}\inferrule{\mbox{{\isacharparenleft}{\isacharparenleft}{\isasymsigma},\ s{\isacharparenright},\ {\isasymnot}unicast\ dst,\ {\isacharparenleft}{\isasymsigma}{\isacharprime},\ s{\isacharprime}{\isacharparenright}{\isacharparenright}{\isasymin}T\isactrlsub A}\\\ \mbox{dst\ {\isasymnotin}\ R}\\\ \mbox{{\isasymforall}j{\isachardot}\ j\ {\isasymnoteq}\ i\ {\isasymlongrightarrow}\ {\isasymsigma}{\isacharprime}\ j\ {\isacharequal}\ {\isasymsigma}\ j}}{\mbox{{\isacharparenleft}{\isacharparenleft}{\isasymsigma},\ \NodeS\ i\ s\ R{\isacharparenright},\ {\isasymtau},\ {\isacharparenleft}{\isasymsigma}{\isacharprime},\ \NodeS\ i\ s{\isacharprime}\ R{\isacharparenright}{\isacharparenright}{\isasymin}\onodesos\ T\isactrlsub A}}}}
\snip{onode_deliver}{0}{0}{\isa{\mbox{}\inferrule{\mbox{{\isacharparenleft}{\isacharparenleft}{\isasymsigma},\ s{\isacharparenright},\ deliver\ d,\ {\isacharparenleft}{\isasymsigma}{\isacharprime},\ s{\isacharprime}{\isacharparenright}{\isacharparenright}{\isasymin}T\isactrlsub A}\\\ \mbox{{\isasymforall}j{\isachardot}\ j\ {\isasymnoteq}\ i\ {\isasymlongrightarrow}\ {\isasymsigma}{\isacharprime}\ j\ {\isacharequal}\ {\isasymsigma}\ j}}{\mbox{{\isacharparenleft}{\isacharparenleft}{\isasymsigma},\ \NodeS\ i\ s\ R{\isacharparenright},\ i{\isacharcolon}deliver{\isacharparenleft}d{\isacharparenright},\ {\isacharparenleft}{\isasymsigma}{\isacharprime},\ \NodeS\ i\ s{\isacharprime}\ R{\isacharparenright}{\isacharparenright}{\isasymin}\onodesos\ T\isactrlsub A}}}}
\snip{onode_receive}{0}{0}{\isa{\mbox{}\inferrule{\mbox{{\isacharparenleft}{\isacharparenleft}{\isasymsigma},\ s{\isacharparenright},\ receive\ m,\ {\isacharparenleft}{\isasymsigma}{\isacharprime},\ s{\isacharprime}{\isacharparenright}{\isacharparenright}{\isasymin}T\isactrlsub A}}{\mbox{{\isacharparenleft}{\isacharparenleft}{\isasymsigma},\ \NodeS\ i\ s\ R{\isacharparenright},\ {\isacharbraceleft}i{\isacharbraceright}{\isasymnot}{\isasymemptyset}{\isacharcolon}arrive{\isacharparenleft}m{\isacharparenright},\ {\isacharparenleft}{\isasymsigma}{\isacharprime},\ \NodeS\ i\ s{\isacharprime}\ R{\isacharparenright}{\isacharparenright}{\isasymin}\onodesos\ T\isactrlsub A}}}}
\snip{onode_tau}{0}{0}{\isa{\mbox{}\inferrule{\mbox{{\isacharparenleft}{\isacharparenleft}{\isasymsigma},\ s{\isacharparenright},\ {\isasymtau},\ {\isacharparenleft}{\isasymsigma}{\isacharprime},\ s{\isacharprime}{\isacharparenright}{\isacharparenright}{\isasymin}T\isactrlsub A}\\\ \mbox{{\isasymforall}j {\isasymnoteq} i{\isachardot}\ {\isasymsigma}{\isacharprime}\ j\ {\isacharequal}\ {\isasymsigma}\ j}}{\mbox{{\isacharparenleft}{\isacharparenleft}{\isasymsigma},\ \NodeS\ i\ s\ R{\isacharparenright},\ {\isasymtau},\ {\isacharparenleft}{\isasymsigma}{\isacharprime},\ \NodeS\ i\ s{\isacharprime}\ R{\isacharparenright}{\isacharparenright}{\isasymin}\onodesos\ T\isactrlsub A}}}}
\snip{onode_arrive}{0}{0}{\isa{\mbox{}\inferrule{\mbox{{\isasymsigma}{\isacharprime}\ i\ {\isacharequal}\ {\isasymsigma}\ i}}{\mbox{{\isacharparenleft}{\isacharparenleft}{\isasymsigma},\ \NodeS\ i\ s\ R{\isacharparenright},\ {\isasymemptyset}{\isasymnot}{\isacharbraceleft}i{\isacharbraceright}{\isacharcolon}arrive{\isacharparenleft}m{\isacharparenright},\ {\isacharparenleft}{\isasymsigma}{\isacharprime},\ \NodeS\ i\ s\ R{\isacharparenright}{\isacharparenright}{\isasymin}\onodesos\ T\isactrlsub A}}}}
\snip{onode_connect1}{0}{0}{\isa{\mbox{}\inferrule{\mbox{{\isasymsigma}{\isacharprime}\ i\ {\isacharequal}\ {\isasymsigma}\ i}}{\mbox{{\isacharparenleft}{\isacharparenleft}{\isasymsigma},\ \NodeS\ i\ s\ R{\isacharparenright},\ connect{\isacharparenleft}i,\ i{\isacharprime}{\isacharparenright},\ {\isacharparenleft}{\isasymsigma}{\isacharprime},\ \NodeS\ i\ s\ {R\ {\isasymunion}\ {\isacharbraceleft}i{\isacharprime}{\isacharbraceright}}{\isacharparenright}{\isacharparenright}{\isasymin}\onodesos\ T\isactrlsub A}}}}
\snip{onode_connect2}{0}{0}{\isa{\mbox{}\inferrule{\mbox{{\isasymsigma}{\isacharprime}\ i\ {\isacharequal}\ {\isasymsigma}\ i}}{\mbox{{\isacharparenleft}{\isacharparenleft}{\isasymsigma},\ \NodeS\ i\ s\ R{\isacharparenright},\ connect{\isacharparenleft}i{\isacharprime},\ i{\isacharparenright},\ {\isacharparenleft}{\isasymsigma}{\isacharprime},\ \NodeS\ i\ s\ {R\ {\isasymunion}\ {\isacharbraceleft}i{\isacharprime}{\isacharbraceright}}{\isacharparenright}{\isacharparenright}{\isasymin}\onodesos\ T\isactrlsub A}}}}
\snip{onode_disconnect1}{0}{0}{\isa{\mbox{}\inferrule{\mbox{{\isasymsigma}{\isacharprime}\ i\ {\isacharequal}\ {\isasymsigma}\ i}}{\mbox{{\isacharparenleft}{\isacharparenleft}{\isasymsigma},\ \NodeS\ i\ s\ R{\isacharparenright},\ disconnect{\isacharparenleft}i,\ i{\isacharprime}{\isacharparenright},\ {\isacharparenleft}{\isasymsigma}{\isacharprime},\ \NodeS\ i\ s\ {R\ {\isacharminus}\ {\isacharbraceleft}i{\isacharprime}{\isacharbraceright}}{\isacharparenright}{\isacharparenright}{\isasymin}\onodesos\ T\isactrlsub A}}}}
\snip{onode_disconnect2}{0}{0}{\isa{\mbox{}\inferrule{\mbox{{\isasymsigma}{\isacharprime}\ i\ {\isacharequal}\ {\isasymsigma}\ i}}{\mbox{{\isacharparenleft}{\isacharparenleft}{\isasymsigma},\ \NodeS\ i\ s\ R{\isacharparenright},\ disconnect{\isacharparenleft}i{\isacharprime},\ i{\isacharparenright},\ {\isacharparenleft}{\isasymsigma}{\isacharprime},\ \NodeS\ i\ s\ {R\ {\isacharminus}\ {\isacharbraceleft}i{\isacharprime}{\isacharbraceright}}{\isacharparenright}{\isacharparenright}{\isasymin}\onodesos\ T\isactrlsub A}}}}
\snip{onode_connect_other}{0}{0}{\isa{\mbox{}\inferrule{\mbox{i\ {\isasymnoteq}\ i{\isacharprime}}\\\ \mbox{i\ {\isasymnoteq}\ i{\isacharprime}{\isacharprime}}\\\ \mbox{{\isasymsigma}{\isacharprime}\ i\ {\isacharequal}\ {\isasymsigma}\ i}}{\mbox{{\isacharparenleft}{\isacharparenleft}{\isasymsigma},\ \NodeS\ i\ s\ R{\isacharparenright},\ connect{\isacharparenleft}i{\isacharprime},\ i{\isacharprime}{\isacharprime}{\isacharparenright},\ {\isacharparenleft}{\isasymsigma}{\isacharprime},\ \NodeS\ i\ s\ R{\isacharparenright}{\isacharparenright}{\isasymin}\onodesos\ T\isactrlsub A}}}}
\snip{onode_disconnect_other}{0}{0}{\isa{\mbox{}\inferrule{\mbox{i\ {\isasymnoteq}\ i{\isacharprime}}\\\ \mbox{i\ {\isasymnoteq}\ i{\isacharprime}{\isacharprime}}\\\ \mbox{{\isasymsigma}{\isacharprime}\ i\ {\isacharequal}\ {\isasymsigma}\ i}}{\mbox{{\isacharparenleft}{\isacharparenleft}{\isasymsigma},\ \NodeS\ i\ s\ R{\isacharparenright},\ disconnect{\isacharparenleft}i{\isacharprime},\ i{\isacharprime}{\isacharprime}{\isacharparenright},\ {\isacharparenleft}{\isasymsigma}{\isacharprime},\ \NodeS\ i\ s\ R{\isacharparenright}{\isacharparenright}{\isasymin}\onodesos\ T\isactrlsub A}}}}

\snip{onode_tau_prem_2}{0}{0}{\isa{{\isasymforall}j {\isasymnoteq} i{\isachardot}\ {\isasymsigma}{\isacharprime}\ j\ {\isacharequal}\ {\isasymsigma}\ j}}

\snip{onode_comp}{0}{0}{\isa{{\isasymlangle}i\ {\isacharcolon}\ onp\ {\isacharcolon}\ \Ri{\isasymrangle}\isactrlsub o\ {\isacharequal}\ {\isasymlparr}init\ {\isacharequal}\ {\isacharbraceleft}{\isacharparenleft}{\isasymsigma},\ \NodeS\ i\ s\ \Ri{\isacharparenright}\ {\isacharbar}\ {\isacharparenleft}{\isasymsigma},\ s{\isacharparenright}{\isasymin}init\ onp{\isacharbraceright},\ trans\ {\isacharequal}\ \onodesos\ {\isacharparenleft}trans\ onp{\isacharparenright}{\isasymrparr}}}
\snip{onode_comp_term}{0}{0}{\isa{{\isasymlangle}i\ {\isacharcolon}\ S\ {\isacharcolon}\ R{\isasymrangle}\isactrlsub o{\isasymColon}{\isacharparenleft}{\isacharparenleft}nat\ {\isasymRightarrow}\ {\isacharprime}a{\isacharparenright}\ {\isasymtimes}\ {\isacharprime}b\ net{\isacharunderscore}state,\ {\isacharprime}c\ node{\isacharunderscore}action{\isacharparenright}\ automaton}}

\snip{opnet_cast1}{0}{0}{\isa{\mbox{}\inferrule{\mbox{{\isacharparenleft}{\isacharparenleft}{\isasymsigma},\ s{\isacharparenright},\ R{\isacharcolon}{\isacharasterisk}cast{\isacharparenleft}m{\isacharparenright},\ {\isacharparenleft}{\isasymsigma}{\isacharprime},\ s{\isacharprime}{\isacharparenright}{\isacharparenright}{\isasymin}T\isactrlsub A}\\\ \mbox{{\isacharparenleft}{\isacharparenleft}{\isasymsigma},\ t{\isacharparenright},\ H{\isasymnot}K{\isacharcolon}arrive{\isacharparenleft}m{\isacharparenright},\ {\isacharparenleft}{\isasymsigma}{\isacharprime},\ t{\isacharprime}{\isacharparenright}{\isacharparenright}{\isasymin}T\isactrlsub B}\\\ \mbox{H\ {\isasymsubseteq}\ R}\\\ \mbox{K\ {\isasyminter}\ R\ {\isacharequal}\ {\isasymemptyset}}}{\mbox{{\isacharparenleft}{\isacharparenleft}{\isasymsigma},\ \subnets{s}{t}{\isacharparenright},\ R{\isacharcolon}{\isacharasterisk}cast{\isacharparenleft}m{\isacharparenright},\ {\isacharparenleft}{\isasymsigma}{\isacharprime},\ \subnets{s{\isacharprime}}{t{\isacharprime}}{\isacharparenright}{\isacharparenright}{\isasymin}\opnetsos\ T\isactrlsub A\ T\isactrlsub B}}}}
\snip{opnet_cast2}{0}{0}{\isa{\mbox{}\inferrule{\mbox{{\isacharparenleft}{\isacharparenleft}{\isasymsigma},\ s{\isacharparenright},\ H{\isasymnot}K{\isacharcolon}arrive{\isacharparenleft}m{\isacharparenright},\ {\isacharparenleft}{\isasymsigma}{\isacharprime},\ s{\isacharprime}{\isacharparenright}{\isacharparenright}{\isasymin}T\isactrlsub A}\\\ \mbox{{\isacharparenleft}{\isacharparenleft}{\isasymsigma},\ t{\isacharparenright},\ R{\isacharcolon}{\isacharasterisk}cast{\isacharparenleft}m{\isacharparenright},\ {\isacharparenleft}{\isasymsigma}{\isacharprime},\ t{\isacharprime}{\isacharparenright}{\isacharparenright}{\isasymin}T\isactrlsub B}\\\ \mbox{H\ {\isasymsubseteq}\ R}\\\ \mbox{K\ {\isasyminter}\ R\ {\isacharequal}\ {\isasymemptyset}}}{\mbox{{\isacharparenleft}{\isacharparenleft}{\isasymsigma},\ \subnets{s}{t}{\isacharparenright},\ R{\isacharcolon}{\isacharasterisk}cast{\isacharparenleft}m{\isacharparenright},\ {\isacharparenleft}{\isasymsigma}{\isacharprime},\ \subnets{s{\isacharprime}}{t{\isacharprime}}{\isacharparenright}{\isacharparenright}{\isasymin}\opnetsos\ T\isactrlsub A\ T\isactrlsub B}}}}
\snip{opnet_arrive}{0}{0}{\isa{\mbox{}\inferrule{\mbox{{\isacharparenleft}{\isacharparenleft}{\isasymsigma},\ s{\isacharparenright},\ H{\isasymnot}K{\isacharcolon}arrive{\isacharparenleft}m{\isacharparenright},\ {\isacharparenleft}{\isasymsigma}{\isacharprime},\ s{\isacharprime}{\isacharparenright}{\isacharparenright}{\isasymin}T\isactrlsub A}\\\ \mbox{{\isacharparenleft}{\isacharparenleft}{\isasymsigma},\ t{\isacharparenright},\ H{\isacharprime}{\isasymnot}K{\isacharprime}{\isacharcolon}arrive{\isacharparenleft}m{\isacharparenright},\ {\isacharparenleft}{\isasymsigma}{\isacharprime},\ t{\isacharprime}{\isacharparenright}{\isacharparenright}{\isasymin}T\isactrlsub B}}{\mbox{{\isacharparenleft}{\isacharparenleft}{\isasymsigma},\ \subnets{s}{t}{\isacharparenright},\ {\isacharparenleft}H\ {\isasymunion}\ H{\isacharprime}{\isacharparenright}{\isasymnot}{\isacharparenleft}K\ {\isasymunion}\ K{\isacharprime}{\isacharparenright}{\isacharcolon}arrive{\isacharparenleft}m{\isacharparenright},\ {\isacharparenleft}{\isasymsigma}{\isacharprime},\ \subnets{s{\isacharprime}}{t{\isacharprime}}{\isacharparenright}{\isacharparenright}{\isasymin}\opnetsos\ T\isactrlsub A\ T\isactrlsub B}}}}
\snip{opnet_deliver1}{0}{0}{\isa{\mbox{}\inferrule{\mbox{{\isacharparenleft}{\isacharparenleft}{\isasymsigma},\ s{\isacharparenright},\ i{\isacharcolon}deliver{\isacharparenleft}d{\isacharparenright},\ {\isacharparenleft}{\isasymsigma}{\isacharprime},\ s{\isacharprime}{\isacharparenright}{\isacharparenright}{\isasymin}T\isactrlsub A}}{\mbox{{\isacharparenleft}{\isacharparenleft}{\isasymsigma},\ \subnets{s}{t}{\isacharparenright},\ i{\isacharcolon}deliver{\isacharparenleft}d{\isacharparenright},\ {\isacharparenleft}{\isasymsigma}{\isacharprime},\ \subnets{s{\isacharprime}}{t}{\isacharparenright}{\isacharparenright}{\isasymin}\opnetsos\ T\isactrlsub A\ T\isactrlsub B}}}}
\snip{opnet_deliver2}{0}{0}{\isa{\mbox{}\inferrule{\mbox{{\isacharparenleft}{\isacharparenleft}{\isasymsigma},\ t{\isacharparenright},\ i{\isacharcolon}deliver{\isacharparenleft}d{\isacharparenright},\ {\isacharparenleft}{\isasymsigma}{\isacharprime},\ t{\isacharprime}{\isacharparenright}{\isacharparenright}{\isasymin}T\isactrlsub B}}{\mbox{{\isacharparenleft}{\isacharparenleft}{\isasymsigma},\ \subnets{s}{t}{\isacharparenright},\ i{\isacharcolon}deliver{\isacharparenleft}d{\isacharparenright},\ {\isacharparenleft}{\isasymsigma}{\isacharprime},\ \subnets{s}{t{\isacharprime}}{\isacharparenright}{\isacharparenright}{\isasymin}\opnetsos\ T\isactrlsub A\ T\isactrlsub B}}}}
\snip{opnet_tau1}{0}{0}{\isa{\mbox{}\inferrule{\mbox{{\isacharparenleft}{\isacharparenleft}{\isasymsigma},\ s{\isacharparenright},\ {\isasymtau},\ {\isacharparenleft}{\isasymsigma}{\isacharprime},\ s{\isacharprime}{\isacharparenright}{\isacharparenright}{\isasymin}T\isactrlsub A}}{\mbox{{\isacharparenleft}{\isacharparenleft}{\isasymsigma},\ \subnets{s}{t}{\isacharparenright},\ {\isasymtau},\ {\isacharparenleft}{\isasymsigma}{\isacharprime},\ \subnets{s{\isacharprime}}{t}{\isacharparenright}{\isacharparenright}{\isasymin}\opnetsos\ T\isactrlsub A\ T\isactrlsub B}}}}
\snip{opnet_tau2}{0}{0}{\isa{\mbox{}\inferrule{\mbox{{\isacharparenleft}{\isacharparenleft}{\isasymsigma},\ t{\isacharparenright},\ {\isasymtau},\ {\isacharparenleft}{\isasymsigma}{\isacharprime},\ t{\isacharprime}{\isacharparenright}{\isacharparenright}{\isasymin}T\isactrlsub B}}{\mbox{{\isacharparenleft}{\isacharparenleft}{\isasymsigma},\ \subnets{s}{t}{\isacharparenright},\ {\isasymtau},\ {\isacharparenleft}{\isasymsigma}{\isacharprime},\ \subnets{s}{t{\isacharprime}}{\isacharparenright}{\isacharparenright}{\isasymin}\opnetsos\ T\isactrlsub A\ T\isactrlsub B}}}}
\snip{opnet_connect}{0}{0}{\isa{\mbox{}\inferrule{\mbox{{\isacharparenleft}{\isacharparenleft}{\isasymsigma},\ s{\isacharparenright},\ connect{\isacharparenleft}i,\ i{\isacharprime}{\isacharparenright},\ {\isacharparenleft}{\isasymsigma}{\isacharprime},\ s{\isacharprime}{\isacharparenright}{\isacharparenright}{\isasymin}T\isactrlsub A}\\\ \mbox{{\isacharparenleft}{\isacharparenleft}{\isasymsigma},\ t{\isacharparenright},\ connect{\isacharparenleft}i,\ i{\isacharprime}{\isacharparenright},\ {\isacharparenleft}{\isasymsigma}{\isacharprime},\ t{\isacharprime}{\isacharparenright}{\isacharparenright}{\isasymin}T\isactrlsub B}}{\mbox{{\isacharparenleft}{\isacharparenleft}{\isasymsigma},\ \subnets{s}{t}{\isacharparenright},\ connect{\isacharparenleft}i,\ i{\isacharprime}{\isacharparenright},\ {\isacharparenleft}{\isasymsigma}{\isacharprime},\ \subnets{s{\isacharprime}}{t{\isacharprime}}{\isacharparenright}{\isacharparenright}{\isasymin}\opnetsos\ T\isactrlsub A\ T\isactrlsub B}}}}
\snip{opnet_disconnect}{0}{0}{\isa{\mbox{}\inferrule{\mbox{{\isacharparenleft}{\isacharparenleft}{\isasymsigma},\ s{\isacharparenright},\ disconnect{\isacharparenleft}i,\ i{\isacharprime}{\isacharparenright},\ {\isacharparenleft}{\isasymsigma}{\isacharprime},\ s{\isacharprime}{\isacharparenright}{\isacharparenright}{\isasymin}T\isactrlsub A}\\\ \mbox{{\isacharparenleft}{\isacharparenleft}{\isasymsigma},\ t{\isacharparenright},\ disconnect{\isacharparenleft}i,\ i{\isacharprime}{\isacharparenright},\ {\isacharparenleft}{\isasymsigma}{\isacharprime},\ t{\isacharprime}{\isacharparenright}{\isacharparenright}{\isasymin}T\isactrlsub B}}{\mbox{{\isacharparenleft}{\isacharparenleft}{\isasymsigma},\ \subnets{s}{t}{\isacharparenright},\ disconnect{\isacharparenleft}i,\ i{\isacharprime}{\isacharparenright},\ {\isacharparenleft}{\isasymsigma}{\isacharprime},\ \subnets{s{\isacharprime}}{t{\isacharprime}}{\isacharparenright}{\isacharparenright}{\isasymin}\opnetsos\ T\isactrlsub A\ T\isactrlsub B}}}}

\snip{opnet}{0}{0}{\isa{opnet\ onp\ {\isasymlangle}i{\isacharsemicolon}\ \Ri{\isasymrangle}\ {\isacharequal}\ {\isasymlangle}i\ {\isacharcolon}\ onp\ i\ {\isacharcolon}\ \Ri{\isasymrangle}\isactrlsub o\isasep\isanewline%
opnet\ onp\ {\isacharparenleft}p\isactrlsub {\isadigit{1}}\parallelcomp{}p\isactrlsub {\isadigit{2}}{\isacharparenright}\ {\isacharequal}\ {\isasymlparr}init\ {\isacharequal}\ {\isacharbraceleft}{\isacharparenleft}{\isasymsigma},\ \subnets{s\isactrlsub {\isadigit{1}}}{s\isactrlsub {\isadigit{2}}}{\isacharparenright}\ {\isacharbar}\ {\isacharparenleft}{\isasymsigma},\ s\isactrlsub {\isadigit{1}}{\isacharparenright}{\isasymin}init\ {\isacharparenleft}opnet\ onp\ p\isactrlsub {\isadigit{1}}{\isacharparenright}\ {\isasymand}\ {\isacharparenleft}{\isasymsigma},\ s\isactrlsub {\isadigit{2}}{\isacharparenright}{\isasymin}init\ {\isacharparenleft}opnet\ onp\ p\isactrlsub {\isadigit{2}}{\isacharparenright}\ {\isasymand}\ net{\isacharunderscore}ips\ s\isactrlsub {\isadigit{1}}\ {\isasyminter}\ net{\isacharunderscore}ips\ s\isactrlsub {\isadigit{2}}\ {\isacharequal}\ {\isasymemptyset}{\isacharbraceright},\ trans\ {\isacharequal}\ \opnetsos\ {\isacharparenleft}trans\ {\isacharparenleft}opnet\ onp\ p\isactrlsub {\isadigit{1}}{\isacharparenright}{\isacharparenright}\ {\isacharparenleft}trans\ {\isacharparenleft}opnet\ onp\ p\isactrlsub {\isadigit{2}}{\isacharparenright}{\isacharparenright}{\isasymrparr}}}

\snip{ocnet_connect}{0}{0}{\isa{\mbox{}\inferrule{\mbox{{\isacharparenleft}{\isacharparenleft}{\isasymsigma},\ s{\isacharparenright},\ connect{\isacharparenleft}i,\ i{\isacharprime}{\isacharparenright},\ {\isacharparenleft}{\isasymsigma}{\isacharprime},\ s{\isacharprime}{\isacharparenright}{\isacharparenright}{\isasymin}T\isactrlsub A}\\\ \mbox{{\isasymforall}j{\isachardot}\ j\ {\isasymnotin}\ net{\isacharunderscore}ips\ s\ {\isasymlongrightarrow}\ {\isasymsigma}{\isacharprime}\ j\ {\isacharequal}\ {\isasymsigma}\ j}}{\mbox{{\isacharparenleft}{\isacharparenleft}{\isasymsigma},\ s{\isacharparenright},\ connect{\isacharparenleft}i,\ i{\isacharprime}{\isacharparenright},\ {\isacharparenleft}{\isasymsigma}{\isacharprime},\ s{\isacharprime}{\isacharparenright}{\isacharparenright}{\isasymin}\ocnetsos\ T\isactrlsub A}}}}
\snip{ocnet_disconnect}{0}{0}{\isa{\mbox{}\inferrule{\mbox{{\isacharparenleft}{\isacharparenleft}{\isasymsigma},\ s{\isacharparenright},\ disconnect{\isacharparenleft}i,\ i{\isacharprime}{\isacharparenright},\ {\isacharparenleft}{\isasymsigma}{\isacharprime},\ s{\isacharprime}{\isacharparenright}{\isacharparenright}{\isasymin}T\isactrlsub A}\\\ \mbox{{\isasymforall}j{\isachardot}\ j\ {\isasymnotin}\ net{\isacharunderscore}ips\ s\ {\isasymlongrightarrow}\ {\isasymsigma}{\isacharprime}\ j\ {\isacharequal}\ {\isasymsigma}\ j}}{\mbox{{\isacharparenleft}{\isacharparenleft}{\isasymsigma},\ s{\isacharparenright},\ disconnect{\isacharparenleft}i,\ i{\isacharprime}{\isacharparenright},\ {\isacharparenleft}{\isasymsigma}{\isacharprime},\ s{\isacharprime}{\isacharparenright}{\isacharparenright}{\isasymin}\ocnetsos\ T\isactrlsub A}}}}
\snip{ocnet_cast}{0}{0}{\isa{\mbox{}\inferrule{\mbox{{\isacharparenleft}{\isacharparenleft}{\isasymsigma},\ s{\isacharparenright},\ R{\isacharcolon}{\isacharasterisk}cast{\isacharparenleft}m{\isacharparenright},\ {\isacharparenleft}{\isasymsigma}{\isacharprime},\ s{\isacharprime}{\isacharparenright}{\isacharparenright}{\isasymin}T\isactrlsub A}\\\ \mbox{{\isasymforall}j{\isachardot}\ j\ {\isasymnotin}\ net{\isacharunderscore}ips\ s\ {\isasymlongrightarrow}\ {\isasymsigma}{\isacharprime}\ j\ {\isacharequal}\ {\isasymsigma}\ j}}{\mbox{{\isacharparenleft}{\isacharparenleft}{\isasymsigma},\ s{\isacharparenright},\ {\isasymtau},\ {\isacharparenleft}{\isasymsigma}{\isacharprime},\ s{\isacharprime}{\isacharparenright}{\isacharparenright}{\isasymin}\ocnetsos\ T\isactrlsub A}}}}
\snip{ocnet_tau}{0}{0}{\isa{\mbox{}\inferrule{\mbox{{\isacharparenleft}{\isacharparenleft}{\isasymsigma},\ s{\isacharparenright},\ {\isasymtau},\ {\isacharparenleft}{\isasymsigma}{\isacharprime},\ s{\isacharprime}{\isacharparenright}{\isacharparenright}{\isasymin}T\isactrlsub A}\\\ \mbox{{\isasymforall}j{\isachardot}\ j\ {\isasymnotin}\ net{\isacharunderscore}ips\ s\ {\isasymlongrightarrow}\ {\isasymsigma}{\isacharprime}\ j\ {\isacharequal}\ {\isasymsigma}\ j}}{\mbox{{\isacharparenleft}{\isacharparenleft}{\isasymsigma},\ s{\isacharparenright},\ {\isasymtau},\ {\isacharparenleft}{\isasymsigma}{\isacharprime},\ s{\isacharprime}{\isacharparenright}{\isacharparenright}{\isasymin}\ocnetsos\ T\isactrlsub A}}}}
\snip{ocnet_deliver}{0}{0}{\isa{\mbox{}\inferrule{\mbox{{\isacharparenleft}{\isacharparenleft}{\isasymsigma},\ s{\isacharparenright},\ i{\isacharcolon}deliver{\isacharparenleft}d{\isacharparenright},\ {\isacharparenleft}{\isasymsigma}{\isacharprime},\ s{\isacharprime}{\isacharparenright}{\isacharparenright}{\isasymin}T\isactrlsub A}\\\ \mbox{{\isasymforall}j{\isachardot}\ j\ {\isasymnotin}\ net{\isacharunderscore}ips\ s\ {\isasymlongrightarrow}\ {\isasymsigma}{\isacharprime}\ j\ {\isacharequal}\ {\isasymsigma}\ j}}{\mbox{{\isacharparenleft}{\isacharparenleft}{\isasymsigma},\ s{\isacharparenright},\ i{\isacharcolon}deliver{\isacharparenleft}d{\isacharparenright},\ {\isacharparenleft}{\isasymsigma}{\isacharprime},\ s{\isacharprime}{\isacharparenright}{\isacharparenright}{\isasymin}\ocnetsos\ T\isactrlsub A}}}}
\snip{ocnet_newpkt}{0}{0}{\isa{\mbox{}\inferrule{\mbox{{\isacharparenleft}{\isacharparenleft}{\isasymsigma},\ s{\isacharparenright},\ {\isacharbraceleft}i{\isacharbraceright}{\isasymnot}K{\isacharcolon}arrive{\isacharparenleft}Newpkt\ d\ dst{\isacharparenright},\ {\isacharparenleft}{\isasymsigma}{\isacharprime},\ s{\isacharprime}{\isacharparenright}{\isacharparenright}{\isasymin}T\isactrlsub A}\\\ \mbox{{\isasymforall}j{\isachardot}\ j\ {\isasymnotin}\ net{\isacharunderscore}ips\ s\ {\isasymlongrightarrow}\ {\isasymsigma}{\isacharprime}\ j\ {\isacharequal}\ {\isasymsigma}\ j}}{\mbox{{\isacharparenleft}{\isacharparenleft}{\isasymsigma},\ s{\isacharparenright},\ i{\isacharcolon}newpkt{\isacharparenleft}d,\ dst{\isacharparenright},\ {\isacharparenleft}{\isasymsigma}{\isacharprime},\ s{\isacharprime}{\isacharparenright}{\isacharparenright}{\isasymin}\ocnetsos\ T\isactrlsub A}}}}

\snip{oclosed}{0}{0}{\isa{oclosed\ {\isacharparenleft}opnet\ {\isacharparenleft}{\isasymlambda}i{\isachardot}\ optoy\ i\ {\isasymlangle}{\isasymlangle}\isactrlbsub i\isactrlesub \ qmsg{\isacharparenright}\ n{\isacharparenright}}}
\snip{oclosed_withtype}{0}{0}{\isa{oclosed\ {\isacharparenleft}opnet\ {\isacharparenleft}{\isasymlambda}i{\isachardot}\ optoy\ i\ {\isasymlangle}{\isasymlangle}\isactrlbsub i\isactrlesub \ qmsg{\isacharparenright}\ n{\isacharparenright}{\isasymColon}{\isacharparenleft}{\isacharparenleft}nat\ {\isasymRightarrow}\ state{\isacharparenright}\ {\isasymtimes}\ {\isacharparenleft}{\isacharparenleft}state,\ msg,\ pseqp,\ pseqp\ label{\isacharparenright}\ seqp\ {\isasymtimes}\ msg\ list\ {\isasymtimes}\ {\isacharparenleft}msg\ list,\ msg,\ unit,\ unit\ label{\isacharparenright}\ seqp{\isacharparenright}\ net{\isacharunderscore}state,\ msg\ node{\isacharunderscore}action{\isacharparenright}\ automaton}}
\snip{microstep_choiceI1}{0}{0}{\isa{{\isacharparenleft}p\ {\isasymoplus}\ q{\isacharparenright}\ {\isasymleadsto}\isactrlbsub {\isasymGamma}\isactrlesub \ p}}
\snip{microstep_choiceI2}{0}{0}{\isa{{\isacharparenleft}p\ {\isasymoplus}\ q{\isacharparenright}\ {\isasymleadsto}\isactrlbsub {\isasymGamma}\isactrlesub \ q}}
\snip{microstep_callI}{0}{0}{\isa{{\isacharparenleft}call{\isacharparenleft}pn{\isacharparenright}{\isacharparenright}\ {\isasymleadsto}\isactrlbsub {\isasymGamma}\isactrlesub \ {\isasymGamma}\ pn}}

\snip{wellformed}{0}{0}{\isa{wellformed\ {\isasymGamma}\ {\isacharequal}\ wf\ {\isacharbraceleft}{\isacharparenleft}q,\ p{\isacharparenright}\ {\isacharbar}\ p\ {\isasymleadsto}\isactrlbsub {\isasymGamma}\isactrlesub \ q{\isacharbraceright}}}

\snip{sterms_termination}{0}{0}{\isa{wellformed\ {\isasymGamma}\ {\isasymLongrightarrow}\ sterms{\isacharunderscore}dom\ {\isacharparenleft}{\isasymGamma},\ p{\isacharparenright}}}
\snip{sterms_termination_concl}{0}{0}{\isa{sterms{\isacharunderscore}dom\ {\isacharparenleft}{\isasymGamma},\ p{\isacharparenright}}}

\snip{sterms_choice}{0}{0}{\isa{\mbox{}\inferrule{\mbox{wellformed\ {\isasymGamma}}}{\mbox{sterms\ {\isasymGamma}\ {\isacharparenleft}p\ {\isasymoplus}\ q{\isacharparenright}\ {\isacharequal}\ sterms\ {\isasymGamma}\ p\ {\isasymunion}\ sterms\ {\isasymGamma}\ q}}}}

\snip{sterms_call}{0}{0}{\isa{\mbox{}\inferrule{\mbox{wellformed\ {\isasymGamma}}}{\mbox{sterms\ {\isasymGamma}\ {\isacharparenleft}call{\isacharparenleft}pn{\isacharparenright}{\isacharparenright}\ {\isacharequal}\ sterms\ {\isasymGamma}\ {\isacharparenleft}{\isasymGamma}\ pn{\isacharparenright}}}}}
\snip{sterms_other1}{0}{0}{\isa{\mbox{}\inferrule{\mbox{wellformed\ {\isasymGamma}}}{\mbox{sterms\ {\isasymGamma}\ {\isacharparenleft}{\isacharbraceleft}v{\isacharbraceright}{\isasymlangle}va{\isasymrangle}\ vb{\isacharparenright}\ {\isacharequal}\ {\isacharbraceleft}{\isacharbraceleft}v{\isacharbraceright}{\isasymlangle}va{\isasymrangle}\ vb{\isacharbraceright}}}}}
\snip{sterms_other2}{0}{0}{\isa{\mbox{}\inferrule{\mbox{wellformed\ {\isasymGamma}}}{\mbox{sterms\ {\isasymGamma}\ {\isacharparenleft}{\isacharbraceleft}v{\isacharbraceright}{\isasymlbrakk}va{\isasymrbrakk}\ vb{\isacharparenright}\ {\isacharequal}\ {\isacharbraceleft}{\isacharbraceleft}v{\isacharbraceright}{\isasymlbrakk}va{\isasymrbrakk}\ vb{\isacharbraceright}}}}}

\snip{sterms_choice_concl}{0}{0}{\isa{sterms\ {\isasymGamma}\ {\isacharparenleft}p\ {\isasymoplus}\ q{\isacharparenright}\ {\isacharequal}\ sterms\ {\isasymGamma}\ p\ {\isasymunion}\ sterms\ {\isasymGamma}\ q}}
\snip{sterms_choice_concl_lhs}{0}{0}{\isa{sterms\ {\isasymGamma}\ {\isacharparenleft}p\ {\isasymoplus}\ q{\isacharparenright}}}
\snip{sterms_choice_concl_rhs}{0}{0}{\isa{sterms\ {\isasymGamma}\ p\ {\isasymunion}\ sterms\ {\isasymGamma}\ q}}

\snip{sterms_other}{0}{0}{\isa{sterms\ {\isasymGamma}\ p\ {\isacharequal}\ {\isacharbraceleft}p{\isacharbraceright}}}
\snip{sterms_other_lhs}{0}{0}{\isa{sterms\ {\isasymGamma}\ p}}
\snip{sterms_other_rhs}{0}{0}{\isa{{\isacharbraceleft}p{\isacharbraceright}}}

\snip{sterms_call_concl}{0}{0}{\isa{sterms\ {\isasymGamma}\ {\isacharparenleft}call{\isacharparenleft}pn{\isacharparenright}{\isacharparenright}\ {\isacharequal}\ sterms\ {\isasymGamma}\ {\isacharparenleft}{\isasymGamma}\ pn{\isacharparenright}}}
\snip{sterms_call_concl_lhs}{0}{0}{\isa{sterms\ {\isasymGamma}\ {\isacharparenleft}call{\isacharparenleft}pn{\isacharparenright}{\isacharparenright}}}
\snip{sterms_call_concl_rhs}{0}{0}{\isa{sterms\ {\isasymGamma}\ {\isacharparenleft}{\isasymGamma}\ pn{\isacharparenright}}}

\snip{sterms_other1_concl}{0}{0}{\isa{sterms\ {\isasymGamma}\ {\isacharparenleft}{\isacharbraceleft}v{\isacharbraceright}{\isasymlangle}va{\isasymrangle}\ vb{\isacharparenright}\ {\isacharequal}\ {\isacharbraceleft}{\isacharbraceleft}v{\isacharbraceright}{\isasymlangle}va{\isasymrangle}\ vb{\isacharbraceright}}}
\snip{sterms_other2_concl}{0}{0}{\isa{sterms\ {\isasymGamma}\ {\isacharparenleft}{\isacharbraceleft}v{\isacharbraceright}{\isasymlbrakk}va{\isasymrbrakk}\ vb{\isacharparenright}\ {\isacharequal}\ {\isacharbraceleft}{\isacharbraceleft}v{\isacharbraceright}{\isasymlbrakk}va{\isasymrbrakk}\ vb{\isacharbraceright}}}

\snip{sterms_maximal_microstep}{0}{0}{\isa{\mbox{}\inferrule{\mbox{wellformed\ {\isasymGamma}}}{\mbox{sterms\ {\isasymGamma}\ p\ {\isacharequal}\ {\isacharbraceleft}q\ {\isacharbar}\ p\ {\isasymleadsto}\isactrlbsub {\isasymGamma}\isactrlesub \isactrlsup {\isacharasterisk}\ q\ {\isasymand}\ {\isacharparenleft}{\isasymnexists}q{\isacharprime}{\isachardot}\ q\ {\isasymleadsto}\isactrlbsub {\isasymGamma}\isactrlesub \ q{\isacharprime}{\isacharparenright}{\isacharbraceright}}}}}
\snip{sterms_maximal_microstep_rhs}{0}{0}{\isa{{\isacharbraceleft}q\ {\isacharbar}\ p\ {\isasymleadsto}\isactrlbsub {\isasymGamma}\isactrlesub \isactrlsup {\isacharasterisk}\ q\ {\isasymand}\ {\isacharparenleft}{\isasymnexists}q{\isacharprime}{\isachardot}\ q\ {\isasymleadsto}\isactrlbsub {\isasymGamma}\isactrlesub \ q{\isacharprime}{\isacharparenright}{\isacharbraceright}}}

\snip{dterms_choice}{0}{0}{\isa{\mbox{}\inferrule{\mbox{wellformed\ {\isasymGamma}}}{\mbox{dterms\ {\isasymGamma}\ {\isacharparenleft}p\ {\isasymoplus}\ q{\isacharparenright}\ {\isacharequal}\ dterms\ {\isasymGamma}\ p\ {\isasymunion}\ dterms\ {\isasymGamma}\ q}}}}
\snip{dterms_call}{0}{0}{\isa{\mbox{}\inferrule{\mbox{wellformed\ {\isasymGamma}}}{\mbox{dterms\ {\isasymGamma}\ {\isacharparenleft}call{\isacharparenleft}pn{\isacharparenright}{\isacharparenright}\ {\isacharequal}\ dterms\ {\isasymGamma}\ {\isacharparenleft}{\isasymGamma}\ pn{\isacharparenright}}}}}
\snip{dterms_other1}{0}{0}{\isa{\mbox{}\inferrule{\mbox{wellformed\ {\isasymGamma}}}{\mbox{dterms\ {\isasymGamma}\ {\isacharparenleft}\gray{{\isacharbraceleft}l{\isacharbraceright}}{\isasymlangle}g{\isasymrangle}\ p{\isacharparenright}\ {\isacharequal}\ sterms\ {\isasymGamma}\ p}}}}
\snip{dterms_other2}{0}{0}{\isa{\mbox{}\inferrule{\mbox{wellformed\ {\isasymGamma}}}{\mbox{dterms\ {\isasymGamma}\ {\isacharparenleft}\gray{{\isacharbraceleft}l{\isacharbraceright}}{\isasymlbrakk}u{\isasymrbrakk}\ p{\isacharparenright}\ {\isacharequal}\ sterms\ {\isasymGamma}\ p}}}}

\snip{dterms_choice_concl}{0}{0}{\isa{dterms\ {\isasymGamma}\ {\isacharparenleft}p\ {\isasymoplus}\ q{\isacharparenright}\ {\isacharequal}\ dterms\ {\isasymGamma}\ p\ {\isasymunion}\ dterms\ {\isasymGamma}\ q}}
\snip{dterms_choice_concl_lhs}{0}{0}{\isa{dterms\ {\isasymGamma}\ {\isacharparenleft}p\ {\isasymoplus}\ q{\isacharparenright}}}
\snip{dterms_choice_concl_rhs}{0}{0}{\isa{dterms\ {\isasymGamma}\ p\ {\isasymunion}\ dterms\ {\isasymGamma}\ q}}

\snip{dterms_call_concl}{0}{0}{\isa{dterms\ {\isasymGamma}\ {\isacharparenleft}call{\isacharparenleft}pn{\isacharparenright}{\isacharparenright}\ {\isacharequal}\ dterms\ {\isasymGamma}\ {\isacharparenleft}{\isasymGamma}\ pn{\isacharparenright}}}
\snip{dterms_call_concl_lhs}{0}{0}{\isa{dterms\ {\isasymGamma}\ {\isacharparenleft}call{\isacharparenleft}pn{\isacharparenright}{\isacharparenright}}}
\snip{dterms_call_concl_rhs}{0}{0}{\isa{dterms\ {\isasymGamma}\ {\isacharparenleft}{\isasymGamma}\ pn{\isacharparenright}}}

\snip{dterms_other1_concl}{0}{0}{\isa{dterms\ {\isasymGamma}\ {\isacharparenleft}\gray{{\isacharbraceleft}l{\isacharbraceright}}{\isasymlangle}g{\isasymrangle}\ p{\isacharparenright}\ {\isacharequal}\ sterms\ {\isasymGamma}\ p}}
\snip{dterms_other1_concl_lhs}{0}{0}{\isa{dterms\ {\isasymGamma}\ {\isacharparenleft}\gray{{\isacharbraceleft}l{\isacharbraceright}}{\isasymlangle}g{\isasymrangle}\ p{\isacharparenright}}}
\snip{dterms_other1_concl_rhs}{0}{0}{\isa{sterms\ {\isasymGamma}\ p}}

\snip{dterms_other2_concl}{0}{0}{\isa{dterms\ {\isasymGamma}\ {\isacharparenleft}\gray{{\isacharbraceleft}l{\isacharbraceright}}{\isasymlbrakk}u{\isasymrbrakk}\ p{\isacharparenright}\ {\isacharequal}\ sterms\ {\isasymGamma}\ p}}
\snip{dterms_other2_concl_lhs}{0}{0}{\isa{dterms\ {\isasymGamma}\ {\isacharparenleft}\gray{{\isacharbraceleft}l{\isacharbraceright}}{\isasymlbrakk}u{\isasymrbrakk}\ p{\isacharparenright}}}
\snip{dterms_other2_concl_rhs}{0}{0}{\isa{sterms\ {\isasymGamma}\ p}}

\snip{dterms_unicast_concl}{0}{0}{\isa{dterms\ {\isasymGamma}\ {\isacharparenleft}\gray{{\isacharbraceleft}l{\isacharbraceright}}unicast{\isacharparenleft}\selip,\ \selmsg{\isacharparenright}\ {\isachardot}\ p\ {\isasymtriangleright}\ q{\isacharparenright}\ {\isacharequal}\ sterms\ {\isasymGamma}\ p\ {\isasymunion}\ sterms\ {\isasymGamma}\ q}}
\snip{dterms_unicast_concl_lhs}{0}{0}{\isa{dterms\ {\isasymGamma}\ {\isacharparenleft}\gray{{\isacharbraceleft}l{\isacharbraceright}}unicast{\isacharparenleft}\selip,\ \selmsg{\isacharparenright}\ {\isachardot}\ p\ {\isasymtriangleright}\ q{\isacharparenright}}}
\snip{dterms_unicast_concl_rhs}{0}{0}{\isa{sterms\ {\isasymGamma}\ p\ {\isasymunion}\ sterms\ {\isasymGamma}\ q}}

\snip{dterms_broadcast_concl}{0}{0}{\isa{dterms\ {\isasymGamma}\ {\isacharparenleft}\gray{{\isacharbraceleft}l{\isacharbraceright}}broadcast{\isacharparenleft}\selmsg{\isacharparenright}\ {\isachardot}\ p{\isacharparenright}\ {\isacharequal}\ sterms\ {\isasymGamma}\ p}}
\snip{dterms_broadcast_concl_lhs}{0}{0}{\isa{dterms\ {\isasymGamma}\ {\isacharparenleft}\gray{{\isacharbraceleft}l{\isacharbraceright}}broadcast{\isacharparenleft}\selmsg{\isacharparenright}\ {\isachardot}\ p{\isacharparenright}}}
\snip{dterms_broadcast_concl_rhs}{0}{0}{\isa{sterms\ {\isasymGamma}\ p}}

\snip{dterms_groupcast_concl}{0}{0}{\isa{dterms\ {\isasymGamma}\ {\isacharparenleft}\gray{{\isacharbraceleft}l{\isacharbraceright}}groupcast{\isacharparenleft}\selips,\ \selmsg{\isacharparenright}\ {\isachardot}\ p{\isacharparenright}\ {\isacharequal}\ sterms\ {\isasymGamma}\ p}}
\snip{dterms_groupcast_concl_lhs}{0}{0}{\isa{dterms\ {\isasymGamma}\ {\isacharparenleft}\gray{{\isacharbraceleft}l{\isacharbraceright}}groupcast{\isacharparenleft}\selips,\ \selmsg{\isacharparenright}\ {\isachardot}\ p{\isacharparenright}}}
\snip{dterms_groupcast_concl_rhs}{0}{0}{\isa{sterms\ {\isasymGamma}\ p}}

\snip{dterms_send_concl}{0}{0}{\isa{dterms\ {\isasymGamma}\ {\isacharparenleft}\gray{{\isacharbraceleft}l{\isacharbraceright}}send{\isacharparenleft}\selmsg{\isacharparenright}\ {\isachardot}\ p{\isacharparenright}\ {\isacharequal}\ sterms\ {\isasymGamma}\ p}}
\snip{dterms_send_concl_lhs}{0}{0}{\isa{dterms\ {\isasymGamma}\ {\isacharparenleft}\gray{{\isacharbraceleft}l{\isacharbraceright}}send{\isacharparenleft}\selmsg{\isacharparenright}\ {\isachardot}\ p{\isacharparenright}}}
\snip{dterms_send_concl_rhs}{0}{0}{\isa{sterms\ {\isasymGamma}\ p}}

\snip{dterms_deliver_concl}{0}{0}{\isa{dterms\ {\isasymGamma}\ {\isacharparenleft}\gray{{\isacharbraceleft}l{\isacharbraceright}}deliver{\isacharparenleft}\seldata{\isacharparenright}\ {\isachardot}\ p{\isacharparenright}\ {\isacharequal}\ sterms\ {\isasymGamma}\ p}}
\snip{dterms_deliver_concl_lhs}{0}{0}{\isa{dterms\ {\isasymGamma}\ {\isacharparenleft}\gray{{\isacharbraceleft}l{\isacharbraceright}}deliver{\isacharparenleft}\seldata{\isacharparenright}\ {\isachardot}\ p{\isacharparenright}}}
\snip{dterms_deliver_concl_rhs}{0}{0}{\isa{sterms\ {\isasymGamma}\ p}}

\snip{dterms_receive_concl}{0}{0}{\isa{dterms\ {\isasymGamma}\ {\isacharparenleft}\gray{{\isacharbraceleft}l{\isacharbraceright}}receive{\isacharparenleft}\updmsg{\isacharparenright}\ {\isachardot}\ p{\isacharparenright}\ {\isacharequal}\ sterms\ {\isasymGamma}\ p}}
\snip{dterms_receive_concl_lhs}{0}{0}{\isa{dterms\ {\isasymGamma}\ {\isacharparenleft}\gray{{\isacharbraceleft}l{\isacharbraceright}}receive{\isacharparenleft}\updmsg{\isacharparenright}\ {\isachardot}\ p{\isacharparenright}}}
\snip{dterms_receive_concl_rhs}{0}{0}{\isa{sterms\ {\isasymGamma}\ p}}

\snip{dterms}{0}{0}{\isa{wellformed\ {\isasymGamma}\ {\isasymLongrightarrow}\ dterms\ {\isasymGamma}\ {\isacharparenleft}\gray{{\isacharbraceleft}l{\isacharbraceright}}{\isasymlangle}g{\isasymrangle}\ p{\isacharparenright}\ {\isacharequal}\ sterms\ {\isasymGamma}\ p\isasep\isanewline%
wellformed\ {\isasymGamma}\ {\isasymLongrightarrow}\ dterms\ {\isasymGamma}\ {\isacharparenleft}\gray{{\isacharbraceleft}l{\isacharbraceright}}{\isasymlbrakk}u{\isasymrbrakk}\ p{\isacharparenright}\ {\isacharequal}\ sterms\ {\isasymGamma}\ p\isasep\isanewline%
wellformed\ {\isasymGamma}\ {\isasymLongrightarrow}\ dterms\ {\isasymGamma}\ {\isacharparenleft}p\ {\isasymoplus}\ q{\isacharparenright}\ {\isacharequal}\ dterms\ {\isasymGamma}\ p\ {\isasymunion}\ dterms\ {\isasymGamma}\ q\isasep\isanewline%
wellformed\ {\isasymGamma}\ {\isasymLongrightarrow}\ dterms\ {\isasymGamma}\ {\isacharparenleft}\gray{{\isacharbraceleft}l{\isacharbraceright}}unicast{\isacharparenleft}\selip,\ \selmsg{\isacharparenright}\ {\isachardot}\ p\ {\isasymtriangleright}\ q{\isacharparenright}\ {\isacharequal}\ sterms\ {\isasymGamma}\ p\ {\isasymunion}\ sterms\ {\isasymGamma}\ q\isasep\isanewline%
wellformed\ {\isasymGamma}\ {\isasymLongrightarrow}\ dterms\ {\isasymGamma}\ {\isacharparenleft}\gray{{\isacharbraceleft}l{\isacharbraceright}}broadcast{\isacharparenleft}\selmsg{\isacharparenright}\ {\isachardot}\ p{\isacharparenright}\ {\isacharequal}\ sterms\ {\isasymGamma}\ p\isasep\isanewline%
wellformed\ {\isasymGamma}\ {\isasymLongrightarrow}\ dterms\ {\isasymGamma}\ {\isacharparenleft}\gray{{\isacharbraceleft}l{\isacharbraceright}}groupcast{\isacharparenleft}\selips,\ \selmsg{\isacharparenright}\ {\isachardot}\ p{\isacharparenright}\ {\isacharequal}\ sterms\ {\isasymGamma}\ p\isasep\isanewline%
wellformed\ {\isasymGamma}\ {\isasymLongrightarrow}\ dterms\ {\isasymGamma}\ {\isacharparenleft}\gray{{\isacharbraceleft}l{\isacharbraceright}}send{\isacharparenleft}\selmsg{\isacharparenright}\ {\isachardot}\ p{\isacharparenright}\ {\isacharequal}\ sterms\ {\isasymGamma}\ p\isasep\isanewline%
wellformed\ {\isasymGamma}\ {\isasymLongrightarrow}\ dterms\ {\isasymGamma}\ {\isacharparenleft}\gray{{\isacharbraceleft}l{\isacharbraceright}}deliver{\isacharparenleft}\seldata{\isacharparenright}\ {\isachardot}\ p{\isacharparenright}\ {\isacharequal}\ sterms\ {\isasymGamma}\ p\isasep\isanewline%
wellformed\ {\isasymGamma}\ {\isasymLongrightarrow}\ dterms\ {\isasymGamma}\ {\isacharparenleft}\gray{{\isacharbraceleft}l{\isacharbraceright}}receive{\isacharparenleft}\updmsg{\isacharparenright}\ {\isachardot}\ p{\isacharparenright}\ {\isacharequal}\ sterms\ {\isasymGamma}\ p\isasep\isanewline%
wellformed\ {\isasymGamma}\ {\isasymLongrightarrow}\ dterms\ {\isasymGamma}\ {\isacharparenleft}call{\isacharparenleft}pn{\isacharparenright}{\isacharparenright}\ {\isacharequal}\ dterms\ {\isasymGamma}\ {\isacharparenleft}{\isasymGamma}\ pn{\isacharparenright}}}

\snip{wf_no_direct_calls}{0}{0}{\isa{\mbox{}\inferrule{\mbox{{\isasymAnd}pn{\isachardot}\ {\isasymforall}pn{\isacharprime}{\isachardot}\ call{\isacharparenleft}pn{\isacharprime}{\isacharparenright}\ {\isasymnotin}\ stermsl\ {\isacharparenleft}{\isasymGamma}\ pn{\isacharparenright}}}{\mbox{wellformed\ {\isasymGamma}}}}}
\snip{wf_no_direct_calls_prem_1}{0}{0}{\isa{call{\isacharparenleft}pn{\isacharprime}{\isacharparenright}\ {\isasymnotin}\ stermsl\ {\isacharparenleft}{\isasymGamma}\ pn{\isacharparenright}}}

\snip{cterms_SI}{0}{0}{\isa{\mbox{}\inferrule{\mbox{p{\isasymin}sterms\ {\isasymGamma}\ {\isacharparenleft}{\isasymGamma}\ pn{\isacharparenright}}}{\mbox{p{\isasymin}cterms\ {\isasymGamma}}}}}
\snip{cterms_DI}{0}{0}{\isa{\mbox{}\inferrule{\mbox{q{\isasymin}cterms\ {\isasymGamma}}\\\ \mbox{p{\isasymin}dterms\ {\isasymGamma}\ q}}{\mbox{p{\isasymin}cterms\ {\isasymGamma}}}}}

\snip{cterms_def'}{0}{0}{\isa{\mbox{}\inferrule{\mbox{wellformed\ {\isasymGamma}}}{\mbox{cterms\ {\isasymGamma}\ {\isacharequal}\ {\isacharbraceleft}p\ {\isacharbar}\ {\isasymexists}pn{\isachardot}\ p{\isasymin}ctermsl\ {\isacharparenleft}{\isasymGamma}\ pn{\isacharparenright}\ {\isasymand}\ not{\isacharunderscore}call\ p{\isacharbraceright}}}}}
\snip{cterms_def'_concl}{0}{0}{\isa{cterms\ {\isasymGamma}\ {\isacharequal}\ {\isacharbraceleft}p\ {\isacharbar}\ {\isasymexists}pn{\isachardot}\ p{\isasymin}ctermsl\ {\isacharparenleft}{\isasymGamma}\ pn{\isacharparenright}\ {\isasymand}\ not{\isacharunderscore}call\ p{\isacharbraceright}}}

\snip{cterms_subterms}{0}{0}{\isa{\mbox{}\inferrule{\mbox{wellformed\ {\isasymGamma}}}{\mbox{cterms\ {\isasymGamma}\ {\isacharequal}\ {\isacharbraceleft}p\ {\isacharbar}\ {\isasymexists}pn{\isachardot}\ p{\isasymin}subterms\ {\isacharparenleft}{\isasymGamma}\ pn{\isacharparenright}\ {\isasymand}\ not{\isacharunderscore}call\ p\ {\isasymand}\ not{\isacharunderscore}choice\ p{\isacharbraceright}}}}}  
\snip{cterms_subterms_concl}{0}{0}{\isa{cterms\ {\isasymGamma}\ {\isacharequal}\ {\isacharbraceleft}p\ {\isacharbar}\ {\isasymexists}pn{\isachardot}\ p{\isasymin}subterms\ {\isacharparenleft}{\isasymGamma}\ pn{\isacharparenright}\ {\isasymand}\ not{\isacharunderscore}call\ p\ {\isasymand}\ not{\isacharunderscore}choice\ p{\isacharbraceright}}}

\snip{ctermsl_choice}{0}{0}{\isa{ctermsl\ {\isacharparenleft}p\ {\isasymoplus}\ q{\isacharparenright}\ {\isacharequal}\ ctermsl\ p\ {\isasymunion}\ ctermsl\ q}}
\snip{ctermsl_choice_lhs}{0}{0}{\isa{ctermsl\ {\isacharparenleft}p\ {\isasymoplus}\ q{\isacharparenright}}}
\snip{ctermsl_choice_rhs}{0}{0}{\isa{ctermsl\ p\ {\isasymunion}\ ctermsl\ q}}

\snip{ctermsl_call}{0}{0}{\isa{ctermsl\ {\isacharparenleft}call{\isacharparenleft}pn{\isacharparenright}{\isacharparenright}\ {\isacharequal}\ {\isacharbraceleft}call{\isacharparenleft}pn{\isacharparenright}{\isacharbraceright}}}
\snip{ctermsl_call_lhs}{0}{0}{\isa{ctermsl\ {\isacharparenleft}call{\isacharparenleft}pn{\isacharparenright}{\isacharparenright}}}
\snip{ctermsl_call_rhs}{0}{0}{\isa{{\isacharbraceleft}call{\isacharparenleft}pn{\isacharparenright}{\isacharbraceright}}}

\snip{ctermsl_other1}{0}{0}{\isa{ctermsl\ {\isacharparenleft}\gray{{\isacharbraceleft}l{\isacharbraceright}}{\isasymlangle}g{\isasymrangle}\ p{\isacharparenright}\ {\isacharequal}\ {\isacharbraceleft}{\isacharbraceleft}l{\isacharbraceright}{\isasymlangle}g{\isasymrangle}\ p{\isacharbraceright}\ {\isasymunion}\ ctermsl\ p}}
\snip{ctermsl_other1_lhs}{0}{0}{\isa{ctermsl\ {\isacharparenleft}\gray{{\isacharbraceleft}l{\isacharbraceright}}{\isasymlangle}g{\isasymrangle}\ p{\isacharparenright}}}
\snip{ctermsl_other1_rhs}{0}{0}{\isa{{\isacharbraceleft}\gray{{\isacharbraceleft}l{\isacharbraceright}}{\isasymlangle}g{\isasymrangle}\ p{\isacharbraceright}\ {\isasymunion}\ ctermsl\ p}}

\snip{ctermsl_other2}{0}{0}{\isa{ctermsl\ {\isacharparenleft}\gray{{\isacharbraceleft}l{\isacharbraceright}}{\isasymlbrakk}u{\isasymrbrakk}\ p{\isacharparenright}\ {\isacharequal}\ {\isacharbraceleft}{\isacharbraceleft}l{\isacharbraceright}{\isasymlbrakk}u{\isasymrbrakk}\ p{\isacharbraceright}\ {\isasymunion}\ ctermsl\ p}}
\snip{ctermsl_other2_lhs}{0}{0}{\isa{ctermsl\ {\isacharparenleft}\gray{{\isacharbraceleft}l{\isacharbraceright}}{\isasymlbrakk}u{\isasymrbrakk}\ p{\isacharparenright}}}
\snip{ctermsl_other2_rhs}{0}{0}{\isa{{\isacharbraceleft}\gray{{\isacharbraceleft}l{\isacharbraceright}}{\isasymlbrakk}u{\isasymrbrakk}\ p{\isacharbraceright}\ {\isasymunion}\ ctermsl\ p}}

\snip{ctermsl_unicast}{0}{0}{\isa{ctermsl\ {\isacharparenleft}\gray{{\isacharbraceleft}l{\isacharbraceright}}broadcast{\isacharparenleft}\selmsg{\isacharparenright}\ {\isachardot}\ p{\isacharparenright}\ {\isacharequal}\ {\isacharbraceleft}{\isacharbraceleft}l{\isacharbraceright}broadcast{\isacharparenleft}\selmsg{\isacharparenright}\ {\isachardot}\ p{\isacharbraceright}\ {\isasymunion}\ ctermsl\ p}}

\snip{ctermsl_unicast_lhs}{0}{0}{\isa{ctermsl\ {\isacharparenleft}\gray{{\isacharbraceleft}l{\isacharbraceright}}unicast{\isacharparenleft}\selip,\ \selmsg{\isacharparenright}\ {\isachardot}\ p\ {\isasymtriangleright}\ q{\isacharparenright}}}
\snip{ctermsl_unicast_rhs}{0}{0}{\isa{{\isacharbraceleft}\gray{{\isacharbraceleft}l{\isacharbraceright}}unicast{\isacharparenleft}\selip,\ \selmsg{\isacharparenright}\ {\isachardot}\ p\ {\isasymtriangleright}\ q{\isacharbraceright}\ {\isasymunion}\ {\isacharparenleft}ctermsl\ p\ {\isasymunion}\ ctermsl\ q{\isacharparenright}}}

\snip{ctermsl_broadcast_lhs}{0}{0}{\isa{ctermsl\ {\isacharparenleft}\gray{{\isacharbraceleft}l{\isacharbraceright}}broadcast{\isacharparenleft}\selmsg{\isacharparenright}\ {\isachardot}\ p{\isacharparenright}}}
\snip{ctermsl_broadcast_rhs}{0}{0}{\isa{{\isacharbraceleft}\gray{{\isacharbraceleft}l{\isacharbraceright}}broadcast{\isacharparenleft}\selmsg{\isacharparenright}\ {\isachardot}\ p{\isacharbraceright}\ {\isasymunion}\ ctermsl\ p}}

\snip{ctermsl_groupcast_lhs}{0}{0}{\isa{ctermsl\ {\isacharparenleft}\gray{{\isacharbraceleft}l{\isacharbraceright}}groupcast{\isacharparenleft}\selips,\ \selmsg{\isacharparenright}\ {\isachardot}\ p{\isacharparenright}}}
\snip{ctermsl_groupcast_rhs}{0}{0}{\isa{{\isacharbraceleft}\gray{{\isacharbraceleft}l{\isacharbraceright}}groupcast{\isacharparenleft}\selips,\ \selmsg{\isacharparenright}\ {\isachardot}\ p{\isacharbraceright}\ {\isasymunion}\ ctermsl\ p}}

\snip{ctermsl_send_lhs}{0}{0}{\isa{ctermsl\ {\isacharparenleft}\gray{{\isacharbraceleft}l{\isacharbraceright}}send{\isacharparenleft}\selmsg{\isacharparenright}\ {\isachardot}\ p{\isacharparenright}}}
\snip{ctermsl_send_rhs}{0}{0}{\isa{{\isacharbraceleft}\gray{{\isacharbraceleft}l{\isacharbraceright}}send{\isacharparenleft}\selmsg{\isacharparenright}\ {\isachardot}\ p{\isacharbraceright}\ {\isasymunion}\ ctermsl\ p}}

\snip{ctermsl_deliver_lhs}{0}{0}{\isa{ctermsl\ {\isacharparenleft}\gray{{\isacharbraceleft}l{\isacharbraceright}}deliver{\isacharparenleft}\seldata{\isacharparenright}\ {\isachardot}\ p{\isacharparenright}}}
\snip{ctermsl_deliver_rhs}{0}{0}{\isa{{\isacharbraceleft}\gray{{\isacharbraceleft}l{\isacharbraceright}}deliver{\isacharparenleft}\seldata{\isacharparenright}\ {\isachardot}\ p{\isacharbraceright}\ {\isasymunion}\ ctermsl\ p}}

\snip{ctermsl_receive_lhs}{0}{0}{\isa{ctermsl\ {\isacharparenleft}\gray{{\isacharbraceleft}l{\isacharbraceright}}receive{\isacharparenleft}\updmsg{\isacharparenright}\ {\isachardot}\ p{\isacharparenright}}}
\snip{ctermsl_receive_rhs}{0}{0}{\isa{{\isacharbraceleft}\gray{{\isacharbraceleft}l{\isacharbraceright}}receive{\isacharparenleft}\updmsg{\isacharparenright}\ {\isachardot}\ p{\isacharbraceright}\ {\isasymunion}\ ctermsl\ p}}

\snip{stermsl_ctermsl}{0}{0}{\isa{\mbox{}\inferrule{\mbox{q{\isasymin}stermsl\ p}}{\mbox{q{\isasymin}ctermsl\ p}}}}
\snip{stermsl_ctermsl_prem}{0}{0}{\isa{q{\isasymin}stermsl\ p}}
\snip{stermsl_ctermsl_concl}{0}{0}{\isa{q{\isasymin}ctermsl\ p}}

\snip{stermsl}{0}{0}{\isa{stermsl}}
\snip{stermsl_choice}{0}{0}{\isa{stermsl\ {\isacharparenleft}p\isactrlsub {\isadigit{1}}\ {\isasymoplus}\ p\isactrlsub {\isadigit{2}}{\isacharparenright}\ {\isacharequal}\ stermsl\ p\isactrlsub {\isadigit{1}}\ {\isasymunion}\ stermsl\ p\isactrlsub {\isadigit{2}}}}
\snip{stermsl_other}{0}{0}{\isa{stermsl\ p\ {\isacharequal}\ {\isacharbraceleft}p{\isacharbraceright}}}

\snip{seq_reachable_in_cterms_prem_1}{0}{0}{\isa{wellformed\ {\isasymGamma}}}
\snip{seq_reachable_in_cterms_prem_2}{0}{0}{\isa{control{\isacharunderscore}within\ {\isasymGamma}\ {\isacharparenleft}init\ A{\isacharparenright}}}
\snip{seq_reachable_in_cterms_prem_3}{0}{0}{\isa{trans\ A\ {\isacharequal}\ \seqpsos\ {\isasymGamma}}}
\snip{seq_reachable_in_cterms_prem_4}{0}{0}{\isa{{\isacharparenleft}{\isasymxi},\ p{\isacharparenright}{\isasymin}reachable\ A\ I}}
\snip{seq_reachable_in_cterms_prem_5}{0}{0}{\isa{q{\isasymin}sterms\ {\isasymGamma}\ p}}
\snip{seq_reachable_in_cterms_concl}{0}{0}{\isa{q{\isasymin}cterms\ {\isasymGamma}}}

\snip{control_within}{0}{0}{\isa{control{\isacharunderscore}within\ {\isasymGamma}\ Z\ {\isacharequal}\ {\isasymforall}{\isacharparenleft}{\isasymxi},\ p{\isacharparenright}{\isasymin}Z{\isachardot}\ {\isasymexists}pn{\isachardot}\ p{\isasymin}subterms\ {\isacharparenleft}{\isasymGamma}\ pn{\isacharparenright}}}
\snip{simple_labels}{0}{0}{\isa{simple{\isacharunderscore}labels\ {\isasymGamma}\ {\isacharequal}\ {\isasymforall}pn{\isachardot}\ {\isasymforall}p{\isasymin}subterms\ {\isacharparenleft}{\isasymGamma}\ pn{\isacharparenright}{\isachardot}\ {\isasymexists}\hspace{-.2em}{\isacharbang}\hspace{.15em}l{\isachardot}\ labels\ {\isasymGamma}\ p\ {\isacharequal}\ \gray{{\isacharbraceleft}l{\isacharbraceright}}}}
\snip{not_simple_labels_eg}{0}{0}{\isa{{\isasymGamma}\isactrlsub c\ Q\ {\isacharequal}\ \gray{{\isacharbraceleft}Q{\isacharminus}{\isacharcolon}{\isadigit{1}}{\isacharbraceright}}{\isasymlbrakk}f{\isasymrbrakk}\ call{\isacharparenleft}Q{\isacharparenright}\ {\isasymoplus}\ \gray{{\isacharbraceleft}Q{\isacharminus}{\isacharcolon}{\isadigit{2}}{\isacharbraceright}}{\isasymlbrakk}g{\isasymrbrakk}\ call{\isacharparenleft}Q{\isacharparenright}}}

\snip{seq_invariant_ctermsI_prem_1}{0}{0}{\isa{wellformed\ {\isasymGamma}}}
\snip{seq_invariant_ctermsI_prem_2}{0}{0}{\isa{control{\isacharunderscore}within\ {\isasymGamma}\ {\isacharparenleft}init\ A{\isacharparenright}}}
\snip{seq_invariant_ctermsI_prem_3}{0}{0}{\isa{simple{\isacharunderscore}labels\ {\isasymGamma}}}
\snip{seq_invariant_ctermsI_prem_4}{0}{0}{\isa{trans\ A\ {\isacharequal}\ \seqpsos\ {\isasymGamma}}}

\snip{seq_invariant_ctermsI_init}{0}{0}{\isa{{\isasymAnd}{\isasymxi}\ p\ l{\isachardot}\ \mbox{}\inferrule{\mbox{{\isacharparenleft}{\isasymxi},\ p{\isacharparenright}{\isasymin}init\ A}\\\ \mbox{l{\isasymin}labels\ {\isasymGamma}\ p}}{\mbox{P\ {\isacharparenleft}{\isasymxi},\ l{\isacharparenright}}}}}
\snip{seq_invariant_ctermsI_init_1}{0}{0}{\isa{{\isacharparenleft}{\isasymxi},\ p{\isacharparenright}{\isasymin}init\ A}}
\snip{seq_invariant_ctermsI_init_2}{0}{0}{\isa{l{\isasymin}labels\ {\isasymGamma}\ p}}
\snip{seq_invariant_ctermsI_init_3}{0}{0}{\isa{P\ {\isacharparenleft}{\isasymxi},\ l{\isacharparenright}}}

\snip{seq_invariant_ctermsI_trans_1}{0}{0}{\isa{p{\isasymin}ctermsl\ {\isacharparenleft}{\isasymGamma}\ pn{\isacharparenright}}}
\snip{seq_invariant_ctermsI_trans_2}{0}{0}{\isa{not{\isacharunderscore}call\ p}}
\snip{seq_invariant_ctermsI_trans_3}{0}{0}{\isa{l{\isasymin}labels\ {\isasymGamma}\ p}}
\snip{seq_invariant_ctermsI_trans_4}{0}{0}{\isa{P\ {\isacharparenleft}{\isasymxi},\ l{\isacharparenright}}}
\snip{seq_invariant_ctermsI_trans_5}{0}{0}{\isa{{\isacharparenleft}{\isacharparenleft}{\isasymxi},\ p{\isacharparenright},\ a,\ {\isacharparenleft}{\isasymxi}{\isacharprime},\ q{\isacharparenright}{\isacharparenright}{\isasymin}\seqpsos\ {\isasymGamma}}}
\snip{seq_invariant_ctermsI_trans_6}{0}{0}{\isa{l{\isacharprime}{\isasymin}labels\ {\isasymGamma}\ q}}
\snip{seq_invariant_ctermsI_trans_7}{0}{0}{\isa{{\isacharparenleft}{\isasymxi},\ pp{\isacharparenright}{\isasymin}reachable\ A\ I}}
\snip{seq_invariant_ctermsI_trans_8}{0}{0}{\isa{p{\isasymin}sterms\ {\isasymGamma}\ pp}}
\snip{seq_invariant_ctermsI_trans_9}{0}{0}{\isa{I\ a}}
\snip{seq_invariant_ctermsI_trans_10}{0}{0}{\isa{P\ {\isacharparenleft}{\isasymxi}{\isacharprime},\ l{\isacharprime}{\isacharparenright}}}

\snip{seq_invariant_ctermsI_concl}{0}{0}{\isa{A\ {\isasymTTurnstile}\ {\isacharparenleft}I\ {\isasymrightarrow}{\isacharparenright}\ onl\ {\isasymGamma}\ P}}

\snip{oseq_invariant_ctermsI_trans_1}{0}{0}{\isa{{\isacharparenleft}{\isasymsigma},\ p{\isacharparenright}{\isasymin}oreachable\ A\ S\ U}}
\snip{oseq_invariant_ctermsI_trans_2}{0}{0}{\isa{l{\isasymin}labels\ {\isasymGamma}\ p}}
\snip{oseq_invariant_ctermsI_trans_3}{0}{0}{\isa{P\ {\isacharparenleft}{\isasymsigma},\ l{\isacharparenright}}}
\snip{oseq_invariant_ctermsI_trans_4}{0}{0}{\isa{U\ {\isasymsigma}\ {\isasymsigma}{\isacharprime}}}
\snip{oseq_invariant_ctermsI_trans_5}{0}{0}{\isa{P\ {\isacharparenleft}{\isasymsigma}{\isacharprime},\ l{\isacharparenright}}}

\snip{oseq_invariant_ctermsI_concl}{0}{0}{\isa{A\ {\isasymTurnstile}\ {\isacharparenleft}S,\ U\ {\isasymrightarrow}{\isacharparenright}\ onl\ {\isasymGamma}\ P}}

\snip{seq_step_invariant_ctermsI_prem_1}{0}{0}{\isa{wellformed\ {\isasymGamma}}}
\snip{seq_step_invariant_ctermsI_prem_2}{0}{0}{\isa{control{\isacharunderscore}within\,{\isasymGamma}\,\!{\isacharparenleft}init\ A{\isacharparenright}}}
\snip{seq_step_invariant_ctermsI_prem_3}{0}{0}{\isa{simple{\isacharunderscore}labels\ {\isasymGamma}}}
\snip{seq_step_invariant_ctermsI_prem_4}{0}{0}{\isa{trans\ A\ {\isacharequal}\ \seqpsos\;{\isasymGamma}}}

\snip{seq_step_invariant_ctermsI_trans_1}{0}{0}{\isa{p{\isasymin}ctermsl\ {\isacharparenleft}{\isasymGamma}\ pn{\isacharparenright}}}
\snip{seq_step_invariant_ctermsI_trans_2}{0}{0}{\isa{not{\isacharunderscore}call\ p}}
\snip{seq_step_invariant_ctermsI_trans_3}{0}{0}{\isa{l{\isasymin}labels\ {\isasymGamma}\ p}}
\snip{seq_step_invariant_ctermsI_trans_4}{0}{0}{\isa{{\isacharparenleft}{\isacharparenleft}{\isasymxi},\ p{\isacharparenright},\ a,\ {\isacharparenleft}{\isasymxi}{\isacharprime},\ q{\isacharparenright}{\isacharparenright}{\isasymin}\seqpsos\ {\isasymGamma}}}
\snip{seq_step_invariant_ctermsI_trans_5}{0}{0}{\isa{l{\isacharprime}{\isasymin}labels\ {\isasymGamma}\ q}}
\snip{seq_step_invariant_ctermsI_trans_6}{0}{0}{\isa{{\isacharparenleft}{\isasymxi},\ pp{\isacharparenright}{\isasymin}reachable\ A\ I}}
\snip{seq_step_invariant_ctermsI_trans_7}{0}{0}{\isa{p{\isasymin}sterms\ {\isasymGamma}\ pp}}
\snip{seq_step_invariant_ctermsI_trans_8}{0}{0}{\isa{I\ a}}
\snip{seq_step_invariant_ctermsI_trans_9}{0}{0}{\isa{P\ {\isacharparenleft}{\isacharparenleft}{\isasymxi},\ l{\isacharparenright},\ a,\ {\isacharparenleft}{\isasymxi}{\isacharprime},\ l{\isacharprime}{\isacharparenright}{\isacharparenright}}}

\snip{seq_step_invariant_ctermsI_concl}{0}{0}{\isa{A\ \stepinv\ {\isacharparenleft}I\ {\isasymrightarrow}{\isacharparenright}\ onll\ {\isasymGamma}\ P}}
\snip{seq_invariant_ctermI}{0}{0}{\begin{isabelle}%
{\isasymlbrakk}wellformed\ {\isasymGamma}{\isacharsemicolon}\ control{\isacharunderscore}within\ {\isasymGamma}\ {\isacharparenleft}init\ A{\isacharparenright}{\isacharsemicolon}\ simple{\isacharunderscore}labels\ {\isasymGamma}{\isacharsemicolon}\isanewline
\isaindent{\ }trans\ A\ {\isacharequal}\ \seqpsos\ {\isasymGamma}{\isacharsemicolon}\isanewline
\isaindent{\ }{\isasymAnd}{\isasymxi}\ p\ l{\isachardot}\ {\isasymlbrakk}{\isacharparenleft}{\isasymxi},\ p{\isacharparenright}{\isasymin}init\ A{\isacharsemicolon}\ l{\isasymin}labels\ {\isasymGamma}\ p{\isasymrbrakk}\ {\isasymLongrightarrow}\ P\ {\isacharparenleft}{\isasymxi},\ l{\isacharparenright}{\isacharsemicolon}\isanewline
\isaindent{\ }{\isasymAnd}p\ l\ {\isasymxi}\ a\ q\ l{\isacharprime}\ {\isasymxi}{\isacharprime}\ pp{\isachardot}\isanewline
\isaindent{\ \ \ \ }{\isasymlbrakk}p{\isasymin}cterms\ {\isasymGamma}{\isacharsemicolon}\ l{\isasymin}labels\ {\isasymGamma}\ p{\isacharsemicolon}\ P\ {\isacharparenleft}{\isasymxi},\ l{\isacharparenright}{\isacharsemicolon}\isanewline
\isaindent{\ \ \ \ \ }{\isacharparenleft}{\isacharparenleft}{\isasymxi},\ p{\isacharparenright},\ a,\ {\isacharparenleft}{\isasymxi}{\isacharprime},\ q{\isacharparenright}{\isacharparenright}{\isasymin}\seqpsos\ {\isasymGamma}{\isacharsemicolon}\isanewline
\isaindent{\ \ \ \ \ }{\isacharparenleft}{\isacharparenleft}{\isasymxi},\ p{\isacharparenright},\ a,\ {\isasymxi}{\isacharprime},\ q{\isacharparenright}{\isasymin}trans\ A{\isacharsemicolon}\ l{\isacharprime}{\isasymin}labels\ {\isasymGamma}\ q{\isacharsemicolon}\isanewline
\isaindent{\ \ \ \ \ }{\isacharparenleft}{\isasymxi},\ pp{\isacharparenright}{\isasymin}reachable\ A\ I{\isacharsemicolon}\ p{\isasymin}sterms\ {\isasymGamma}\ pp{\isacharsemicolon}\ {\isacharparenleft}{\isasymxi}{\isacharprime},\ q{\isacharparenright}{\isasymin}reachable\ A\ I{\isacharsemicolon}\ I\ a{\isasymrbrakk}\isanewline
\isaindent{\ \ \ \ }{\isasymLongrightarrow}\ P\ {\isacharparenleft}{\isasymxi}{\isacharprime},\ l{\isacharprime}{\isacharparenright}{\isasymrbrakk}\isanewline
{\isasymLongrightarrow}\ A\ {\isasymTTurnstile}\ {\isacharparenleft}I\ {\isasymrightarrow}{\isacharparenright}\ onl\ {\isasymGamma}\ P%
\end{isabelle}}
\snip{seq_invariant_ctermI_wf}{0}{0}{\isa{wellformed\ {\isasymGamma}}}
\snip{seq_invariant_ctermI_cw}{0}{0}{\isa{control{\isacharunderscore}within\ {\isasymGamma}\ {\isacharparenleft}init\ A{\isacharparenright}}}
\snip{seq_invariant_ctermI_sl}{0}{0}{\isa{simple{\isacharunderscore}labels\ {\isasymGamma}}}
\snip{seq_invariant_ctermI_init}{0}{0}{\begin{isabelle}%
trans\ A\ {\isacharequal}\ \seqpsos\ {\isasymGamma}%
\end{isabelle}}
\snip{seq_invariant_ctermI_step}{0}{0}{\begin{isabelle}%
{\isasymAnd}{\isasymxi}\ p\ l{\isachardot}\ {\isasymlbrakk}{\isacharparenleft}{\isasymxi},\ p{\isacharparenright}{\isasymin}init\ A{\isacharsemicolon}\ l{\isasymin}labels\ {\isasymGamma}\ p{\isasymrbrakk}\ {\isasymLongrightarrow}\ P\ {\isacharparenleft}{\isasymxi},\ l{\isacharparenright}%
\end{isabelle}}
\snip{seq_invariant_ctermI_concl}{0}{0}{\isa{A\ {\isasymTTurnstile}\ {\isacharparenleft}I\ {\isasymrightarrow}{\isacharparenright}\ onl\ {\isasymGamma}\ P}}

\snip{seq_reachable_in_cterms}{0}{0}{\isa{\mbox{}\inferrule{\mbox{wellformed\ {\isasymGamma}}\\\ \mbox{control{\isacharunderscore}within\ {\isasymGamma}\ {\isacharparenleft}init\ A{\isacharparenright}}\\\ \mbox{trans\ A\ {\isacharequal}\ \seqpsos\ {\isasymGamma}}\\\ \mbox{{\isacharparenleft}{\isasymxi},\ p{\isacharparenright}{\isasymin}reachable\ A\ I}\\\ \mbox{p{\isacharprime}{\isasymin}sterms\ {\isasymGamma}\ p}}{\mbox{p{\isacharprime}{\isasymin}cterms\ {\isasymGamma}}}}}
\snip{open_seq_step_invariant}{0}{0}{\isa{\mbox{}\inferrule{\mbox{A\ \stepinv\ {\isacharparenleft}I\ {\isasymrightarrow}{\isacharparenright}\ P}\\\ \mbox{initiali\ i\ {\isacharparenleft}init\ OA{\isacharparenright}\ {\isacharparenleft}init\ A{\isacharparenright}}\\\ \mbox{trans\ OA\ {\isacharequal}\ \oseqpsos\ {\isasymGamma}\ i}\\\ \mbox{trans\ A\ {\isacharequal}\ \seqpsos\ {\isasymGamma}}}{\mbox{OA\ \ostepinv\ {\isacharparenleft}act\ I,\ other\ ANY\ {\isacharbraceleft}i{\isacharbraceright}\ {\isasymrightarrow}{\isacharparenright}\ seqll\ i\ P}}}}

\snip{open_seq_invariant}{0}{0}{\isa{\mbox{}\inferrule{\mbox{A\ {\isasymTTurnstile}\ {\isacharparenleft}I\ {\isasymrightarrow}{\isacharparenright}\ P}\\\ \mbox{initiali\ i\ {\isacharparenleft}init\ OA{\isacharparenright}\ {\isacharparenleft}init\ A{\isacharparenright}}\\\ \mbox{trans\ OA\ {\isacharequal}\ \oseqpsos\ {\isasymGamma}\ i}\\\ \mbox{trans\ A\ {\isacharequal}\ \seqpsos\ {\isasymGamma}}}{\mbox{OA\ {\isasymTurnstile}\ {\isacharparenleft}act\ I,\ other\ ANY\ {\isacharbraceleft}i{\isacharbraceright}\ {\isasymrightarrow}{\isacharparenright}\ seql\ i\ P}}}}
\snip{open_seq_invariant_prem_1}{0}{0}{\isa{A\ {\isasymTTurnstile}\ {\isacharparenleft}I\ {\isasymrightarrow}{\isacharparenright}\ P}}
\snip{open_seq_invariant_prem_2}{0}{0}{\isa{initiali\ i\ {\isacharparenleft}init\ A{\isacharprime}{\isacharparenright}\ {\isacharparenleft}init\ A{\isacharparenright}}}
\snip{open_seq_invariant_prem_2'}{0}{0}{\isa{init\ A\ {\isacharequal}\ {\isacharbraceleft}{\isacharparenleft}{\isasymsigma}\ i,\ p{\isacharparenright}\ {\isacharbar}\ {\isacharparenleft}{\isasymsigma},\ p{\isacharparenright}{\isasymin}init\ A{\isacharprime}{\isacharbraceright}}}
\snip{open_seq_invariant_prem_3}{0}{0}{\isa{trans\ A{\isacharprime}\ {\isacharequal}\ \oseqpsos\ {\isasymGamma}\ i}}
\snip{open_seq_invariant_prem_4}{0}{0}{\isa{trans\ A\ {\isacharequal}\ \seqpsos\ {\isasymGamma}}}
\snip{open_seq_invariant_concl}{0}{0}{\isa{A{\isacharprime}\ {\isasymTurnstile}\ {\isacharparenleft}act\ I,\ other\ F\ {\isacharbraceleft}i{\isacharbraceright}\ {\isasymrightarrow}{\isacharparenright}\ seql\ i\ P}}
\snip{open_seq_invariant_concl'}{0}{0}{\isa{A{\isacharprime}\ {\isasymTurnstile}\ {\isacharparenleft}{\isasymlambda}{\isacharunderscore}\ {\isacharunderscore}{\isachardot}\ I,\ other\ F\ {\isacharbraceleft}i{\isacharbraceright}\ {\isasymrightarrow}{\isacharparenright}\ {\isacharparenleft}{\isasymlambda}{\isacharparenleft}{\isasymsigma},\ p{\isacharparenright}{\isachardot}\ P\ {\isacharparenleft}{\isasymsigma}\ i,\ p{\isacharparenright}{\isacharparenright}}}

\snip{oinvariant}{0}{0}{\begin{isabelle}%
{\isacharparenleft}A\ {\isasymTurnstile}\ {\isacharparenleft}S,\ U\ {\isasymrightarrow}{\isacharparenright}\ P{\isacharparenright}\ {\isacharequal}\ {\isacharparenleft}{\isasymforall}s{\isasymin}oreachable\ A\ S\ U{\isachardot}\ P\ s{\isacharparenright}%
\end{isabelle}}
\snip{oinvariant_lhs}{0}{0}{\isa{A\ {\isasymTurnstile}\ {\isacharparenleft}S,\ U\ {\isasymrightarrow}{\isacharparenright}\ P}}
\snip{oinvariant_rhs}{0}{0}{\isa{{\isasymforall}s{\isasymin}oreachable\ A\ S\ U{\isachardot}\ P\ s}}
\snip{ostep_invariant}{0}{0}{\begin{isabelle}%
{\isacharparenleft}A\ \ostepinv\ {\isacharparenleft}S,\ U\ {\isasymrightarrow}{\isacharparenright}\ P{\isacharparenright}\ {\isacharequal}\isanewline
{\isacharparenleft}{\isasymforall}s{\isasymin}oreachable\ A\ S\ U{\isachardot}\isanewline
\isaindent{{\isacharparenleft}\ \ \ }{\isasymforall}a\ s{\isacharprime}{\isachardot}\isanewline
\isaindent{{\isacharparenleft}\ \ \ \ \ \ }{\isacharparenleft}s,\ a,\ s{\isacharprime}{\isacharparenright}{\isasymin}trans\ A\ {\isasymand}\ S\ {\isacharparenleft}fst\ s{\isacharparenright}\ {\isacharparenleft}fst\ s{\isacharprime}{\isacharparenright}\ a\ {\isasymlongrightarrow}\isanewline
\isaindent{{\isacharparenleft}\ \ \ \ \ \ }P\ {\isacharparenleft}s,\ a,\ s{\isacharprime}{\isacharparenright}{\isacharparenright}%
\end{isabelle}}
\snip{ostep_invariant_lhs}{0}{0}{\isa{A\ \ostepinv\ {\isacharparenleft}S,\ U\ {\isasymrightarrow}{\isacharparenright}\ P}}
\snip{ostep_invariant_rhs}{0}{0}{\isa{{\isasymforall}s{\isasymin}oreachable\ A\ S\ U{\isachardot}\ {\isasymforall}a\ s{\isacharprime}{\isachardot}\ {\isacharparenleft}s,\ a,\ s{\isacharprime}{\isacharparenright}{\isasymin}trans\ A\ {\isasymand}\ S\ {\isacharparenleft}fst\ s{\isacharparenright}\ {\isacharparenleft}fst\ s{\isacharprime}{\isacharparenright}\ a\ {\isasymlongrightarrow}\ P\ {\isacharparenleft}s,\ a,\ s{\isacharprime}{\isacharparenright}}}

\snip{invariant}{0}{0}{\begin{isabelle}%
{\isacharparenleft}A\ {\isasymTTurnstile}\ {\isacharparenleft}I\ {\isasymrightarrow}{\isacharparenright}\ P{\isacharparenright}\ {\isacharequal}\ {\isacharparenleft}{\isasymforall}s{\isasymin}reachable\ A\ I{\isachardot}\ P\ s{\isacharparenright}%
\end{isabelle}}
\snip{invariant_lhs}{0}{0}{\isa{A\ {\isasymTTurnstile}\ {\isacharparenleft}I\ {\isasymrightarrow}{\isacharparenright}\ P}}
\snip{invariant_rhs}{0}{0}{\isa{{\isasymforall}s{\isasymin}reachable\ A\ I{\isachardot}\ P\ s}}
\snip{invariant_TT_lhs}{0}{0}{\isa{A\ {\isasymTTurnstile}\ P}}

\snip{step_invariant}{0}{0}{\begin{isabelle}%
{\isacharparenleft}A\ \stepinv\ {\isacharparenleft}I\ {\isasymrightarrow}{\isacharparenright}\ P{\isacharparenright}\ {\isacharequal}\isanewline
{\isacharparenleft}{\isasymforall}a{\isachardot}\ I\ a\ {\isasymlongrightarrow}\isanewline
\isaindent{{\isacharparenleft}{\isasymforall}a{\isachardot}\ }{\isacharparenleft}{\isasymforall}s{\isasymin}reachable\ A\ I{\isachardot}\isanewline
\isaindent{{\isacharparenleft}{\isasymforall}a{\isachardot}\ {\isacharparenleft}\ \ \ }{\isasymforall}s{\isacharprime}{\isachardot}\ {\isacharparenleft}s,\ a,\ s{\isacharprime}{\isacharparenright}{\isasymin}trans\ A\ {\isasymlongrightarrow}\ P\ {\isacharparenleft}s,\ a,\ s{\isacharprime}{\isacharparenright}{\isacharparenright}{\isacharparenright}%
\end{isabelle}}
\snip{step_invariant_lhs}{0}{0}{\isa{A\ \stepinv\ {\isacharparenleft}I\ {\isasymrightarrow}{\isacharparenright}\ P}}
\snip{step_invariant_rhs}{0}{0}{\isa{{\isasymforall}a{\isachardot}\ I\ a\ {\isasymlongrightarrow}\ {\isacharparenleft}{\isasymforall}s{\isasymin}reachable\ A\ I{\isachardot}\ {\isasymforall}s{\isacharprime}{\isachardot}\ {\isacharparenleft}s,\ a,\ s{\isacharprime}{\isacharparenright}{\isasymin}trans\ A\ {\isasymlongrightarrow}\ P\ {\isacharparenleft}s,\ a,\ s{\isacharprime}{\isacharparenright}{\isacharparenright}}}

\snip{TT}{0}{0}{\isa{TT\ {\isacharequal}\ {\isacharparenleft}{\isasymlambda}{\isacharunderscore}{\isachardot}\ True{\isacharparenright}}}
\snip{TT_rhs}{0}{0}{\isa{{\isasymlambda}{\isacharunderscore}{\isachardot}\ True}}

\snip{open_closed_invariant}{0}{0}{\isa{\mbox{}\inferrule{\mbox{A\ {\isasymTTurnstile}\ {\isacharparenleft}I\ {\isasymrightarrow}{\isacharparenright}\ P}\\\ \mbox{subreachable\ A\ U\ J}\\\ \mbox{{\isasymAnd}{\isasymsigma}\ {\isasymsigma}{\isacharprime}\ s{\isachardot}\ \mbox{}\inferrule{\mbox{{\isasymforall}j{\isasymin}J{\isachardot}\ {\isasymsigma}{\isacharprime}\ j\ {\isacharequal}\ {\isasymsigma}\ j}\\\ \mbox{P\ {\isacharparenleft}{\isasymsigma}{\isacharprime},\ s{\isacharparenright}}}{\mbox{P\ {\isacharparenleft}{\isasymsigma},\ s{\isacharparenright}}}}}{\mbox{A\ {\isasymTurnstile}\ {\isacharparenleft}act\ I,\ U\ {\isasymrightarrow}{\isacharparenright}\ P}}}}  

\snip{oseq_invariant_ctermI}{0}{0}{\begin{isabelle}%
{\isasymlbrakk}wellformed\ {\isasymGamma}{\isacharsemicolon}\ control{\isacharunderscore}within\ {\isasymGamma}\ {\isacharparenleft}init\ A{\isacharparenright}{\isacharsemicolon}\ simple{\isacharunderscore}labels\ {\isasymGamma}{\isacharsemicolon}\isanewline
\isaindent{\ }trans\ A\ {\isacharequal}\ \oseqpsos\ {\isasymGamma}\ i{\isacharsemicolon}\isanewline
\isaindent{\ }{\isasymAnd}{\isasymsigma}\ p\ l{\isachardot}\ {\isasymlbrakk}{\isacharparenleft}{\isasymsigma},\ p{\isacharparenright}{\isasymin}init\ A{\isacharsemicolon}\ l{\isasymin}labels\ {\isasymGamma}\ p{\isasymrbrakk}\ {\isasymLongrightarrow}\ P\ {\isacharparenleft}{\isasymsigma},\ l{\isacharparenright}{\isacharsemicolon}\isanewline
\isaindent{\ }{\isasymAnd}{\isasymsigma}\ {\isasymsigma}{\isacharprime}\ p\ l{\isachardot}\isanewline
\isaindent{\ \ \ \ }{\isasymlbrakk}{\isacharparenleft}{\isasymsigma},\ p{\isacharparenright}{\isasymin}oreachable\ A\ S\ U{\isacharsemicolon}\ l{\isasymin}labels\ {\isasymGamma}\ p{\isacharsemicolon}\ P\ {\isacharparenleft}{\isasymsigma},\ l{\isacharparenright}{\isacharsemicolon}\ U\ {\isasymsigma}\ {\isasymsigma}{\isacharprime}{\isasymrbrakk}\isanewline
\isaindent{\ \ \ \ }{\isasymLongrightarrow}\ P\ {\isacharparenleft}{\isasymsigma}{\isacharprime},\ l{\isacharparenright}{\isacharsemicolon}\isanewline
\isaindent{\ }{\isasymAnd}p\ l\ {\isasymsigma}\ a\ q\ l{\isacharprime}\ {\isasymsigma}{\isacharprime}\ pp{\isachardot}\isanewline
\isaindent{\ \ \ \ }{\isasymlbrakk}p{\isasymin}cterms\ {\isasymGamma}{\isacharsemicolon}\ l{\isasymin}labels\ {\isasymGamma}\ p{\isacharsemicolon}\ P\ {\isacharparenleft}{\isasymsigma},\ l{\isacharparenright}{\isacharsemicolon}\isanewline
\isaindent{\ \ \ \ \ }{\isacharparenleft}{\isacharparenleft}{\isasymsigma},\ p{\isacharparenright},\ a,\ {\isacharparenleft}{\isasymsigma}{\isacharprime},\ q{\isacharparenright}{\isacharparenright}{\isasymin}\oseqpsos\ {\isasymGamma}\ i{\isacharsemicolon}\isanewline
\isaindent{\ \ \ \ \ }{\isacharparenleft}{\isacharparenleft}{\isasymsigma},\ p{\isacharparenright},\ a,\ {\isasymsigma}{\isacharprime},\ q{\isacharparenright}{\isasymin}trans\ A{\isacharsemicolon}\ l{\isacharprime}{\isasymin}labels\ {\isasymGamma}\ q{\isacharsemicolon}\isanewline
\isaindent{\ \ \ \ \ }{\isacharparenleft}{\isasymsigma},\ pp{\isacharparenright}{\isasymin}oreachable\ A\ S\ U{\isacharsemicolon}\ p{\isasymin}sterms\ {\isasymGamma}\ pp{\isacharsemicolon}\isanewline
\isaindent{\ \ \ \ \ }{\isacharparenleft}{\isasymsigma}{\isacharprime},\ q{\isacharparenright}{\isasymin}oreachable\ A\ S\ U{\isacharsemicolon}\ S\ {\isasymsigma}\ {\isasymsigma}{\isacharprime}\ a{\isasymrbrakk}\isanewline
\isaindent{\ \ \ \ }{\isasymLongrightarrow}\ P\ {\isacharparenleft}{\isasymsigma}{\isacharprime},\ l{\isacharprime}{\isacharparenright}{\isasymrbrakk}\isanewline
{\isasymLongrightarrow}\ A\ {\isasymTurnstile}\ {\isacharparenleft}S,\ U\ {\isasymrightarrow}{\isacharparenright}\ onl\ {\isasymGamma}\ P%
\end{isabelle}}
\snip{oseq_invariant_ctermI_wf}{0}{0}{\isa{wellformed\ {\isasymGamma}}}
\snip{oseq_invariant_ctermI_cw}{0}{0}{\isa{control{\isacharunderscore}within\ {\isasymGamma}\ {\isacharparenleft}init\ A{\isacharparenright}}}
\snip{oseq_invariant_ctermI_sl}{0}{0}{\isa{simple{\isacharunderscore}labels\ {\isasymGamma}}}
\snip{oseq_invariant_ctermI_init}{0}{0}{\begin{isabelle}%
trans\ A\ {\isacharequal}\ \oseqpsos\ {\isasymGamma}\ i%
\end{isabelle}}
\snip{oseq_invariant_ctermI_other}{0}{0}{\isa{{\isasymAnd}{\isasymsigma}\ p\ l{\isachardot}\ {\isasymlbrakk}{\isacharparenleft}{\isasymsigma},\ p{\isacharparenright}{\isasymin}init\ A{\isacharsemicolon}\ l{\isasymin}labels\ {\isasymGamma}\ p{\isasymrbrakk}\ {\isasymLongrightarrow}\ P\ {\isacharparenleft}{\isasymsigma},\ l{\isacharparenright}}}
\snip{oseq_invariant_ctermI_step}{0}{0}{\begin{isabelle}%
{\isasymAnd}{\isasymsigma}\ {\isasymsigma}{\isacharprime}\ p\ l{\isachardot}\isanewline
\isaindent{\ \ \ }{\isasymlbrakk}{\isacharparenleft}{\isasymsigma},\ p{\isacharparenright}{\isasymin}oreachable\ A\ S\ U{\isacharsemicolon}\ l{\isasymin}labels\ {\isasymGamma}\ p{\isacharsemicolon}\ P\ {\isacharparenleft}{\isasymsigma},\ l{\isacharparenright}{\isacharsemicolon}\ U\ {\isasymsigma}\ {\isasymsigma}{\isacharprime}{\isasymrbrakk}\isanewline
\isaindent{\ \ \ }{\isasymLongrightarrow}\ P\ {\isacharparenleft}{\isasymsigma}{\isacharprime},\ l{\isacharparenright}%
\end{isabelle}}
\snip{oseq_invariant_ctermI_concl}{0}{0}{\isa{A\ {\isasymTurnstile}\ {\isacharparenleft}S,\ U\ {\isasymrightarrow}{\isacharparenright}\ onl\ {\isasymGamma}\ P}}

\snip{oseq_step_invariant_ctermI}{0}{0}{\begin{isabelle}%
{\isasymlbrakk}wellformed\ {\isasymGamma}{\isacharsemicolon}\ control{\isacharunderscore}within\ {\isasymGamma}\ {\isacharparenleft}init\ A{\isacharparenright}{\isacharsemicolon}\ simple{\isacharunderscore}labels\ {\isasymGamma}{\isacharsemicolon}\isanewline
\isaindent{\ }trans\ A\ {\isacharequal}\ \oseqpsos\ {\isasymGamma}\ i{\isacharsemicolon}\isanewline
\isaindent{\ }{\isasymAnd}p\ l\ {\isasymsigma}\ a\ q\ l{\isacharprime}\ {\isasymsigma}{\isacharprime}\ pp{\isachardot}\isanewline
\isaindent{\ \ \ \ }{\isasymlbrakk}p{\isasymin}cterms\ {\isasymGamma}{\isacharsemicolon}\ l{\isasymin}labels\ {\isasymGamma}\ p{\isacharsemicolon}\ {\isacharparenleft}{\isacharparenleft}{\isasymsigma},\ p{\isacharparenright},\ a,\ {\isacharparenleft}{\isasymsigma}{\isacharprime},\ q{\isacharparenright}{\isacharparenright}{\isasymin}\oseqpsos\ {\isasymGamma}\ i{\isacharsemicolon}\isanewline
\isaindent{\ \ \ \ \ }{\isacharparenleft}{\isacharparenleft}{\isasymsigma},\ p{\isacharparenright},\ a,\ {\isasymsigma}{\isacharprime},\ q{\isacharparenright}{\isasymin}trans\ A{\isacharsemicolon}\ l{\isacharprime}{\isasymin}labels\ {\isasymGamma}\ q{\isacharsemicolon}\isanewline
\isaindent{\ \ \ \ \ }{\isacharparenleft}{\isasymsigma},\ pp{\isacharparenright}{\isasymin}oreachable\ A\ S\ U{\isacharsemicolon}\ p{\isasymin}sterms\ {\isasymGamma}\ pp{\isacharsemicolon}\isanewline
\isaindent{\ \ \ \ \ }{\isacharparenleft}{\isasymsigma}{\isacharprime},\ q{\isacharparenright}{\isasymin}oreachable\ A\ S\ U{\isacharsemicolon}\ S\ {\isasymsigma}\ {\isasymsigma}{\isacharprime}\ a{\isasymrbrakk}\isanewline
\isaindent{\ \ \ \ }{\isasymLongrightarrow}\ P\ {\isacharparenleft}{\isacharparenleft}{\isasymsigma},\ l{\isacharparenright},\ a,\ {\isasymsigma}{\isacharprime},\ l{\isacharprime}{\isacharparenright}{\isasymrbrakk}\isanewline
{\isasymLongrightarrow}\ A\ \ostepinv\ {\isacharparenleft}S,\ U\ {\isasymrightarrow}{\isacharparenright}\ onll\ {\isasymGamma}\ P%
\end{isabelle}}

\snip{reachable_init}{0}{0}{\isa{\mbox{}\inferrule{\mbox{s{\isasymin}init\ A}}{\mbox{s{\isasymin}reachable\ A\ I}}}}
\snip{reachable_step}{0}{0}{\isa{\mbox{}\inferrule{\mbox{s{\isasymin}reachable\ A\ I}\\\ \mbox{{\isacharparenleft}s,\ a,\ s{\isacharprime}{\isacharparenright}{\isasymin}trans\ A}\\\ \mbox{I\ a}}{\mbox{s{\isacharprime}{\isasymin}reachable\ A\ I}}}}

\snip{oreachable_init}{0}{0}{\isa{\mbox{}\inferrule{\mbox{{\isacharparenleft}{\isasymsigma},\ s{\isacharparenright}{\isasymin}init\ A}}{\mbox{{\isacharparenleft}{\isasymsigma},\ s{\isacharparenright}{\isasymin}oreachable\ A\ S\ U}}}}
\snip{oreachable_local}{0}{0}{\isa{\mbox{}\inferrule{\mbox{s{\isasymin}oreachable\ A\ S\ U}\\\ \mbox{{\isacharparenleft}s,\ {\isacharparenleft}a,\ s{\isacharprime}{\isacharparenright}{\isacharparenright}{\isasymin}trans\ A}\\\ \mbox{S\ {\isacharparenleft}fst\ s{\isacharparenright}\ {\isacharparenleft}fst\ s{\isacharprime}{\isacharparenright}\ a}}{\mbox{s{\isacharprime}{\isasymin}oreachable\ A\ S\ U}}}}
\snip{oreachable_other}{0}{0}{\isa{\mbox{}\inferrule{\mbox{s{\isasymin}oreachable\ A\ S\ U}\\\ \mbox{U\ {\isacharparenleft}fst\ s{\isacharparenright}\ {\isasymsigma}{\isacharprime}}}{\mbox{{\isacharparenleft}{\isasymsigma}{\isacharprime},\ snd\ s{\isacharparenright}{\isasymin}oreachable\ A\ S\ U}}}}
\snip{oreachable_local'}{0}{0}{\isa{\mbox{}\inferrule{\mbox{{\isacharparenleft}{\isasymsigma},\ s{\isacharparenright}{\isasymin}oreachable\ A\ S\ U}\\\ \mbox{{\isacharparenleft}{\isacharparenleft}{\isasymsigma},\ s{\isacharparenright},\ a,\ {\isacharparenleft}{\isasymsigma}{\isacharprime},\ s{\isacharprime}{\isacharparenright}{\isacharparenright}{\isasymin}trans\ A}\\\ \mbox{S\ {\isasymsigma}\ {\isasymsigma}{\isacharprime}\ a}}{\mbox{{\isacharparenleft}{\isasymsigma}{\isacharprime},\ s{\isacharprime}{\isacharparenright}{\isasymin}oreachable\ A\ S\ U}}}}
\snip{oreachable_other'}{0}{0}{\isa{\mbox{}\inferrule{\mbox{{\isacharparenleft}{\isasymsigma},\ s{\isacharparenright}{\isasymin}oreachable\ A\ S\ U}\\\ \mbox{U\ {\isasymsigma}\ {\isasymsigma}{\isacharprime}}}{\mbox{{\isacharparenleft}{\isasymsigma}{\isacharprime},\ s{\isacharparenright}{\isasymin}oreachable\ A\ S\ U}}}}

\snip{otherwith}{0}{0}{\isa{otherwith\ {\isacharparenleft}op\ {\isacharequal}{\isacharparenright}\ {\isacharbraceleft}i{\isacharbraceright}\ {\isacharparenleft}orecvmsg\ P{\isacharparenright}}}
\snip{other}{0}{0}{\isa{other\ quality{\isacharunderscore}increases\ {\isacharbraceleft}i{\isacharbraceright}}}
\snip{otherwith_def_lhs}{0}{0}{\isa{otherwith\ E\ N\ I\ {\isasymsigma}\ {\isasymsigma}{\isacharprime}\ a}}
\snip{otherwith_def_rhs}{0}{0}{\isa{{\isacharparenleft}{\isasymforall}i{\isachardot}\ i\ {\isasymnotin}\ N\ {\isasymlongrightarrow}\ E\ {\isacharparenleft}{\isasymsigma}\ i{\isacharparenright}\ {\isacharparenleft}{\isasymsigma}{\isacharprime}\ i{\isacharparenright}{\isacharparenright}\ {\isasymand}\ I\ {\isasymsigma}\ a}}
\snip{other_def_lhs}{0}{0}{\isa{other\ E\ N\ {\isasymsigma}\ {\isasymsigma}{\isacharprime}}}
\snip{other_def_rhs}{0}{0}{\isa{{\isasymforall}i{\isachardot}\ \textsf{if}\ i{\isasymin}N\ \textsf{then}\ {\isasymsigma}{\isacharprime}\ i\ {\isacharequal}\ {\isasymsigma}\ i\ \textsf{else}\ E\ {\isacharparenleft}{\isasymsigma}\ i{\isacharparenright}\ {\isacharparenleft}{\isasymsigma}{\isacharprime}\ i{\isacharparenright}}}
\snip{otherwith_def}{0}{0}{\isa{otherwith\ E\ N\ I\ {\isasymsigma}\ {\isasymsigma}{\isacharprime}\ a\ {\isacharequal}\ {\isacharparenleft}{\isasymforall}i{\isachardot}\ i\ {\isasymnotin}\ N\ {\isasymlongrightarrow}\ E\ {\isacharparenleft}{\isasymsigma}\ i{\isacharparenright}\ {\isacharparenleft}{\isasymsigma}{\isacharprime}\ i{\isacharparenright}{\isacharparenright}\ {\isasymand}\ I\ {\isasymsigma}\ a}}
\snip{other_def}{0}{0}{\isa{other\ E\ N\ {\isasymsigma}\ {\isasymsigma}{\isacharprime}\ {\isacharequal}\ {\isasymforall}i{\isachardot}\ \textsf{if}\ i{\isasymin}N\ \textsf{then}\ {\isasymsigma}{\isacharprime}\ i\ {\isacharequal}\ {\isasymsigma}\ i\ \textsf{else}\ E\ {\isacharparenleft}{\isasymsigma}\ i{\isacharparenright}\ {\isacharparenleft}{\isasymsigma}{\isacharprime}\ i{\isacharparenright}}}
\snip{qmsg_msgs_not_empty}{0}{0}{\begin{isabelle}%
qmsg\ {\isasymTTurnstile}\ onl\ \gammaqmsg\ {\isacharparenleft}{\isasymlambda}{\isacharparenleft}msgs,\ l{\isacharparenright}{\isachardot}\ l\ {\isacharequal}\ {\isacharparenleft}{\isacharparenright}{\isacharminus}{\isacharcolon}{\isadigit{1}}\ {\isasymlongrightarrow}\ msgs\ {\isasymnoteq}\ {\isacharbrackleft}\,{\isacharbrackright}{\isacharparenright}%
\end{isabelle}}
\snip{qmsg_send_receive_or_tau}{0}{0}{\begin{isabelle}%
qmsg\ \stepinv\ {\isacharparenleft}{\isasymlambda}{\isacharparenleft}uu{\isacharunderscore},\ a,\ {\isacharunderscore}{\isacharparenright}{\isachardot}\ {\isasymexists}m{\isachardot}\ a\ {\isacharequal}\ send\ m\ {\isasymor}\ a\ {\isacharequal}\ receive\ m\ {\isasymor}\ a\ {\isacharequal}\ {\isasymtau}{\isacharparenright}%
\end{isabelle}}

\snip{qmsg_send_from_queue}{0}{0}{\isa{%
qmsg\ \stepinv\ {\isacharparenleft}{\isasymlambda}{\isacharparenleft}{\isacharparenleft}msgs,\ q{\isacharparenright},\ a,\ {\isacharunderscore}{\isacharparenright}{\isachardot}\ sendmsg\ {\isacharparenleft}{\isasymlambda}m{\isachardot}\ m{\isasymin}set\ msgs{\isacharparenright}\ a{\isacharparenright}%
}}
\snip{qmsg_queue_contents}{0}{0}{\begin{isabelle}%
qmsg\ \stepinv\ {\isacharparenleft}{\isasymlambda}{\isacharparenleft}{\isacharparenleft}msgs,\ q{\isacharparenright},\ a,\ {\isacharparenleft}msgs{\isacharprime},\ q{\isacharprime}{\isacharparenright}{\isacharparenright}{\isachardot}\isanewline
\isaindent{qmsg\ \stepinv\ {\isacharparenleft}\ \ \ }\textsf{case}\ a\ \textsf{of}\ receive\ m\ {\isasymRightarrow}\ msgs{\isacharprime}\ {\isacharequal}\ msgs\ {\isacharat}\ {\isacharbrackleft}m{\isacharbrackright}\isanewline
\isaindent{qmsg\ \stepinv\ {\isacharparenleft}\ \ \ }{\isacharbar}\ {\isacharunderscore}\ {\isasymRightarrow}\ set\ msgs{\isacharprime}\ {\isasymsubseteq}\ set\ msgs{\isacharparenright}%
\end{isabelle}}

\snip{par_qmsg_oreachable}{0}{0}{\begin{isabelle}%
{\isasymlbrakk}{\isacharparenleft}{\isasymsigma},\ {\isasymzeta}{\isacharparenright}\isanewline
\isaindent{{\isasymlbrakk}}{\isasymin}\ oreachable\ {\isacharparenleft}A\ {\isasymlangle}{\isasymlangle}\isactrlbsub i\isactrlesub \ qmsg{\isacharparenright}\ {\isacharparenleft}otherwith\ S\ {\isacharbraceleft}i{\isacharbraceright}\ {\isacharparenleft}orecvmsg\ R{\isacharparenright}{\isacharparenright}\ {\isacharparenleft}other\ U\ {\isacharbraceleft}i{\isacharbraceright}{\isacharparenright}{\isacharsemicolon}\isanewline
\isaindent{\ }A\ \ostepinv\ {\isacharparenleft}otherwith\ S\ {\isacharbraceleft}i{\isacharbraceright}\ {\isacharparenleft}orecvmsg\ R{\isacharparenright},\ other\ U\ {\isacharbraceleft}i{\isacharbraceright}\ {\isasymrightarrow}{\isacharparenright}\isanewline
\isaindent{\ A\ \ostepinv\ \ }globala\ {\isacharparenleft}{\isasymlambda}{\isacharparenleft}{\isasymsigma},\ {\isacharunderscore},\ {\isasymsigma}{\isacharprime}{\isacharparenright}{\isachardot}\ U\ {\isacharparenleft}{\isasymsigma}\ i{\isacharparenright}\ {\isacharparenleft}{\isasymsigma}{\isacharprime}\ i{\isacharparenright}{\isacharparenright}{\isacharsemicolon}\isanewline
\isaindent{\ }{\isasymAnd}{\isasymxi}{\isachardot}\ U\ {\isasymxi}\ {\isasymxi}{\isacharsemicolon}\ {\isasymAnd}{\isasymxi}\ {\isasymxi}{\isacharprime}{\isachardot}\ S\ {\isasymxi}\ {\isasymxi}{\isacharprime}\ {\isasymLongrightarrow}\ U\ {\isasymxi}\ {\isasymxi}{\isacharprime}{\isacharsemicolon}\isanewline
\isaindent{\ }{\isasymAnd}{\isasymsigma}\ {\isasymsigma}{\isacharprime}\ m{\isachardot}\ {\isasymlbrakk}{\isasymforall}j{\isachardot}\ U\ {\isacharparenleft}{\isasymsigma}\ j{\isacharparenright}\ {\isacharparenleft}{\isasymsigma}{\isacharprime}\ j{\isacharparenright}{\isacharsemicolon}\ R\ {\isasymsigma}\ m{\isasymrbrakk}\ {\isasymLongrightarrow}\ R\ {\isasymsigma}{\isacharprime}\ m{\isasymrbrakk}\isanewline
{\isasymLongrightarrow}\ {\isacharparenleft}{\isasymsigma},\ fst\ {\isasymzeta}{\isacharparenright}{\isasymin}oreachable\ A\ {\isacharparenleft}otherwith\ S\ {\isacharbraceleft}i{\isacharbraceright}\ {\isacharparenleft}orecvmsg\ R{\isacharparenright}{\isacharparenright}\ {\isacharparenleft}other\ U\ {\isacharbraceleft}i{\isacharbraceright}{\isacharparenright}\ {\isasymand}\isanewline
\isaindent{{\isasymLongrightarrow}\ }snd\ {\isasymzeta}{\isasymin}reachable\ qmsg\ {\isacharparenleft}recvmsg\ {\isacharparenleft}R\ {\isasymsigma}{\isacharparenright}{\isacharparenright}\ {\isasymand}\ {\isacharparenleft}{\isasymforall}m{\isasymin}set\ {\isacharparenleft}fst\ {\isacharparenleft}snd\ {\isasymzeta}{\isacharparenright}{\isacharparenright}{\isachardot}\ R\ {\isasymsigma}\ m{\isacharparenright}%
\end{isabelle}}
\snip{par_qmsg_oreachable_upreservesq}{0}{0}{\begin{isabelle}%
{\isasymAnd}{\isasymsigma}\ {\isasymsigma}{\isacharprime}\ m{\isachardot}\ {\isasymlbrakk}{\isasymforall}j{\isachardot}\ U\ {\isacharparenleft}{\isasymsigma}\ j{\isacharparenright}\ {\isacharparenleft}{\isasymsigma}{\isacharprime}\ j{\isacharparenright}{\isacharsemicolon}\ R\ {\isasymsigma}\ m{\isasymrbrakk}\ {\isasymLongrightarrow}\ R\ {\isasymsigma}{\isacharprime}\ m%
\end{isabelle}}
\snip{par_qmsg_oreachable_concl3}{0}{0}{\begin{isabelle}%
{\isasymforall}m{\isasymin}set\ {\isacharparenleft}fst\ {\isacharparenleft}snd\ {\isasymzeta}{\isacharparenright}{\isacharparenright}{\isachardot}\ M\ {\isasymsigma}\ m%
\end{isabelle}}

\snip{lift_into_qmsg}{0}{0}{\begin{isabelle}%
{\isasymlbrakk}A\ {\isasymTurnstile}\ {\isacharparenleft}otherwith\ S\ {\isacharbraceleft}i{\isacharbraceright}\ {\isacharparenleft}orecvmsg\ R{\isacharparenright},\ other\ U\ {\isacharbraceleft}i{\isacharbraceright}\ {\isasymrightarrow}{\isacharparenright}\ global\ P{\isacharsemicolon}\ {\isasymAnd}{\isasymxi}{\isachardot}\ U\ {\isasymxi}\ {\isasymxi}{\isacharsemicolon}\isanewline
\isaindent{\ }{\isasymAnd}{\isasymxi}\ {\isasymxi}{\isacharprime}{\isachardot}\ S\ {\isasymxi}\ {\isasymxi}{\isacharprime}\ {\isasymLongrightarrow}\ U\ {\isasymxi}\ {\isasymxi}{\isacharprime}{\isacharsemicolon}\ {\isasymAnd}{\isasymsigma}\ {\isasymsigma}{\isacharprime}\ m{\isachardot}\ {\isasymlbrakk}{\isasymforall}j{\isachardot}\ U\ {\isacharparenleft}{\isasymsigma}\ j{\isacharparenright}\ {\isacharparenleft}{\isasymsigma}{\isacharprime}\ j{\isacharparenright}{\isacharsemicolon}\ R\ {\isasymsigma}\ m{\isasymrbrakk}\ {\isasymLongrightarrow}\ R\ {\isasymsigma}{\isacharprime}\ m{\isacharsemicolon}\isanewline
\isaindent{\ }A\ \ostepinv\ {\isacharparenleft}otherwith\ S\ {\isacharbraceleft}i{\isacharbraceright}\ {\isacharparenleft}orecvmsg\ R{\isacharparenright},\ other\ U\ {\isacharbraceleft}i{\isacharbraceright}\ {\isasymrightarrow}{\isacharparenright}\isanewline
\isaindent{\ A\ \ostepinv\ \ }globala\ {\isacharparenleft}{\isasymlambda}{\isacharparenleft}{\isasymsigma},\ {\isacharunderscore},\ {\isasymsigma}{\isacharprime}{\isacharparenright}{\isachardot}\ U\ {\isacharparenleft}{\isasymsigma}\ i{\isacharparenright}\ {\isacharparenleft}{\isasymsigma}{\isacharprime}\ i{\isacharparenright}{\isacharparenright}{\isasymrbrakk}\isanewline
{\isasymLongrightarrow}\ A\ {\isasymlangle}{\isasymlangle}\isactrlbsub i\isactrlesub \ qmsg\ {\isasymTurnstile}\ {\isacharparenleft}otherwith\ S\ {\isacharbraceleft}i{\isacharbraceright}\ {\isacharparenleft}orecvmsg\ R{\isacharparenright},\ other\ U\ {\isacharbraceleft}i{\isacharbraceright}\ {\isasymrightarrow}{\isacharparenright}\ global\ P%
\end{isabelle}}
\snip{lift_step_into_qmsg}{0}{0}{\begin{isabelle}%
{\isasymlbrakk}A\ \ostepinv\ {\isacharparenleft}otherwith\ S\ {\isacharbraceleft}i{\isacharbraceright}\ {\isacharparenleft}orecvmsg\ R{\isacharparenright},\ other\ U\ {\isacharbraceleft}i{\isacharbraceright}\ {\isasymrightarrow}{\isacharparenright}\ globala\ P{\isacharsemicolon}\ {\isasymAnd}{\isasymxi}{\isachardot}\ U\ {\isasymxi}\ {\isasymxi}{\isacharsemicolon}\isanewline
\isaindent{\ }{\isasymAnd}{\isasymxi}\ {\isasymxi}{\isacharprime}{\isachardot}\ S\ {\isasymxi}\ {\isasymxi}{\isacharprime}\ {\isasymLongrightarrow}\ U\ {\isasymxi}\ {\isasymxi}{\isacharprime}{\isacharsemicolon}\ {\isasymAnd}{\isasymsigma}\ {\isasymsigma}{\isacharprime}\ m{\isachardot}\ {\isasymlbrakk}{\isasymforall}j{\isachardot}\ U\ {\isacharparenleft}{\isasymsigma}\ j{\isacharparenright}\ {\isacharparenleft}{\isasymsigma}{\isacharprime}\ j{\isacharparenright}{\isacharsemicolon}\ R\ {\isasymsigma}\ m{\isasymrbrakk}\ {\isasymLongrightarrow}\ R\ {\isasymsigma}{\isacharprime}\ m{\isacharsemicolon}\isanewline
\isaindent{\ }A\ \ostepinv\ {\isacharparenleft}otherwith\ S\ {\isacharbraceleft}i{\isacharbraceright}\ {\isacharparenleft}orecvmsg\ R{\isacharparenright},\ other\ U\ {\isacharbraceleft}i{\isacharbraceright}\ {\isasymrightarrow}{\isacharparenright}\isanewline
\isaindent{\ A\ \ostepinv\ \ }globala\ {\isacharparenleft}{\isasymlambda}{\isacharparenleft}{\isasymsigma},\ {\isacharunderscore},\ {\isasymsigma}{\isacharprime}{\isacharparenright}{\isachardot}\ U\ {\isacharparenleft}{\isasymsigma}\ i{\isacharparenright}\ {\isacharparenleft}{\isasymsigma}{\isacharprime}\ i{\isacharparenright}{\isacharparenright}{\isacharsemicolon}\isanewline
\isaindent{\ }{\isasymAnd}{\isasymsigma}\ {\isasymsigma}{\isacharprime}\ m{\isachardot}\ {\isasymlbrakk}{\isasymforall}j{\isachardot}\ U\ {\isacharparenleft}{\isasymsigma}\ j{\isacharparenright}\ {\isacharparenleft}{\isasymsigma}{\isacharprime}\ j{\isacharparenright}{\isacharsemicolon}\ {\isasymsigma}{\isacharprime}\ i\ {\isacharequal}\ {\isasymsigma}\ i{\isasymrbrakk}\ {\isasymLongrightarrow}\ P\ {\isacharparenleft}{\isasymsigma},\ receive\ m,\ {\isasymsigma}{\isacharprime}{\isacharparenright}{\isacharsemicolon}\isanewline
\isaindent{\ }{\isasymAnd}{\isasymsigma}\ {\isasymsigma}{\isacharprime}\ m{\isachardot}\ P\ {\isacharparenleft}{\isasymsigma},\ receive\ m,\ {\isasymsigma}{\isacharprime}{\isacharparenright}\ {\isasymLongrightarrow}\ P\ {\isacharparenleft}{\isasymsigma},\ {\isasymtau},\ {\isasymsigma}{\isacharprime}{\isacharparenright}{\isasymrbrakk}\isanewline
{\isasymLongrightarrow}\ A\ {\isasymlangle}{\isasymlangle}\isactrlbsub i\isactrlesub \ qmsg\ \ostepinv\ {\isacharparenleft}otherwith\ S\ {\isacharbraceleft}i{\isacharbraceright}\ {\isacharparenleft}orecvmsg\ R{\isacharparenright},\ other\ U\ {\isacharbraceleft}i{\isacharbraceright}\ {\isasymrightarrow}{\isacharparenright}\ globala\ P%
\end{isabelle}}

\snip{node_lift}{0}{0}{\begin{isabelle}%
{\isasymlbrakk}T\ {\isasymTurnstile}\ {\isacharparenleft}otherwith\ S\ {\isacharbraceleft}ii{\isacharbraceright}\ {\isacharparenleft}orecvmsg\ I{\isacharparenright},\ other\ U\ {\isacharbraceleft}ii{\isacharbraceright}\ {\isasymrightarrow}{\isacharparenright}\ global\ P{\isacharsemicolon}\isanewline
\isaindent{\ }{\isasymAnd}{\isasymxi}\ {\isasymxi}{\isacharprime}{\isachardot}\ S\ {\isasymxi}\ {\isasymxi}{\isacharprime}\ {\isasymLongrightarrow}\ U\ {\isasymxi}\ {\isasymxi}{\isacharprime}{\isasymrbrakk}\isanewline
{\isasymLongrightarrow}\ {\isasymlangle}ii\ {\isacharcolon}\ T\ {\isacharcolon}\ \Ri{\isasymrangle}\isactrlsub o\ {\isasymTurnstile}\ {\isacharparenleft}otherwith\ S\ {\isacharbraceleft}ii{\isacharbraceright}\ {\isacharparenleft}oarrivemsg\ I{\isacharparenright},\ other\ U\ {\isacharbraceleft}ii{\isacharbraceright}\ {\isasymrightarrow}{\isacharparenright}\isanewline
\isaindent{{\isasymLongrightarrow}\ {\isasymlangle}ii\ {\isacharcolon}\ T\ {\isacharcolon}\ \Ri{\isasymrangle}\isactrlsub o\ {\isasymTurnstile}\ \ }global\ P%
\end{isabelle}}
\snip{node_lift_step}{0}{0}{\begin{isabelle}%
{\isasymlbrakk}T\ \ostepinv\ {\isacharparenleft}otherwith\ S\ {\isacharbraceleft}i{\isacharbraceright}\ {\isacharparenleft}orecvmsg\ I{\isacharparenright},\ other\ U\ {\isacharbraceleft}i{\isacharbraceright}\ {\isasymrightarrow}{\isacharparenright}\isanewline
\isaindent{{\isasymlbrakk}T\ \ostepinv\ \ }globala\ {\isacharparenleft}{\isasymlambda}{\isacharparenleft}{\isasymsigma},\ {\isacharunderscore},\ {\isasymsigma}{\isacharprime}{\isacharparenright}{\isachardot}\ Q\ {\isasymsigma}\ {\isasymsigma}{\isacharprime}{\isacharparenright}{\isacharsemicolon}\isanewline
\isaindent{\ }{\isasymAnd}{\isasymsigma}\ {\isasymsigma}{\isacharprime}{\isachardot}\ other\ U\ {\isacharbraceleft}i{\isacharbraceright}\ {\isasymsigma}\ {\isasymsigma}{\isacharprime}\ {\isasymLongrightarrow}\ Q\ {\isasymsigma}\ {\isasymsigma}{\isacharprime}{\isacharsemicolon}\ {\isasymAnd}{\isasymxi}\ {\isasymxi}{\isacharprime}{\isachardot}\ S\ {\isasymxi}\ {\isasymxi}{\isacharprime}\ {\isasymLongrightarrow}\ U\ {\isasymxi}\ {\isasymxi}{\isacharprime}{\isasymrbrakk}\isanewline
{\isasymLongrightarrow}\ {\isasymlangle}i\ {\isacharcolon}\ T\ {\isacharcolon}\ \Ri{\isasymrangle}\isactrlsub o\ \ostepinv\ {\isacharparenleft}otherwith\ S\ {\isacharbraceleft}i{\isacharbraceright}\ {\isacharparenleft}oarrivemsg\ I{\isacharparenright},\ other\ U\ {\isacharbraceleft}i{\isacharbraceright}\ {\isasymrightarrow}{\isacharparenright}\isanewline
\isaindent{{\isasymLongrightarrow}\ {\isasymlangle}i\ {\isacharcolon}\ T\ {\isacharcolon}\ \Ri{\isasymrangle}\isactrlsub o\ \ostepinv\ \ }globala\ {\isacharparenleft}{\isasymlambda}{\isacharparenleft}{\isasymsigma},\ {\isacharunderscore},\ {\isasymsigma}{\isacharprime}{\isacharparenright}{\isachardot}\ Q\ {\isasymsigma}\ {\isasymsigma}{\isacharprime}{\isacharparenright}%
\end{isabelle}}

\snip{optoy_qmsg_term}{0}{0}{\isa{optoy\ i\ {\isasymlangle}{\isasymlangle}\isactrlbsub i\isactrlesub \ qmsg}}

\snip{otherwith_r_term}{0}{0}{\isa{otherwith\ S\ {\isacharbraceleft}i{\isacharbraceright}\ {\isacharparenleft}orecvmsg\ M{\isacharparenright}}}
\snip{eg1_nettree}{0}{0}{\isa{{\isacharparenleft}{\isacharparenleft}{\isasymlangle}{\isadigit{1}}{\isacharsemicolon}\ {\isacharbraceleft}{\isadigit{2}}{\isacharbraceright}{\isasymrangle}\parallelcomp{}{\isasymlangle}{\isadigit{2}}{\isacharsemicolon}\ {\isacharbraceleft}{\isadigit{1}},\ {\isadigit{3}}{\isacharbraceright}{\isasymrangle}{\isacharparenright}\parallelcomp{}{\isasymlangle}{\isadigit{3}}{\isacharsemicolon}\ {\isacharbraceleft}{\isadigit{2}}{\isacharbraceright}{\isasymrangle}{\isacharparenright}}}
\snip{eg1_pnet}{0}{0}{\isa{{\isacharparenleft}pnet\ {\isacharparenleft}{\isasymlambda}i{\isachardot}\ {\isacharparenleft}{\isacharparenleft}ptoy\ i{\isacharparenright}\ {\isasymlangle}{\isasymlangle}\ qmsg{\isacharparenright}{\isacharparenright}\ {\isacharparenleft}{\isacharparenleft}{\isasymlangle}{\isadigit{1}}{\isacharsemicolon}\ {\isacharbraceleft}{\isadigit{2}}{\isacharbraceright}{\isasymrangle}\parallelcomp{}{\isasymlangle}{\isadigit{2}}{\isacharsemicolon}\ {\isacharbraceleft}{\isadigit{1}},\ {\isadigit{3}}{\isacharbraceright}{\isasymrangle}{\isacharparenright}\parallelcomp{}{\isasymlangle}{\isadigit{3}}{\isacharsemicolon}\ {\isacharbraceleft}{\isadigit{2}}{\isacharbraceright}{\isasymrangle}{\isacharparenright}{\isacharparenright}}}
\snip{eg1_all}{0}{0}{\isa{{\isacharparenleft}closed\ {\isacharparenleft}pnet\ {\isacharparenleft}{\isasymlambda}i{\isachardot}\ {\isacharparenleft}{\isacharparenleft}ptoy\ i{\isacharparenright}\ {\isasymlangle}{\isasymlangle}\ qmsg{\isacharparenright}{\isacharparenright}\ {\isacharparenleft}{\isasymlangle}{\isadigit{1}}{\isacharsemicolon}\ {\isacharbraceleft}{\isadigit{2}},\ {\isadigit{4}}{\isacharbraceright}{\isasymrangle}\parallelcomp{}{\isacharparenleft}{\isasymlangle}{\isadigit{2}}{\isacharsemicolon}\ {\isacharbraceleft}{\isadigit{1}},\ {\isadigit{3}}{\isacharbraceright}{\isasymrangle}\parallelcomp{}{\isacharparenleft}{\isasymlangle}{\isadigit{3}}{\isacharsemicolon}\ {\isacharbraceleft}{\isadigit{2}}{\isacharbraceright}{\isasymrangle}\parallelcomp{}{\isasymlangle}{\isadigit{4}}{\isacharsemicolon}\ {\isacharbraceleft}{\isadigit{1}}{\isacharbraceright}{\isasymrangle}{\isacharparenright}{\isacharparenright}{\isacharparenright}{\isacharparenright}{\isacharparenright}}}
\snip{stmt_guard}{0}{0}{\isa{{\isasymlangle}{\isasymlambda}{\isasymxi}{\isachardot}\ \textsf{if}\ h\ {\isasymxi}\ \textsf{then}\ {\isacharbraceleft}{\isasymxi}{\isacharbraceright}\ \textsf{else}\ {\isasymemptyset}{\isasymrangle}\ p}}
\snip{stmt_bind}{0}{0}{\isa{{\isasymlangle}{\isasymlambda}{\isasymxi}{\isachardot}\ {\isacharbraceleft}{\isasymxi}{\isasymlparr}no\ {\isacharcolon}{\isacharequal}\ n{\isasymrparr}\ {\isacharbar}\ n\ {\isacharless}\ {\isadigit{5}}{\isacharbraceright}{\isasymrangle}\ p}}

\snip{act_unicast}{0}{0}{\isa{unicast\ i\ m}}
\snip{act_not_unicast}{0}{0}{\isa{{\isasymnot}unicast\ i}}

\snip{eg1_node_a'}{0}{0}{\isa{{\isasymlangle}{\isadigit{1}}\ {\isacharcolon}\ ptoy\ {\isadigit{1}}\ {\isasymlangle}{\isasymlangle}\ qmsg\ {\isacharcolon}\ {\isacharbraceleft}{\isadigit{2}},\ {\isadigit{4}}{\isacharbraceright}{\isasymrangle}}}
\snip{eg1_node_b'}{0}{0}{\isa{{\isasymlangle}{\isadigit{2}}\ {\isacharcolon}\ ptoy\ {\isadigit{2}}\ {\isasymlangle}{\isasymlangle}\ qmsg\ {\isacharcolon}\ {\isacharbraceleft}{\isadigit{1}},\ {\isadigit{3}}{\isacharbraceright}{\isasymrangle}}}
\snip{eg1_node_c'}{0}{0}{\isa{{\isasymlangle}{\isadigit{3}}\ {\isacharcolon}\ ptoy\ {\isadigit{3}}\ {\isasymlangle}{\isasymlangle}\ qmsg\ {\isacharcolon}\ {\isacharbraceleft}{\isadigit{2}}{\isacharbraceright}{\isasymrangle}}}
\snip{eg1_node_d'}{0}{0}{\isa{{\isasymlangle}{\isadigit{4}}\ {\isacharcolon}\ ptoy\ {\isadigit{4}}\ {\isasymlangle}{\isasymlangle}\ qmsg\ {\isacharcolon}\ {\isacharbraceleft}{\isadigit{1}}{\isacharbraceright}{\isasymrangle}}}

\snip{eg1_node_a}{0}{0}{\isa{{\isasymlangle}{\isadigit{1}}{\isacharsemicolon}\ {\isacharbraceleft}{\isadigit{2}},\ {\isadigit{4}}{\isacharbraceright}{\isasymrangle}}}
\snip{eg1_node_b}{0}{0}{\isa{{\isasymlangle}{\isadigit{2}}{\isacharsemicolon}\ {\isacharbraceleft}{\isadigit{1}},\ {\isadigit{3}}{\isacharbraceright}{\isasymrangle}}}
\snip{eg1_node_c}{0}{0}{\isa{{\isasymlangle}{\isadigit{3}}{\isacharsemicolon}\ {\isacharbraceleft}{\isadigit{2}}{\isacharbraceright}{\isasymrangle}}}
\snip{eg1_node_d}{0}{0}{\isa{{\isasymlangle}{\isadigit{4}}{\isacharsemicolon}\ {\isacharbraceleft}{\isadigit{1}}{\isacharbraceright}{\isasymrangle}}}

\snip{dummy_par}{0}{0}{\isa{\_\parallelcomp{}\_}}

\snip{eg_sterms_foo}{0}{0}{\isa{sterms\ {\isasymGamma}\ {\isacharparenleft}{\isasymGamma}\ foo{\isacharparenright}}}
\snip{eg_sterms_groupcast}{0}{0}{\isa{dterms\ {\isasymGamma}\ {\isacharparenleft}groupcast{\isacharparenleft}\_,\ \_{\isacharparenright}\ {\isachardot}\ p{\isacharparenright}}}
\snip{eg_sterms_p}{0}{0}{\isa{sterms\ {\isasymGamma}\ p}}
\snip{gammap}{0}{0}{\isa{\gammatoy}}

\snip{gammap_ptoy}{0}{0}{\begin{isabelle}%
\gammatoy\ PToy\ {\isacharequal}\ {\isacharbraceleft}PToy{\isacharminus}{\isacharcolon}{\isadigit{0}}{\isacharbraceright}receive{\isacharparenleft}{\isasymlambda}msg{\isacharprime}{\isachardot}\ msg{\isacharunderscore}update\ {\isacharparenleft}{\isasymlambda}{\isacharunderscore}{\isachardot}\ msg{\isacharprime}{\isacharparenright}{\isacharparenright}\ {\isachardot}\isanewline
{\isacharbraceleft}PToy{\isacharminus}{\isacharcolon}{\isadigit{1}}{\isacharbraceright}{\isasymlbrakk}{\isasymlambda}{\isasymxi}{\isachardot}\ {\isasymxi}{\isasymlparr}nhid\ {\isacharcolon}{\isacharequal}\ id\ {\isasymxi}{\isasymrparr}{\isasymrbrakk}\isanewline
{\isacharparenleft}{\isacharbraceleft}PToy{\isacharminus}{\isacharcolon}{\isadigit{2}}{\isacharbraceright}{\isasymlangle}is{\isacharunderscore}newpkt{\isasymrangle}\isanewline
\isaindent{{\isacharparenleft}}{\isacharbraceleft}PToy{\isacharminus}{\isacharcolon}{\isadigit{3}}{\isacharbraceright}{\isasymlbrakk}{\isasymlambda}{\isasymxi}{\isachardot}\ {\isasymxi}{\isasymlparr}no\ {\isacharcolon}{\isacharequal}\ max\ {\isacharparenleft}no\ {\isasymxi}{\isacharparenright}\ {\isacharparenleft}num\ {\isasymxi}{\isacharparenright}{\isasymrparr}{\isasymrbrakk}\isanewline
\isaindent{{\isacharparenleft}}{\isacharbraceleft}PToy{\isacharminus}{\isacharcolon}{\isadigit{4}}{\isacharbraceright}broadcast{\isacharparenleft}{\isasymlambda}{\isasymxi}{\isachardot}\ Pkt\ {\isacharparenleft}no\ {\isasymxi}{\isacharparenright}\ {\isacharparenleft}id\ {\isasymxi}{\isacharparenright}{\isacharparenright}\ {\isachardot}\isanewline
\isaindent{{\isacharparenleft}}{\isacharbraceleft}PToy{\isacharminus}{\isacharcolon}{\isadigit{5}}{\isacharbraceright}{\isasymlbrakk}clear{\isacharunderscore}locals{\isasymrbrakk}\isanewline
\isaindent{{\isacharparenleft}}call{\isacharparenleft}PToy{\isacharparenright}\isanewline
\isaindent{{\isacharparenleft}}{\isasymoplus}\isanewline
\isaindent{{\isacharparenleft}}{\isacharbraceleft}PToy{\isacharminus}{\isacharcolon}{\isadigit{2}}{\isacharbraceright}{\isasymlangle}is{\isacharunderscore}pkt{\isasymrangle}\isanewline
\isaindent{{\isacharparenleft}}{\isacharparenleft}{\isacharbraceleft}PToy{\isacharminus}{\isacharcolon}{\isadigit{6}}{\isacharbraceright}{\isasymlangle}{\isasymlambda}{\isasymxi}{\isachardot}\ \textsf{if}\ no\ {\isasymxi}\ {\isasymle}\ num\ {\isasymxi}\ \textsf{then}\ {\isacharbraceleft}{\isasymxi}{\isacharbraceright}\ \textsf{else}\ {\isasymemptyset}{\isasymrangle}\isanewline
\isaindent{{\isacharparenleft}{\isacharparenleft}}{\isacharbraceleft}PToy{\isacharminus}{\isacharcolon}{\isadigit{7}}{\isacharbraceright}{\isasymlbrakk}{\isasymlambda}{\isasymxi}{\isachardot}\ {\isasymxi}{\isasymlparr}no\ {\isacharcolon}{\isacharequal}\ num\ {\isasymxi}{\isasymrparr}{\isasymrbrakk}\isanewline
\isaindent{{\isacharparenleft}{\isacharparenleft}}{\isacharbraceleft}PToy{\isacharminus}{\isacharcolon}{\isadigit{8}}{\isacharbraceright}{\isasymlbrakk}{\isasymlambda}{\isasymxi}{\isachardot}\ {\isasymxi}{\isasymlparr}nhid\ {\isacharcolon}{\isacharequal}\ sid\ {\isasymxi}{\isasymrparr}{\isasymrbrakk}\isanewline
\isaindent{{\isacharparenleft}{\isacharparenleft}}{\isacharbraceleft}PToy{\isacharminus}{\isacharcolon}{\isadigit{9}}{\isacharbraceright}broadcast{\isacharparenleft}{\isasymlambda}{\isasymxi}{\isachardot}\ Pkt\ {\isacharparenleft}no\ {\isasymxi}{\isacharparenright}\ {\isacharparenleft}id\ {\isasymxi}{\isacharparenright}{\isacharparenright}\ {\isachardot}\isanewline
\isaindent{{\isacharparenleft}{\isacharparenleft}}{\isacharbraceleft}PToy{\isacharminus}{\isacharcolon}{\isadigit{1}}{\isadigit{0}}{\isacharbraceright}{\isasymlbrakk}clear{\isacharunderscore}locals{\isasymrbrakk}\isanewline
\isaindent{{\isacharparenleft}{\isacharparenleft}}call{\isacharparenleft}PToy{\isacharparenright}\isanewline
\isaindent{{\isacharparenleft}{\isacharparenleft}}{\isasymoplus}\isanewline
\isaindent{{\isacharparenleft}{\isacharparenleft}}{\isacharbraceleft}PToy{\isacharminus}{\isacharcolon}{\isadigit{6}}{\isacharbraceright}{\isasymlangle}{\isasymlambda}{\isasymxi}{\isachardot}\ \textsf{if}\ num\ {\isasymxi}\ {\isacharless}\ no\ {\isasymxi}\ \textsf{then}\ {\isacharbraceleft}{\isasymxi}{\isacharbraceright}\ \textsf{else}\ {\isasymemptyset}{\isasymrangle}\isanewline
\isaindent{{\isacharparenleft}{\isacharparenleft}}{\isacharbraceleft}PToy{\isacharminus}{\isacharcolon}{\isadigit{1}}{\isadigit{1}}{\isacharbraceright}{\isasymlbrakk}clear{\isacharunderscore}locals{\isasymrbrakk}\isanewline
\isaindent{{\isacharparenleft}{\isacharparenleft}}call{\isacharparenleft}PToy{\isacharparenright}{\isacharparenright}{\isacharparenright}%
\end{isabelle}}

\snip{sigmap}{0}{0}{\isa{\sigmatoy\ i\ {\isacharequal}\ {\isacharbraceleft}{\isacharparenleft}toy{\isacharunderscore}init\ i,\ \gammatoy\ PToy{\isacharparenright}{\isacharbraceright}}}
\snip{ptoy}{0}{0}{\isa{ptoy}}
\snip{ptoy_term}{0}{0}{\isa{{\isasymlparr}init\ {\isacharequal}\ {\isacharbraceleft}{\isacharparenleft}toy{\isacharunderscore}init\ i,\ \gammatoy\ PToy{\isacharparenright}{\isacharbraceright},\ trans\ {\isacharequal}\ \seqpsos\ \gammatoy{\isasymrparr}}}
\snip{ptoy_term'}{0}{0}{\isa{{\isasymlparr}init\ {\isacharequal}\ \sigmatoy\ i,\ trans\ {\isacharequal}\ \seqpsos\ \gammatoy{\isasymrparr}}}
\snip{ptoy_type}{0}{0}{\isa{nat\ {\isasymRightarrow}\ {\isacharparenleft}state\ {\isasymtimes}\ {\isacharparenleft}state,\ msg,\ pseqp,\ pseqp\ label{\isacharparenright}\ seqp,\ msg\ seq{\isacharunderscore}action{\isacharparenright}\ automaton}}

\snip{sigmap'}{0}{0}{\isa{{\isasymsigma}\isactrlsub O\isactrlsub T\isactrlsub O\isactrlsub Y\ {\isacharequal}\ {\isacharbraceleft}{\isacharparenleft}toy{\isacharunderscore}init,\ \gammatoy\ PToy{\isacharparenright}{\isacharbraceright}}}
\snip{optoy}{0}{0}{\isa{optoy}}
\snip{optoy_term}{0}{0}{\isa{{\isasymlparr}init\ {\isacharequal}\ {\isacharbraceleft}{\isacharparenleft}toy{\isacharunderscore}init,\ \gammatoy\ PToy{\isacharparenright}{\isacharbraceright},\ trans\ {\isacharequal}\ \oseqpsos\ \gammatoy\ i{\isasymrparr}}}

\snip{gammaq}{0}{0}{\isa{\gammaqmsg\ {\isacharparenleft}{\isacharparenright}}}
\snip{gammaq_unit}{0}{0}{\begin{isabelle}%
\gammaqmsg\ {\isacharparenleft}{\isacharparenright}\ {\isacharequal}\isanewline
{\isacharbraceleft}{\isacharparenleft}{\isacharparenright}{\isacharminus}{\isacharcolon}{\isadigit{0}}{\isacharbraceright}receive{\isacharparenleft}{\isasymlambda}msg\ msgs{\isachardot}\ msgs\ {\isacharat}\ {\isacharbrackleft}msg{\isacharbrackright}{\isacharparenright}\ {\isachardot}\isanewline
call{\isacharparenleft}{\isacharparenleft}{\isacharparenright}{\isacharparenright}\isanewline
{\isasymoplus}\isanewline
{\isacharbraceleft}{\isacharparenleft}{\isacharparenright}{\isacharminus}{\isacharcolon}{\isadigit{0}}{\isacharbraceright}{\isasymlangle}{\isasymlambda}msgs{\isachardot}\ \textsf{if}\ msgs\ {\isasymnoteq}\ {\isacharbrackleft}\,{\isacharbrackright}\ \textsf{then}\ {\isacharbraceleft}msgs{\isacharbraceright}\ \textsf{else}\ {\isasymemptyset}{\isasymrangle}\isanewline
{\isacharparenleft}{\isacharbraceleft}{\isacharparenleft}{\isacharparenright}{\isacharminus}{\isacharcolon}{\isadigit{1}}{\isacharbraceright}send{\isacharparenleft}hd{\isacharparenright}\ {\isachardot}\isanewline
\isaindent{{\isacharparenleft}}{\isacharbraceleft}{\isacharparenleft}{\isacharparenright}{\isacharminus}{\isacharcolon}{\isadigit{2}}{\isacharbraceright}{\isasymlbrakk}tl{\isasymrbrakk}\isanewline
\isaindent{{\isacharparenleft}}call{\isacharparenleft}{\isacharparenleft}{\isacharparenright}{\isacharparenright}\isanewline
\isaindent{{\isacharparenleft}}{\isasymoplus}\isanewline
\isaindent{{\isacharparenleft}}{\isacharbraceleft}{\isacharparenleft}{\isacharparenright}{\isacharminus}{\isacharcolon}{\isadigit{1}}{\isacharbraceright}receive{\isacharparenleft}{\isasymlambda}msg\ msgs{\isachardot}\ msgs\ {\isacharat}\ {\isacharbrackleft}msg{\isacharbrackright}{\isacharparenright}\ {\isachardot}\isanewline
\isaindent{{\isacharparenleft}}call{\isacharparenleft}{\isacharparenleft}{\isacharparenright}{\isacharparenright}{\isacharparenright}%
\end{isabelle}}
\snip{gammaq_unit'}{0}{0}{\begin{isabelle}%
\gammaqmsg\ {\isacharparenleft}{\isacharparenright}\ {\isacharequal}\isanewline
labelled\ {\isacharparenleft}{\isacharparenright}\isanewline
\isaindent{\ }{\isacharparenleft}receive{\isacharparenleft}{\isasymlambda}msg\ msgs{\isachardot}\ msgs\ {\isacharat}\ {\isacharbrackleft}msg{\isacharbrackright}{\isacharparenright}\ {\isachardot}\isanewline
\isaindent{\ {\isacharparenleft}}call{\isacharparenleft}{\isacharparenleft}{\isacharparenright}{\isacharparenright}\isanewline
\isaindent{\ {\isacharparenleft}}{\isasymoplus}\isanewline
\isaindent{\ {\isacharparenleft}}{\isasymlangle}{\isasymlambda}msgs{\isachardot}\ \textsf{if}\ msgs\ {\isasymnoteq}\ {\isacharbrackleft}\,{\isacharbrackright}\ \textsf{then}\ {\isacharbraceleft}msgs{\isacharbraceright}\ \textsf{else}\ {\isasymemptyset}{\isasymrangle}\isanewline
\isaindent{\ {\isacharparenleft}}{\isacharparenleft}send{\isacharparenleft}hd{\isacharparenright}\ {\isachardot}\isanewline
\isaindent{\ {\isacharparenleft}{\isacharparenleft}}{\isasymlbrakk}tl{\isasymrbrakk}\isanewline
\isaindent{\ {\isacharparenleft}{\isacharparenleft}}call{\isacharparenleft}{\isacharparenleft}{\isacharparenright}{\isacharparenright}\isanewline
\isaindent{\ {\isacharparenleft}{\isacharparenleft}}{\isasymoplus}\isanewline
\isaindent{\ {\isacharparenleft}{\isacharparenleft}}receive{\isacharparenleft}{\isasymlambda}msg\ msgs{\isachardot}\ msgs\ {\isacharat}\ {\isacharbrackleft}msg{\isacharbrackright}{\isacharparenright}\ {\isachardot}\isanewline
\isaindent{\ {\isacharparenleft}{\isacharparenleft}}call{\isacharparenleft}{\isacharparenleft}{\isacharparenright}{\isacharparenright}{\isacharparenright}{\isacharparenright}%
\end{isabelle}}
\snip{sigmaq}{0}{0}{\isa{\gammaqmsg\ {\isacharequal}\ {\isacharbraceleft}{\isacharparenleft}{\isacharbrackleft}\,{\isacharbrackright},\ \gammaqmsg\ {\isacharparenleft}{\isacharparenright}{\isacharparenright}{\isacharbraceright}}}
\snip{qmsg}{0}{0}{\isa{qmsg}}
\snip{qmsg_term}{0}{0}{\isa{{\isasymlparr}init\ {\isacharequal}\ \gammaqmsg,\ trans\ {\isacharequal}\ \seqpsos\ \gammaqmsg{\isasymrparr}}}

\snip{clear_locals_sip}{0}{0}{\isa{{\isasymxi}{\isasymlparr}sid\ {\isacharcolon}{\isacharequal}\ SOME\ x{\isachardot}\ x\ {\isasymnoteq}\ id\ {\isasymxi}{\isasymrparr}}}

\snip{ptoy_qmsg_term}{0}{0}{\isa{ptoy\ i\ {\isasymlangle}{\isasymlangle}\ qmsg}}
\snip{opnet_p}{0}{0}{\isa{opnet\ {\isasymPsi}}}
\snip{opnet_node}{0}{0}{\isa{opnet\ {\isacharparenleft}{\isasymlambda}i{\isachardot}\ optoy\ i\ {\isasymlangle}{\isasymlangle}\isactrlbsub i\isactrlesub \ qmsg{\isacharparenright}\ {\isasymlangle}i{\isacharsemicolon}\ R{\isasymrangle}}}

\snip{net_tree_ips_node}{0}{0}{\isa{net{\isacharunderscore}tree{\isacharunderscore}ips\ {\isasymlangle}i{\isacharsemicolon}\ R{\isasymrangle}\ {\isacharequal}\ {\isacharbraceleft}i{\isacharbraceright}}}
\snip{net_tree_ips_node_lhs}{0}{0}{\isa{net{\isacharunderscore}tree{\isacharunderscore}ips\ {\isasymlangle}i{\isacharsemicolon}\ \Ri{\isasymrangle}}}
\snip{net_tree_ips_node_rhs}{0}{0}{\isa{{\isacharbraceleft}i{\isacharbraceright}}}

\snip{net_tree_ips_par}{0}{0}{\isa{net{\isacharunderscore}tree{\isacharunderscore}ips\ {\isacharparenleft}{\isasymPsi}\isactrlsub {\isadigit{1}}\parallelcomp{}{\isasymPsi}\isactrlsub {\isadigit{2}}{\isacharparenright}\ {\isacharequal}\ net{\isacharunderscore}tree{\isacharunderscore}ips\ {\isasymPsi}\isactrlsub {\isadigit{1}}\ {\isasymunion}\ net{\isacharunderscore}tree{\isacharunderscore}ips\ {\isasymPsi}\isactrlsub {\isadigit{2}}}}
\snip{net_tree_ips_par_lhs}{0}{0}{\isa{net{\isacharunderscore}tree{\isacharunderscore}ips\ {\isacharparenleft}{\isasymPsi}\isactrlsub {\isadigit{1}}\!\parallelcomp{}{\isasymPsi}\isactrlsub {\isadigit{2}}{\isacharparenright}}}
\snip{net_tree_ips_par_rhs}{0}{0}{\isa{net{\isacharunderscore}tree{\isacharunderscore}ips\ {\isasymPsi}\isactrlsub {\isadigit{1}}\ {\isasymunion}\ net{\isacharunderscore}tree{\isacharunderscore}ips\ {\isasymPsi}\isactrlsub {\isadigit{2}}}}

\snip{opnet_p1_par_p2}{0}{0}{\isa{opnet\ {\isacharparenleft}{\isasymlambda}i{\isachardot}\ optoy\ i\ {\isasymlangle}{\isasymlangle}\isactrlbsub i\isactrlesub \ qmsg{\isacharparenright}\ {\isacharparenleft}{\isasymPsi}\isactrlsub {\isadigit{1}}\parallelcomp{}{\isasymPsi}\isactrlsub {\isadigit{2}}{\isacharparenright}}}

\snip{pnet_lift_prem1}{0}{0}{\isa{opnet\ onp\ {\isasymlangle}i{\isacharsemicolon}\ R{\isasymrangle}\ {\isasymTurnstile}\ {\isacharparenleft}otherwith\ S\ {\isacharbraceleft}i{\isacharbraceright}\ {\isacharparenleft}oarrivemsg\ I{\isacharparenright},\ other\ U\ {\isacharbraceleft}i{\isacharbraceright}\ {\isasymrightarrow}{\isacharparenright}\ global\ {\isacharparenleft}P\ i{\isacharparenright}}}

\snip{pnet_lift_assume_s}{0}{0}{\isa{otherwith\ S\ {\isacharparenleft}net{\isacharunderscore}tree{\isacharunderscore}ips\ p{\isacharparenright}\ {\isacharparenleft}oarrivemsg\ I{\isacharparenright}}}
\snip{pnet_lift_assume_u}{0}{0}{\isa{other\ U\ {\isacharparenleft}net{\isacharunderscore}tree{\isacharunderscore}ips\ p{\isacharparenright}}}
\snip{pnet_lift_shows}{0}{0}{\isa{global\ {\isacharparenleft}{\isasymlambda}{\isasymsigma}{\isachardot}\ {\isasymforall}i{\isasymin}net{\isacharunderscore}tree{\isacharunderscore}ips\ p{\isachardot}\ P\ i\ {\isasymsigma}{\isacharparenright}}}

\snip{pnet_lift_castmsg_pred}{0}{0}{\isa{castmsg\ {\isacharparenleft}I\ {\isasymsigma}{\isacharparenright}}}
\snip{pnet_lift_arrivemsg_pred}{0}{0}{\isa{oarrivemsg\ I\ {\isasymsigma}}}
\snip{castmsg_cast}{0}{0}{\isa{castmsg\ M\ {\isasymsigma}\ {\isacharparenleft}R{\isacharcolon}{\isacharasterisk}cast{\isacharparenleft}m{\isacharparenright}{\isacharparenright}}}
\snip{castmsg_other}{0}{0}{\isa{castmsg\ M\ \isasymsigma\ a}}

\snip{pnet_lift_castmsg}{0}{0}{\isa{opnet\ onp\ {\isasymlangle}i{\isacharsemicolon}\ R{\isasymrangle}\ \ostepinv\ {\isacharparenleft}{\isasymlambda}{\isasymsigma}\ {\isacharunderscore}{\isachardot}\ oarrivemsg\ I\ {\isasymsigma},\ other\ U\ {\isacharbraceleft}i{\isacharbraceright}\ {\isasymrightarrow}{\isacharparenright}\ globala\ {\isacharparenleft}{\isasymlambda}{\isacharparenleft}{\isasymsigma},\ a,\ {\isasymsigma}{\isacharprime}{\isacharparenright}{\isachardot}\ castmsg\ {\isacharparenleft}I\ {\isasymsigma}{\isacharparenright}\ a{\isacharparenright}}}
\snip{pnet_lift_s}{0}{0}{\isa{opnet\ onp\ {\isasymlangle}i{\isacharsemicolon}\ R{\isasymrangle}\ \ostepinv\ {\isacharparenleft}{\isasymlambda}{\isasymsigma}\ {\isacharunderscore}{\isachardot}\ oarrivemsg\ I\ {\isasymsigma},\ other\ U\ {\isacharbraceleft}i{\isacharbraceright}\ {\isasymrightarrow}{\isacharparenright}\ globala\ {\isacharparenleft}{\isasymlambda}{\isacharparenleft}{\isasymsigma},\ a,\ {\isasymsigma}{\isacharprime}{\isacharparenright}{\isachardot}\ S\ {\isacharparenleft}{\isasymsigma}\ i{\isacharparenright}\ {\isacharparenleft}{\isasymsigma}{\isacharprime}\ i{\isacharparenright}{\isacharparenright}}}
\snip{pnet_lift_u}{0}{0}{\isa{opnet\ onp\ {\isasymlangle}i{\isacharsemicolon}\ R{\isasymrangle}\ \ostepinv\ {\isacharparenleft}{\isasymlambda}{\isasymsigma}\ {\isacharunderscore}{\isachardot}\ oarrivemsg\ I\ {\isasymsigma},\ other\ U\ {\isacharbraceleft}i{\isacharbraceright}\ {\isasymrightarrow}{\isacharparenright}\ globala\ {\isacharparenleft}{\isasymlambda}{\isacharparenleft}{\isasymsigma},\ a,\ {\isasymsigma}{\isacharprime}{\isacharparenright}{\isachardot}\ U\ {\isacharparenleft}{\isasymsigma}\ i{\isacharparenright}\ {\isacharparenleft}{\isasymsigma}{\isacharprime}\ i{\isacharparenright}{\isacharparenright}}}

\snip{pnet_lift_concl}{0}{0}{\isa{opnet\ onp\ p\ {\isasymTurnstile}\ {\isacharparenleft}otherwith\ S\ {\isacharparenleft}net{\isacharunderscore}tree{\isacharunderscore}ips\ p{\isacharparenright}\ {\isacharparenleft}oarrivemsg\ I{\isacharparenright},\ other\ U\ {\isacharparenleft}net{\isacharunderscore}tree{\isacharunderscore}ips\ p{\isacharparenright}\ {\isasymrightarrow}{\isacharparenright}\ global\ {\isacharparenleft}{\isasymlambda}{\isasymsigma}{\isachardot}\ {\isasymforall}i{\isasymin}net{\isacharunderscore}tree{\isacharunderscore}ips\ p{\isachardot}\ P\ i\ {\isasymsigma}{\isacharparenright}}}
\snip{other_net_tree_ips_par_left}{0}{0}{\isa{\mbox{}\inferrule{\mbox{other\ U\ {\isacharparenleft}net{\isacharunderscore}tree{\isacharunderscore}ips\ {\isacharparenleft}p\isactrlsub {\isadigit{1}}\parallelcomp{}p\isactrlsub {\isadigit{2}}{\isacharparenright}{\isacharparenright}\ {\isasymsigma}\ {\isasymsigma}{\isacharprime}}\\\ \mbox{{\isasymAnd}{\isasymxi}{\isachardot}\ U\ {\isasymxi}\ {\isasymxi}}}{\mbox{other\ U\ {\isacharparenleft}net{\isacharunderscore}tree{\isacharunderscore}ips\ p\isactrlsub {\isadigit{1}}{\isacharparenright}\ {\isasymsigma}\ {\isasymsigma}{\isacharprime}}}}}
\snip{net_tree_ips_p1_par_p2}{0}{0}{\isa{i{\isasymin}net{\isacharunderscore}tree{\isacharunderscore}ips\ {\isacharparenleft}p\isactrlsub {\isadigit{1}}\parallelcomp{}p\isactrlsub {\isadigit{2}}{\isacharparenright}}}
\snip{net_tree_ips_not_p1_par_p2}{0}{0}{\isa{i\ {\isasymnotin}\ net{\isacharunderscore}tree{\isacharunderscore}ips\ {\isacharparenleft}p\isactrlsub {\isadigit{1}}\parallelcomp{}p\isactrlsub {\isadigit{2}}{\isacharparenright}}}
\snip{net_tree_ips_p1}{0}{0}{\isa{i{\isasymin}net{\isacharunderscore}tree{\isacharunderscore}ips\ p\isactrlsub {\isadigit{1}}}}
\snip{net_tree_ips_p2}{0}{0}{\isa{i{\isasymin}net{\isacharunderscore}tree{\isacharunderscore}ips\ p\isactrlsub {\isadigit{2}}}}
\snip{net_tree_ips_not_p1_sigma}{0}{0}{\isa{{\isasymforall}j{\isachardot}\ j\ {\isasymnotin}\ net{\isacharunderscore}tree{\isacharunderscore}ips\ p\isactrlsub {\isadigit{1}}\ {\isasymlongrightarrow}\ {\isasymsigma}{\isacharprime}\ j\ {\isacharequal}\ {\isasymsigma}\ j}}
\snip{node_proc_reachable_prem_1}{0}{0}{\isa{{\isacharparenleft}{\isasymsigma},\ \NodeS\ i\ s\ R{\isacharparenright}{\isasymin}oreachable\ {\isacharparenleft}{\isasymlangle}i\ {\isacharcolon}\ A\ {\isacharcolon}\ \Ri{\isasymrangle}\isactrlsub o{\isacharparenright}\ {\isacharparenleft}otherwith\ E\ {\isacharbraceleft}i{\isacharbraceright}\ {\isacharparenleft}oarrivemsg\ M{\isacharparenright}{\isacharparenright}\ {\isacharparenleft}other\ F\ {\isacharbraceleft}i{\isacharbraceright}{\isacharparenright}}}

\snip{node_proc_reachable_prem_2}{0}{0}{\isa{{\isasymAnd}{\isasymxi}\ {\isasymxi}{\isacharprime}{\isachardot}\ E\ {\isasymxi}\ {\isasymxi}{\isacharprime}\ {\isasymLongrightarrow}\ F\ {\isasymxi}\ {\isasymxi}{\isacharprime}}}
\snip{node_proc_reachable_prem_2_prem}{0}{0}{\isa{E\ {\isasymxi}\ {\isasymxi}{\isacharprime}}}
\snip{node_proc_reachable_prem_2_concl}{0}{0}{\isa{F\ {\isasymxi}\ {\isasymxi}{\isacharprime}}}
\snip{node_proc_reachable_concl}{0}{0}{\isa{{\isacharparenleft}{\isasymsigma},\ s{\isacharparenright}{\isasymin}oreachable\ A\ {\isacharparenleft}otherwith\ E\ {\isacharbraceleft}i{\isacharbraceright}\ {\isacharparenleft}orecvmsg\ M{\isacharparenright}{\isacharparenright}\ {\isacharparenleft}other\ F\ {\isacharbraceleft}i{\isacharbraceright}{\isacharparenright}}}

\snip{node_proc_lift_prem_1}{0}{0}{\isa{A\ {\isasymTurnstile}\ {\isacharparenleft}otherwith\ E\ {\isacharbraceleft}i{\isacharbraceright}\ {\isacharparenleft}orecvmsg\ M{\isacharparenright},\ other\ F\ {\isacharbraceleft}i{\isacharbraceright}\ {\isasymrightarrow}{\isacharparenright}\ {\isacharparenleft}{\isasymlambda}{\isacharparenleft}{\isasymsigma},\ {\isacharunderscore}{\isacharparenright}{\isachardot}\ P\ {\isasymsigma}{\isacharparenright}}}
\snip{node_proc_lift_prem_2}{0}{0}{\isa{{\isasymAnd}{\isasymxi}\ {\isasymxi}{\isacharprime}{\isachardot}\ E\ {\isasymxi}\ {\isasymxi}{\isacharprime}\ {\isasymLongrightarrow}\ F\ {\isasymxi}\ {\isasymxi}{\isacharprime}}}
\snip{node_proc_lift_concl}{0}{0}{\isa{{\isasymlangle}i\ {\isacharcolon}\ A\ {\isacharcolon}\ \Ri{\isasymrangle}\isactrlsub o\ {\isasymTurnstile}\ {\isacharparenleft}otherwith\ E\ {\isacharbraceleft}i{\isacharbraceright}\ {\isacharparenleft}oarrivemsg\ M{\isacharparenright},\ other\ F\ {\isacharbraceleft}i{\isacharbraceright}\ {\isasymrightarrow}{\isacharparenright}\ {\isacharparenleft}{\isasymlambda}{\isacharparenleft}{\isasymsigma},\ {\isacharunderscore}{\isacharparenright}{\isachardot}\ P\ {\isasymsigma}{\isacharparenright}}}

\snip{subnet_oreachable_prem_1}{0}{0}{\isa{{\isacharparenleft}{\isasymsigma}\!,\!\ \subnetsless{s}{t}{\isacharparenright}{\isasymin}oreachable\ {\isacharparenleft}opnet\;onp\;{\isacharparenleft}{\isasymPsi}\isactrlsub {\isadigit{1}}\!\parallelcomp{}{\isasymPsi}\isactrlsub {\isadigit{2}}{\isacharparenright}{\isacharparenright}\ S\ U}}
\snip{subnet_oreachable_prem_1_L}{0}{0}{\isa{S\ {\isacharequal}\ otherwith\ E\ {\isacharparenleft}net{\isacharunderscore}tree{\isacharunderscore}ips\ {\isacharparenleft}{\isasymPsi}\isactrlsub {\isadigit{1}}\!\parallelcomp{}{\isasymPsi}\isactrlsub {\isadigit{2}}{\isacharparenright}{\isacharparenright}\ {\isacharparenleft}oarrivemsg\ M{\isacharparenright}}}
\snip{subnet_oreachable_prem_1_E}{0}{0}{\isa{U\ {\isacharequal}\ other\ F\ {\isacharparenleft}net{\isacharunderscore}tree{\isacharunderscore}ips\ {\isacharparenleft}{\isasymPsi}\isactrlsub {\isadigit{1}}\!\parallelcomp{}{\isasymPsi}\isactrlsub {\isadigit{2}}{\isacharparenright}{\isacharparenright}}}

\snip{subnet_oreachable_prem_2}{0}{0}{\isa{{\isasymAnd}{\isasymxi}{\isachardot}\ E\ {\isasymxi}\ {\isasymxi}}}
\snip{subnet_oreachable_prem_3}{0}{0}{\isa{{\isasymAnd}{\isasymxi}{\isachardot}\ F\ {\isasymxi}\ {\isasymxi}}}
\snip{subnet_oreachable_prem_4}{0}{0}{\isa{{\isasymAnd}i\ R{\isachardot}\ {\isasymlangle}i\ {\isacharcolon}\ onp\ i\ {\isacharcolon}\ R{\isasymrangle}\isactrlsub o\ \ostepinv\ {\isacharparenleft}{\isasymlambda}{\isasymsigma}\ {\isacharunderscore}{\isachardot}\ oarrivemsg\ I\ {\isasymsigma},\ other\ F\ {\isacharbraceleft}i{\isacharbraceright}\ {\isasymrightarrow}{\isacharparenright}\ globala\ {\isacharparenleft}{\isasymlambda}{\isacharparenleft}{\isasymsigma},\ a,\ {\isacharunderscore}{\isacharparenright}{\isachardot}\ castmsg\ {\isacharparenleft}I\ {\isasymsigma}{\isacharparenright}\ a{\isacharparenright}}}

\snip{subnet_oreachable_prem_4_p}{0}{0}{\isa{{\isasymlangle}i\ {\isacharcolon}\ onp\ i\ {\isacharcolon}\ R{\isasymrangle}\isactrlsub o\ \ostepinv\ {\isacharparenleft}{\isasymlambda}{\isasymsigma}\ {\isacharunderscore}{\isachardot}\ oarrivemsg\ I\ {\isasymsigma},\ other\ F\ {\isacharbraceleft}i{\isacharbraceright}\ {\isasymrightarrow}{\isacharparenright}\ {\isacharparenleft}{\isasymlambda}{\isacharparenleft}{\isacharparenleft}{\isasymsigma},\ {\isacharunderscore}{\isacharparenright},\ a,\ ({\isasymsigma}{\isacharprime},\ {\isacharunderscore}){\isacharparenright}{\isachardot}\ castmsg\ {\isacharparenleft}I\ {\isasymsigma}{\isacharparenright}\ a{\isacharparenright}}}
\snip{subnet_oreachable_prem_4_p'}{0}{0}{\isa{A\ i\ {\isacharequal}\ {\isasymlangle}i\ {\isacharcolon}\ onp\ i\ {\isacharcolon}\ R{\isasymrangle}\isactrlsub o}}
\snip{subnet_oreachable_prem_5_p}{0}{0}{\isa{{\isasymlangle}i\ {\isacharcolon}\ onp\ i\ {\isacharcolon}\ R{\isasymrangle}\isactrlsub o\ \ostepinv\ {\isacharparenleft}{\isasymlambda}{\isasymsigma}\ {\isacharunderscore}{\isachardot}\ oarrivemsg\ I\ {\isasymsigma},\ other\ F\ {\isacharbraceleft}i{\isacharbraceright}\ {\isasymrightarrow}{\isacharparenright}\ {\isacharparenleft}{\isasymlambda}{\isacharparenleft}{\isacharparenleft}{\isasymsigma},\ {\isacharunderscore}{\isacharparenright},\ a,\ ({\isasymsigma}{\isacharprime},\ {\isacharunderscore}){\isacharparenright}{\isachardot}\ a\ {\isasymnoteq}\ {\isasymtau}\ {\isasymand}\ {\isacharparenleft}{\isasymforall}i\ d{\isachardot}\ a\ {\isasymnoteq}\ i{\isacharcolon}deliver{\isacharparenleft}d{\isacharparenright}{\isacharparenright}\ {\isasymlongrightarrow}\ E\ {\isacharparenleft}{\isasymsigma}\ i{\isacharparenright}\ {\isacharparenleft}{\isasymsigma}{\isacharprime}\ i{\isacharparenright}{\isacharparenright}}}
\snip{subnet_oreachable_prem_6_p}{0}{0}{\isa{{\isasymlangle}i\ {\isacharcolon}\ onp\ i\ {\isacharcolon}\ R{\isasymrangle}\isactrlsub o\ \ostepinv\ {\isacharparenleft}{\isasymlambda}{\isasymsigma}\ {\isacharunderscore}{\isachardot}\ oarrivemsg\ I\ {\isasymsigma},\ other\ F\ {\isacharbraceleft}i{\isacharbraceright}\ {\isasymrightarrow}{\isacharparenright}\ {\isacharparenleft}{\isasymlambda}{\isacharparenleft}{\isacharparenleft}{\isasymsigma},\ {\isacharunderscore}{\isacharparenright},\ a,\ ({\isasymsigma}{\isacharprime},\ {\isacharunderscore}){\isacharparenright}{\isachardot}\ a\ {\isacharequal}\ {\isasymtau}\ {\isasymor}\ {\isacharparenleft}{\isasymexists}d{\isachardot}\ a\ {\isacharequal}\ i{\isacharcolon}deliver{\isacharparenleft}d{\isacharparenright}{\isacharparenright}\ {\isasymlongrightarrow}\ F\ {\isacharparenleft}{\isasymsigma}\ i{\isacharparenright}\ {\isacharparenleft}{\isasymsigma}{\isacharprime}\ i{\isacharparenright}{\isacharparenright}}}

\snip{subnet_oreachable_prem_5}{0}{0}{\isa{{\isasymAnd}i\ R{\isachardot}\ {\isasymlangle}i\ {\isacharcolon}\ onp\ i\ {\isacharcolon}\ R{\isasymrangle}\isactrlsub o\ \ostepinv\ {\isacharparenleft}{\isasymlambda}{\isasymsigma}\ {\isacharunderscore}{\isachardot}\ oarrivemsg\ I\ {\isasymsigma},\ other\ F\ {\isacharbraceleft}i{\isacharbraceright}\ {\isasymrightarrow}{\isacharparenright}\ globala\ {\isacharparenleft}{\isasymlambda}{\isacharparenleft}{\isasymsigma},\ a,\ {\isasymsigma}{\isacharprime}{\isacharparenright}{\isachardot}\ a\ {\isasymnoteq}\ {\isasymtau}\ {\isasymand}\ {\isacharparenleft}{\isasymforall}i\ d{\isachardot}\ a\ {\isasymnoteq}\ i{\isacharcolon}deliver{\isacharparenleft}d{\isacharparenright}{\isacharparenright}\ {\isasymlongrightarrow}\ E\ {\isacharparenleft}{\isasymsigma}\ i{\isacharparenright}\ {\isacharparenleft}{\isasymsigma}{\isacharprime}\ i{\isacharparenright}{\isacharparenright}}}
\snip{subnet_oreachable_prem_5_pred}{0}{0}{\isa{E\ {\isacharparenleft}{\isasymsigma}\ i{\isacharparenright}\ {\isacharparenleft}{\isasymsigma}{\isacharprime}\ i{\isacharparenright}}}

\snip{subnet_oreachable_prem_6}{0}{0}{\isa{{\isasymAnd}i\ R{\isachardot}\ {\isasymlangle}i\ {\isacharcolon}\ onp\ i\ {\isacharcolon}\ R{\isasymrangle}\isactrlsub o\ \ostepinv\ {\isacharparenleft}{\isasymlambda}{\isasymsigma}\ {\isacharunderscore}{\isachardot}\ oarrivemsg\ I\ {\isasymsigma},\ other\ F\ {\isacharbraceleft}i{\isacharbraceright}\ {\isasymrightarrow}{\isacharparenright}\ globala\ {\isacharparenleft}{\isasymlambda}{\isacharparenleft}{\isasymsigma},\ a,\ {\isasymsigma}{\isacharprime}{\isacharparenright}{\isachardot}\ a\ {\isacharequal}\ {\isasymtau}\ {\isasymor}\ {\isacharparenleft}{\isasymexists}d{\isachardot}\ a\ {\isacharequal}\ i{\isacharcolon}deliver{\isacharparenleft}d{\isacharparenright}{\isacharparenright}\ {\isasymlongrightarrow}\ F\ {\isacharparenleft}{\isasymsigma}\ i{\isacharparenright}\ {\isacharparenleft}{\isasymsigma}{\isacharprime}\ i{\isacharparenright}{\isacharparenright}}}
\snip{subnet_oreachable_prem_6_pred}{0}{0}{\isa{F\ {\isacharparenleft}{\isasymsigma}\ i{\isacharparenright}\ {\isacharparenleft}{\isasymsigma}{\isacharprime}\ i{\isacharparenright}}}

\snip{subnet_oreachable_prem_9}{0}{0}{\isa{S\isactrlsub {\isadigit{1}}\ {\isacharequal}\ otherwith\ E\ {\isacharparenleft}net{\isacharunderscore}tree{\isacharunderscore}ips\ {\isasymPsi}\isactrlsub {\isadigit{1}}{\isacharparenright}\ {\isacharparenleft}oarrivemsg\ M{\isacharparenright}}}
\snip{subnet_oreachable_prem_10}{0}{0}{\isa{U\isactrlsub {\isadigit{1}}\ {\isacharequal}\ other\ F\ {\isacharparenleft}net{\isacharunderscore}tree{\isacharunderscore}ips\ {\isasymPsi}\isactrlsub {\isadigit{1}}{\isacharparenright}}}
\snip{subnet_oreachable_prem_11}{0}{0}{\isa{S\isactrlsub {\isadigit{2}}\ {\isacharequal}\ otherwith\ E\ {\isacharparenleft}net{\isacharunderscore}tree{\isacharunderscore}ips\ {\isasymPsi}\isactrlsub {\isadigit{2}}{\isacharparenright}\ {\isacharparenleft}oarrivemsg\ M{\isacharparenright}}}
\snip{subnet_oreachable_prem_12}{0}{0}{\isa{U\isactrlsub {\isadigit{2}}\ {\isacharequal}\ other\ F\ {\isacharparenleft}net{\isacharunderscore}tree{\isacharunderscore}ips\ {\isasymPsi}\isactrlsub {\isadigit{2}}{\isacharparenright}}}

\snip{subnet_oreachable_concl_1}{0}{0}{\isa{{\isacharparenleft}{\isasymsigma},\ s{\isacharparenright}{\isasymin}oreachable\ {\isacharparenleft}opnet\ onp\ {\isasymPsi}\isactrlsub {\isadigit{1}}{\isacharparenright}\ S\isactrlsub {\isadigit{1}}\ U\isactrlsub {\isadigit{1}}}}
\snip{subnet_oreachable_concl_2}{0}{0}{\isa{{\isacharparenleft}{\isasymsigma},\ t{\isacharparenright}{\isasymin}oreachable\ {\isacharparenleft}opnet\ onp\ {\isasymPsi}\isactrlsub {\isadigit{2}}{\isacharparenright}\ S\isactrlsub {\isadigit{2}}\ U\isactrlsub {\isadigit{2}}}}
\snip{subnet_oreachable_concl_3}{0}{0}{\isa{net{\isacharunderscore}tree{\isacharunderscore}ips\ {\isasymPsi}\isactrlsub {\isadigit{1}}\ {\isasyminter}\ net{\isacharunderscore}tree{\isacharunderscore}ips\ {\isasymPsi}\isactrlsub {\isadigit{2}}\ {\isacharequal}\ {\isasymemptyset}}}

\snip{S1}{0}{0}{\isa{S\isactrlsub {\isadigit{1}}}}
\snip{U1}{0}{0}{\isa{U\isactrlsub {\isadigit{1}}}}
\snip{S2}{0}{0}{\isa{S\isactrlsub {\isadigit{2}}}}
\snip{U2}{0}{0}{\isa{U\isactrlsub {\isadigit{2}}}}

\snip{subnet_oreachable_proof_step}{0}{0}{\isa{{\isacharparenleft}{\isacharparenleft}{\isasymsigma},\ \subnets{s}{t}{\isacharparenright},\ a,\ {\isacharparenleft}{\isasymsigma}{\isacharprime},\ \subnetsadj{s{\isacharprime}}{t{\isacharprime}}{\isacharparenright}{\isacharparenright}{\isasymin}trans\ {\isacharparenleft}opnet\ onp\ {\isacharparenleft}{\isasymPsi}\isactrlsub {\isadigit{1}}\parallelcomp{}{\isasymPsi}\isactrlsub {\isadigit{2}}{\isacharparenright}{\isacharparenright}}}
\snip{subnet_oreachable_proof_jL}{0}{0}{\isa{j\ {\isasymnotin}\ net{\isacharunderscore}tree{\isacharunderscore}ips\ {\isacharparenleft}{\isasymPsi}\isactrlsub {\isadigit{1}}\!\parallelcomp{}{\isasymPsi}\isactrlsub {\isadigit{2}}{\isacharparenright}}}
\snip{subnet_oreachable_proof_oarrivemsg}{0}{0}{\isa{oarrivemsg\ M\ {\isasymsigma}}}

\snip{pnet_lift2_prem_1}{0}{0}{\isa{{\isasymAnd}ii\ \Ri{\isachardot}\ {\isasymlangle}ii\ {\isacharcolon}\ onp\ ii\ {\isacharcolon}\ \Ri{\isasymrangle}\isactrlsub o\ {\isasymTurnstile}\ {\isacharparenleft}otherwith\ E\ {\isacharbraceleft}ii{\isacharbraceright}\ {\isacharparenleft}oarrivemsg\ I{\isacharparenright},\ other\ F\ {\isacharbraceleft}ii{\isacharbraceright}\ {\isasymrightarrow}{\isacharparenright}\ {\isacharparenleft}{\isasymlambda}{\isacharparenleft}{\isasymsigma},\ {\isacharunderscore}{\isacharparenright}{\isachardot}\ P\ ii\ {\isasymsigma}{\isacharparenright}}}
\snip{pnet_lift2_prem_4}{0}{0}{\isa{{\isasymAnd}i\ R{\isachardot}\ {\isasymlangle}i\ {\isacharcolon}\ onp\ i\ {\isacharcolon}\ R{\isasymrangle}\isactrlsub o\ \ostepinv\ {\isacharparenleft}{\isasymlambda}{\isasymsigma}\ {\isacharunderscore}{\isachardot}\ oarrivemsg\ I\ {\isasymsigma},\ other\ F\ {\isacharbraceleft}i{\isacharbraceright}\ {\isasymrightarrow}{\isacharparenright}\ {\isacharparenleft}{\isasymlambda}{\isacharparenleft}{\isacharparenleft}{\isasymsigma},\ {\isacharunderscore}{\isacharparenright},\ a,\ {\isacharunderscore}{\isacharparenright}{\isachardot}\ castmsg\ {\isacharparenleft}I\ {\isasymsigma}{\isacharparenright}\ a{\isacharparenright}}}
\snip{pnet_lift2_prem_5}{0}{0}{\isa{{\isasymAnd}i\ R{\isachardot}\ {\isasymlangle}i\ {\isacharcolon}\ onp\ i\ {\isacharcolon}\ R{\isasymrangle}\isactrlsub o\ \ostepinv\ {\isacharparenleft}{\isasymlambda}{\isasymsigma}\ {\isacharunderscore}{\isachardot}\ oarrivemsg\ I\ {\isasymsigma},\ other\ F\ {\isacharbraceleft}i{\isacharbraceright}\ {\isasymrightarrow}{\isacharparenright}\ {\isacharparenleft}{\isasymlambda}{\isacharparenleft}{\isacharparenleft}{\isasymsigma},\ {\isacharunderscore}{\isacharparenright},\ a,\ {\isasymsigma}{\isacharprime},\ {\isacharunderscore}{\isacharparenright}{\isachardot}\ a\ {\isasymnoteq}\ {\isasymtau}\ {\isasymand}\ {\isacharparenleft}{\isasymforall}i\ d{\isachardot}\ a\ {\isasymnoteq}\ i{\isacharcolon}deliver{\isacharparenleft}d{\isacharparenright}{\isacharparenright}\ {\isasymlongrightarrow}\ E\ {\isacharparenleft}{\isasymsigma}\ i{\isacharparenright}\ {\isacharparenleft}{\isasymsigma}{\isacharprime}\ i{\isacharparenright}{\isacharparenright}}}
\snip{pnet_lift2_prem_6}{0}{0}{\isa{{\isasymAnd}i\ R{\isachardot}\ {\isasymlangle}i\ {\isacharcolon}\ onp\ i\ {\isacharcolon}\ R{\isasymrangle}\isactrlsub o\ \ostepinv\ {\isacharparenleft}{\isasymlambda}{\isasymsigma}\ {\isacharunderscore}{\isachardot}\ oarrivemsg\ I\ {\isasymsigma},\ other\ F\ {\isacharbraceleft}i{\isacharbraceright}\ {\isasymrightarrow}{\isacharparenright}\ {\isacharparenleft}{\isasymlambda}{\isacharparenleft}{\isacharparenleft}{\isasymsigma},\ {\isacharunderscore}{\isacharparenright},\ a,\ {\isasymsigma}{\isacharprime},\ {\isacharunderscore}{\isacharparenright}{\isachardot}\ a\ {\isacharequal}\ {\isasymtau}\ {\isasymor}\ {\isacharparenleft}{\isasymexists}d{\isachardot}\ a\ {\isacharequal}\ i{\isacharcolon}deliver{\isacharparenleft}d{\isacharparenright}{\isacharparenright}\ {\isasymlongrightarrow}\ F\ {\isacharparenleft}{\isasymsigma}\ i{\isacharparenright}\ {\isacharparenleft}{\isasymsigma}{\isacharprime}\ i{\isacharparenright}{\isacharparenright}}}
\snip{pnet_lift2_concl}{0}{0}{\isa{opnet\ onp\ p\ {\isasymTurnstile}\ {\isacharparenleft}otherwith\ E\ {\isacharparenleft}net{\isacharunderscore}tree{\isacharunderscore}ips\ p{\isacharparenright}\ {\isacharparenleft}oarrivemsg\ I{\isacharparenright},\ other\ F\ {\isacharparenleft}net{\isacharunderscore}tree{\isacharunderscore}ips\ p{\isacharparenright}\ {\isasymrightarrow}{\isacharparenright}\ {\isacharparenleft}{\isasymlambda}{\isacharparenleft}{\isasymsigma},\ {\isacharunderscore}{\isacharparenright}{\isachardot}\ {\isasymforall}i{\isasymin}net{\isacharunderscore}tree{\isacharunderscore}ips\ p{\isachardot}\ P\ i\ {\isasymsigma}{\isacharparenright}}}

\snip{oclosed_oreachable_inclosed_1}{0}{0}{\isa{{\isacharparenleft}{\isasymsigma},\ {s}{\isacharparenright}{\isasymin}oreachable\ {\isacharparenleft}oclosed\ {\isacharparenleft}opnet\ onp\ {\isasymPsi}{\isacharparenright}{\isacharparenright}\ {\isacharparenleft}{\isasymlambda}{\isacharunderscore}\ {\isacharunderscore}\ {\isacharunderscore}{\isachardot}\ True{\isacharparenright}\ U}}
\snip{oclosed_oreachable_inclosed_concl}{0}{0}{\isa{{\isacharparenleft}{\isasymsigma},\ {s}{\isacharparenright}{\isasymin}oreachable\ {\isacharparenleft}opnet\ onp\ {\isasymPsi}{\isacharparenright}\ {\isacharparenleft}otherwith\ {\isacharparenleft}op\ {\isacharequal}{\isacharparenright}\ {\isacharparenleft}net{\isacharunderscore}tree{\isacharunderscore}ips\ {\isasymPsi}{\isacharparenright}\ inoclosed{\isacharparenright}\ U}}

\snip{inclosed_closed_1}{0}{0}{\isa{opnet\ np\ {\isasymPsi}\ {\isasymTurnstile}\ {\isacharparenleft}otherwith\ {\isacharparenleft}op\ {\isacharequal}{\isacharparenright}\ {\isacharparenleft}net{\isacharunderscore}tree{\isacharunderscore}ips\ {\isasymPsi}{\isacharparenright}\ inoclosed,\ U\ {\isasymrightarrow}{\isacharparenright}\ P}}
\snip{inclosed_closed_concl}{0}{0}{\isa{oclosed\ {\isacharparenleft}opnet\ np\ {\isasymPsi}{\isacharparenright}\ {\isasymTurnstile}\ {\isacharparenleft}{\isasymlambda}{\isacharunderscore}\ {\isacharunderscore}\ {\isacharunderscore}{\isachardot}\ True,\ U\ {\isasymrightarrow}{\isacharparenright}\ P}}

\snip{closed_reachable_oreachable'_prem_2}{0}{0}{\isa{s{\isasymin}reachable\ {\isacharparenleft}closed\ {\isacharparenleft}pnet\ {\isacharparenleft}{\isasymlambda}i{\isachardot}\ ptoy\ i\ {\isasymlangle}{\isasymlangle}\ qmsg{\isacharparenright}\ n{\isacharparenright}{\isacharparenright}\ {\isacharparenleft}{\isasymlambda}{\isacharunderscore}{\isachardot}\ True{\isacharparenright}}}
\snip{closed_reachable_oreachable'_concl}{0}{0}{\isa{netgmap\ {\isacharparenleft}{\isasymlambda}{\isacharparenleft}p,\ q{\isacharparenright}{\isachardot}\ {\isacharparenleft}fst\ {\isacharparenleft}Fun{\isachardot}id\ p{\isacharparenright},\ snd\ {\isacharparenleft}Fun{\isachardot}id\ p{\isacharparenright},\ q{\isacharparenright}{\isacharparenright}\ s{\isasymin}netmask\ {\isacharparenleft}net{\isacharunderscore}tree{\isacharunderscore}ips\ n{\isacharparenright}\ {\isacharbackquote}\ oreachable\ {\isacharparenleft}oclosed\ {\isacharparenleft}opnet\ {\isacharparenleft}{\isasymlambda}i{\isachardot}\ optoy\ i\ {\isasymlangle}{\isasymlangle}\isactrlbsub i\isactrlesub \ qmsg{\isacharparenright}\ n{\isacharparenright}{\isacharparenright}\ {\isacharparenleft}{\isasymlambda}{\isacharunderscore}\ {\isacharunderscore}\ {\isacharunderscore}{\isachardot}\ True{\isacharparenright}\ U}}

\snip{inoclosed_arrive_newpkt}{0}{0}{\isa{inoclosed\ {\isasymsigma}\ {\isacharparenleft}H{\isasymnot}K{\isacharcolon}arrive{\isacharparenleft}Newpkt\ d\ dst{\isacharparenright}{\isacharparenright}}}
\snip{inoclosed_arrive_other}{0}{0}{\isa{{\isasymnot}\ inoclosed\ {\isasymsigma}\ {\isacharparenleft}H{\isasymnot}K{\isacharcolon}arrive{\isacharparenleft}m{\isacharparenright}{\isacharparenright}}}
\snip{inoclosed_newpkt}{0}{0}{\isa{{\isasymnot}\ inoclosed\ {\isasymsigma}\ {\isacharparenleft}i{\isacharcolon}newpkt{\isacharparenleft}d,\ dst{\isacharparenright}{\isacharparenright}}}
\snip{inoclosed_other}{0}{0}{\isa{inoclosed\ {\isasymsigma}\ a}}

\snip{close_opnet_prem_1}{0}{0}{\isa{openproc\ np\ onp\ sr}}
\snip{close_opnet_prem_2}{0}{0}{\isa{wf{\isacharunderscore}net{\isacharunderscore}tree\ {\isasymPsi}}}
\snip{close_opnet_prem_3}{0}{0}{\isa{oclosed\ {\isacharparenleft}opnet\ onp\ {\isasymPsi}{\isacharparenright}\ {\isasymTurnstile}\ {\isacharparenleft}{\isasymlambda}{\isacharunderscore}\ {\isacharunderscore}\ {\isacharunderscore}{\isachardot}\ True,\ U\ {\isasymrightarrow}{\isacharparenright}\ {\isacharparenleft}{\isasymlambda}{\isacharparenleft}{\isasymsigma},{\isacharunderscore}{\isacharparenright}{\isachardot}\ P\ {\isasymsigma}{\isacharparenright}}}
\snip{close_opnet_concl}{0}{0}{\isa{closed\ {\isacharparenleft}pnet\ np\ {\isasymPsi}{\isacharparenright}\ {\isasymTTurnstile}\ netglobal\ np\ sr\ P}}
\snip{pnet_reachable_transfer}{0}{0}{\isa{{\isasymlbrakk}openproc\ np\ onp\ sr{\isacharsemicolon}\ wf{\isacharunderscore}net{\isacharunderscore}tree\ {\isasymPsi}{\isacharsemicolon}\ s{\isasymin}reachable\ {\isacharparenleft}closed\ {\isacharparenleft}pnet\ np\ {\isasymPsi}{\isacharparenright}{\isacharparenright}\ TT{\isasymrbrakk}\ {\isasymLongrightarrow}\ {\isacharparenleft}default\ {\isacharparenleft}someinit\ np\ sr{\isacharparenright}\ {\isacharparenleft}netlift\ sr\ s{\isacharparenright},\ netliftc\ sr\ s{\isacharparenright}{\isasymin}oreachable\ {\isacharparenleft}oclosed\ {\isacharparenleft}opnet\ onp\ {\isasymPsi}{\isacharparenright}{\isacharparenright}\ {\isacharparenleft}{\isasymlambda}{\isacharunderscore}\ {\isacharunderscore}\ {\isacharunderscore}{\isachardot}\ True{\isacharparenright}\ U}}
\snip{pnet_reachable_transfer_prem_1}{0}{0}{\isa{openproc\ np\ onp\ sr}}
\snip{pnet_reachable_transfer_prem_2}{0}{0}{\isa{wf{\isacharunderscore}net{\isacharunderscore}tree\ {\isasymPsi}}}
\snip{pnet_reachable_transfer_prem_3}{0}{0}{\isa{s{\isasymin}reachable\ {\isacharparenleft}closed\ {\isacharparenleft}pnet\ np\ {\isasymPsi}{\isacharparenright}{\isacharparenright}\ {\isacharparenleft}{\isasymlambda}{\isacharunderscore}{\isachardot}\ True{\isacharparenright}}}
\snip{pnet_reachable_transfer_concl}{0}{0}{\isa{{\isacharparenleft}default\ {\isacharparenleft}someinit\ np\ sr{\isacharparenright}\ {\isacharparenleft}netlift\ sr\ s{\isacharparenright},\ netliftc\ sr\ s{\isacharparenright}\\\mbox{}\hfill{}{\isasymin}oreachable\ {\isacharparenleft}oclosed\ {\isacharparenleft}opnet\ onp\ {\isasymPsi}{\isacharparenright}{\isacharparenright}\ {\isacharparenleft}{\isasymlambda}{\isacharunderscore}\ {\isacharunderscore}\ {\isacharunderscore}{\isachardot}\ True{\isacharparenright}\ U}}

\snip{someinit}{0}{0}{\isa{openproc\ np\ onp\ sr\ {\isasymLongrightarrow}\ someinit\ np\ sr\ i\ {\isacharequal}\ SOME\ x{\isachardot}\ x{\isasymin}{\isacharparenleft}fst\ {\isasymcirc}\ sr{\isacharparenright}\ {\isacharbackquote}\ init\ {\isacharparenleft}np\ i{\isacharparenright}}}
\snip{someinit_concl}{0}{0}{\isa{someinit\ np\ sr\ i\ {\isacharequal}\ SOME\ x{\isachardot}\ x{\isasymin}{\isacharparenleft}fst\ {\isasymcirc}\ sr{\isacharparenright}\ {\isacharbackquote}\ init\ {\isacharparenleft}np\ i{\isacharparenright}}}
\snip{someinit_concl_lhs}{0}{0}{\isa{someinit\ np\ sr\ i}}
\snip{someinit_concl_rhs}{0}{0}{\isa{SOME\ x{\isachardot}\ x{\isasymin}{\isacharparenleft}fst\ {\isasymcirc}\ sr{\isacharparenright}\ {\isacharbackquote}\ init\ {\isacharparenleft}np\ i{\isacharparenright}}}

\snip{openproc_init}{0}{0}{\isa{openproc\ np\ onp\ sr\ {\isasymLongrightarrow}\ {\isacharbraceleft}uu{\isacharunderscore}\ {\isacharbar}\ {\isasymexists}{\isasymsigma}\ {\isasymzeta}\ s{\isachardot}\ {\isacharunderscore}\ {\isacharequal}\ {\isacharparenleft}{\isasymsigma},\ {\isasymzeta}{\isacharparenright}\ {\isasymand}\ s{\isasymin}init\ {\isacharparenleft}np\ i{\isacharparenright}\ {\isasymand}\ {\isacharparenleft}{\isasymsigma}\ i,\ {\isasymzeta}{\isacharparenright}\ {\isacharequal}\ sr\ s\ {\isasymand}\ {\isacharparenleft}{\isasymforall}j{\isachardot}\ j\ {\isasymnoteq}\ i\ {\isasymlongrightarrow}\ {\isasymsigma}\ j{\isasymin}{\isacharparenleft}fst\ {\isasymcirc}\ sr{\isacharparenright}\ {\isacharbackquote}\ init\ {\isacharparenleft}np\ j{\isacharparenright}{\isacharparenright}{\isacharbraceright}\ {\isasymsubseteq}\ init\ {\isacharparenleft}onp\ i{\isacharparenright}}}
\snip{openproc_init_notempty}{0}{0}{\isa{openproc\ np\ onp\ sr\ {\isasymLongrightarrow}\ {\isasymforall}j{\isachardot}\ init\ {\isacharparenleft}np\ j{\isacharparenright}\ {\isasymnoteq}\ {\isasymemptyset}}}
\snip{openproc_trans}{0}{0}{\isa{{\isasymlbrakk}openproc\ np\ onp\ sr{\isacharsemicolon}\ {\isasymsigma}\ i\ {\isacharequal}\ fst\ {\isacharparenleft}sr\ s{\isacharparenright}{\isacharsemicolon}\ {\isasymsigma}{\isacharprime}\ i\ {\isacharequal}\ fst\ {\isacharparenleft}sr\ s{\isacharprime}{\isacharparenright}{\isacharsemicolon}\ {\isacharparenleft}s,\ a,\ s{\isacharprime}{\isacharparenright}{\isasymin}trans\ {\isacharparenleft}np\ i{\isacharparenright}{\isasymrbrakk}\ {\isasymLongrightarrow}\ {\isacharparenleft}{\isacharparenleft}{\isasymsigma},\ snd\ {\isacharparenleft}sr\ s{\isacharparenright}{\isacharparenright},\ a,\ ({\isasymsigma}{\isacharprime},\ snd\ {\isacharparenleft}sr\ s{\isacharprime}{\isacharparenright}){\isacharparenright}{\isasymin}trans\ {\isacharparenleft}onp\ i{\isacharparenright}}}

\snip{openproc_trans_prem_1}{0}{0}{\isa{openproc\ np\ onp\ sr}}
\snip{openproc_trans_prem_2}{0}{0}{\isa{{\isasymsigma}\ i\ {\isacharequal}\ fst\ {\isacharparenleft}sr\ s{\isacharparenright}}}
\snip{openproc_trans_prem_3}{0}{0}{\isa{{\isasymsigma}{\isacharprime}\ i\ {\isacharequal}\ fst\ {\isacharparenleft}sr\ s{\isacharprime}{\isacharparenright}}}
\snip{openproc_trans_prem_4}{0}{0}{\isa{{\isacharparenleft}s,\ a,\ s{\isacharprime}{\isacharparenright}{\isasymin}trans\ {\isacharparenleft}np\ i{\isacharparenright}}}
\snip{openproc_trans_concl}{0}{0}{\isa{{\isacharparenleft}{\isacharparenleft}{\isasymsigma},\ snd\ {\isacharparenleft}sr\ s{\isacharparenright}{\isacharparenright},\ a,\ ({\isasymsigma}{\isacharprime},\ snd\ {\isacharparenleft}sr\ s{\isacharprime}{\isacharparenright}){\isacharparenright}{\isasymin}trans\ {\isacharparenleft}onp\ i{\isacharparenright}}}

\snip{old_openproc_np_type}{0}{0}{\isa{ip\ {\isasymRightarrow}\ {\isacharparenleft}{\isacharprime}s,\ {\isacharprime}m\ proc{\isacharunderscore}action{\isacharparenright}\ automaton}}
\snip{openproc_np_type}{0}{0}{\isa{ip\ {\isasymRightarrow}\ {\isacharparenleft}{\isacharprime}s,\ proc{\isacharunderscore}action{\isacharparenright}\ automaton}}
\snip{old_openproc_onp_type}{0}{0}{\isa{ip\ {\isasymRightarrow}\ {\isacharparenleft}{\isacharparenleft}ip\ {\isasymRightarrow}\ {\isacharprime}g{\isacharparenright}\ {\isasymtimes}\ {\isacharprime}l,\ {\isacharprime}m\ proc{\isacharunderscore}action{\isacharparenright}\ automaton}}
\snip{openproc_onp_type}{0}{0}{\isa{ip\ {\isasymRightarrow}\ {\isacharparenleft}{\isacharparenleft}ip\ {\isasymRightarrow}\ {\isacharprime}k{\isacharparenright}\ {\isasymtimes}\ {\isacharprime}t,\ proc{\isacharunderscore}action{\isacharparenright}\ automaton}}
\snip{openproc_sr_type}{0}{0}{\isa{{\isacharprime}s\ {\isasymRightarrow}\ {\isacharprime}k\ {\isasymtimes}\ {\isacharprime}t}}
\snip{par_qmsg_oreachable_prem_1}{0}{0}{\isa{{\isacharparenleft}{\isasymsigma},\ (s,\ (msgs,\ q)){\isacharparenright}{\isasymin}oreachable\ {\isacharparenleft}A\ {\isasymlangle}{\isasymlangle}\isactrlbsub i\isactrlesub \ qmsg{\isacharparenright}\ S\ U}}
\snip{par_qmsg_oreachable_prem_2}{0}{0}{\isa{A\ \ostepinv\ {\isacharparenleft}S,\ U\ {\isasymrightarrow}{\isacharparenright}\ {\isacharparenleft}{\isasymlambda}{\isacharparenleft}{\isacharparenleft}{\isasymsigma},\ {\isacharunderscore}{\isacharparenright},\ {\isacharunderscore},\ ({\isasymsigma}{\isacharprime},\ {\isacharunderscore}){\isacharparenright}{\isachardot}\ F\ {\isacharparenleft}{\isasymsigma}\ i{\isacharparenright}\ {\isacharparenleft}{\isasymsigma}{\isacharprime}\ i{\isacharparenright}{\isacharparenright}}}
\snip{par_qmsg_oreachable_prem_3}{0}{0}{\isa{{\isasymAnd}{\isasymxi}{\isachardot}\ F\ {\isasymxi}\ {\isasymxi}}}
\snip{par_qmsg_oreachable_prem_4}{0}{0}{\isa{{\isasymAnd}{\isasymxi}\ {\isasymxi}{\isacharprime}{\isachardot}\ E\ {\isasymxi}\ {\isasymxi}{\isacharprime}\ {\isasymLongrightarrow}\ F\ {\isasymxi}\ {\isasymxi}{\isacharprime}}}
\snip{par_qmsg_oreachable_prem_5}{0}{0}{\isa{{\isasymAnd}{\isasymsigma}\ {\isasymsigma}{\isacharprime}\ m{\isachardot}\ {\isasymlbrakk}{\isasymforall}j{\isachardot}\ F\ {\isacharparenleft}{\isasymsigma}\ j{\isacharparenright}\ {\isacharparenleft}{\isasymsigma}{\isacharprime}\ j{\isacharparenright}{\isacharsemicolon}\ M\ {\isasymsigma}\ m{\isasymrbrakk}\ {\isasymLongrightarrow}\ M\ {\isasymsigma}{\isacharprime}\ m}}

\snip{par_qmsg_oreachable_L}{0}{0}{\isa{S\ {\isacharequal}\ otherwith\ E\ {\isacharbraceleft}i{\isacharbraceright}\ {\isacharparenleft}orecvmsg\ M{\isacharparenright}}}
\snip{par_qmsg_oreachable_E}{0}{0}{\isa{U\ {\isacharequal}\ other\ F\ {\isacharbraceleft}i{\isacharbraceright}}}

\snip{par_qmsg_oreachable_conj1}{0}{0}{\isa{{\isacharparenleft}{\isasymsigma},\ s{\isacharparenright}{\isasymin}oreachable\ A\ S\ U}}
\snip{par_qmsg_oreachable_conj2}{0}{0}{\isa{{\isacharparenleft}msgs,\ q{\isacharparenright}{\isasymin}reachable\ qmsg\ {\isacharparenleft}recvmsg\ {\isacharparenleft}M\ {\isasymsigma}{\isacharparenright}{\isacharparenright}}}
\snip{par_qmsg_oreachable_conj3}{0}{0}{\isa{{\isasymforall}m{\isasymin}set\ msgs{\isachardot}\ M\ {\isasymsigma}\ m}}

\snip{par_qmsg_oreachable_Exixi'}{0}{0}{\isa{E\ {\isasymxi}\ {\isasymxi}{\isacharprime}}}
\snip{par_qmsg_oreachable_Fxixi'}{0}{0}{\isa{F\ {\isasymxi}\ {\isasymxi}{\isacharprime}}}
\snip{par_qmsg_oreachable_Fsigmaj}{0}{0}{\isa{{\isasymforall}j{\isachardot}\ F\ {\isacharparenleft}{\isasymsigma}\ j{\isacharparenright}\ {\isacharparenleft}{\isasymsigma}{\isacharprime}\ j{\isacharparenright}}}
\snip{par_qmsg_oreachable_Rsigmam}{0}{0}{\isa{M\ {\isasymsigma}\ m}}
\snip{par_qmsg_oreachable_Rsigma'm}{0}{0}{\isa{M\ {\isasymsigma}{\isacharprime}\ m}}

\snip{par_qmsg_lift_prem_1}{0}{0}{\isa{A\ {\isasymTurnstile}\ {\isacharparenleft}S,\ U\ {\isasymrightarrow}{\isacharparenright}\ {\isacharparenleft}{\isasymlambda}{\isacharparenleft}{\isasymsigma},\ {\isacharunderscore}{\isacharparenright}{\isachardot}\ P\ {\isasymsigma}{\isacharparenright}}}
\snip{par_qmsg_lift_prem_2}{0}{0}{\isa{A\ \ostepinv\ {\isacharparenleft}S,\ U\ {\isasymrightarrow}{\isacharparenright}\ {\isacharparenleft}{\isasymlambda}{\isacharparenleft}{\isacharparenleft}{\isasymsigma},\ {\isacharunderscore}{\isacharparenright},\ {\isacharunderscore},\ {\isasymsigma}{\isacharprime},\ {\isacharunderscore}{\isacharparenright}{\isachardot}\ F\ {\isacharparenleft}{\isasymsigma}\ i{\isacharparenright}\ {\isacharparenleft}{\isasymsigma}{\isacharprime}\ i{\isacharparenright}{\isacharparenright}}}
\snip{par_qmsg_lift_prem_3}{0}{0}{\isa{{\isasymAnd}{\isasymxi}{\isachardot}\ F\ {\isasymxi}\ {\isasymxi}}}
\snip{par_qmsg_lift_prem_4}{0}{0}{\isa{{\isasymAnd}{\isasymxi}\ {\isasymxi}{\isacharprime}{\isachardot}\ E\ {\isasymxi}\ {\isasymxi}{\isacharprime}\ {\isasymLongrightarrow}\ F\ {\isasymxi}\ {\isasymxi}{\isacharprime}}}
\snip{par_qmsg_lift_prem_5}{0}{0}{\isa{{\isasymAnd}{\isasymsigma}\ {\isasymsigma}{\isacharprime}\ m{\isachardot}\ {\isasymlbrakk}{\isasymforall}j{\isachardot}\ F\ {\isacharparenleft}{\isasymsigma}\ j{\isacharparenright}\ {\isacharparenleft}{\isasymsigma}{\isacharprime}\ j{\isacharparenright}{\isacharsemicolon}\ M\ {\isasymsigma}\ m{\isasymrbrakk}\ {\isasymLongrightarrow}\ M\ {\isasymsigma}{\isacharprime}\ m}}
\snip{par_qmsg_lift_concl}{0}{0}{\isa{A\ {\isasymlangle}{\isasymlangle}\isactrlbsub i\isactrlesub \ qmsg\ {\isasymTurnstile}\ {\isacharparenleft}S,\ U\ {\isasymrightarrow}{\isacharparenright}\ {\isacharparenleft}{\isasymlambda}{\isacharparenleft}{\isasymsigma},\ {\isacharunderscore}{\isacharparenright}{\isachardot}\ P\ {\isasymsigma}{\isacharparenright}}}
\snip{closed_reachable_oreachable}{0}{0}{\isa{\mbox{}\inferrule{\mbox{wf{\isacharunderscore}net{\isacharunderscore}tree\ n}\\\ \mbox{s{\isasymin}reachable\ {\isacharparenleft}closed\ {\isacharparenleft}pnet\ {\isacharparenleft}{\isasymlambda}i{\isachardot}\ ptoy\ i\ {\isasymlangle}{\isasymlangle}\ qmsg{\isacharparenright}\ n{\isacharparenright}{\isacharparenright}\ TT}}{\mbox{initmissing\ {\isacharparenleft}netgmap\ {\isacharparenleft}{\isasymlambda}{\isacharparenleft}p,\ q{\isacharparenright}{\isachardot}\ {\isacharparenleft}fst\ {\isacharparenleft}Fun{\isachardot}id\ p{\isacharparenright},\ snd\ {\isacharparenleft}Fun{\isachardot}id\ p{\isacharparenright},\ q{\isacharparenright}{\isacharparenright}\ s{\isacharparenright}{\isasymin}oreachable\ {\isacharparenleft}oclosed\ {\isacharparenleft}opnet\ {\isacharparenleft}{\isasymlambda}i{\isachardot}\ optoy\ i\ {\isasymlangle}{\isasymlangle}\isactrlbsub i\isactrlesub \ qmsg{\isacharparenright}\ n{\isacharparenright}{\isacharparenright}\ {\isacharparenleft}{\isasymlambda}{\isacharunderscore}\ {\isacharunderscore}\ {\isacharunderscore}{\isachardot}\ True{\isacharparenright}\ U}}}}

\snip{close_toy_opnet}{0}{0}{\isa{\mbox{}\inferrule{\mbox{wf{\isacharunderscore}net{\isacharunderscore}tree\ n}\\\ \mbox{oclosed\ {\isacharparenleft}opnet\ optoy\ n{\isacharparenright}\ {\isasymTurnstile}\ {\isacharparenleft}{\isasymlambda}{\isacharunderscore}\ {\isacharunderscore}\ {\isacharunderscore}{\isachardot}\ True,\ U\ {\isasymrightarrow}{\isacharparenright}\ global\ P}}{\mbox{closed\ {\isacharparenleft}pnet\ ptoy\ n{\isacharparenright}\ {\isasymTTurnstile}\ netglobal\ ptoy\ Fun{\isachardot}id\ P}}}}
\snip{close_toy_opnet_prem_1}{0}{0}{\isa{wf{\isacharunderscore}net{\isacharunderscore}tree\ n}}
\snip{close_toy_opnet_prem_2}{0}{0}{\isa{oclosed\ {\isacharparenleft}opnet\ optoy\ n{\isacharparenright}\ {\isasymTurnstile}\ {\isacharparenleft}{\isasymlambda}{\isacharunderscore}\ {\isacharunderscore}\ {\isacharunderscore}{\isachardot}\ True,\ U\ {\isasymrightarrow}{\isacharparenright}\ {\isacharparenleft}{\isasymlambda}{\isacharparenleft}{\isasymsigma},\ {\isacharunderscore}{\isacharparenright}{\isachardot}\ P\ {\isasymsigma}{\isacharparenright}}}
\snip{close_toy_opnet_concl}{0}{0}{\isa{closed\ {\isacharparenleft}pnet\ ptoy\ n{\isacharparenright}\ {\isasymTTurnstile}\ netglobal\ ptoy\ Fun{\isachardot}id\ P}}

\snip{nos_increase}{0}{0}{\isa{nos{\isacharunderscore}inc\ {\isasymxi}\ {\isasymxi}{\isacharprime}\ {\isacharequal}\ no\ {\isasymxi}\ {\isasymle}\ no\ {\isasymxi}{\isacharprime}}}
\snip{nos_increase_lhs}{0}{0}{\isa{nos{\isacharunderscore}inc\ {\isasymxi}\ {\isasymxi}{\isacharprime}}}
\snip{nos_increase_rhs}{0}{0}{\isa{no\ {\isasymxi}\ {\isasymle}\ no\ {\isasymxi}{\isacharprime}}}

\snip{msg_num_ok}{0}{0}{\isa{msg{\isacharunderscore}ok\ {\isasymsigma}\ m\ {\isacharequal}\ \textsf{case}\ m\ \textsf{of}\ Pkt\ num{\isacharprime}\ sid{\isacharprime}\ {\isasymRightarrow}\ num{\isacharprime}\ {\isasymle}\ no\ {\isacharparenleft}{\isasymsigma}\ sid{\isacharprime}{\isacharparenright}\ {\isacharbar}\ Newpkt\ nat{\isadigit{1}}\ nat{\isadigit{2}}\ {\isasymRightarrow}\ True}}
\snip{msg_num_ok_pkt}{0}{0}{\isa{msg{\isacharunderscore}ok\ {\isasymsigma}\ {\isacharparenleft}Pkt\ data src{\isacharparenright}\ {\isacharequal}\ {\isacharparenleft}data\ {\isasymle}\ no\ {\isacharparenleft}{\isasymsigma}\ src{\isacharparenright}{\isacharparenright}}}
\snip{msg_num_ok_newpkt}{0}{0}{\isa{msg{\isacharunderscore}ok\ {\isasymsigma}\ {\isacharparenleft}Newpkt\ data dst{\isacharparenright}}}
\snip{msg_num_ok_newpkt2}{0}{0}{\isa{msg{\isacharunderscore}ok\ {\isasymsigma}\ {\isacharparenleft}Newpkt\ d\ dst{\isacharparenright}\ {\isacharequal}\ True}}

\snip{oseq_msg_num_ok}{0}{0}{\isa{optoy\ i\ \ostepinv\ {\isacharparenleft}act\ TT,\ other\ U\ {\isacharbraceleft}i{\isacharbraceright}\ {\isasymrightarrow}{\isacharparenright}\ globala\ {\isacharparenleft}{\isasymlambda}{\isacharparenleft}{\isasymsigma},\ a,\ {\isacharunderscore}{\isacharparenright}{\isachardot}\ anycast\ {\isacharparenleft}msg{\isacharunderscore}ok\ {\isasymsigma}{\isacharparenright}\ a{\isacharparenright}}}
\snip{seq_nos_increases}{0}{0}{\isa{ptoy\ i\ \stepinv\ onll\ \gammatoy\ {\isacharparenleft}{\isasymlambda}{\isacharparenleft}{\isacharparenleft}{\isasymxi},\ {\isacharunderscore}{\isacharparenright},\ {\isacharunderscore},\ {\isacharparenleft}{\isasymxi}{\isacharprime},\ {\isacharunderscore}{\isacharparenright}{\isacharparenright}{\isachardot}\ no\ {\isasymxi}\ {\isasymle}\ no\ {\isasymxi}{\isacharprime}{\isacharparenright}}}
\snip{seq_nos_increases'}{0}{0}{\isa{ptoy\ i\ \stepinv\ {\isacharparenleft}{\isasymlambda}{\isacharparenleft}{\isacharparenleft}{\isasymxi},\ {\isacharunderscore}{\isacharparenright},\ {\isacharunderscore},\ {\isacharparenleft}{\isasymxi}{\isacharprime},\ {\isacharunderscore}{\isacharparenright}{\isacharparenright}{\isachardot}\ no\ {\isasymxi}\ {\isasymle}\ no\ {\isasymxi}{\isacharprime}{\isacharparenright}}}
\snip{nhip_eq_ip}{0}{0}{\isa{ptoy\ i\ {\isasymTTurnstile}\ onl\ \gammatoy\ {\isacharparenleft}{\isasymlambda}{\isacharparenleft}{\isasymxi},\ l{\isacharparenright}{\isachardot}\ l{\isasymin}{\isacharbraceleft}PToy{\isacharminus}{\isacharcolon}{\isadigit{2}}{\isachardot}{\isachardot}PToy{\isacharminus}{\isacharcolon}{\isadigit{8}}{\isacharbraceright}\ {\isasymlongrightarrow}\ nhid\ {\isasymxi}\ {\isacharequal}\ id\ {\isasymxi}{\isacharparenright}}}

\snip{net_global}{0}{0}{\isa{netglobal\ P\ {\isacharequal}\ {\isasymlambda}s{\isachardot}\ P\ {\isacharparenleft}default\ toy{\isacharunderscore}init\ {\isacharparenleft}netlift\ fst\ s{\isacharparenright}{\isacharparenright}}}
\snip{net_global_lhs}{0}{0}{\isa{netglobal\ P}}
\snip{net_global_rhs}{0}{0}{\isa{{\isasymlambda}s{\isachardot}\ P\ {\isacharparenleft}default\ toy{\isacharunderscore}init\ {\isacharparenleft}netlift\ fst\ s{\isacharparenright}{\isacharparenright}}}

\snip{oseq_bigger_than_next}{0}{0}{\isa{optoy\ i\ {\isasymTurnstile}\ {\isacharparenleft}otherwith\ nos{\isacharunderscore}inc\ {\isacharbraceleft}i{\isacharbraceright}\ {\isacharparenleft}orecvmsg\ msg{\isacharunderscore}ok{\isacharparenright},\ other\ nos{\isacharunderscore}inc\ {\isacharbraceleft}i{\isacharbraceright}\ {\isasymrightarrow}{\isacharparenright}\ }\\\isa{\phantom{optoy\ i\ {\isasymTurnstile}\ }{\isacharparenleft}{\isasymlambda}{\isacharparenleft}{\isasymsigma},\ {\isacharunderscore}{\isacharparenright}{\isachardot}\ no\ {\isacharparenleft}{\isasymsigma}\ i{\isacharparenright}\ {\isasymle}\ no\ {\isacharparenleft}{\isasymsigma}\ {\isacharparenleft}nhid\ {\isacharparenleft}{\isasymsigma}\ i{\isacharparenright}{\isacharparenright}{\isacharparenright}{\isacharparenright}}}
\snip{pnet_bigger_than_next}{0}{0}{\isa{opnet\ {\isacharparenleft}{\isasymlambda}i{\isachardot}\ optoy\ i\ {\isasymlangle}{\isasymlangle}\isactrlbsub i\isactrlesub \ qmsg{\isacharparenright}\ n\ {\isasymTurnstile}\ {\isacharparenleft}otherwith\ nos{\isacharunderscore}inc\ {\isacharparenleft}net{\isacharunderscore}tree{\isacharunderscore}ips\ n{\isacharparenright}\ {\isacharparenleft}oarrivemsg\ msg{\isacharunderscore}ok{\isacharparenright},\ other\ nos{\isacharunderscore}inc\ {\isacharparenleft}net{\isacharunderscore}tree{\isacharunderscore}ips\ n{\isacharparenright}\ {\isasymrightarrow}{\isacharparenright}\ global\ {\isacharparenleft}{\isasymlambda}{\isasymsigma}{\isachardot}\ {\isasymforall}i{\isasymin}net{\isacharunderscore}tree{\isacharunderscore}ips\ n{\isachardot}\ no\ {\isacharparenleft}{\isasymsigma}\ i{\isacharparenright}\ {\isasymle}\ no\ {\isacharparenleft}{\isasymsigma}\ {\isacharparenleft}nhid\ {\isacharparenleft}{\isasymsigma}\ i{\isacharparenright}{\isacharparenright}{\isacharparenright}{\isacharparenright}}}
\snip{bigger_than_next}{0}{0}{\mbox{\isa{wf{\isacharunderscore}net{\isacharunderscore}tree\ {\isasymPsi}\ {\isasymLongrightarrow}\ closed\ {\isacharparenleft}pnet\ {\isacharparenleft}{\isasymlambda}i{\isachardot}\ ptoy\ i\ {\isasymlangle}{\isasymlangle}\ qmsg{\isacharparenright}\ {\isasymPsi}{\isacharparenright}\ {\isasymTTurnstile}\ }}\\\mbox{\isa{\phantom{wf{\isacharunderscore}net{\isacharunderscore}tree\ n\ }netglobal\ {\isacharparenleft}{\isasymlambda}{\isasymsigma}{\isachardot}\ {\isasymforall}i{\isachardot}\ no\ {\isacharparenleft}{\isasymsigma}\ i{\isacharparenright}\ {\isasymle}\ no\ {\isacharparenleft}{\isasymsigma}\ {\isacharparenleft}nhid\ {\isacharparenleft}{\isasymsigma}\ i{\isacharparenright}{\isacharparenright}{\isacharparenright}{\isacharparenright}}}}
\snip{bigger_than_next_prem}{0}{0}{\isa{wf{\isacharunderscore}net{\isacharunderscore}tree\ {\isasymPsi}}}
\snip{bigger_than_next_concl}{0}{0}{\isa{closed\ {\isacharparenleft}pnet\ {\isacharparenleft}{\isasymlambda}i{\isachardot}\ ptoy\ i\ {\isasymlangle}{\isasymlangle}\ qmsg{\isacharparenright}\ {\isasymPsi}{\isacharparenright}\ {\isasymTTurnstile}\ netglobal\ {\isacharparenleft}{\isasymlambda}{\isasymsigma}{\isachardot}\ {\isasymforall}i{\isachardot}\ no\ {\isacharparenleft}{\isasymsigma}\ i{\isacharparenright}\ {\isasymle}\ no\ {\isacharparenleft}{\isasymsigma}\ {\isacharparenleft}nhid\ {\isacharparenleft}{\isasymsigma}\ i{\isacharparenright}{\isacharparenright}{\isacharparenright}{\isacharparenright}}}

\snip{seq_no_leq_num_restricted}{0}{0}{\isa{ptoy\ i\ {\isasymTTurnstile}\ onl\ \gammatoy\ {\isacharparenleft}{\isasymlambda}{\isacharparenleft}{\isasymxi},\ l{\isacharparenright}{\isachardot}\ l{\isasymin}{\isacharbraceleft}PToy{\isacharminus}{\isacharcolon}{\isadigit{7}}..PToy{\isacharminus}{\isacharcolon}{\isadigit{8}}{\isacharbraceright}\ {\isasymlongrightarrow}\ no\ {\isasymxi}\ {\isasymle}\ num\ {\isasymxi}{\isacharparenright}}}